\documentstyle[preprint,aps,eqsecnum]{revtex}

\newcommand{\beq}{\begin{equation}}
\newcommand{\eeq}{\end{equation}}
\newcommand{\bea}{\begin{eqnarray}}
\newcommand{\eea}{\end{eqnarray}}

\begin{document}
\title{The Semi-Classical Relativistic Darwin Potential for Spinning Particles in
the Rest-Frame Instant Form: 2-Body Bound States with Spin 1/2 Constituents.}
\author{David Alba}
\address{Dipartimento di Fisica\\
Universita' di Firenze\\
L.go E.Fermi 2 (Arcetri)\\
50125 Firenze, Italy\\
E-mail: ALBA@FI.INFN.IT}
\author{and}
\author{Horace Crater}
\address{The University of Tennessee Space Institute\\
Tullahoma, Tennessee\\
37388 USA\\
E-mail: HCrater@utsi.edu}
\author{and}
\author{Luca Lusanna}
\address{Sezione INFN di Firenze\\
L.go E.Fermi 2 (Arcetri)\\
50125 Firenze, Italy\\
E-mail: LUSANNA@FI.INFN.IT}
\maketitle

\begin{abstract}
We extend previous results on the extraction of the Darwin potential to all
orders in $c^{-2}$ from the radiation gauge Lienard-Wiechert solution for
the system of N positive-energy scalar particles plus the electromagnetic
field in the Wigner-covariant rest-frame instant form of dynamics to the
case of N positive-energy spinning particles. This is done in the
semi-classical approximation of using {\it Grassmann-valued electric charges}
for {\it regularizing} the Coulomb self-energies and {\it extracting} the
{\it unique semi-classical} action-at-a-distance interaction hidden in any
Green function used for the Lienard-Wiechert solution. By describing
semi-classically also the {\it spin} of the particles with
Grassmann-variables, by means of a semi-classical Foldy-Wouthuysen
transformation applied the the Dirac-like constraints of the manifestly
Lorentz covariant spinning particles, we determine the coupling of
positive-energy spinning particles to the electric field in the
semi-classical approximation. Then we follow the same procedure developed
for scalar particles and, in the sector where there is no {\it in}%
-radiation, we determine the effective semi-classical interparticle
potential. Besides the relativistic {\it Darwin} term there are {\it %
spin-orbit} and {\it spin-spin} terms in the potential. Quantization of the
lowest order (in $c^{-2}$) part of the closed form of the effective
Hamiltonian in the case N=2 reproduces {\it exactly} the standard result of
the reduction of the {\it Bethe-Salpeter equation} for the bound states of
two spin 1/2 constituents of arbitrary mass (hydrogen atom, positronium,
muonium).

\vskip 1truecm

\today

\vskip 1truecm
\end{abstract}


\tightenlines

\vfill\eject

\vfill\eject

\section{Introduction.}

\bigskip

At the classical level, \ the accepted mathematical description of the four
basic interactions starts with an action principle. Manifest Lorentz
invariance and local gauge invariance force these interactions to make use
of {\it singular Lagrangians}.\ This implies that their Hamiltonian
description must rely on the {\it Dirac-Bergmann theory of constraints}\cite
{dd,brg}. \ Constraint dynamics provides a natural formalism for any program
of a unified description of these interactions in terms of {\it %
Dirac-Bergmann observables}. \ Those observables are gauge invariant and
deterministic variables which describe a canonical basis of measurable
quantities. When one begins with a singular Lagrangian, the canonical
momenta $p_{i}=\partial L/\partial \dot{q}_{i}$ are not independent. \ The
relations among them, $\phi _{\alpha }(q,p)\approx 0,$ are called {\it %
primary} constraints (the weak inequality $\approx $ means that the equality
sign cannot be used inside Poisson brackets). \ The canonical Hamiltonian $%
H_{c}(q,p)$ has to be replaced by the Dirac Hamiltonian $H_{D}=H_{c}+\sum_{%
\alpha }\lambda _{\alpha }(t)\phi _{a}$ which, by way of the arbitrary Dirac
multipliers, then accounts for the restriction to the submanifold defined by
the constraints.\ The time constancy of the primary constraints, $\partial
_{t}\phi _{\alpha }=\{\phi _{\alpha },H_{D}\}\approx 0,$ either produces
{\it secondary} Hamiltonian constraints or determines some of the Lagrange
multipliers. \ This procedure is repeated for the secondary constraints and
again if necessary (this is the {\it Dirac-Bergmann algorithm})\cite{wein}.
\ When the procedure is finished, there is a final set of constraints, $\chi
_{\alpha }\approx 0,$ defining the final submanifold on which the dynamics
is consistently restricted, and a final Dirac Hamiltonian with a reduced set
of arbitrary Dirac multipliers describing the remaining arbitrariness of the
time evolution. \ Dirac divides the constraints into two subgroups: \ i) \
the {\it first class} constraints $\chi _{\alpha }^{(1)}\approx 0$ having
weakly zero Poisson brackets with all constraints and being the generators
of the {\it Hamiltonian gauge transformations} of the theory and \ ii) the
{\it second class} constraints $\chi _{\alpha }^{(2)}\approx 0$ \footnote{%
They always appear in pairs when there are only bosonic degrees of freedom.}
with $\det (\{\chi _{n_{1}}^{(2)},\chi _{n_{2}}^{(2)}\})\neq 0$,
corresponding to pairs of {\it inessential} variables that can be
eliminated. \ The vehicle that Dirac invented to eliminate, in a systematic
and symmetrical way, from the dynamical equations those variables
constrained by the second class constraints are known as {\it Dirac brackets}%
. Second class constraints may be primary or secondary constraints but they
can also be introduced by hand. \ This can be done in two different ways. \
Firstly one can explicitly break the gauge freedom by the introduction of
gauge fixing conditions on the gauge variables, so that the original first
class constraints and the gauge fixing ones become a system of second class
constraints. \ Secondly one can introduce pairs of solutions to the
equations of motions ( for example, field equations whose solutions express
fields in terms of canonical variables for particles serving as sources for
the fields). \ In general this procedure may break the original manifest
Lorentz covariance of the theory ( but maintaining Wigner covariance, when
we work in the {\it rest-frame instant form of dynamics}\cite{lu1}).

If the end aim is a unified description of the four interactions, then in
the application of the Dirac-Bergmann algorithm to special relativistic
theories it is advantageous to reformulate the theory in such a way \ as to
allow a natural transition to the coupling of gravity. \ With this in mind,
\ recent studies (see \cite{lu1,india} and references cited therein) have
taken advantage of work by Dirac\cite{dd} on general relativity arriving to
a description of special relativistic systems on arbitrary spacelike
hypersurfaces ({\it parametrized Minkowski theories}). Then, for every
configuration of the system with timelike total 4-momentum the description
can be restricted to the so-called {\it Wigner hyperplanes} orthogonal to
the total 4-momentum. This is the intrinsic rest frame of the configuration.
On it a new instant form of dynamics (the {\it Wigner-covariant rest-frame
instant form}) can be introduced.

In a recent paper \cite{ap} we presented the technical completion of one
segment of this program\cite{india}, that of $N$ scalar charged particles
plus the electromagnetic field when the charges of the particles are
described by bilinears in Grassmann variables. \ There we analyzed how to
extract the action-at-a-distance interparticle potential hidden in the
semi-classical Lienard-Wiechert solution of the electromagnetic field
equations, a subset of the solutions of the equations of motion for the
isolated system formed by $N$ positive-energy scalar charged particles plus
the electromagnetic field. \ The problem is formulated in the
Wigner-covariant rest-frame instant form of dynamics\cite{lu1,india}, which
is defined on the Wigner hyperplanes orthogonal to the total time-like
four-momentum of the isolated system and which requires the choice of the
sign of the energy of the particles (we considered only positive energies)

This was possible due to the {\it semiclassical approximation} of using
Grassmann-valued electric charges ($Q_{i}^{2}=0$, $Q_{i}Q_{j}\not=0$ for $i
\not=j$) as an alternative to the extended electron models used for the
regularization of the Coulomb self energies. How this happens was shown in
Ref.\cite{lu1}, where the Coulomb potential was extracted from the
electromagnetic potential by making the canonical reduction of the
electromagnetic gauge freedom via the Shanmugadhasan canonical
transformation. This is equivalent to the use of a {\it Wigner-covariant
radiation} (or Coulomb) {\it gauge} in the rest-frame instant form.

Ref.\cite{lu4} presented the retarded Lienard-Wiechert solution for the
transverse electromagnetic field in the rest-frame instant form radiation
gauge: in this gauge, due to the transversality, the retarded
Lienard-Wiechert potential associated with each charged particle depends on
the whole past history of the other particles. At the semi-classical level a
single accelerated charged positive-energy particle with Grassmann-valued
electric charge does {\it not} radiate even if it has a non-trivial
Lienard-Wiechert potential, avoiding therefore the acausal features of the
Abraham-Lorentz-Dirac equations, and has {\it no} mass renormalization. \
However, a system of $N$ charged particles produces, by virtue of the {\it %
interference terms} from the various retarded Lienard-Wiechert potentials of
the particles, a radiation which reproduces the standard Larmor expression
for radiation in the wave zone, when the particles are considered as {\it %
external} sources of the electromagnetic field and their equations of motion
are {\it not} used.

If instead the particles are considered {\it dynamical}, the use of their
equations of motion and of the semi-classical approximation leads to a
drastic simplification of the Lienard-Wiechert potentials and fields.
Indeed, if we make an equal time expansion of the delay by expressing these
potentials and fields in terms of particle coordinates, velocities and
accelerations of every order, it turns out that {\it all the accelerations
decouple} at the semi-classical level due to the particle equations of
motion. Therefore, at the semi-classical level the retarded, advanced and
symmetric Lienard-Wiechert potentials and electric and magnetic fields
coincide and {\it depend only} on the positions and velocities of the
particles, so that we can find their phase space expression in terms of
particle positions and momenta.

In this way the semi-classical Lienard-Wiechert potential and fields can be
reinterpreted as {\it scalar and vector interparticle instantaneous
action-at-a-distance potentials}. It is then possible to identify a
semiclassical reduced phase space containing only particles by eliminating
the electromagnetic field by adding by hand second class constraints which
force the transverse potential and electric field canonical variables to
coincide with the semi-classical Lienard-Wiechert ones in the absence of
incoming radiation: $\vec{A}_{\perp }(\tau ,\vec{\sigma})-\vec{A}_{\perp
LW}(\tau ,\vec{\sigma})\approx 0$, $\vec{\pi}_{\perp }(\tau ,\vec{\sigma})-%
\vec{\pi}_{\perp LW}(\tau ,\vec{\sigma})\approx 0$. Let us remark that this
could be done also in presence of an arbitrary incoming radiation ${\vec{A}}%
_{\perp (rad)}(\tau ,\vec{\sigma})$, ${\vec{\pi}}_{\perp (rad)}(\tau ,\vec{%
\sigma})=-{\frac{{\partial }}{{\partial \tau }}}{\vec{A}}_{\perp (rad)}(\tau
,\vec{\sigma})$ \footnote{%
It is an arbitrary solution of the homogeneous wave equation and must not be
interpreted as a pair of canonical variables.} by modifying the constraints
to the form $\vec{A}_{\perp }(\tau ,\vec{\sigma})-\vec{A}_{\perp LW}(\tau ,%
\vec{\sigma})-{\vec{A}}_{\perp (rad)}(\tau ,\vec{\sigma})\approx 0$, $\vec{%
\pi}_{\perp }(\tau ,\vec{\sigma})-\vec{\pi}_{\perp LW}(\tau ,\vec{\sigma})-{%
\ \vec{\pi}}_{\perp (rad)}(\tau ,\vec{\sigma})\approx 0$.

The reduced phase space is obtained by means of the introduction of the
Dirac brackets associated with these second class constraints. Since the old
particle positions and momenta are no longer canonical in this reduced phase
space, we had to find the new (Darboux) basis of particle canonical
variables. The generators of the {\it internal} Poincar\'{e}` group inside
the Wigner hyperplanes in the rest-frame instant form of dynamics can be
reexpressed in terms of these new variables: the 3-momentum ${\cal \vec{P}}%
_{(int)}$ and the angular momentum ${\cal \vec{J}}_{(int)}$ \ become equal
to those for $N$ free scalar particles (as expected in an instant form). \
The interaction dependent boosts ${\vec{{\cal K}}}_{(int)}$ are proportional
to the {\it internal} canonical center of mass $\vec{q}_{+}$ inside the
Wigner hyperplane: $\vec{q}_{+}\approx 0$ are the gauge-fixings to be added
to the rest-frame conditions ${\cal \vec{P}}_{(int)}\approx 0$ , if one
wishes to re-express the dynamics only in terms of particle {\it internal }
relative variables. Therefore we have a unfaithful representation of the
{\it internal} Poincar\'{e} group. Also the energy-momentum tensor has been
evaluated in the new canonical variables and there is a suggestion on how to
find the M\o ller center of energy of a cluster of $n$ particles contained
in the N particle isolated system.

The Hamiltonian in the rest frame instant form, generating the evolution in
the rest-frame time of the decoupled {\it external} canonical center of
mass, is the {\it internal} energy generator $M={\cal P}_{(int)}^{\tau }$
(the invariant mass of the isolated $N$ particle system). \ The
semi-classical Lienard-Wiechert solution implies the existence of
interparticle action-at-a-distance potentials of two types: {\it vector
potentials}, minimally coupled to the Wigner spin 1 particle three-momentum,
under the square root associated with the kinetic energies; ii) a {\it %
scalar potential} (including the Coulomb potential) outside the square
roots. \ In the {\it semi-classical approximation} all these potentials can
be replaced by a {\it unique scalar potential}, which is the sum of the {\it %
Coulomb potential} and of a {\it generalized Darwin one} for arbitrary $N$.
It is the (semi-classical) non-static complete potential corresponding to
the \ one photon exchange tree Feynman diagrams of scalar electrodynamics
and is a completely new result. The expression we find contains no $N$-body
forces, being simply a sum of two particle interactions. \ This is a
consequence of our use of Grassmann charges and of the equal time
description on the Wigner hyperplanes of the rest-frame instant form.

In the $N=2$ case we obtain a closed form of the solution by evaluating it
in the rest frame after the gauge fixing ${\vec q}_{+}\approx 0$: the lowest
order in $1/c^{2}$ contribution of the generalized Darwin potential agrees
with the expression of the standard Darwin potential. We then show that in a
\ semi-classical sense a special solution of the Hamilton equations is the
Schild solution \cite{schild} in which the two particles move in concentric
circular orbits.

${}$

${}$

In this paper we will extend these results to positive-energy charged
spinning particles, with both the electric charges and the spins described
by Grassmann variables (pseudo- or semi-classical approximation), whose
rest-frame instant form description was given in Ref.\cite{lu3}.

The resulting generalized Darwin potential will be the (semiclassical)
static and non-static complete potential corresponding to the one-photon
exchange tree Feynman diagrams of spinor electrodynamics, after the
restriction to the subsector of positive-energy fermions.

The fact that the positive-energy spinning particles are a pseudo-classical
description of positive-energy fermions immediately forces us to face a
problem absent with scalar particles. In the manifestly Lorentz covariant
approach, relativistic scalar particles are described by means of the
mass-shell first class constraints

\begin{equation}
\chi_i = p^2_i-m^2_i \approx 0.  \label{I1}
\end{equation}

\noindent The positive- and negative-energy sheets of the mass hyperboloid
correspond to the particle energies $p_{i}^{0}\approx \pm \sqrt{m_{i}^{2}+{\
\vec{p}}_{i}^{2}}$. When an {\it external} electromagnetic field is present
we have

\begin{equation}
\chi_i = [p_i - Q_i A(x_i)]^2-m^2_i \approx 0,  \label{I2}
\end{equation}

\noindent so that $p_{i}^{0}\approx Q_{i}A^{0}(x_{i})\pm \sqrt{m_{i}^{2}+[{\
\vec{p}}_{i}-Q_{i}\vec{A}(x_{i})]^{2}}$. We see that at the classical level
the positive- and negative-energy sheets of the mass hyperboloid of scalar
particles never intersect even in presence of external electromagnetic
fields \footnote{%
As shown in Ref.\cite{mate}, Appendix D, scalar particles give the
semiclassical description only of those solutions of the Klein-Gordon
equation coupled to an external electromagnetic field for which the
Feshbach-Villar approach to the Klein-Gordon field allows a separation of
positive-energies from negative-energies in an eikonal approximation.
Otherwise the non-diagonalizability of the Feshbach-Villar Hamiltonian
implies that generic solutions cannot be described only in terms of scalar
particles when external fields are present, but require more elaborate
quasi-particle concepts.}. Therefore, there is no problem in the rest-frame
description of scalar particles with a fixed sign of the energy, since these
particles will be always interpreted as the classical basis for
Tomonaga-Schwinger asymptotic states.

On the other hand, the manifestly Lorentz covariant description of spinning
particles requires the introduction of extra Dirac-like first class
pseudo-classical constraints\cite{cas}

\begin{equation}
\chi_{Di} = p_{i\mu} \xi^{\mu}_i- m_i \xi_{i5} \approx 0,  \label{I3}
\end{equation}

\noindent whose quantization generates the one-particle Dirac equation
\footnote{%
The Grassmann variables $\xi^{\mu}_i$, $\xi_{i5}$ describe the {\it spin
structure} and, after quantization, generate the Clifford algebra of Dirac
matrices. We assume that the Grassmann variables describing the spin of
different particles commute: $\xi^{\mu}_i\xi^{\nu}_j=\xi^{\nu}_j\xi^{\mu}_i$
for $i\not= j$, $\xi^{\mu}_i\xi^{\nu}_i+\xi^{\nu}_i\xi^{\mu}_i=0$ and so on.}
(for each particle we have $\xi_{i5} \mapsto \gamma_5$, $\xi^{\mu}_i \mapsto
\gamma_5\gamma^{\mu}$, $\chi_{Di}\approx 0 \mapsto
\gamma_5(p_{i\mu}\gamma^{\mu}-m_i)\psi (p_i)=0$).

The mass-shell constraints $\chi _{i}=p_{i}^{2}-m_{i}^{2}\approx 0$ must be
consistent with the pseudo-classical version $\{\chi _{Di},\chi
_{Dj}\}=i\delta _{ij}\chi _{i}$ . In this sense {\it the Dirac equation is
the square root of the Klein-Gordon one}.

When there is an {\it external} electromagnetic field, the Dirac-like
constraints become

\begin{equation}
\chi_{Di} = [p_{i\mu} - Q_i A_{\mu}(x_i)] \xi^{\mu}_i - m_i \xi_{i5} \approx
0,  \label{I4}
\end{equation}

\noindent and the resulting mass-shell constraints

\begin{equation}
\chi_i=-i \{ \chi_{Di}, \chi_{Di} \} = [p_i- Q_i A(x_i)]^2-m^2_i +iQ_i
F_{\mu\nu}(x_i) \xi^{\mu}_i\xi^{\nu}_i \approx 0,  \label{I5}
\end{equation}

\noindent contain a {\it non-minimal term} connected with the {\it %
spin-magnetic field coupling in the rest frame}.

Since we know that we cannot separate positive from negative energies in the
Dirac equation coupled to an external electromagnetic field with an exact
Foldy-Wouthuysen transformation \footnote{%
See Ref.\cite{thaller} and its rich bibliography for the special cases in
which this is possible and for the lack of a mathematical justification of
the validity of the Foldy-Wouthuysen transformation.}, we expect to have the
same problems present also to the pseudo-classical level, where the
pseudo-classical Foldy-Wouthuysen transformation is known\cite{pfw} in
absence of external electric fields \footnote{%
It produces forms of the Dirac-like constraint in which there are only
polynomials in the Grassmann variables which after quantization become {\it %
even} operators.}. Now in the present case this pseudo-classical
transformation has to be reformulated by using Grassmann-valued electric
charges: this further semiclassical approximation will allow to get an exact
semi-classical separation of the positive and negative energies. Therefore,
due to spin, differently from the case of scalar particles, already at the
semi-classical level we must make a pseudo-classical Foldy-Wouthuysen
transformation which puts the Dirac-like constraint at order $Q_{i}$ in a
form producing only {\it even} operators after quantization. In this way we
shall identify the kind of semiclassical couplings of the spinning particle
to an {\it external} electric field which are compatible with the separation
of positive energies from negative ones. These couplings will appear in the
semiclassical approximation of the solutions of the one-particle Dirac
equation in all those cases in which an eikonal approximation plus a
standard Foldy-Wouthuysen transformation will allow us to identify positive
(or negative) energy spinning particle as a realistic approximation\footnote{%
In particular the approximation will be valid as long as the electric fields
are not too strong.}.

The next step will be to take the parametrized Minkowski action describing
charged spinning particles and the electromagnetic field on arbitrary
spacelike hypersurfaces, in which the minimal coupling describes correctly
the magnetic couplings, and to {\it add non-minimally} these electric
couplings of the spinning particles with the ({\it now dynamical})
electromagnetic field. Since we have not found the needed covariant
modification of the Lagrangian, we limit ourselves to present the form of
these couplings after the reduction to the Wigner hyperplanes, viz. in the
rest-frame instant form of dynamics.

Therefore, we will have a modified invariant mass for the isolated system
plus the ordinary rest-frame conditions, namely the vanishing of the total
3-momentum of the isolated system inside the Wigner hyperplane. This will
produce a modification of the equations of motion. In particular the
electromagnetic field equations will have a more complex particle source
term, one dependent on the pseudoclassical spin variables. Therefore, the
radiation gauge Lienard-Wiechert solutions of Ref.\cite{lu3} will be
modified.

At this point we have only to repeat the whole procedure, identified in the
previous paper on charged scalar particles, to get the semi-classical phase
space version of the {\it unique} semi-classical Lienard-Wiechert potential
and electric field. Then we will make the canonical reduction to only
particle degrees of freedom by imposing conditions on the electromagnetic
field in the radiation gauge such that it coincide with this Hamiltonian
Lienard-Wiechert electromagnetic field. After the identification of the new
particle canonical variables, the final form of the Hamiltonian will
identify the relativistic form of the {\it Darwin potential for spinning
particles}.

In the 2-body case, the quantization of the lowest order in $c^{-2}$ of the $%
N$=2 final Hamiltonian has to be compared with the positive-energy sector of
the 3-dimensional positive-energy reduction of the {\it Bethe-Salpeter
equation} \cite{bethe,das,breit,alstine,stroscio} for 2-body bound states
with spin 1/2 constituents of arbitrary mass (hydrogen atom, positronium,
muonium). We find that our {\it Darwin}, {\it spin-orbit} and {\it spin-spin}
terms reproduce {\it exactly} the standard Bethe-Salpeter result: this is
the first time that, by making a relativistic separation of the
center-of-mass motion and a subsequent study and canonical reduction of the
relative motion, {\it results coming from quantum field theory can be
exactly reproduced}.

The paper is organized as follows. In Section II we study the
pseudo-classical Foldy-Wouthuysen transformation applied to the manifestly
Lorentz-covariant spinning particle in an external electromagnetic field. A
review of the theory of the positive-energy spinning particle interacting
with a dynamical electromagnetic field in the rest-frame instant form of
dynamics is given in Section III. In Section IV we introduce the non-minimal
coupling of the electric field to the positive-energy spinning particles on
the Wigner hyperplane using the results of Section II. In Section V we give
the equations of motion for the particles and the field in the radiation
gauge on the Wigner hyperplane. In Section VI we determine the unique
semi-classical Lienard-Wiechert solution, we find its phase space
expression, we add second class constraints to eliminate the radiation
field, we find the Dirac brackets and the new canonical basis for the
particles. The semi-classical Hamiltonian with the Darwin and spin-dependent
potentials for the $N$-body problem is found in Section VII. In Section VIII
we study the 2-body problem: we find the semiclassical Hamiltonian for
muonium-, hydrogen- and positronium-like systems and then we quantize its
lowest order part, reproducing the Bethe-Salpeter result. In the Conclusions
after some comments we delineate how the research program could be developed
to treat the non-Abelian case of the quark model.

In Appendix A there is the determination of some functions defined in
Section II. In Appendix B, after a review on spacelike hypersurfaces, there
is an attempt to extend the results of Section IV outside Wigner hyperplanes
and outside the radiation gauge. In Appendix C there is the computation of
the electromagnetic energy and 3-momentum when the Lienard-Wiechert solution
is inserted in them. In Appendix D there the evaluation of some of the
potentials of Section VII. In Appendix E there is the summation of the
2-body rest energy to a closed form.

\vfill\eject

\section{The Pseudo-classical Foldy-Wouthuysen transformation of the
manifestly Lorentz covariant spinning particle.}

In Ref.\cite{cas} there is the coupling of the manifestly Lorentz covariant
spinning particle to external electromagnetic fields. It is based on the
singular Lagrangian

\begin{eqnarray}
&L& = - \frac {i}{2} \xi_5 \dot{\xi}_5 - \frac {i}{2} \xi_\mu \dot{\xi}^\mu +
\nonumber \\
&-& \sqrt{m^2 - i e F_{\mu \nu}(x) \xi^\mu \xi^\nu} \sqrt{\Big( \dot{x}_\mu
- \frac im \xi_\mu \dot{\xi}_5 \Big)^2} - e \dot{x}_\mu A^\mu(x)=  \nonumber
\\
&=& - \frac {i}{2} \xi_5 \dot{\xi}_5 - \frac {i}{2} \xi_\mu \dot{\xi}^\mu -e
\dot{x}_\mu A^\mu(x)-  \nonumber \\
&-& \Big[ m -\frac{i e}{2m} F_{\mu \nu}(x) \xi^\mu \xi^\nu - \frac {e^2}{m^3}
F_{\mu \nu}(x) F_{\rho \lambda}(x) \xi^\mu \xi^\nu \xi^\rho \xi^\lambda %
\Big] \sqrt{\Big( \dot{x}_\mu - \frac im \xi_\mu \dot{\xi}_5 \Big)^2},
\label{II1}
\end{eqnarray}

\noindent which, besides the standard minimal coupling, has a {\it %
non-minimal mass renormalization} $-ieF_{\mu\nu}\xi^{\mu}\xi^{\nu}=eF_{\mu%
\nu}S^{\mu\nu}$. The $ie$ coefficient in front of $F_{\mu\nu}\xi^{\mu}\xi^{%
\nu}$, corresponding to the absence of an anomalous magnetic moment of the
electron, is the only one ensuring that the two constraints remain first
class even in presence of an external electromagnetic field.

Indeed, besides the second class constraints $\pi_{\mu}-{\frac{i}{2}}%
\xi_{\mu} \approx 0$ and the added one (like in the free case) $\pi_5+{\frac{%
i}{2}}\xi_5 \approx 0$ \footnote{%
After the elimination of the Grassmann momenta by going to Dirac brackets
with respect to these second class constraints, the original Poisson
brackets $\{ x^{\mu}, p^{\nu} \} = \{ \xi^{\mu}, \pi^{\nu} \}
=-\eta^{\mu\nu} $, $\{ \xi_5, \pi_5 \} =-1$ become the following non-null
Dirac bracket: $\{x^{\mu },p^{\nu }\}^{*}=-\eta ^{\mu \nu }$, $\{ \xi^{\mu},
\xi^{\nu} \}^{*} =i\eta^{\mu\nu}$, $\{ \xi_5, \xi_5 \}^{*} =-i$; in what
follows we will use the notation $\{ .,. \}$ for these Dirac brackets.}, one
gets the first class constraints

\begin{eqnarray}
\chi_D &=& (p_\mu - e A_\mu(x)) \xi^\mu - m \xi_5 \approx 0  \nonumber \\
\chi &=& (p-e A(x))^2 - m^2 + i e F_{\mu \nu}(x) \xi^\mu \xi^\nu \approx 0 ,
\nonumber \\
&&{}  \nonumber \\
&&\{ \chi_D,\chi_D \} {}^{*} =i \chi ,\quad\quad \{ \chi ,\chi \} {}^{*}= \{
\chi ,\chi_D \} {}^{*} =0.  \label{II2}
\end{eqnarray}

As said in the Introduction, if we have many particles we assume that the
spin Grassmann variables of each particle commute with those of the other
particles.

Following Ref.\cite{bar}, we now describe also the electric charge of each
particle in a semi-classical way by means of a pair of complex conjugate
Grassmann variables \footnote{%
They are assumed to commute with the spin Grassmann variables.} $\theta
_{i}(\tau ),\theta _{i}^{\ast }(\tau )$ \cite{bar} satisfying ($%
I_{i}=I_{i}^{\ast }=\theta _{i}^{\ast }\theta _{i}$ is the generator of the $%
U_{em}(1)$ group of particle $i$)

\begin{eqnarray}
&&\theta _{i}^{2}=\theta _{i}^{{\ast }2}=0,\quad \quad \theta _{i}\theta
_{i}^{\ast }+\theta _{i}^{\ast }\theta _{i}=0,  \nonumber \\
&&\theta _{i}\theta _{j}=\theta _{j}\theta _{i},\quad \quad \theta
_{i}\theta _{j}^{\ast }=\theta _{j}^{\ast }\theta _{i},\quad \quad \theta
_{i}^{\ast }\theta _{j}^{\ast }=\theta _{j}^{\ast }\theta _{i}^{\ast },\quad
\quad i\not=j,  \nonumber \\
&&{}  \nonumber \\
Q_i &=& e_i \theta^{*}_i \theta_i,\qquad Q^2_i=0,\qquad Q_iQ_j=Q_jQ_i.
\label{II3}
\end{eqnarray}

The action now depends also on the configuration variables $\theta _{i}(\tau
)$ and $\theta _{i}^{\ast }(\tau )$,$i=1,..,N$, through an extra {\it kinetic%
} piece for the complex Grassmann charges $\int \frac{i}{2}[\theta_{i}^{\ast
}(\tau ){\dot{\theta}}_{i}(\tau )- {\dot{\theta}}_{i}^{\ast }(\tau )\theta
_{i}(\tau )]d\tau $,

The Grassmann momenta associated to these extra variables give rise to the
second class constraints $\pi_{\theta \, i}+{\frac{i}{2}}\theta^{*}_i\approx
0$, $\pi_{\theta^{*}\, i}+{\frac{i}{2} } \theta_i\approx 0$ \footnote{$\{
\theta_i(\tau ) ,\pi_{\theta j}(\tau ) \} = \{ \theta^{*}_i(\tau ),
\pi_{\theta^{*} j}(\tau ) \} =-\delta_{ij}$; $\lbrace \pi_{\theta \, i}+{%
\frac{i}{2}}\theta^{*}_i, \pi_{\theta^{*}\, j}+{\frac{i}{2}}\theta_j\rbrace
=-i\delta_{ij}$.}; $\pi _{\theta \, i}$ and $\pi_{\theta^{*}\, i}$ are then
eliminated with the help of Dirac brackets

\begin{equation}
\lbrace A,B\rbrace {}^{*}=\lbrace A,B\rbrace -i[\lbrace A,\pi_{\theta \, i}+
{\frac{i}{2}}\theta^{*}_i\rbrace \lbrace \pi_{\theta^{*}\, i}+{\frac{i}{2}}
\theta_i,B\rbrace +\lbrace A,\pi_{\theta^{*}\, i}+{\frac{i}{2}} \theta_i
\rbrace \lbrace \pi_{\theta \, i}+{\frac{i}{2}}\theta^{*}_i,B\rbrace ],
\label{II4}
\end{equation}

\noindent so that the remaining Grassmann variables have the fundamental
Dirac brackets (which we will still denote $\lbrace .,.\rbrace$ for the sake
of simplicity)

\begin{eqnarray}
&&\{\theta _{i}(\tau ),\theta _{j}(\tau )\}=\{\theta _{i}^{\ast }(\tau
),\theta _{j}^{\ast }(\tau )\}=0,  \nonumber \\
&&\{\theta _{i}(\tau ),\theta_{j}^{\ast }(\tau )\}=-i\delta _{ij}.
\label{II5}
\end{eqnarray}

In Ref.\cite{pfw} a pseudo-classical Foldy-Wouthuysen canonical
transformation was introduced, which, after quantization, realizes an exact
Foldy-Wouthuysen transformation of the Dirac equation in the case $%
A_{0}(x)=0 $, $\frac{\partial \vec{A}(x)}{\partial x^{0}}=0$. However in
that paper the electric charge was not Grassmann-valued.

The infinitesimal generator $S$ of the pseudo-classical FW transformation of
Ref.\cite{pfw} is

\begin{equation}
S = 2i(\vec{\Pi}\cdot \vec{\xi})\xi _{5}\theta (\alpha ),  \label{II6}
\end{equation}

\noindent where [in the free case we get $\alpha ={\vec p}^2$, $\theta
(\alpha )=\theta (p)$]

\begin{eqnarray}
p &=&\mid \vec{p}\mid ,  \nonumber \\
\vec{\Pi} &=&\vec{p}-Q\vec{A}(x),  \nonumber \\
\alpha &=&{\vec{\Pi}}^{2}-iQF_{hk}(x)\xi ^{h}\xi ^{k},  \nonumber \\
\sqrt{\alpha }\theta (\alpha ) &=&(1/2)%
\mathop{\rm arctg}%
\left( \frac{\sqrt{\alpha }}{m}\right) .  \label{II7}
\end{eqnarray}

The action $T_S$ of the canonical transformation generated by $S$ on a
generic function $F$ in phase space is

\begin{equation}
T_{S}:F\rightarrow F+\{F,S\}+(1/2!)\{\{F,S\},S\}+....  \label{II8}
\end{equation}

Let us now see what is the effect of the canonical transformation (\ref{II8}%
) when the electric charge $Q$ is Grassmann-valued, so that we have $Q^2=0$.
If we consider its action on the Dirac-like constraint

\begin{equation}
\chi _D =(p_{0}-QA_{0}(x))\xi _{0}-(\vec{p}-Q\vec{A}(x))\cdot \vec{\xi}-m\xi
_{5}\approx 0,  \label{II9}
\end{equation}

\noindent we get from Ref.\cite{pfw} the result

\begin{equation}
T_{S}:(-\vec{\Pi}\cdot \vec{\xi}-m\xi _{5})\rightarrow -\xi _{5}\sqrt{%
m^{2}+\alpha }.  \label{II10}
\end{equation}

We have now to evaluate the following two terms at order $Q$

\begin{eqnarray}
T_{S} &:&-QA_{0}(x)\xi _{0}\rightarrow -Q[A_{0}(x)\xi _{0}+\{A_{0}(x)\xi_o
,S\}+(1/2!)\{\{A_{0}(x)\xi_o,S\},S\}+....,  \nonumber \\
&&{}  \nonumber \\
T_{S} &:&p_{0}\xi _{0}\rightarrow p_{0}\xi _{0}+\{p_{0}\xi
_{0},S\}+(1/2!)\{\{p_{0}\xi _{0},S\},S\}+....  \label{II11}
\end{eqnarray}
Using $\xi _{5}^{2}=0$, we can write:

\begin{equation}
\{\{...\{QA_{0}(x)\xi _{0},\underbrace{S\},S\},...\},S\}}_{n}=\left\{
\begin{array}{lr}
Q(2i)^{n}\Phi _{n}(p,x,\vec{\xi})\xi _{5}\xi _{0} & n-odd \\
&  \\
Q(2i)^{n}\Phi _{n}(p,x,\vec{\xi})\xi _{0} & n-even
\end{array}
\right. +{\cal O}(Q^{2}),  \label{II12}
\end{equation}

\noindent with the odd functions $\Phi _{n}$ defined in Appendix A. Dropping
the $Q^{2}$ term we have:

\begin{equation}
T_{S}:-QA_{0}(x)\xi _{0}\rightarrow -QF_{1}(p,x,\vec{\xi})\xi
_{0}-QG_{1}(p,x,\vec{\xi})\xi _{5}\xi _{0},  \label{II13}
\end{equation}

\noindent with the following two functions (they are odd, i.e. linear in the
three Grassmann variables $\vec{\xi}$)

\begin{equation}
\left\{
\begin{array}{lcl}
F_{1}(p,x,\vec{\xi}) & = & \sum_{k=0}^{\infty }\frac{(2i)^{2k}}{2k!}\Phi
_{2k}(p,x,\vec{\xi}), \nonumber \\
&  & \nonumber \\
G_{1}(p,x,\vec{\xi}) & = & \sum_{k=0}^{\infty}\frac{(2i)^{2k+1}}{(2k+1)!}
\Phi _{2k+1}(p,x,\vec{\xi}).
\end{array}
\right.  \label{II14}
\end{equation}

Equally, we can write:

\begin{equation}
\{\{...\{p_{0}\xi _{0},\underbrace{S\},S\},...\},S\}}_{n}=\left\{
\begin{array}{lr}
Q(2i)^{n}\Psi _{n}(p,x,\vec{\xi})\xi _{5}\xi _{0} & n-odd \\
&  \\
Q(2i)^{n}\Psi _{n}(p,x,\vec{\xi})\xi _{0} & n-even,n\neq 0,
\end{array}
\right.  \label{II15}
\end{equation}

\noindent with the odd functions $\Psi _{n}$ defined in Appendix A. Then we
get

\begin{equation}
T_{S}:p_{0}(x)\xi _{0}\rightarrow p_{0}\xi _{0}+QF_{2}(p,x,\vec{\xi})\xi
_{0}+QG_{2}(p,x,\vec{\xi})\xi _{5}\xi _{0},  \label{II16}
\end{equation}

\noindent with the following two functions (they are odd, i.e. linear, in
the three Grassmann variables $\vec \xi$)

\begin{equation}
\left\{
\begin{array}{lcl}
F_{2}(p,x,\vec{\xi}) & = & \sum_{k=1}^{\infty }\frac{(2i)^{2k}}{2k!}\Psi
_{2k}(p,x,\vec{\xi}),\nonumber \\
&  & \nonumber \\
G_{2}(p,x,\vec{\xi}) & = & \sum_{k=0}^{\infty}\frac{(2i)^{2k+1}}{(2k+1)!}
\Psi _{2k+1}(p,x,\vec{\xi}).
\end{array}
\right.  \label{II17}
\end{equation}

In total, we can sum the two term to get

\begin{equation}
T_{S}:(p_{0}-QA_{0}(x))\xi _{0}\rightarrow \Big[p_{0}+QF(p,x,\vec{\xi})\Big]%
\xi _{0}+QG(p,x,\vec{\xi})\xi _{5}\xi _{0},  \label{II18}
\end{equation}

\noindent where $F=F_{2}-F_{1}$ and $G=G_{2}-G_{1}$. See Appendix A for the
determination of the functions $F$ and $G$. From Eqs. (\ref{a26}), (\ref{a32}%
) their final expression is [$\theta (p)=(1/2p)%
\mathop{\rm arctg}%
(p/m)$]

\begin{eqnarray}
F(p,x,\vec{\xi}) &=&A_{0}(x)-i{\frac{{(\vec{p}\cdot \vec{\xi})\vec{\xi}\cdot
\vec{E}(x)}}{{(m+\sqrt{m^{2}+{\vec{p}}^{2}})\sqrt{m^{2}+{\vec{p}}^{2}}}}},
\nonumber \\
&&{}  \nonumber \\
G(p,x,\vec{\xi}) &=&i{\frac{{sin\,[2p\theta (p)]}}{p}}\vec{\xi}\cdot \vec{%
\partial}A_{o}(x)+  \nonumber \\
&+&i\Big[2\sum_u{\frac{{\partial \theta (p)}}{{\partial p^{u}}}}{\frac{{%
\partial A_{o}(x)}}{{\partial x^{u}}}}-{\frac{{sin\,[2p\theta (p)]-2p\theta
(p)}}{{p^{3}}}}\vec{p}\cdot \vec{\partial}A_{o}(x)\Big]\vec{p}\cdot \vec{\xi}%
.  \label{II19}
\end{eqnarray}

The action of the canonical transformation $T_{S}$'s on the Dirac-like
constraint $\chi _{D}$ thus turns out to be

\begin{equation}
T_{S}:\chi _{D}\rightarrow \chi _{D}^{\prime }=\Big[p_{0}+QF(p,x,\vec{\xi})%
\Big]\xi _{0}-\xi _{5}\sqrt{\alpha +m^{2}}+QG(p,x,\vec{\xi})\xi _{5}\xi _{0}.
\label{II20}
\end{equation}
The new constraint $\chi _{D}^{\prime }$ contains some terms, linear in $\xi
_{0}$ or $\xi _{5}$, corresponding to {\it even} diagonal terms in quantum
case and some terms, linear in $\xi _{5}\xi _{0}$, corresponding to {\it odd}
skew-diagonal terms, according to the rule: $\xi _{5}\rightarrow \gamma
_{5},\xi _{0}\rightarrow \gamma _{5}\gamma _{0}$:

\begin{equation}
\chi _{5}^{\prime }\rightarrow \gamma _{5}\left[ \underbrace{
(p_{0}+QF)\gamma _{0}-\sqrt{m^2+\alpha}}_{diagonal}+\underbrace{QG\gamma
_{5}\gamma _{0}}_{skew-diagonal}\right].  \label{II21}
\end{equation}

In order to cancel the {\it odd} skew-diagonal terms at order $Q$, we define
a unknown infinitesimal generator $W=Qw(p,x,\vec{\xi})$ for a new canonical
transformation $T_W$

\begin{eqnarray}
T_{W}:\chi _{D}^{\prime }\rightarrow \chi _{D}^{\prime \prime } &=&\chi
_{D}^{\prime }+Q\{\chi _{D}^{\prime },w\}=  \nonumber \\
&=&\chi _{D}^{\prime }+Q\{p_{0}\xi _{0}-\xi _{5}\sqrt{m^{2}+\vec{p}^{2}},w\}.
\label{II22}
\end{eqnarray}
By imposing the Ansatz:

\begin{equation}
w=w_{1}(p,x,\vec{\xi})\xi _{0}+w_{2}(p,x,\vec{\xi})\xi _{5},  \label{II23}
\end{equation}

\noindent the condition for the cancellation of the skew-diagonal terms is

\begin{equation}
\{p_{0}\xi _{0}-\xi _{5}\sqrt{m^{2}+\vec{p}^{2}},w_{1}(p,x,\vec{\xi})\xi
_{0}+w_{2}(p,x,\vec{\xi})\xi _{5}\}+G(p,x,\vec{\xi})\xi _{5}\xi _{0}=0.
\label{II24}
\end{equation}

This condition is equivalent to a partial differential equation for the
unknown functions $w_{1},w_{2}$:

\begin{eqnarray}
&&(-i)w_{1}p_{0}-(-i)w_{2}\sqrt{m^{2}+\vec{p}^{2}}-\{p_{0},w_{2}\}\xi
_{5}\xi _{0}-\{\sqrt{m^{2}+\vec{p}^{2}},w_{1}\}\xi _{5}\xi _{0}+G\xi _{5}\xi
_{0}=0,  \nonumber \\
&&{}  \nonumber \\
&\Rightarrow &\left\{
\begin{array}{l}
w_{1}p_{0}-w_{2}\sqrt{m^{2}+\vec{p}^{2}}=0, \\
\\
-\{p_{0},w_{2}\}-\{\sqrt{m^{2}+\vec{p}^{2}},w_{1}\}+G=0,
\end{array}
\right.  \nonumber \\
&&{}  \nonumber \\
&\Rightarrow &\left\{
\begin{array}{l}
w_{2}=\frac{p_{0}}{\sqrt{m^{2}+\vec{p}^{2}}}w_{1}, \\
\\
-\frac{\partial w_{2}}{\partial x^{o}}-\frac{1}{\sqrt{m^{2}+\vec{p}^{2}}}%
\vec{p}\cdot \partial w_{1}+G=0,
\end{array}
\right.  \nonumber \\
&&{}  \nonumber \\
&\Rightarrow &-\frac{p_{0}}{\sqrt{m^{2}+\vec{p}^{2}}}\frac{\partial w_{1}}{%
\partial x^{o}}-\frac{1}{\sqrt{\vec{p}^{2}+m^{2}}}\vec{p}\cdot \vec \partial
w_{1}+G=0.  \label{II25}
\end{eqnarray}

By introducing the Green function $g_p(x)$ for $p^{\mu}\partial_{\mu}$ we
find the following solution for $w_1$, $w_2$ [see Appendix B for the tetrads
$\epsilon^{\mu}_A(u(p))$]

\begin{eqnarray}
w_2&=&\frac{p_0}{\sqrt{m^2+\vec{p}^2}}w_1,  \nonumber \\
&&{}  \nonumber \\
p^\mu\partial_\mu w_1(p,x,\vec{\xi})&=&\sqrt{m^2+\vec{p}^2}G(p,x, \vec{\xi}),
\nonumber \\
&&\Downarrow  \nonumber \\
w_1(p,x,\vec \xi )&=& \int d^4y\, g_p(x-y)\, \sqrt{m^2+\vec{p}^2}G(p,y, \vec{%
\xi}),  \nonumber \\
&&{}  \nonumber \\
&&p_{\mu}\partial^{\mu} g_p(x) = \delta^4(x),  \nonumber \\
&&g_p(x) = \pm {\frac{{\theta (\pm p^{\mu}x_{\mu})\,
\delta^3(\epsilon^{\mu}_r(u(p)) x_{\mu})}}{\sqrt{{\vec p}^2}}}.  \label{II26}
\end{eqnarray}

\noindent This shows the existence of the canonical transformation $T_{W}$
and the non-locality of the separation of positive and negative energies.
Therefore, the {\it odd} term $QG\xi _{5}\xi _{0}$ of Eq.(\ref{II20}) can be
eliminated at the order Q with the canonical transformation $T_{W}$.

After these two canonical transformations $T_W \circ T_S$, we get the
following form of the Dirac-like constraint at order $Q$

\begin{eqnarray}
\chi _{D}^{\prime \prime }&=&\Big[p_{0}-QA_{0}(x)+iQ(\vec{p}\cdot \xi )(\vec{%
\xi}\cdot \vec{E}(x))\frac{1}{(m+\sqrt{m^{2}+{\vec{p}}^{2}})\sqrt{m^{2}+{%
\vec{p}}^{2}}}\Big]\xi _{0}-\sqrt{m^{2}+\alpha }\xi _{5}=  \nonumber \\
&=&[p_{0}-QA_{0}(x)-Q\frac{\vec p \cdot \vec E(x)\times \vec S} {(m+\sqrt{%
m^{2}+{\vec{p}}^{2}})\sqrt{m^{2}+{\vec{p}}^{2}}}\Big]\xi _{0}-\sqrt{%
m^{2}+\alpha }\xi _{5}\approx 0,  \label{II27}
\end{eqnarray}

\noindent where we introduced the spin $\vec S=-{\frac{i}{2}} \vec \xi
\times \vec \xi$.

To find the final form $\chi^{\prime \prime}$ ($T_{W}\circ T_{S}:\chi
\rightarrow \chi ^{\prime \prime }$) of the mass-shell constraint we use the
fact that $T_W \circ T_S$ is a canonical transformation, so that

\begin{equation}
\{\chi _D^{\prime \prime },\chi _D^{\prime \prime }\}=i\chi ^{\prime \prime
} .  \label{II28}
\end{equation}

We get ($F_{hk}=\epsilon_{hkr}B_r$)

\begin{eqnarray}
\chi ^{\prime \prime } &=&\left( p_{0}-QA_{0}(x)-Q\vec{p}\cdot \vec E%
(x)\times \vec S \frac{m-\sqrt{m^{2}+{\vec p}^{2}}}{{\vec p}^{2}\sqrt{m^{2}+{%
\vec p}^{2}}}\right) ^{2}-  \nonumber \\
&-&\Big[ m^2+(\vec p - Q\vec A(x))^2 +2Q\vec S\cdot \vec B(x) \Big]\approx 0,
\label{II29}
\end{eqnarray}

\noindent after having discarded terms proportional to $\xi _{0}\xi _{5}$,
since they vanish due to $\chi _D^{\prime \prime }\approx 0$, which implies $%
\xi_5$ weakly proportional to $\xi_o$ at order $Q$.

The mass-shell constraint can to be resolved at order $Q$ in two
constraints, corresponding to the two signs of the energy:

\begin{eqnarray}
\chi ^{(\pm )}&=&p_{0}-QA_{0}(x)+iQ(\vec{p}\cdot \xi )(\vec{\xi}\cdot \vec{E}
(x))\frac{1}{(m+\sqrt{m^{2}+{\vec p}^{2}})\sqrt{m^{2}+{\vec p}^{2}}}\mp
\nonumber \\
&\mp& \sqrt{ m^2+(\vec p - Q\vec A(x))^2 -iQ\xi^h\xi^k F_{hk}(x)}=  \nonumber
\\
&=&p_o-QA_o(x)-{\frac{{Q \vec p\cdot \vec E(x)\times \vec S}}{{(m+\sqrt{m^2+{%
\vec p}^2})\sqrt{m^2+{\vec p}^2}}}} \mp  \nonumber \\
&\mp& \sqrt{m^2+(\vec p-Q\vec A(x))^2+2Q\vec S\cdot \vec B(x)} \approx 0.
\label{II30}
\end{eqnarray}

In this way we have identified the extra coupling at order $Q$ to the
electric field $\vec{E}(x)=-{\frac{{\partial \vec{A}(x)}}{{\partial x^{0}}}}-%
\vec{\partial}A_{0}(x)$ of a spinning particle with a definite sign of the
energy. {\it Only} the semi-classical approximation $Q^{2}=0$ allows us to
get this result in closed form.

Now the problem is to interpret this result. Since $A_{0}(x)$ and $\vec{A}%
(x) $ describe an {\it external} electromagnetic field, we are in an
arbitrary fixed gauge. But we will need the result in the radiation gauge
with both transverse radiation fields and action-at-a-distance Coulomb
potentials from other charges acting simultaneously on the given charged
spinning particle. It seems reasonable to interpret $A_{0}(x)$ as the scalar
Coulomb potential generated by the other charges. Then the term $-{\frac{{%
\partial \vec{A}(x)}}{{\partial x^{0}}}}-\vec{\partial}A_{0}(x)$ should be
interpreted as the {\it sum of the transverse radiation electric field and
of the gradient of the Coulomb potential} (action-at-a-distance electric
field generated by the other charges), since this is the total electric
field acting on the charged spinning particle.

\vfill\eject

\section{The Positive-energy spinning particle.}

In this Section we shall review the description of the positive-energy
charged spinning particle given in Ref.\cite{lu3}. To define a charged
spinning particle with a definite sign of the energy on spacelike
hypersurfaces\footnote{%
See Appendix B for some notions on spacelike hypersurfaces.}, the starting
point was the Lagrangian description of a charged scalar particle (see Refs.
\cite{lu1,lu6}) with only a real Grassmann 4-vector $\xi ^{\mu }(\tau )$ for
the description of spin\footnote{%
To describe a positive-energy spinning particle we have to solve the first
class constraint $\chi _{D}\approx 0$ of Eq.(\ref{I3}) to express $\xi _{5}$
in terms of the Grassmann 4-vector $\xi ^{\mu }$. The needed gauge fixing to
$\chi _{D}\approx 0$ is a constraint eliminating one of the four components
of $\xi ^{\mu }$: as shown in Ref.\cite{lu3} in the free case it is $p_{\mu
}\xi ^{\mu }+m\xi ^{5}\approx 0$, so that $\xi _{5}\approx 0$ and $p_{\mu
}\xi ^{\mu }\approx 0$ hold simultaneously. The mass-shell constraint $\chi
\approx 0$ of Eq.(\ref{I1}) is eliminated by the choice of the energy like
for scalar particles.}. After the Legendre transformation to the Hamiltonian
formalism, a Hamiltonian odd second class constraint \footnote{%
It is of the transversality type $p_{\mu }\xi ^{\mu }\approx 0$, but with $%
p_{\mu }$ being the conserved total momentum of the isolated system and {\it %
not} the particle momentum.} is added to eliminate one of the components of $%
\xi ^{\mu }$. In this way only three Grassmann variables will survive for
each particle and, after quantization, they will generate Pauli matrices
acting on the 2-spinors describing the positive-energy wave functions. It
turns out that the addition by hand of this constraint gives a consistent
set of constraints.

Since, due to this last constraint, the Lagrangian description is too
complicated \footnote{%
In Eq.(29) of Ref.\cite{lu3} there is the Lagrangian generating all the
constraints ( \ref{III3}) except the transversality ones $\phi_i(\tau
)=p_{\mu}\xi_i^{\mu}\approx 0$. The Lagrangian generating all the
constraints (\ref{III3}) could be recovered by inverse Legendre
transformation. This has been done in Ref.\cite{lu3} [see its Eq.(50)] only
in absence of electromagnetic field, because the general case is very
complicated and not particularly interesting, except for the determination
of the energy-momentum tensor. In any case, the Lagrangian contains both a
minimal coupling of the positive-energy particles to the electromagnetic
field and a non-minimal coupling like in Eq.(\ref{II1}).} the model was
defined directly in phase space by means of a set of constraints. As usual
in relativistic particle mechanics, only the Hamiltonian description is
tractable, because the Lagrangian one is too involved and very often it is
impossible to get it in closed form.

Given a 3+1 splitting of Minkowski spacetime with a foliation whose
spacelike leaves are defined by the embeddings $\Sigma \mapsto \Sigma _{\tau
}$ , $(\tau ,\vec{\sigma})\mapsto z^{\mu }(\tau ,\vec{\sigma})$, the
embeddings become new configuration variables describing all possible
hypersurfaces (all possible congruences of timelike accelerated observers).
Since, due to the separate $\tau $- and $\vec{\sigma}$-reprametrization
invariances of the action, there are first class constraints implying the
independence of the description from the choice of the 3+1 splitting, the $%
z^{\mu }(\tau ,\vec{\sigma})$'s are the {\it gauge} variables of this type
of general covariance\footnote{%
The descriptions given by arbitrary congruences of timelike observers are
gauge equivalent.} See Appendix B for the definition of the notations and of
the induced metric. Each positive energy particle is described by the
coordinates ${\vec{\eta}}_{i}(\tau )$ such that $x_{i}^{\mu }(\tau )=z^{\mu
}(\tau ,{\vec{\eta}}_{i}(\tau ))$ . Moreover, for each particle there will
be a Grassmann 4-vector $\xi _{i}^{\mu }(\tau )$ for the spin and a pair $%
\theta _{i}(\tau )$, $\theta _{i}^{\ast }(\tau )$ of complex scalar
Grassmann variables for the electric charge.

On the hypersurface $\Sigma _{\tau }$, we describe the electromagnetic
potential and field strength with Lorentz-scalar variables $A_{\check{A}%
}(\tau ,\vec{\sigma})$ and $F_{{\check{A}}{\check{B}}}(\tau ,\vec{\sigma})$
respectively, which have the {\it equal time} concept introduced by the
embedding, defined by [$\tilde{A}$, ${\tilde{F}}_{\mu \nu }$ are the
standard electromagnetic potentials and filed strengths, respectively]

\begin{eqnarray}
&&A_{\check A}(\tau ,\vec \sigma )=z^{\mu}_{\check A}(\tau ,\vec \sigma ) {%
\tilde A}_{\mu}(z(\tau ,\vec \sigma )),  \nonumber \\
&&F_{{\check A}{\check B}}(\tau ,\vec \sigma )={\partial}_{\check A}A_{%
\check B}(\tau ,\vec \sigma )-{\partial}_{\check B}A_{\check A}(\tau ,\vec %
\sigma )= z^{\mu}_{\check A}(\tau ,\vec \sigma )z^{\nu}_{\check B}(\tau ,%
\vec \sigma ) {\tilde F}_{\mu\nu}(z(\tau ,\vec \sigma )).  \label{III1}
\end{eqnarray}

The model of Ref.\cite{lu3} is defined in a phase space spanned by the
canonically conjugate pairs of variables $z^{\mu }(\tau ,\vec{\sigma})$, $%
\rho _{\mu }(\tau ,\vec{\sigma})$ ; $A_{\check{A}}(\tau ,\vec{\sigma})$, $%
\pi ^{\check{A}}(\tau ,\vec{\sigma})$ ${}$\footnote{$E_{\check{r}}=F_{{%
\check{r}}\tau }$ and $B_{\check{r}}={\frac{1}{2}}\epsilon _{{\check{r}}{%
\check{s}}{\check{t}}}F_{{\check{s}}{\check{t}}}$ ($\epsilon _{{\check{r}}{%
\check{s}}{\check{t}}}=\epsilon ^{{\check{r}}{\check{s}}{\check{t}}}$) are
the electric and magnetic fields respectively; for $g_{\check{A}\check{B}%
}\rightarrow \eta _{\check{A}\check{B}}$ one gets $\pi ^{\check{r}}=-E_{%
\check{r}}=E^{\check{r}}$.}; ${\vec{\eta}}_{i}(\tau )$, ${\vec{\kappa}}%
_{i}(\tau )$; $\xi _{i}^{\mu }(\tau )$, $\pi _{i}^{\mu }(\tau )$; $\theta
_{i}(\tau )$, $\pi _{\theta i}(\tau )$; $\theta _{i}^{\ast }(\tau )$, $\pi
_{\theta ^{\ast }i}(\tau )$ with the following Poisson brackets

\begin{eqnarray}
&&\lbrace z^{\mu}(\tau ,\vec \sigma ),\rho_{\nu}(\tau ,{\vec \sigma}%
^{^{\prime}}\rbrace =-\eta^{\mu}_{\nu}\delta^3(\vec \sigma -{\vec \sigma}%
^{^{\prime}}),  \nonumber \\
&&\lbrace A_{\check A}(\tau ,\vec \sigma ),\pi^{\check B}(\tau ,\vec \sigma%
^{^{\prime}} )\rbrace =\eta^{\check B}_{\check A} \delta^3(\vec \sigma -\vec %
\sigma^{^{\prime}}),  \nonumber \\
&&\lbrace \eta^{\check r}_i(\tau ),\kappa_{j{\check s}}(\tau )\rbrace =-
\delta_{ij}\delta^{\check r}_{\check s},  \nonumber \\
&&\lbrace \theta_i(\tau ),\pi_{\theta \, j}(\tau )\rbrace =-\delta_{ij},
\nonumber \\
&&\lbrace \theta^{*}_i(\tau ),\pi_{\theta^{*} \, j}(\tau )\rbrace
=-\delta_{ij},  \nonumber \\
&&\lbrace \xi^{\mu}_i,\pi_j^{\nu}\rbrace =-\delta_{ij}\eta^{\mu\nu}.
\label{III2}
\end{eqnarray}

The total conserved 4-momentum of the system is $p^{\mu}_s=\int d^3\sigma
\rho^{\mu}(\tau ,\vec \sigma )$ [see Eqs.(\ref{III6})].

The model is defined by the following set of constraints (only positive
energy particles are considered)

\begin{eqnarray}
&&\pi _{\theta \,i}+{\frac{i}{2}}\theta _{i}^{\ast }\approx 0,  \nonumber \\
&&\pi _{\theta ^{\ast }\,i}+{\frac{i}{2}}\theta _{i}\approx 0,  \nonumber \\
&&{}  \nonumber \\
\phi _{i}(\tau )&=&[\pi _{i}^{\mu }(\tau )+{\frac{i}{2}}\xi _{i}^{\mu }(\tau
)]\int d^{3}\sigma \,\rho _{\mu }(\tau ,\vec{\sigma})= [\pi _{i}^{\mu }(\tau
)+{\frac{i}{2}}\xi _{i}^{\mu }(\tau )] p_{s\mu}\approx 0,  \nonumber \\
\chi _{i}^{\mu }(\tau ) &=&\pi _{i}^{\mu }(\tau )-\frac{i}{2}\xi _{i}^{\mu
}(\tau )\approx 0,\quad \Rightarrow \phi_i(\tau )\approx
p_{s\mu}\xi^{\mu}_i(\tau )\approx 0,  \nonumber \\
&&{}  \nonumber \\
\pi ^{\tau }(\tau ,\vec{\sigma}) &\approx &0  \nonumber \\
\Gamma (\tau ,\vec{\sigma}) &=&\partial _{\breve{r}}\pi ^{\breve{r}}(\tau ,%
\vec{\sigma})-\sum_{i=1}^{N}Q_{i}(\tau )\delta ^{3}(\vec{\sigma}-\vec{\eta}%
_{i}(\tau ))\approx 0,  \nonumber \\
&&{}  \nonumber \\
{\cal H}_{\mu }(\tau ,\vec{\sigma}) &=&\rho _{\mu }(\tau ,\vec{\sigma}%
)-l_{\mu }(\tau ,\vec{\sigma})\Big[-\frac{1}{2\sqrt{\gamma (\tau ,\vec{\sigma%
})}}\pi ^{\breve{r}}(\tau ,\vec{\sigma})g_{\breve{r}\breve{s}}(\tau ,\vec{%
\sigma})\pi ^{s}(\tau ,\vec{\sigma})+  \nonumber \\
&+&\frac{\sqrt{\gamma (\tau ,\vec{\sigma})}}{4}\gamma ^{\breve{r}\breve{s}%
}(\tau ,\vec{\sigma})\gamma ^{\breve{u}\breve{v}}(\tau ,\vec{\sigma})F_{%
\breve{r}\breve{u}}(\tau ,\vec{\sigma})F_{\breve{s}\breve{v}}(\tau ,\vec{%
\sigma})+  \nonumber \\
&+&\sum_{i=1}^{N}\delta ^{3}(\vec{\sigma}-\vec{\eta}_{i}(\tau ))\cdot
\nonumber \\
&\cdot &\sqrt{m_{i}^{2}-\gamma ^{\breve{r}\breve{s}}(\tau ,\vec{\sigma})\big(%
\kappa _{i\breve{r}}(\tau )-Q_{i}(\tau )A_{\breve{r}}(\tau ,\vec{\sigma})%
\big)\big(\kappa _{i\breve{s}}(\tau )-Q_{i}(\tau )A_{\breve{s}}(\tau ,\vec{%
\sigma})\big)}+  \nonumber \\
&+&\frac{1}{2\sqrt{\gamma (\tau ,\vec{\sigma})}}\sum_{i,j=1}^{N}\delta ^{3}(%
\vec{\sigma}-\vec{\eta}_{i}(\tau ))\delta ^{3}(\vec{\sigma}-\vec{\eta}%
_{j}(\tau ))\cdot  \nonumber \\
&\cdot &\frac{Q_{i}(\tau )Q_{j}(\tau )\xi _{i}^{\gamma }(\tau )\xi
_{i}^{\delta }(\tau )l_{\gamma }(\tau ,\vec{\sigma})\eta _{\delta \beta }\xi
_{j}^{\alpha }(\tau )\xi _{j}^{\beta }(\tau )l_{\alpha }(\tau ,\vec{\sigma})%
}{\sqrt{m_{i}^{2}-\gamma ^{\breve{r}\breve{s}}(\tau ,\vec{\sigma})\kappa _{i%
\breve{r}}(\tau )\kappa _{i\breve{s}}(\tau )}\sqrt{m_{j}^{2}-\gamma ^{\breve{%
r}\breve{s}}(\tau ,\vec{\sigma})\kappa _{j\breve{r}}(\tau )\kappa _{j\breve{s%
}}(\tau )}}+  \nonumber \\
&+&\frac{i}{\sqrt{\gamma (\tau ,\vec{\sigma})}}\sum_{i=1}^{N}\delta ^{3}(%
\vec{\sigma}-\vec{\eta}_{i}(\tau ))\cdot  \nonumber \\
&\cdot &\frac{Q_{i}(\tau )\xi _{i}^{\alpha }(\tau )\xi _{i}^{\beta }(\tau )}{%
\sqrt{m_{i}^{2}-\gamma ^{\breve{r}\breve{s}}(\tau ,\vec{\sigma})\kappa _{i%
\breve{r}}(\tau )\kappa _{i\breve{s}}(\tau )}}l_{\alpha }(\tau ,\vec{\sigma}%
)z_{\breve{s}\beta }(\tau ,\vec{\sigma})\pi ^{\breve{s}}(\tau ,\vec{\sigma})-
\nonumber \\
&-&\frac{i}{2}\sum_{i=1}^{N}\delta ^{3}(\vec{\sigma}-\vec{\eta}_{i}(\tau ))%
\frac{Q_{i}(\tau )\xi _{i}^{\alpha }(\tau )\xi _{i}^{\beta }(\tau )}{\sqrt{%
m_{i}^{2}-\gamma ^{\breve{r}\breve{s}}(\tau ,\vec{\sigma})\kappa _{i\breve{r}%
}(\tau )\kappa _{i\breve{s}}(\tau )}}\cdot  \nonumber \\
&\cdot &z_{\breve{u}\alpha }(\tau ,\vec{\sigma})z_{\breve{v}\beta }(\tau ,%
\vec{\sigma})\gamma ^{\breve{r}\breve{u}}(\tau ,\vec{\sigma})\gamma ^{\breve{%
s}\breve{v}}(\tau ,\vec{\sigma})F_{\breve{r}\breve{s}}(\tau ,\vec{\sigma})%
\Big]-  \nonumber \\
&&{}  \nonumber \\
&&{}  \nonumber \\
&-&\gamma ^{\breve{r}\breve{s}}(\tau ,\vec{\sigma})z_{\breve{s}\mu }(\tau ,%
\vec{\sigma})\Big[F_{\breve{r}\breve{u}}(\tau ,\vec{\sigma})\pi ^{\breve{u}%
}(\tau ,\vec{\sigma})+  \nonumber \\
&+&\sum_{i=1}^{N}\delta ^{3}(\vec{\sigma}-\vec{\eta}_{i}(\tau ))\Big(\kappa
_{\breve{r}i}-Q_{i}(\tau )A_{\breve{r}}(\tau ,\vec{\sigma})\Big)\Big]\approx
0.  \label{III3}
\end{eqnarray}

\noindent\ The Lagrangian density that generates all these constraints
except for the combination $\phi _{i}(\tau )=\int d^{3}\sigma \,\rho _{\mu
}(\tau ,\vec{\sigma})\xi _{i}^{\mu }(\tau )\approx 0$ of the third and
fourth set \ above is given by [see Eq.(29) of Ref.\cite{lu3}]

\begin{eqnarray}
&{\cal L}&(\tau ,\vec{\sigma})=\sum_{i=1}^{N}\delta ^{3}(\vec{\sigma}-\vec{%
\eta}_{i}(\tau ))\Big\{\frac{i}{2}\big(\theta _{i}^{\ast }(\tau )\dot{\theta}%
_{i}(\tau )-\dot{\theta}_{i}^{\ast }(\tau )\theta _{i}(\tau )\big)-\frac{i}{2%
}\xi _{\mu i}(\tau )\dot{\xi}_{i}^{\mu }(\tau )-  \nonumber \\
&-&\sqrt{m_{i}^{2}-iQ_{i}(\tau )\xi _{i}^{\mu }(\tau )\xi _{i}^{\nu }(\tau
)z_{\mu }^{\breve{A}}(\tau ,\vec{\sigma})z_{\nu }^{\breve{B}}(\tau ,\vec{%
\sigma})F_{\breve{A}\breve{B}}(\tau ,\vec{\sigma})}\cdot  \nonumber \\
&&\sqrt{g_{\tau \tau }(\tau ,\vec{\sigma})+2g_{\tau \breve{r}}(\tau ,\vec{%
\sigma})\dot{\eta}_{i}^{\breve{r}}(\tau )+g_{\breve{r}\breve{s}}(\tau ,\vec{%
\sigma})\dot{\eta}_{i}^{\breve{r}}(\tau )\dot{\eta}_{i}^{\breve{s}}(\tau )}-
\nonumber \\
&-&Q_{i}(\tau )\Big(A_{\tau }(\tau ,\vec{\sigma})+\dot{\eta}_{i}^{\breve{r}%
}A_{\breve{r}}(\tau ,\vec{\sigma})\Big)\Big\}-\frac{\sqrt{g(\tau ,\vec{\sigma%
})}}{4}F_{\breve{A}\breve{B}}(\tau ,\vec{\sigma})F_{\breve{C}\breve{D}}(\tau
,\vec{\sigma})g^{\breve{A}\breve{C}}(\tau ,\vec{\sigma})g^{\breve{B}\breve{D}%
}(\tau ,\vec{\sigma}).
\end{eqnarray}

The following Dirac Hamiltonian (see Eqs.(46) and (47) of Ref.\cite{lu3} for
its determination) completes the definition of our model

\begin{eqnarray}
H^F_D&=&\int{d^3\sigma \Big[\lambda^\mu (\tau, \vec{\sigma}) {\cal H}%
^\ast_\mu(\tau, \vec{\sigma}) - A_{\tau}(\tau, \vec{\sigma}) \Gamma (\tau,
\vec{\sigma}) + \mu_\tau (\tau, \vec{\sigma}) \pi^\tau (\tau, \vec{\sigma}) %
\Big] } ,  \nonumber \\
&&{}  \nonumber \\
{\cal H}^\ast_\mu(\tau, \vec{\sigma})&=& {\cal H}_\mu(\tau, \vec{\sigma}) +
i \sum_{i=1}^N \{{\cal H}_\mu(\tau, \vec{\sigma}), \chi^\nu_i(\tau)\}
\chi_{i \nu} (\tau) +  \nonumber \\
&-& \frac {i}{p_s^2} \sum_{i=1}^N \{ {\cal H}_\mu(\tau, \vec{\sigma}),
\phi_i(\tau)\}\phi_i(\tau) \approx 0.  \label{III4}
\end{eqnarray}

See Ref.\cite{lu3} for their Poisson brackets. The constraints $\pi _{\theta
\,i}+{\frac{i}{2}}\theta _{i}^{\ast }\approx 0$, $\pi _{\theta ^{\ast }\,i}+{%
\frac{i}{2}}\theta _{i}\approx 0$, $\phi _{i}\approx 0$, $\chi _{i}^{\mu
}\approx 0$ are second class, while the other are first class.

The Dirac multipliers $\lambda ^{\mu }(\tau ,\vec{\sigma})$, $\mu _{\tau
}(\tau ,\vec{\sigma})$ in the Dirac Hamiltonian imply that the constraints $%
{\cal H}_{\mu }^{\ast }(\tau ,\vec{\sigma})\approx 0$, $\pi ^{\tau }(\tau ,%
\vec{\sigma})\approx 0$ are primary constraints of the unknown Lagrangian;
the other primary constraints are the second class ones, whose associated
Dirac multipliers are determined by the Dirac algorithm as shown in Ref.\cite
{lu3}. The Gauss law $\Gamma (\tau ,\vec{\sigma})\approx 0$ is the only
secondary constraints and it appears in the Dirac Hamiltonian with $A_{\tau
}(\tau ,\vec{\sigma})$ as a multiplier not determined by the theory (it is a
gauge variable).

As said in Section II the Grassmann momenta $\pi_{\theta i}(\tau )$, $%
\pi_{\theta^{*} i}(\tau )$ may be eliminated by using the Dirac brackets (%
\ref{II5}).

The conserved Poincar\'e generators are

\begin{eqnarray}
p_s^\mu &=& \int{d^3\sigma \rho^\mu (\tau, \vec{\sigma}) }  \nonumber \\
J^{\mu \nu} &=& \int{d^3\sigma \Big[ z^\mu (\tau, \vec{\sigma}) \rho^\nu
(\tau, \vec{\sigma}) - z^\nu (\tau, \vec{\sigma}) \rho^\mu (\tau, \vec{%
\sigma }) \Big] - \sum_{i=1}^N \Big[ \xi_i^\mu (\tau) \pi_i^\nu (\tau) -
\xi_i^\nu (\tau) \pi_i^\mu (\tau) \Big]}.  \label{III5}
\end{eqnarray}

Since $p_{s}^{\mu }$ is a constant of the motion independent of the isolated
system under investigation, we may write $\phi _{i}=(\pi _{i}^{\mu }+{\frac{i%
}{2}}\xi _{i}^{\mu })p_{s\mu }\approx 0$. Due to the transversality to the
total conserved 4-momentum\footnote{%
This is a highly non-local property, since the whole hypersurface $\Sigma
_{\tau }$ is involved in the reduction. Note that $p_{s}^{\mu }$ weakly
coincides with the total 4-momentum of the isolated system by using ${\cal H}%
_{\mu }(\tau ,\vec{\sigma})\approx 0$.}, we have the possibility of reducing
the $\xi _{i}^{\mu }$'s from 4 to 3 for each particle independently from the
interactions.

As we see, the component of ${\cal H}_\mu(\tau, \vec{\sigma})$ along $%
l^{\mu}(\tau ,\vec \sigma )$ (i.e. orthogonal to $\Sigma_{\tau}$) contains
the electromagnetic energy density and also {\it spin-spin}, {\it %
spin-electric field} and {\it spin-magnetic field} interactions\footnote{%
Only the last one will survive in the rest frame.}. All of them are
necessary to get the first class property for these constraints. Instead the
components of ${\cal H}_\mu(\tau, \vec{\sigma})$ along $z^{\mu}_{\check r
}(\tau ,\vec \sigma )$ )(i.e. tangent to $\Sigma_{\tau}$) contain only the
electromagnetic Poynting vector as with scalar particles \cite{lu1}.

\subsection{The restriction to Wigner's hyperplanes}

As shown in Ref.\cite{lu1}, the restriction from arbitrary hypersurfaces $%
\Sigma _{\tau }$ to hyperplanes $\Sigma _{H\tau }$ is done by introducing
the gauge-fixings by pairing up the first class constraints ${\cal H}_{\nu
}^{\ast }(\tau ,\vec{\sigma}^{\prime })$ with

\begin{eqnarray}
&&\zeta^\mu (\tau, \vec{\sigma}) = z^\mu (\tau, \vec{\sigma}) -x_s^\mu(\tau)
-b_{\breve{r}}^\mu (\tau) \sigma^{\breve{r}} \approx 0,  \nonumber \\
&&{}  \nonumber \\
&&\{ \zeta^\mu (\tau, \vec{\sigma}) , {\cal H}^\ast_\nu(\tau, \vec{\sigma}%
^\prime) \} = - \eta^\mu_\nu \delta^3 (\vec{\sigma} - \vec{\sigma}^\prime),
\label{III6}
\end{eqnarray}

\noindent and the Dirac brackets which eliminate these variables are

\begin{equation}
\{ A, B\}^\ast = \{ A , B \} - \int{d^3\sigma \{ A , \zeta^\mu (\tau, \vec{%
\sigma}) \} \{ {\cal H}^\ast_\mu(\tau, \vec{\sigma}) , B \} } + \int{%
d^3\sigma \{ A , {\cal H}^\ast_\mu (\tau, \vec{\sigma}) \} \{
\zeta^\mu(\tau, \vec{\sigma}) , B \} } .  \label{III7}
\end{equation}

The hyperplane $\Sigma _{H\tau }$ is now described by just 10 configuration
variables: an origin $x_{s}^{\mu }(\tau )$ and the 6 independent degrees of
freedom in an orthonormal tetrad $b_{\check{A}}^{\mu }(\tau )$ [${}b_{\check{%
A}}^{\mu }\,\eta _{\mu \nu }b_{\check{B}}^{\nu }=\eta _{\check{A}\check{B}}$%
] with $b_{\tau }^{\mu }=l^{\mu }$, where $l^{\mu }$ is the $\tau $%
-independent normal to the hyperplane. We have $z_{\check{r}}^{\mu }(\tau ,%
\vec{\sigma})\equiv b_{\check{r}}^{\mu }(\tau )$, $z_{\tau }^{\mu }(\tau ,%
\vec{\sigma})\equiv {\dot{x}}_{s}^{\mu }(\tau )+b_{\check{r}}^{\mu }(\tau
)\sigma ^{\check{r}}$, $g_{\check{r}\check{s}}(\tau ,\vec{\sigma})\equiv
-\delta _{\check{r}\check{s}}$, $\gamma ^{\check{r}\check{s}}(\tau ,\vec{%
\sigma})\equiv -\delta ^{\check{r}\check{s}}$, $\gamma (\tau ,\vec{\sigma}%
)=det\,g_{\check{r}\check{s}}(\tau ,\vec{\sigma})\equiv 1$. The
non-vanishing Dirac brackets of the variables $x_{s}^{\mu }$, $p_{s}^{\mu }$%
, $b_{\breve{A}}^{\mu }$, $S_{s}^{\mu \nu }$, $A_{\breve{A}}$, $\pi ^{\breve{%
A}}$, $\xi _{i}^{\mu }$, $\pi _{j}^{\nu }$ are

\begin{eqnarray}
\{ x_s^\mu (\tau), p_s^\nu \}^\ast &=& -\eta^{\mu \nu},  \nonumber \\
\{\eta_i^{\breve{r}}(\tau), \kappa^{\breve{s}}_j (\tau) \}^\ast &=&
\delta_{ij} \delta^{\breve{r} \breve{s}},  \nonumber \\
\{ S^{\mu \nu}_s(\tau) , b^\rho_{\breve{A}}\}^\ast &=& \eta^{\rho \nu}
b^\mu_{\breve{A}} (\tau) - \eta^{\rho \mu} b^\nu_{\breve{A}} (\tau),
\nonumber \\
\{ S^{\mu \nu}_s(\tau) , S^{\alpha \beta}_s(\tau) \}^\ast &=& C^{\mu \nu
\alpha \beta}_{\gamma \delta} S^{\gamma \delta}_s (\tau),  \nonumber \\
\{ \xi^\mu_i (\tau) , \pi^\nu_j (\tau) \}^\ast &=& - \eta^{\mu \nu}
\delta_{ij}.  \label{III8}
\end{eqnarray}

While $p^{\mu}_s$ is the momentum conjugate to $x^{\mu}_s$, the 6
independent momenta conjugate to the 6 degrees of freedom in the $b^{\mu}_{%
\check A}$'s are hidden in $S_s^{\mu\nu}$, which is a component of the
angular momentum tensor

\begin{eqnarray}
J^{\mu \nu} &=& L^{\mu \nu}_s + S^{\mu \nu}_s + S^{\mu \nu}_\xi ,  \nonumber
\\
L^{\mu \nu}_s &=& x^\mu_s(\tau) p^\nu_s - x^\nu_s(\tau) p^\mu_s ,  \nonumber
\\
S^{\mu \nu}_s &=& b^\mu_{\breve{r}} (\tau) \int{d^3\sigma \sigma^{\breve{r}}
\rho^\nu (\tau, \vec{\sigma})} - b^\nu_{\breve{r}} (\tau) \int{d^3\sigma
\sigma^{\breve{r}} \rho^\mu (\tau, \vec{\sigma})} ,  \nonumber \\
S^{\mu \nu}_\xi &=& - \sum_{i=1}^N \Big( \xi_i^\mu (\tau) \pi_i^\nu (\tau) -
\xi_i^\nu (\tau) \pi_i^\mu (\tau) \Big) ,  \nonumber \\
&&{}  \nonumber \\
&&\{ J^{\mu\nu},J^{\alpha\beta} \}^{*} =
C^{\mu\nu\alpha\beta}_{\gamma\delta} J^{\gamma\delta},\quad\quad \{
L_s^{\mu\nu},L_s^{\alpha\beta} \}^{*} = C
^{\mu\nu\alpha\beta}_{\gamma\delta} L_s^{\gamma\delta},  \nonumber \\
&&\{ S_s^{\mu\nu},S_s^{\alpha\beta} \}^{*} =
C^{\mu\nu\alpha\beta}_{\gamma\delta} S_s^{\gamma\delta},\quad\quad \{
S_{\xi}^{\mu\nu},S_{\xi}^{\alpha\beta} \}^{*} =
C^{\mu\nu\alpha\beta}_{\gamma\delta} S_{\xi}^{\gamma\delta},  \nonumber \\
&&C^{\mu\nu\alpha\beta}_{\gamma\delta}=\eta^{\nu}_{\gamma}\eta^{\alpha}
_{\delta}\eta^{\mu\beta}+\eta^{\mu}_{\gamma}\eta^{\beta}_{\delta}\eta
^{\nu\alpha}-\eta^{\nu}_{\gamma}\eta^{\beta}_{\delta}\eta^{\mu\alpha}-
\eta^{\mu}_{\gamma}\eta^{\alpha}_{\delta}\eta^{\nu\beta}.  \label{III9}
\end{eqnarray}

Next, we eliminate the second class constraints $\chi _{i}^{\mu }(\tau
)\approx 0$, $\phi _{i}(\tau )\approx 0$ with the new Dirac brackets

\begin{equation}
\{A,B\}_{D}^{\ast }=\{A,B\}^{\ast }+i\sum_{i=1}^{N}\{A,\chi _{i}^{\mu }(\tau
)\}^{\ast }\eta _{\mu \nu }\{\chi _{i}^{\nu }(\tau ),B\}^{\ast }-\frac{i}{%
p_{s}^{2}}\sum_{i=1}^{N}\{A,\phi _{i}(\tau )\}^{\ast }\{\phi _{i}(\tau
),B\}^{\ast }.  \label{III10}
\end{equation}
Now we have

\begin{eqnarray}
&&\pi _{i}^{\mu }(\tau )\equiv \frac{i}{2}\xi _{i}^{\mu }(\tau ),  \nonumber
\\
&&\xi _{i}^{\mu }(\tau ){p_{s}}_{\mu }\equiv 0,  \nonumber \\
&\Rightarrow &\,\xi _{i}^{\mu }(\tau )\equiv \xi _{i\perp }^{\mu }(\tau
)\equiv \Pi ^{\mu \nu }\xi _{i\nu }(\tau )=\Big(\eta ^{\mu \nu }-\frac{%
p_{s}^{\mu }p_{s}^{\nu }}{p_{s}^{2}}\Big)\xi _{i\nu }(\tau ),  \nonumber \\
&&S_{\xi }^{\mu \nu }\equiv -i\sum_{i=1}^{N}\xi _{i}^{\mu }\xi _{i}^{\nu
},\quad \quad p_{s\mu }S_{\xi }^{\mu \nu }\equiv 0.  \label{III11}
\end{eqnarray}
However, now we get the following non-canonical Dirac brackets on $\Sigma
_{H\tau }$

\begin{eqnarray}
\{ x_s^\mu(\tau), x_s^\nu(\tau) \}^\ast_D &=& - i \sum_{i=1}^N \frac{%
\xi^\mu_i(\tau) \xi^\nu_i(\tau)}{p_s^2} \equiv \frac{S^{\mu \nu}_\xi (\tau)}{%
p_s^2},  \nonumber \\
\{ x_s^\mu(\tau), \xi_i^\nu(\tau) \}^\ast_D &=& \frac{\xi^\mu_i(\tau) p_s^\nu%
}{p_s^2},  \nonumber \\
\{ \xi_i^\mu(\tau), \xi_j^\nu(\tau) \}^\ast_D &=& i \Big( \eta^{\mu \nu} -
\frac{p_s^\mu p_s^\nu}{p_s^2}\Big) \delta_{ij} \equiv i \Pi^{\mu \nu}
\delta_{ij}.  \label{III12}
\end{eqnarray}

In this way, we have eliminated the components of $\xi^{\mu}_i$ parallel to $%
p^{\mu}_s$ in a Lorentz-invariant way. The spin of each particle is
described only by 3 Grassmann variables and the spin tensor $S_{\xi}
^{\mu\nu}$ satisfies a Weyssenhoff condition. The angular momentum tensor
becomes

\begin{eqnarray}
&&J^{\mu \nu} = L^{\mu \nu}_s + S^{\mu \nu}_s + S^{\mu \nu}_\xi,  \nonumber
\\
&&S^{\mu \nu}_\xi = - i \sum_{i=1}^N\xi_i^\mu (\tau) \xi_i^\nu (\tau) ,
\nonumber \\
&&{}  \nonumber \\
&&\{ J^{\mu \nu} , J^{\alpha \beta} \}^\ast_D = \{ J^{\mu \nu} , J^{\alpha
\beta} \}^\ast = C^{\mu \nu \alpha \beta}_{\gamma \delta} J^{\gamma \delta} ,
\nonumber \\
&&\{ L^{\mu \nu}_s , L^{\alpha \beta}_s \}^\ast_D = C^{\mu \nu \alpha
\beta}_{\gamma \delta} L^{\gamma \delta}_s - P^{\mu \nu \alpha
\beta}_{\gamma \delta} S^{\gamma \delta}_\xi ,  \nonumber \\
&&\{ S^{\mu \nu}_\xi , S^{\alpha \beta}_\xi \}^\ast_D = C^{\mu \nu \alpha
\beta}_{\gamma \delta} S^{\gamma \delta}_\xi - P^{\mu \nu \alpha
\beta}_{\gamma \delta} S^{\gamma \delta}_\xi ,  \nonumber \\
&&\{ L^{\mu \nu}_s , S^{\alpha \beta}_\xi \}^\ast_D = P^{\mu \nu \alpha
\beta}_{\gamma \delta} S^{\gamma \delta}_\xi ,  \nonumber \\
&&\{ L^{\mu \nu}_s + S_\xi^{\mu \nu}, L^{\alpha \beta}_s + S^{\alpha
\beta}_\xi\}^\ast_D = C^{\mu \nu \alpha \beta}_{\gamma \delta} \Big(%
L_s^{\gamma \delta} + S_\xi^{\gamma \delta} \Big) ,  \nonumber \\
&&{}  \nonumber \\
&&P^{\mu \nu \alpha \beta}_{\gamma \delta} \equiv \frac{p_s^\mu p_s^\beta}{%
p_s^2} \eta^\nu_\gamma \eta^\alpha_\delta + \frac{p_s^\nu p_s^\alpha}{p_s^2}
\eta^\mu_\gamma \eta^\beta_\delta - \frac{p_s^\mu p_s^\alpha}{p_s^2}
\eta^\nu_\gamma \eta^\beta_\delta - \frac{p_s^\nu p_s^\beta}{p_s^2}
\eta^\mu_\gamma \eta^\alpha_\delta .  \label{III13}
\end{eqnarray}

Since by asking the time constancy of the gauge fixings (\ref{III6}) we get
\cite{lu1} $\lambda^{\mu}(\tau ,\vec \sigma )={\tilde \lambda} ^{\mu}(\tau )+%
{\tilde \lambda}^{\mu}{}_{\nu}(\tau )b^{\nu}_{\check r}(\tau ) \sigma^{%
\check r}$, ${\tilde \lambda}^{\mu}(\tau )=-{\dot x}^{\mu}_s(\tau )$, ${%
\tilde \lambda}^{\mu\nu}(\tau )=-{\tilde \lambda}^{\nu\mu}(\tau )={\frac{1}{2%
}} \sum_{\check r}[{\dot b}^{\mu}_{\check r}b^{\nu}_{\check r}-b^{\mu}_{%
\check r} {\dot b}^{\nu}_{\check r}](\tau )$, the Dirac Hamiltonian becomes

\begin{equation}
H^F_D = \tilde{\lambda}^\mu(\tau) {\cal H}_\mu(\tau) - \frac 12 \tilde{
\lambda}^{\mu \nu}(\tau) {\cal H}_{\mu \nu}(\tau) + \int{d^3\sigma \Big[ -
A_\tau (\tau, \vec{\sigma}) \Gamma (\tau, \vec{\sigma}) + \mu_\tau (\tau,
\vec{\sigma}) \pi^\tau (\tau, \vec{\sigma}) \Big] },  \label{III14}
\end{equation}

\noindent and we are left with only 12 first class constraints

\begin{eqnarray}
\pi^\tau (\tau, \vec{\sigma}) &\approx & 0,  \nonumber \\
\Gamma (\tau, \vec{\sigma}) &=& \partial_{\breve{r}} \pi^{\breve{r}} (\tau,
\vec{\sigma}) - \sum_{i=1}^N \delta^3 (\vec{\sigma} -\vec{\eta}_i(\tau))
Q_i(\tau) \approx 0,  \nonumber \\
{\cal H}_\mu(\tau) &=& \int{d^3\sigma {\cal H}_\mu (\tau, \vec{\sigma}) }= {%
\ p_s}_\mu - b_{\mu \tau} \Big\{ {\frac{1}{2}}\int d^3\sigma \Big(\vec{\pi}%
^2 (\tau, \vec{\sigma}) + \vec{B}^2(\tau, \vec{\sigma}) \Big) +  \nonumber \\
&+& \sum_{i=1}^N \sqrt{m_i^2 +\Big[\vec{\kappa}_i(\tau) - Q_i(\tau) \vec{A}
(\tau, \vec{\eta}_i(\tau)) \big]^2} +  \nonumber \\
&+& \frac 12 \sum_{i,j=1}^N \delta^3 (\vec{\eta}_i(\tau) - \vec{\eta}%
_j(\tau)) \cdot  \nonumber \\
&\cdot & \frac{ Q_i(\tau) Q_j(\tau) \xi^\gamma_i(\tau)\xi^\delta_i(\tau)} {%
\sqrt{m_i^2 + \vec{\kappa}_i^2(\tau)} \sqrt{m_j^2 + \vec{\kappa}_j^2(\tau)}}
b_{\tau \gamma} \eta_{\delta \beta} \xi^\alpha_j(\tau)\xi^\beta_j(\tau)
b_{\tau \alpha} +  \nonumber \\
&+& i \sum_{i=1}^N \frac{ Q_i(\tau) \xi^\alpha_i(\tau)\xi^\beta_i(\tau)} {%
\sqrt{m_i^2 + \vec{\kappa}_i^2(\tau)}}b_{\tau \alpha} b_{\breve{s} \beta}
(\tau)\pi^{\breve{s}} (\tau, \vec{\eta}_i(\tau)) +  \nonumber \\
&-& \frac i2 \sum_{i=1}^N \frac{ Q_i(\tau) \xi^\alpha_i(\tau)
\xi^\beta_i(\tau)}{\sqrt{m_i^2 + \vec{\kappa}_i^2(\tau)}} b_{\breve{u}
\alpha} (\tau) b_{\breve{v}\beta} (\tau) F_{\breve{u} \breve{v}} (\tau, \vec{%
\eta}_i(\tau))\Big\} +  \nonumber \\
&&{}  \nonumber \\
&+& b_{\breve{r} \mu} (\tau) \Big\{\int {d^3\sigma \big[\vec{\pi} \times
\vec{B}\big]_{\breve{r}}(\tau, \vec{\sigma}) + \sum_{i=1}^N \big[\kappa_{i
\breve{r}}(\tau) - Q_i(\tau) A_{\breve{r}}(\tau, \vec{\eta}_i(\tau))}\big] %
\Big\} \approx 0,  \nonumber \\
&&{}  \nonumber \\
{\cal H}^{\mu \nu}(\tau) &=& b_{\breve{r}}^\mu (\tau) \int{d^3\sigma \sigma^{%
\breve{r}}{\cal H}^\nu (\tau, \vec{\sigma}) } - b_{\breve{r}}^\nu (\tau) \int%
{d^3\sigma \sigma^{\breve{r}}{\cal H}^\mu (\tau, \vec{\sigma}) } =  \nonumber
\\
&=& S_s^{\mu \nu}(\tau )-\big( b_{\breve{r}}^\mu (\tau) b^\nu_\tau - b_{%
\breve{r}}^\nu (\tau) b^\mu_\tau \big) \Big\{ {\frac{1}{2}}\int d^3\sigma
\sigma^{\breve{r}} \Big( \vec{\pi}^2 (\tau, \vec{\sigma}) + \vec{B}^2(\tau,
\vec{\sigma}) \Big) +  \nonumber \\
&+& \frac 12 \sum_{i,j=1}^N \delta^3 (\vec{\eta}_i(\tau) - \vec{\eta}%
_j(\tau)) \cdot  \nonumber \\
&\cdot & \frac{\eta_i^{\check r} Q_i(\tau) Q_j(\tau)
\xi^\gamma_i(\tau)\xi^\delta_i(\tau)} {\sqrt{m_i^2 + \vec{\kappa}_i^2(\tau)}
\sqrt{m_j^2 + \vec{\kappa}_j^2(\tau)}} b_{\tau \gamma} \eta_{\delta \beta}
\xi^\alpha_j(\tau)\xi^\beta_j(\tau) b_{\tau \alpha} +  \nonumber \\
&+& \sum_{i=1}^N \eta_i^{\breve{r}} \sqrt{m_i^2 + \big[\vec{\kappa}_i(\tau)
- Q_i(\tau) \vec{A} (\tau, \vec{\eta}_i(\tau))\big]^2} +  \nonumber \\
&+& i \sum_{i=1}^N \eta_i^{\breve{r}}(\tau) \frac{ Q_i(\tau)
\xi^\alpha_i(\tau)\xi^\beta_i(\tau)} {\sqrt{m_i^2 + \vec{\kappa}_i^2(\tau)}}%
b_{\tau \alpha} b_{\breve{s} \beta} (\tau)\pi^{\breve{s}} (\tau, \vec{\eta}%
_i(\tau)) +  \nonumber \\
&-& \frac i2 \sum_{i=1}^N \eta_i^{\breve{r}}(\tau) \frac{ Q_i(\tau)
\xi^\alpha_i(\tau)\xi^\beta_i(\tau)} {\sqrt{m_i^2 + \vec{\kappa}_i^2(\tau)}}
b_{\breve{u} \alpha} (\tau) b_{\breve{v}\beta} (\tau) F_{\breve{u} \breve{v}%
} (\tau, \vec{\eta}_i(\tau))\Big\} +  \nonumber \\
&&{}  \nonumber \\
&+& \big( b_{\breve{r}}^\mu (\tau) b_{\breve{s}}^\nu (\tau) - b_{\breve{r}%
}^\nu (\tau) b_{\breve{s}}^\mu (\tau) \big) \Big\{ \int{\ d^3\sigma \sigma^{%
\breve{r}}\big[\vec{\pi} \times \vec{B}\big]_{\breve{s}}(\tau, \vec{\sigma})}
+  \nonumber \\
&+&\sum_{i=1}^N \eta_i^{\breve{r}}(\tau) \big[\kappa_{i \breve{s}}(\tau) -
Q_i(\tau) A_{\breve{s}}(\tau, \vec{\eta}_i(\tau))\big] \Big\} \approx 0.
\label{III15}
\end{eqnarray}

The next step \cite{lu1} is to select all the configurations of the isolated
system which are timelike, namely with $p_{s}^{2}>0$. For them we can boost
to the rest system with the standard Wigner boost $L_{.\nu }^{\mu }(%
\stackrel{o}{p}_{s},p_{s})$ for timelike Poincar\'{e} orbits (see Appendix
B) all the variables of the non-canonical basis $x_{s}^{\mu }(\tau )$, $%
p_{s}^{\mu }$ , $b_{\breve{A}}^{\mu }(\tau )$ , $S_{s}^{\mu \nu }(\tau )$ , $%
A_{\breve{A}}(\tau ,\vec{\sigma})$ , $\pi ^{\breve{A}}(\tau ,\vec{\sigma})$,
$\vec{\eta}_{i}(\tau )$ , $\vec{\kappa}_{i}(\tau )$ , $\theta _{i}(\tau )$ ,
$\theta _{i}^{\ast }(\tau )$ , $\xi _{i}^{\mu }(\tau )$ with Lorentz indices
(except $p_{s}^{\mu }$). This is a canonical transformation generated by $%
e^{\{.,{\cal F}(\tau )\}}$ with generating function

\begin{eqnarray}
{\cal F}(\tau ) &=& \frac 12 \omega(p_s) I_{\mu \nu} (p_s) S_s^{\mu
\nu}(\tau ) ,  \nonumber \\
&&{}  \nonumber \\
I(p) &\equiv & \parallel I(p)^\mu_{.\nu} \parallel = \pmatrix{0&
-\frac{p_j}{|\vec{p}|}\cr \frac{p^i}{|\vec{p}|}& 0},  \nonumber \\
I_{\mu \nu}(p) &=& - I_{\nu \mu}(p),~~~~~~~~~~~~~~~~~~I^3(p) = I(p) ,
\nonumber \\
\cosh \omega(p) &=& \frac{\eta p_0}{\sqrt{p^2}},~~~~~~~~~ \sinh \omega(p) =
\eta \frac{|\vec{p}|}{\sqrt{p^2}},  \nonumber \\
L^\mu_{.\nu} (p, \stackrel{o}{p}) &=& \exp \big[ \omega (p) I (p) \big]%
^\mu_{.\nu} =  \nonumber \\
&=& \big[ \cosh\big(\omega (p) I (p)\big) + \sinh \big(\omega (p) I (p)\big) %
\big]^\mu_{.\nu} =  \nonumber \\
&=& \big[ \openone - I^2(p) + I^2(p) \cosh \omega (p) + I(p) \sinh \omega
(p) \big]^\mu_{.\nu},  \nonumber \\
L^\mu_{.\nu}(\stackrel{o}{p}, p) &=& \exp \big[ - \omega (p) I (p) \big]%
^\mu_{.\nu}.  \label{III16}
\end{eqnarray}

\noindent Since we have $\xi^\mu_i {p_s}_\mu \equiv 0$, we get $I_{\mu
\nu}(p_s) S_\xi^{\mu \nu} = 0$, so that the addition of $S^{\mu\nu}_{\xi}$
to $S^{\mu\nu} _s$ in ${\cal F}$ is irrelevant.

The new non-canonical basis (with the same Dirac brackets) is

\begin{eqnarray}
\tilde{x}^\mu_s &=& x_s^\mu - \frac 12 \epsilon_\nu(u(p_s)) \eta_{AB} \frac{%
\partial \epsilon^B_\rho(u(p_s))}{\partial {p_s}_\mu} S_s^{\nu \rho} =
\nonumber \\
&=& x^\mu_s - \frac{1}{\eta_s \sqrt{p^2_s} (p_s^0 +\eta_s \sqrt{p_s^2})} %
\Big[{p_s}_\nu S_s^{\nu \mu} + \eta_s \sqrt{p^2_s} \Big( S_s^{0 \mu} -
S_s^{0 \nu} \frac{{p_s}_\nu p_s^\mu}{p_s^2} \Big) \Big] =  \nonumber \\
&=& x^\mu_s - \frac{1}{\eta_s \sqrt{p^2_s}} \Big[ \eta^\mu_A \Big(\bar{S}_s^{%
\bar{o} A} - \frac{\bar{S}_s^{Ar} p_s^r} {p_s^0 +\eta_s \sqrt{p_s^2}} \Big) %
+ \frac{p_s^\mu + 2 \eta_s \sqrt{p_s^2} \eta^{\mu 0}} {\eta_s \sqrt{p^2_s}
(p_s^0 +\eta_s \sqrt{p_s^2})} \bar{S}_s^{\bar{o} r} p_s^r \Big]  \nonumber \\
p_s^\mu &=& p_s^\mu,~~~\eta_i^{\breve{r}} = \eta_i^{\breve{r}}, ~~~ k_i^{%
\breve{r}} = k_i^{\breve{r}},~~~ A_{\breve{A}} = A_{\breve{A}} ,~~~ \pi^{%
\breve{A}} = \pi^{\breve{A}}  \nonumber \\
\xi_i^\mu &=& \xi_i^\mu ,~~~ \theta_i^\ast = \theta_i^\ast ,~~~ \theta_i =
\theta_i  \nonumber \\
b^A_{\breve{B}} &=& \epsilon^A_\mu (u(p_s)) b_{\breve{B}}^\mu  \nonumber \\
\tilde{S}_s^{\mu \nu} &=& S_s^{\mu \nu} + \frac 12 \epsilon^A_\rho (u(p_s))
\eta_{AB} \Big[ \frac{\partial \epsilon^B_\sigma (u(p_s))}{\partial {p_s}_\mu%
} p_s^\nu - \frac{\partial \epsilon^B_\sigma (u(p_s))}{\partial {p_s}_\nu}
p_s^\mu \Big] S_s^{\rho \sigma} =  \nonumber \\
&=& S_s^{\mu \nu} + \frac{1}{\eta_s \sqrt{p^2_s} (p_s^0 +\eta_s \sqrt{p_s^2})%
} \Big[{p_s}_\beta (S_s^{\beta \mu} p_s^\nu - S_s^{\beta \nu} p_s^\mu) +
\eta_s \sqrt{p^2_s} (S_s^{0 \mu} p_s^\nu - S_s^{0 \nu} p_s^\mu)\Big] ,
\label{III17}
\end{eqnarray}

\noindent where $u^{\mu}(p_s)=p^{\mu}_s/\eta_s\sqrt{p^2_s}=L^\mu_o (%
\stackrel{o}{p}_s, p_s)$ [$\eta_s=\pm1$; from now on we will select only the
positive-energy branch $\eta_s=+1$].

For later use, let us introduce the spin tensors

\begin{eqnarray}
\bar{S}_{s}^{AB} &=&\epsilon _{\mu }^{A}(u(p_{s}))\epsilon _{\nu
}^{B}(u(p_{s}))S_{s}^{\mu \nu },  \nonumber \\
\bar{S}_{\xi }^{AB} &=&\epsilon _{\mu }^{A}(u(p_{s}))\epsilon _{\nu
}^{B}(u(p_{s}))S_{\xi }^{\mu \nu }.  \label{III18}
\end{eqnarray}
Since $\xi _{i\mu }p_{s}^{\mu }\equiv 0$ implies $\xi _{i\tau }=\xi _{i\mu
}u^{\mu }(p_{s})=\xi _{i\mu }L_{o}^{\mu }(\stackrel{o}{p}_{s},p_{s})\equiv 0$%
, we can reduce to 3 for each particle the Grassmann variables describing
the spin

\begin{equation}
\xi _{i}^{r}(\tau )\,:=\,\epsilon _{\mu }^{r}(u(p_{s}))\xi _{i}^{\mu }(\tau
)~~~~~r=1,2,3.  \label{III19}
\end{equation}
We get

\begin{eqnarray}
&&\bar{S}_{\xi }^{\bar{o}B}=0,  \nonumber \\
&&\bar{S}_{\xi }^{rs}=-i\sum_{i=1}^{N}\epsilon _{\mu }^{r}(u(p_{s}))\epsilon
_{\nu }^{s}(u(p_{s}))\xi _{i}^{\mu }\xi _{i}^{\nu }=-i\sum_{i=1}^{N}\xi
_{i}^{r}\xi _{i}^{s},  \nonumber \\
&&{\bar{S}}_{\xi }^{r}={\frac{1}{2}}\epsilon ^{ruv}{\bar{S}}_{\xi
}^{uv}=\sum_{i=1}^{N}{\bar{S}}_{i\xi }^{r},  \nonumber \\
&&{\bar{S}}_{i\xi }^{r}=-{\frac{i}{2}}\epsilon ^{ruv}\xi _{i}^{u}\xi
_{i}^{v}.  \label{III20}
\end{eqnarray}
The $\xi _{i}^{r}(\tau )$'s satisfy

\begin{eqnarray}
\{\xi _{i}^{r},\xi _{j}^{s}\}_{D}^{\ast } &=&-i\delta ^{rs}\delta _{ij},
\nonumber \\
\{\tilde{x}_{s}^{\mu },\xi _{i}^{r}\}_{D}^{\ast } &=&-\frac{\partial
\epsilon _{\nu }^{r}(u(p_{s}))}{\partial {p_{s}}_{\mu }}\xi _{i}^{\nu }.
\label{III21}
\end{eqnarray}
If we define

\begin{eqnarray}
&\hat{x}^\mu_s& \equiv \tilde{x}^\mu_s - \frac 12 \epsilon^A_\nu (u(p_s))
\eta_{AB} \frac{\partial \epsilon^B_\rho (u(p_s))}{\partial {p_s}_\mu}
S_\xi^{\nu \rho} =  \nonumber \\
&=& {x}^\mu_s - \frac 12 \epsilon^A_\nu (u(p_s)) \eta_{AB} \frac{\partial
\epsilon^B_\rho (u(p_s))}{\partial {p_s}_\mu} \Big( S_s^{\nu \rho} +
S_\xi^{\nu \rho} \Big) ,  \label{III22}
\end{eqnarray}

\noindent we get

\begin{eqnarray}
\{ \hat{x}^\mu_s , p^\nu_s \}^\ast_D &=& - \eta^{\mu \nu},  \nonumber \\
\{ \hat{x}^\mu_s , \xi^r_i \}^\ast_D &=& 0,  \nonumber \\
\{ \hat{x}^\mu_s , \hat{x}^\nu_s \}^\ast_D &=& 0.  \label{III23}
\end{eqnarray}

Therefore, with respect to the Dirac brackets $\{ .,. \}^\ast_D$ we have
obtained a basis in which ${\hat x}^{\mu}_s(\tau )$, $p^{\mu}_s$, $A_{\check %
A}(\tau ,\vec \sigma )$, $\pi^{\check A}(\tau ,\vec \sigma )$, ${\vec \eta}
_i(\tau )$, ${\vec \kappa}_i(\tau )$, $\xi^{\check r}_i(\tau )$, $\theta
_i(\tau )$, $\theta_i^{*}(\tau )$, are canonical variables and only $b^{\mu}
_{\check A}(\tau )$, $S^{\mu\nu}_s(\tau )$, are not canonical. The new
canonical origin ${\hat x}_s^{\mu}$ of the hyperplane has the same
noncovariance of the Newton-Wigner position operator and describes the {\it %
external} decoupled canonical center of mass of the isolated system. In
terms of this variable we get

\begin{eqnarray}
J^{\mu \nu} &=& \tilde{x}^\mu_s p_s^\nu - \tilde{x}^\nu_s p_s^\mu + \tilde{S}%
_s^{\mu \nu} + S_\xi^{\mu \nu}= \hat{L}_s^{\mu \nu} + \tilde{S}_s ^{\mu \nu}
+\tilde{S}_\xi^{\mu \nu},  \nonumber \\
&&{}  \nonumber \\
\hat{L}_s^{\mu \nu} &=& \hat{x}^\mu_s p_s^\nu - \hat{x}^\nu_s p_s^\mu ,
\nonumber \\
\tilde{S}_\xi^{\mu \nu} &=& S_\xi^{\mu \nu} + \frac 12 \epsilon^A_\rho
(u(p_s)) \eta_{AB} \Big[ \frac{\partial \epsilon^B_\sigma (u(p_s))}{\partial
{p_s}_\mu} p_s^\nu - \frac{\partial \epsilon^B_\sigma (u(p_s))}{\partial {p_s%
}_\nu} p_s^\mu \Big] S_\xi^{\rho \sigma},  \nonumber \\
&&{}  \nonumber \\
\{ \hat{L}^{\mu \nu}_s , \hat{L}^{\alpha \beta}_s \}^\ast_D &=& C^{\mu \nu
\alpha \beta}_{\gamma \delta} \hat{L}^{\gamma \delta}_s ,  \nonumber \\
\{ \tilde{S}^{\mu \nu}_\xi , \tilde{S}^{\alpha \beta}_\xi \}^\ast_D &=&
C^{\mu \nu \alpha \beta}_{\gamma \delta} \tilde{S}^{\gamma \delta}_\xi,
\nonumber \\
\{ \tilde{S}^{\mu \nu}_s , \tilde{S}^{\alpha \beta}_s \}^\ast_D &=& C^{\mu
\nu \alpha \beta}_{\gamma \delta} \tilde{S}^{\gamma \delta}_s,  \nonumber \\
\{ \tilde{S}^{\mu \nu}_s , \tilde{S}^{\alpha \beta}_\xi \}^\ast_D &=& \{
\tilde{S}^{\mu \nu}_s , \hat{L}^{\alpha \beta}_s \}^\ast_D = \{ \tilde{S}%
^{\mu \nu}_\xi , \hat{L}^{\alpha \beta}_s \}^\ast_D = 0,  \nonumber \\
\{ J^{\mu \nu} , J^{\alpha \beta} \}^\ast_D &=& C^{\mu \nu \alpha
\beta}_{\gamma \delta} J^{\gamma \delta}.  \label{III24}
\end{eqnarray}

As shown in Ref.\cite{lu1}, we can restrict ourselves to the Wigner
hyperplanes $\Sigma_{W\tau}$, orthogonal to $p^{\mu}_s$, i.e. with normals $%
l^{\mu}=u^{\mu}(p_s)$, with the gauge-fixings

\begin{eqnarray}
T^\mu_{\breve{A}}(\tau) = b^\mu_{\breve{A}} (\tau) - \epsilon^\mu_{A=\breve{A%
}}(u(p_s)) &\approx & 0  \nonumber \\
\Rightarrow b^A_{\breve{A}} (\tau) = \epsilon_\mu^A(u(p_s)) b^\mu_{\breve{A}%
} (\tau) &\approx & \eta^A_{\breve{A}},  \label{III25}
\end{eqnarray}

\noindent which imply the new Dirac brackets

\begin{eqnarray}
\{A,B\}_{D}^{\ast \ast } &=&\{A,B\}_{D}^{\ast }-\frac{1}{4}\Big[\{A,\tilde{H}%
^{\gamma \delta }\}_{D}^{\ast }\Big(\eta _{\gamma \sigma }\epsilon _{\gamma
}^{D}(u(p_{s}))-\eta _{\delta \sigma }\epsilon _{\gamma }^{D}(u(p_{s}))\Big)%
\{T_{D}^{\sigma },B\}_{D}^{\ast }+  \nonumber \\
&+&\{A,T_{D}^{\sigma }\}_{D}^{\ast }\Big(\eta _{\sigma \nu }\epsilon _{\mu
}^{B}(u(p_{s}))-\eta _{\sigma \mu }\epsilon _{\nu }^{B}(u(p_{s}))\Big)\{%
\tilde{H}^{\mu \nu },B\}_{D}^{\ast }\Big].  \label{III26}
\end{eqnarray}

The time constancy of the gauge-fixings (\ref{III25}) implies ${\tilde %
\lambda}^{\mu\nu}(\tau )\approx 0$ , $b^{\mu}_{\check A}(\tau )\equiv L^\mu{}%
_A(p_s,\stackrel{o}{p}_s)$ and ${\tilde {{\cal H}}}^{\mu\nu}(\tau )\equiv 0$
(namely the determination of $S^{\mu\nu}_s$ in terms of the variables of the
system). The remaining variables form a canonical basis

\begin{eqnarray}
\{ \hat{x}^\mu_s(\tau), p_s^\nu \}^{\ast \ast}_D &=& - \eta^{\mu \nu},
\nonumber \\
\{ \eta^r_i(\tau) , \kappa^s_j (\tau) \}^{\ast \ast}_D &=& \delta_{ij}
\delta^{rs},  \nonumber \\
\{ \xi^r_i(\tau) , \xi^s_j (\tau) \}^{\ast \ast}_D &=& - i \delta^{rs}
\delta_{ij} ,  \nonumber \\
\{ A_A(\tau ,\vec \sigma ),\pi^B(\tau ,{\vec \sigma}^{^{\prime}}) \}^{\ast
\ast}_D&=& \eta^B_A \delta^3(\vec \sigma -{\vec \sigma}^{^{\prime}}).
\label{III27}
\end{eqnarray}

As shown in Ref.\cite{lu1}, the dependence of the gauge-fixing (\ref{III25})
on $p^{\mu}_s$ implies that the Lorentz-scalar indices $\check A=\{ \tau ,%
\check r \}$ become Wigner indices $A=\{ \tau , r\}$: i) $A_{A=\tau}(\tau ,%
\vec \sigma )$ is a Lorentz-scalar field; ii) $A_{A=r}(\tau ,\vec \sigma )$,
$\xi^r_i(\tau )$, $\eta^r_i(\tau )$ , $\kappa _{ir}(\tau )$, are Wigner spin
1 ${}{}$ 3-vectors which transform with Wigner rotations under the action of
{\it external} Minkowski Lorentz boosts.

On $\Sigma_{W\tau}$ the Poincar\'e generators are

\begin{eqnarray}
p_s^\mu,&&~~~~~J^{\mu \nu}_s = \hat{x}^\mu_s p_s^\nu - \hat{x}^\nu_s p_s^\mu
+ \tilde{S}^{\mu \nu},  \nonumber \\
&&{}  \nonumber \\
\tilde{S}^{\mu \nu} &\equiv & \tilde{S}_s^{\mu \nu} + \tilde{S}_\xi^{\mu
\nu},  \nonumber \\
\tilde{S}^{0i} &=& -\frac{\delta^{ir} \bar{S}^{rs} p_s^s} {p_s^0 + \eta_s
\sqrt{p_s^2}},~~~~~\tilde{S}^{ij} = \delta^{ir} \delta^{js} \bar{S}^{rs} ,
\label{III28}
\end{eqnarray}

\noindent because one can express ${\tilde S}^{\mu\nu}$ in terms of ${\bar S}
^{AB}=\epsilon^A_{\mu}(u(p_s))\epsilon^B_{\nu}(u(p_s)) S^{\mu\nu}$.

Since ${\tilde H}^{\mu\nu}(\tau )\equiv 0$ implies

\begin{eqnarray}
S_s^{\mu \nu} &=& \Big(\epsilon^\mu_r(u(p_s)) u^\nu(p_s) -
\epsilon^\nu_r(u(p_s)) u^\mu(p_s)\Big) \Big\{ {\frac{1}{2}}\int d^3\sigma
\sigma^r \Big( \vec{\pi}^2 (\tau, \vec{\sigma}) + \vec{B}^2 (\tau, \vec{%
\sigma})\Big) +  \nonumber \\
&+& \sum_{i=1}^N \eta^r_i(\tau) \sqrt{m_i^2 - i Q_i(\tau) \xi^u_i(\tau)
\xi^v_i(\tau) F_{uv}(\tau, \vec{\eta}_i(\tau )) + \big[\vec{\kappa}_i(\tau)
-Q_i \vec{A}(\tau, \vec{\eta}_i(\tau ))\big]^2} \Big\} +  \nonumber \\
&-& \Big(\epsilon^\mu_r(u(p_s)) \epsilon^\nu_s(u(p_s)) -
\epsilon^\nu_r(u(p_s)) \epsilon^\mu_s(u(p_s)) \Big) \Big\{ \int{d^3\sigma
\sigma^r (\vec{\pi} \times \vec{B})_s (\tau, \vec{\sigma})} +  \nonumber \\
&+& \sum_{i=1}^N \eta_i^r(\tau) \big(\kappa_{is}(\tau) - Q_i(\tau) A_s(\tau,
\vec{\eta}_i)\big) \Big\} ,  \label{III29}
\end{eqnarray}

\noindent we get

\begin{eqnarray}
\bar{S}_s^{AB} &=& \big( \delta^A_r \delta^B_{\bar{o}} - \delta^B_r
\delta^A_{\bar{o}} \big) \Big[ {\frac{1}{2}}\int d^3\sigma \sigma^r \Big(%
\vec{\pi}^2 (\tau, \vec{\sigma}) + \vec{B}^2 (\tau, \vec{\sigma})\Big) +
\nonumber \\
&+& \sum_{i=1}^N \eta^r_i(\tau) \sqrt{m_i^2 - i Q_i(\tau)\xi^u_i(\tau)
\xi^v_i(\tau) F_{uv}(\tau, \vec{\eta}_i(\tau )) + \big[\vec{k}_i(\tau) -Q_i
\vec{A}(\tau, \vec{\eta}_i(\tau ))\big]^2} \Big] +  \nonumber \\
&-& \Big( \delta^A_r \delta^B_s - \delta^B_r \delta^A_s \big) \Big[\int{%
d^3\sigma \sigma^r (\vec{\pi} \times \vec{B})_s (\tau, \vec{\sigma})} +
\nonumber \\
&+& \sum_{i=1}^N \eta_i^r(\tau) \big(k_{is}(\tau) -Q_i(\tau) A_s(\tau, \vec{%
\eta}_i(\tau ))\Big) \Big] .  \label{III30}
\end{eqnarray}

\noindent However, the boosts ${\bar S}^{or}_s\equiv {\bar S}^{or}$ [${\bar S%
}^{or}_{\xi}=0$] do not contribute to the previous realization of the
Poincar\'e generators: this is the {\it external} Poincar\'e algebra in the
rest-frame Wigner-covariant instant form of dynamics.

The original variables $z^{\mu}(\tau ,\vec \sigma )$, $\rho_{\mu}(\tau ,\vec %
\sigma )$, are reduced only to ${\hat x}^{\mu}_s$, $p^{\mu}_s$ on the Wigner
hyperplane $\Sigma_{W\tau}$. On it only 6 first class constraints survive

\begin{eqnarray}
\pi ^{\tau }(\tau ,\vec{\sigma})&\approx& 0,  \nonumber \\
\Gamma (\tau ,\vec{\sigma})&=&\partial _{r}\pi ^{r}(\tau ,\vec{\sigma}%
)-\sum_{i=1}^{N}Q_{i}\delta ^{3}(\vec{\sigma}-\vec{\eta}_{i}(\tau ))\approx
0,  \nonumber \\
&&{}  \nonumber \\
{\cal H}^{\mu }(\tau )&=&p_{s}^{\mu }-[u^{\mu }(p_{s}){\cal H}_{rel}(\tau
)+\epsilon _{r}^{\mu }(u(p_{s})){\cal H}_{p\,r}(\tau )]=  \nonumber \\
&=&u^{\mu }(p_{s}){\cal H}(\tau )+\epsilon _{r}^{\mu }(u(p_{s})){\cal H}%
_{pr}(\tau )\approx 0,  \nonumber \\
&&{}  \nonumber \\
&&\text{where}  \nonumber \\
&&{}  \nonumber \\
{\cal H}(\tau )&=&\sqrt{p_{s}^{2}}-{\cal H}_{rel}(\tau )=\eta _{s}\sqrt{%
p_{s}^{2}}-\Big[{\frac{1}{2}}\int {\ d^{3}\sigma \Big(\vec{\pi}^{2}+\vec{B}%
^{2}\Big)(\tau ,\vec{\sigma})}+  \nonumber \\
&+&\sum_{i=1}^{N}\cdot \sqrt{m_{i}^{2}-iQ_{i}(\tau )\xi _{i}^{r}(\tau )\xi
_{i}^{s}(\tau )F_{rs}(\tau ,\vec{\eta}_{i}(\tau ))+\big[\vec{\kappa}%
_{i}(\tau )-Q_{i}(\tau )\vec{A}(\tau ,\vec{\eta}_{i}(\tau ))\big]^{2}}\Big]%
\approx 0,  \nonumber \\
&&{}  \nonumber \\
{\cal H}_{pr}(\tau )&=&\int {\ d^{3}\sigma \Big[\vec{\pi}\times \vec{B}\Big]%
_{r}(\tau ,\vec{\sigma})}+\sum_{i=1}^{N}\Big[{\kappa }_{ir}(\tau
)-Q_{i}(\tau ){A}_{r}(\tau ,\vec{\eta}_{i}(\tau ))\Big]\approx 0,  \nonumber
\\
&&{}  \nonumber \\
&&{}  \nonumber \\
&&\{{\cal H}^{\mu }(\tau ),{\cal H}^{\nu }(\tau )\}_{D}^{\ast \ast }=\int {\
d^{3}\sigma \Big\{\Big[u^{\mu }(p_{s})\epsilon _{r}^{\nu }(u(p_{s}))-u^{\nu
}(p_{s})\epsilon _{r}^{\mu }(u(p_{s}))\Big]\pi ^{r}(\tau ,\vec{\sigma})}+
\nonumber \\
&-&\epsilon _{r}^{\mu }(u(p_{s}))F_{rs}(\tau ,\vec{\sigma})\epsilon
_{s}^{\nu }(u(p_{s}))\Big\}\Gamma (\tau ,\vec{\sigma})\approx 0.
\label{III31}
\end{eqnarray}

Let us remark that in ${\cal H}(\tau )\approx 0$ the {\it spin-spin} and
{\it spin-electric field} interactions have disappeared on $\Sigma_{W\tau}$,
i.e. in the inertial systems associated with the Wigner hyperplane. There is
only the {\it spin-magnetic field} interaction

\begin{equation}
- i Q_i \xi^r_i \xi^s_i F_{rs} (\tau, \vec{\eta}_i(\tau )) = - 2 Q_i {\vec{%
\bar S}}_{i\xi} \cdot \vec{B} (\tau, \vec{\eta}_i(\tau )), \quad\quad {\vec{%
\bar S}}_{i\xi} \equiv - \frac i2 \vec{\xi}_i \times \vec{\xi}_i,
\label{III32}
\end{equation}

\noindent like in the non-relativistic Pauli equation. This term has been
put inside the square roots of the kinetic terms in analogy with the results
of Section II.

Therefore, we get a kind of {\it relativistic Pauli Hamiltonian} describing
the interaction of a positive-energy massive spinning particle belonging to
the $({\frac{1}{2}} ,0)$ representation of the Lorentz group with the
electromagnetic field, whose non-relativistic limit is the pseudo-classical
form of the ordinary Pauli Hamiltonian.

The constraints ${\vec {{\cal H}}}_p(\tau )\approx 0$ identify the Wigner
hyperplane $\Sigma_{W\tau}$ (whose embedding in Minkowski spacetime is $%
z^{\mu}(\tau ,\vec \sigma )=x^{\mu}_s(\tau )+\epsilon^{\mu}_r(u(p_s))
\sigma^r$) with the {\it intrinsic rest frame} (vanishing of the total
Wigner spin 1 ${}$ 3-momentum of the isolated system) and mean that the
conjugate 3-coordinate $\vec \sigma ={\vec q}_{+}$ of the {\it internal}
3-center of mass of the isolated system on $\Sigma_{W\tau}$ is a {\it gauge}
variable. The natural gauge-fixing for ${\vec {{\cal H}}}_p(\tau )\approx 0$
is ${\vec q}_{+}\approx 0$; in this way the {\it internal} center of mass
coincides with the origin of $\Sigma_{W\tau}$: ${}x^{\mu}_s(\tau
)=z^{\mu}(\tau ,\vec \sigma =0)$.

On $\Sigma_{W\tau}$ the Dirac Hamiltonian becomes

\begin{equation}
H_D = \lambda(\tau){\cal H}(\tau ) - \vec{\lambda}(\tau) \cdot {\vec {{\cal H%
}}}_p(\tau) + \int{d^3\sigma \Big[-A_\tau (\tau, \vec{\sigma}) \Gamma (\tau,
\vec{\sigma}) + \mu_\tau(\tau, \vec{\sigma}) \pi^\tau(\tau, \vec{\sigma})%
\Big] } ,  \label{III33}
\end{equation}

\noindent so that the {\it external} canonical 4-center of mass $\hat{x}%
_{s}^{\mu }$ has a velocity parallel to ${p_{s}}^{\mu }$, namely it has no
classical zitterbewegung just as with the Foldy-Wouthuysen mean position.

See Ref.\cite{iten1} for the relativistic kinematics of the N-body problem
in the rest-frame instant form of dynamics and Ref.\cite{lu3} for the
non-relativistic limit of ${\cal H}_{rel}$.

\subsection{Dirac's observables and equations of motion.}

As shown in Ref.\cite{lusa}, the Dirac observables of the electromagnetic
field are the transverse quantities ${\vec A}_{\perp r}(\tau, \vec{\sigma})$
, ${\vec \pi}^r_{\perp}(\tau, \vec{\sigma})$, defined by the decomposition [$%
\triangle =-{\vec \partial}^2$]

\begin{eqnarray}
A_r (\tau, \vec{\sigma}) &=& \partial_r \eta_{em}(\tau, \vec{\sigma}) +
A_{\perp r} (\tau, \vec{\sigma}),\quad \vec \partial \cdot {\vec A}%
_{\perp}(\tau ,\vec \sigma )\equiv 0,  \nonumber \\
\pi^r (\tau, \vec{\sigma}) &=& \pi^r_{\perp} (\tau, \vec{\sigma}) + \frac{%
\partial^r}{\triangle_\sigma} \Big[\Gamma (\tau, \vec{\sigma}) -
\sum_{i=1}^N Q_i(\tau) \delta^3(\vec{\sigma} - \vec{\eta}_i(\tau)) \Big] ,
\nonumber \\
\eta_{em}(\tau, \vec{\sigma}) &=& - \frac{\vec{\partial}}{\triangle_\sigma}
\cdot \vec{A} (\tau, \vec{\sigma}),  \label{III34}
\end{eqnarray}

\noindent while the gauge variables are $A_\tau (\tau, \vec{\sigma})$ and $%
\eta_{em}(\tau, \vec{\sigma})$, being conjugated to the first class
constraints $\pi^{\tau}(\tau, \vec{\sigma})\approx 0$, $\Gamma (\tau, \vec{%
\sigma})\approx 0$.

Concerning the particle variables, we have that $\kappa^r_i(\tau)$, $%
\theta_i(\tau)$, $\theta_i^\ast(\tau)$, are not gauge invariant because

\begin{eqnarray}
\{ \kappa^r_i(\tau) , \Gamma (\tau, \vec{\sigma}) \}^{\ast \ast}_D &=& Q_i
\frac{\partial}{\partial \eta_i^r} \delta^3(\vec{\sigma} - \vec{\eta}%
_i(\tau)),  \nonumber \\
\{ \theta_i(\tau) , \Gamma (\tau, \vec{\sigma}) \}^{\ast \ast}_D &=& i e_i
\theta_i(\tau) \delta^3(\vec{\sigma} - \vec{\eta}_i(\tau)),  \nonumber \\
\{ \theta_i^\ast(\tau) , \Gamma (\tau, \vec{\sigma}) \}^{\ast \ast}_D &=& -
i e_i \theta_i^\ast(\tau) \delta^3(\vec{\sigma} - \vec{\eta}_i(\tau)) .
\label{III35}
\end{eqnarray}

However, the position variables $\eta _{i}^{r}(\tau )$ and the spin
variables $\xi _{i}^{r}(\tau )$ are gauge invariant. The Dirac observables
for the particles are obtained through a {\it dressing with a Coulomb cloud}

\begin{eqnarray}
\check{\theta}_i(\tau) &=& e^{i e_i \eta_{em}(\tau, \vec{\eta}_i)}
\theta_i(\tau),  \nonumber \\
\check{\theta}_i^\ast(\tau) &=& e^{- i e_i \eta_{em}(\tau, \vec{\eta}_i)}
\theta_i^\ast(\tau),  \nonumber \\
\check{\vec{\kappa}_i}(\tau) &=& \vec{\kappa}_i(\tau) - Q_i(\tau) \vec{%
\partial}\eta_{em}(\tau, \vec{\sigma})\quad \Rightarrow \check{\vec{\kappa}_i%
}(\tau) - Q_i \vec{A}_\perp (\tau, \vec{\eta}_i) = \vec{\kappa}_i (\tau) -
Q_i \vec{A} (\tau, \vec{\eta}_i).  \nonumber \\
&&  \label{III36}
\end{eqnarray}

The electric charges are gauge invariant

\begin{equation}
{\check{Q}}_{i}=e_{i}\check{\theta}_{i}^{\ast }(\tau )\check{\theta}%
_{i}(\tau )=Q_{i}=e_{i}\theta _{i}^{\ast }(\tau )\theta _{i}(\tau )~~~~~\Big[%
\dot{Q}_{i}(\tau )=0\Rightarrow Q_{i}(\tau )\equiv Q_{i}\Big].  \label{III37}
\end{equation}
When the Gauss law is satisfied, $\Gamma (\tau ,\vec{\sigma})=0$, Eq.(\ref
{III34}) implies

\begin{equation}
\int{d^3\sigma \vec{\pi}^2 (\tau, \vec{\sigma}) } = \int{d^3\sigma \vec{\pi}%
^2_\perp (\tau, \vec{\sigma}) } + \sum_{i \neq j}^{1...N} \frac{Q_i Q_j}{%
4\pi \mid \vec{\eta}_i(\tau) -\vec{\eta}_j(\tau) \mid} ,  \label{III38}
\end{equation}

\noindent so that we get

\begin{eqnarray}
{\cal H}(\tau) &=& \sqrt{p^2_s} - {\cal H}_{rel}=  \nonumber \\
&=& \sqrt{p_s^2} - \Big\{ {\frac{1}{2}}\int d^3\sigma \Big(\vec{\pi}
_{\perp}^2 (\tau, \vec{\sigma}) + \vec{B}^2 (\tau, \vec{\sigma})\Big) +
\sum_{i \neq j}^{1...N} \frac{Q_i Q_j}{4\pi \mid \vec{\eta}_i(\tau) -\vec{%
\eta}_j(\tau) \mid} +  \nonumber \\
&+& \sum_{i =1}^{N} \sqrt{m_i^2 - i Q_i \xi^r_i(\tau ) \xi^s_i(\tau )
F_{rs}(\tau, \vec{\eta}_i(\tau )) + \big[\check{\vec{\kappa}}_i(\tau) - Q_i%
\vec{A}_\perp (\tau, \vec{\eta}_i(\tau ))\big]^2} \Big\} \approx 0 .
\label{III39}
\end{eqnarray}

We see \cite{lu1} the {\it emergence of the Coulomb potential} from field
theory and the {\it regularization of the Coulomb self-energy} (the $\sum_{i%
\not= j}$ rule) due to the Grassmann character of the electric charges, $%
Q^2_i=0$. In this way all the effects of order $Q^2_i$ are eliminated, but
not those of order $Q_iQ_j$, $i\not= j$ \cite{lu6}.

Let us remark that even if we have not fixed the electromagnetic gauge but
simply decoupled the gauge variables $A_{\tau}(\tau ,\vec \sigma )$ and $%
\eta_{em}(\tau ,\vec \sigma )$, this procedure is equivalent to a
Wigner-covariant radiation gauge\footnote{%
The associated natural gauge fixings would be $A_{\tau}(\tau ,\vec \sigma
)\approx 0$ and $\eta_{em}(\tau ,\vec \sigma )\approx 0$.}.

There is no odd first class constraint, because massive 2-spinors do not
satisfy any spinor equation. The quantization of this Hamiltonian in the
free case gives a non-local Schr\"{o}dinger equation $i{\frac{{\partial \psi
_{\pm }}}{{\partial \tau }}}=\pm \sqrt{m^{2}+\triangle }\psi _{\pm }$, with
the kinetic square root operator \cite{lamme}, for a 2-spinor, which
corresponds to the upper (or lower) part of positive (or negative) energy
Dirac spinors boosted at rest \footnote{%
So that they also coincide with the corresponding parts of the positive (or
negative) energy Foldy-Wouthuysen spinors boosted at rest.}. These 2-spinors
are parity eigenstates in the rest frame.

The three constraints defining the rest frame become

\begin{equation}
{\vec {{\cal H}}}_p(\tau) = \sum_{i=1}^N \check{\vec{\kappa}_i}(\tau) + \int{%
d^3\sigma \vec{\pi}_\perp (\tau, \vec{\sigma}) \times \vec{B} (\tau, \vec{%
\sigma}) }\approx 0,  \label{III40}
\end{equation}

\noindent and are independent of the interactions, as expected in an instant
form of dynamics, like the 3-spin

\begin{eqnarray}
{\bar S}^{rs}_s&=&\sum_{i=1}^N[\eta_i^r(\tau ){\check \kappa}^s_i(\tau )-
\eta^s_i(\tau ){\check \kappa}^r_i(\tau )]+\int d^3\sigma [\sigma^r({\vec \pi%
} _{\perp}\times \vec B)^s-\sigma^s({\vec \pi}_{\perp}\times \vec B)^r](\tau
,\vec \sigma ),  \nonumber \\
{\bar S}^{rs}&=&{\bar S}^{rs}_s+{\bar S}^{rs}_{\xi}={\bar S}^{rs}_s-i
\sum_{i=1}^N \xi^r_i\xi^s_i.  \label{III41}
\end{eqnarray}

The Pauli-Lubanski 4-vector and the spin Poincar\'e Casimir are [$%
\epsilon_s= \sqrt{p^2_s}$]

\begin{eqnarray}
W^{\mu}_s&=&{\frac{1}{2}} \epsilon^{\mu\nu\rho\sigma}p_{s \nu} J_{s
\rho\sigma}= {\frac{1}{2}} \epsilon^{\mu\nu\rho\sigma}p_{s \nu} {\tilde S}%
_{\rho\sigma}=  \nonumber \\
&=& ({\vec p}_s\cdot {\vec {\bar S}} ; \epsilon_s {\vec {\bar S}}+{\frac{{\ {%
\vec p}_s \cdot {\vec {\bar S}} }}{{p^o_s+\epsilon_s}}} {\vec p}_s )=
\nonumber \\
&=&{\frac{1}{2}} \epsilon^{\mu\nu\rho\sigma}p_{s \nu} ({\tilde S}_{s
\rho\sigma}+ {\tilde S}_{\xi \rho\sigma})\, :=\, W^{(L) \mu}_s +
\Sigma^{\mu}_s,  \nonumber \\
W^2_s&=&-{\frac{1}{2}} p^2_s {\tilde S}_{\mu\nu}{\tilde S}^{\mu\nu}=-p^2_s {%
\vec {\bar S}}^2.  \label{III42}
\end{eqnarray}

\noindent This shows that ${\vec {\bar S}}={\vec {\bar S}}_s+{\vec {\bar S}}
_{\xi}$ is the {\it rest-frame Thomas spin}: $W_{s \mu}(p_s)=W_{s \nu}
L^{\nu}{}_{\mu}(p_s;\stackrel{\circ }{p}_s)= (0; \epsilon_s {\vec {\bar S}})$%
.

As shown in Ref.\cite{lu1}, it is convenient to replace the {\it external}
center-of-mass canonical coordinates ${\hat x}^{\mu}_s$, $p^{\mu}_s$ with a
new basis defined by the following canonical transformation ($T_s$ is the
Lorentz-scalar time of the rest frame)

\begin{eqnarray}
T_s &=& \frac{p_s^\mu x_{\mu s}}{\sqrt{p_s^2}},~~~~~~~~ \epsilon_s = \sqrt{%
p_s^2},  \nonumber \\
\vec{z}_s &=& \sqrt{p_s^2} \Big( \vec{\hat{x}_s} - \frac{\vec{p_s}}{ p_s^0}
\hat{x}_s^0 \Big) ,~~~~~ \vec{k}_s = \frac{\vec{p}_s}{\sqrt{ p_s^2}}.
\label{III43}
\end{eqnarray}

The inverse canonical transformation is

\begin{eqnarray}
\hat{x}^0_s &=& \sqrt{1-\vec{k}_s} \Big( T_s + \frac{\vec{k}_s \cdot \vec{z}%
_s}{\epsilon_s} \Big) ,  \nonumber \\
\vec{\hat{x}_s} &=& \frac{\vec{z}_s}{\epsilon_s} + \Big( T_s + \frac{\vec{k}%
_s \cdot \vec{z}_s}{\epsilon_s} \Big) \vec{k}_s,  \nonumber \\
p_s^0 &=& \epsilon_s \sqrt{1+\vec{k}_s^2}.  \label{III44}
\end{eqnarray}

By adding the gauge-fixing $T_s - \tau \approx 0$ \footnote{%
It identifies the rest-frame time $T_s$ with the parameter $\tau$ of the
foliation of Minkowski spacetime with the Wigner hyperplanes associated with
the isolated system.}, whose time constancy implies $\lambda(\tau) = -1$, we
get a frozen phase space with the Dirac Hamiltonian $H_D=-\vec \lambda (\tau
)\cdot {\vec {{\cal H}}}_p(\tau )$. Like in the Hamilton-Jacobi theory, to
reintroduce an evolution in $\tau \equiv T_s$ one uses the energy as
Hamiltonian\footnote{%
See Ref.\cite{pons} for a different demonstration of this result.} so that
the Dirac Hamiltonian for the rest-frame instant form of dynamics is

\begin{equation}
\hat{H}_D = {\cal H}_{rel}(\tau) - \vec{\lambda}(\tau) \cdot {\vec {{\cal H}}%
}_p(\tau).  \label{III45}
\end{equation}

The effective Hamiltonian is the invariant mass $M$ of the isolated system

\begin{eqnarray}
M &=&{\cal H}_{rel}(\tau )=\sum_{i=1}^{N}\sqrt{m_{i}^{2}-iQ_{i}\xi
_{i}^{r}(\tau )\xi _{i}^{s}(\tau )F_{rs}(\tau ,\vec{\eta}_{i}(\tau ))+\big[%
\check{\vec{\kappa}}_{i}(\tau )-Q_{i}\vec{A}_{\perp }(\tau ,\vec{\eta}%
_{i}(\tau ))\big]^{2}}+  \nonumber \\
&+&\sum_{i\neq j}^{1...N}Q_{i}Q_{j}\frac{1}{4\pi \mid \vec{\eta}_{i}(\tau )-%
\vec{\eta}_{j}(\tau )\mid }+\int {d^{3}\sigma \Big(\frac{\vec{\pi}_{\perp
}^{2}(\tau ,\vec{\sigma})+\vec{B}^{2}(\tau ,\vec{\sigma})}{2}\Big)}=
\nonumber \\
&=&\sum_{i=1}^{N}\sqrt{m_{i}^{2}-2Q_{i}{\vec{\bar{S}}}_{i\xi }\cdot \vec{B}%
(\tau ,\vec{\eta}_{i})+\big[\check{\vec{\kappa}_{i}}(\tau )-Q_{i}\vec{A}%
_{\perp }(\tau ,\vec{\eta}_{i})\big]^{2}}+  \nonumber \\
&+&\sum_{i\neq j}^{1...N}Q_{i}Q_{j}\frac{1}{4\pi \mid \vec{\eta}_{i}(\tau )-%
\vec{\eta}_{j}(\tau )\mid }+\int {d^{3}\sigma \frac{\big(\vec{\pi}_{\perp
}^{2}+\vec{B}^{2}\big)}{2}(\tau ,\vec{\sigma})}.  \label{III46}
\end{eqnarray}
See Ref.\cite{lu3} for the associated Hamilton equations.

In the next Section we will modify this invariant mass to include the
non-minimal couplings of the electric field identified in Section II.

\vfill\eject

\section{Non-minimal coupling to the electric field of the positive-energy
spinning particle.}

In this Section we modify the invariant mass (\ref{III46}) on the Wigner
hyperplane to include the non-minimal coupling to the electric field of
Section II. Then we define the {\it external} and {\it internal}
realizations of the Poincar\'e group.

\subsection{The Non-Minimal Coupling to the Electric Field.}

As shown in the previous Section on the Wigner hyperplane of the rest-frame
instant form we have the following four first class constraints for the
description of positive-energy charged spinning particles coupled to the
electromagnetic field in terms of Dirac observables corresponding to the
Wigner-covariant radiation gauge

\begin{eqnarray}
{\cal H}(\tau ) &=&\epsilon _{s}-\Big\{{\frac{1}{2}}\int d^{3}\sigma \Big(%
\vec{\pi}_{\perp }^{2}(\tau ,\vec{\sigma})+\vec{B}^{2}(\tau ,\vec{\sigma})%
\Big)+\sum_{i\neq j}^{1...N}\frac{Q_{i}Q_{j}}{4\pi \mid \vec{\eta}_{i}(\tau
)-\vec{\eta}_{j}(\tau )\mid }+  \nonumber \\
&+&\sum_{i=1}^{N}\sqrt{m_{i}^{2}-iQ_{i}\xi _{i}^{r}(\tau )\xi _{i}^{s}(\tau
)F_{rs}(\tau ,\vec{\eta}_{i}(\tau ))+\big[\check{\vec{\kappa}_{i}}(\tau
)-Q_{i}\vec{A}_{\perp }(\tau ,\vec{\eta}_{i}(\tau ))\big]^{2}}\Big\}\approx
0,  \nonumber \\
&&{}  \nonumber \\
{\vec{{\cal H}}}_{p}(\tau ) &=&\sum_{i=1}^{N}\check{\vec{\kappa}_{i}}(\tau
)+\int {d^{3}\sigma \vec{\pi}_{\perp }(\tau ,\vec{\sigma})\times \vec{B}%
(\tau ,\vec{\sigma})}\approx 0,  \label{IV1}
\end{eqnarray}

The original minimal and non-minimal couplings to the electromagnetic field
produced the Coulomb potential and the non-minimal coupling to the magnetic
field. Now we have to introduce the modification suggested by the
pseudo-classical Foldy-Wouthuysen canonical transformation of Section II to
include the semiclassical non-minimal coupling to the electric field.

We see that the kinetic term of each particle is the same in Eq.(\ref{IV1})
and (\ref{II30}). In Eq.(\ref{II30}) there was no choice of the
electromagnetic gauge. Instead Eq.(\ref{IV1}) is in the radiation gauge, so
that instead of the vector potential we now have the transverse vector
potential of the radiation field.

On each particle $i$ acts: i) the radiation field; ii) the scalar Coulomb
potential of the other charges. As said at the end of Section II, we have to
replace $A_{0}(x)$ of Eq.(\ref{II30}) with just the scalar Coulomb potential

\begin{equation}
Q_i V_i({\vec \eta}_i(\tau ))=Q_i \sum_{j\not= i}^{1..N} {\frac{{Q_j}}{{4\pi
|{\vec \eta}_i(\tau )-{\vec \eta}_j(\tau )|}}},  \label{IV2}
\end{equation}

\noindent acting on particle $i$ (the total Coulomb potential is $V=\frac{1}{%
2}\sum_{i=1}^{N}Q_{i}V_{i}$) while the term $\vec{\xi}\cdot ({\frac{{%
\partial \vec{A}(x)}}{{\partial x}^{0}}}+\vec{\partial}A_{0}(x))$ of Eq.(\ref
{II30}) will be replaced by two terms: i) the transverse electric field $%
\vec{\xi}\cdot {\vec{\pi}}_{\perp }(\tau ,\vec{\sigma}={\vec{\eta}}_{i}(\tau
))$ of the radiation field at the position of the particle and ii) the
spatial variation of the Coulomb field of the other particles when particle $%
i$ moves appearing in the combination\footnote{%
As shown in Section VIII it is just the quantization of the consequences of
this term that will produce the {\it spin-orbit} and {\it Darwin} terms in
the final semiclassical Hamiltonian.}

\begin{equation}
{\frac{{iQ_{i}{\check{\vec{\kappa}}}_{i}(\tau )\cdot \vec{\xi}(\tau )\,\,%
\vec{\xi}(\tau )\cdot {\vec{\partial}}_{{\vec{\eta}}_{i}}V_{i}({\vec{\eta}}%
_{i}(\tau ))}}{{(m_{i}+\sqrt{m_{i}^{2}+{\check{\vec{\kappa}}}_{i}^{2}(\tau )}%
)\sqrt{m_{i}^{2}+{\check{\vec{\kappa}}}_{i}^{2}(\tau )}}}}.  \label{IV3}
\end{equation}

\noindent On the other hand, the rest-frame conditions ${\vec{{\cal H}}}%
_{p}(\tau )\approx 0$ remain the same, since they do not depend on the
interaction in an instant form of the dynamics.

Therefore the modified constraints on the Wigner hyperplane, written in
terms of the Dirac observables, are

\begin{eqnarray}
{\cal H}^{^{\prime }} &=&\epsilon _{s}-M^{^{\prime }}=  \nonumber \\
&=&\epsilon _{s}-\Big(\sum_{i=1}^{N}\sqrt{m_{i}^{2}-iQ_{i}\xi _{i}^{r}(\tau
)\xi _{i}^{s}(\tau )F_{rs}(\tau ,{\vec{\eta}}_{i}(\tau ))+\check{\vec{\kappa}%
}_{i}(\tau )-Q_{i}{\check{\vec{A}}}_{\perp }(\tau ,\vec{\eta}_{i}(\tau
)))^{2}}-  \nonumber \\
&-&i\sum_{i=1}^{N}\frac{Q_{i}\check{\vec{\kappa}}_{i}(\tau )\cdot \vec{\xi}%
_{i}(\tau )\,\vec{\xi}_{i}(\tau )\cdot \check{\vec{\pi}}_{\perp }(\tau ,\vec{%
\eta}_{i}(\tau ))}{(m_{i}+\sqrt{m_{i}^{2}+\check{\vec{\kappa}}_{i}^{2}(\tau )%
})\sqrt{m_{i}^{2}+\check{\vec{\kappa}}_{i}^{2}(\tau )}}+  \nonumber \\
&+&\sum_{i\neq j}\Big[\frac{Q_{i}Q_{j}}{4\pi \mid \vec{\eta}_{i}(\tau )-\vec{%
\eta}_{j}(\tau )\mid }-  \nonumber \\
&-&i\frac{Q_{i}Q_{j}\check{\vec{\kappa}}_{i}(\tau )\cdot \vec{\xi}_{i}(\tau
)\,\vec{\xi}_{i}(\tau )\cdot (\vec{\eta}_{i}(\tau )-\vec{\eta}_{j}(\tau ))}{%
4\pi \mid \vec{\eta}_{i}(\tau )-\vec{\eta}_{j}(\tau )\mid ^{3}(m_{i}+\sqrt{%
m_{i}^{2}+\check{\vec{\kappa}}_{i}^{2}(\tau )})\sqrt{m_{i}^{2}+\check{\vec{%
\kappa}}_{i}^{2}(\tau )}}\Big]+  \nonumber \\
&+&\int d^{3}\sigma {\frac{1}{2}}[\check{\vec{\pi}}_{\perp }^{2}+\check{\vec{%
B}}^{2}](\tau ,\vec{\sigma})\Big)=  \nonumber \\
&&{}  \nonumber \\
&=&\epsilon _{s}-\Big(\sum_{i=1}^{N}\Big[\sqrt{m_{i}^{2}+(\check{\vec{\kappa}%
}_{i}(\tau )-Q_{i}{\check{\vec{A}}}_{\perp }(\tau ,\vec{\eta}_{i}(\tau
)))^{2}}-  \nonumber \\
&+&i\frac{Q_{i}\vec{\xi}_{i}(\tau )\times \vec{\xi}_{i}(\tau )\cdot \check{%
\vec{B}}(\tau ,\vec{\eta}_{i}(\tau ))}{2\sqrt{m_{i}^{2}+\check{\vec{\kappa}}%
_{i}^{2}(\tau )}}-i\frac{Q_{i}\check{\vec{\kappa}}_{i}(\tau )\cdot \vec{\xi}%
_{i}(\tau )\,\vec{\xi}_{i}(\tau )\cdot \check{\vec{\pi}}_{\perp }(\tau ,\vec{%
\eta}_{i}(\tau ))}{(m_{i}+\sqrt{m_{i}^{2}+\check{\vec{\kappa}}_{i}^{2}(\tau )%
})\sqrt{m_{i}^{2}+\check{\vec{\kappa}}_{i}^{2}(\tau )}}\Big]+  \nonumber \\
&+&\sum_{i\neq j}\Big[\frac{Q_{i}Q_{j}}{4\pi \mid \vec{\eta}_{i}(\tau )-\vec{%
\eta}_{j}(\tau )\mid }-  \nonumber \\
&-&i\frac{Q_{i}Q_{j}\check{\vec{\kappa}}_{i}(\tau )\cdot \vec{\xi}_{i}(\tau
)\,\vec{\xi}_{i}(\tau )\cdot (\vec{\eta}_{i}(\tau )-\vec{\eta}_{j}(\tau ))}{%
4\pi \mid \vec{\eta}_{i}(\tau )-\vec{\eta}_{j}(\tau )\mid ^{3}(m_{i}+\sqrt{%
m_{i}^{2}+\check{\vec{\kappa}}_{i}^{2}(\tau )})\sqrt{m_{i}^{2}+\check{\vec{%
\kappa}}_{i}^{2}(\tau )}}\Big]+  \nonumber \\
&+&\int d^{3}\sigma {\frac{1}{2}}[\check{\vec{\pi}}_{\perp }^{2}+\check{\vec{%
B}}^{2}](\tau ,\vec{\sigma})\Big)\approx 0,  \nonumber \\
&&{}  \nonumber \\
{\vec{{\cal H}}}_{p}(\tau ) &=&\sum_{i=1}^{N}\check{\vec{\kappa}_{i}}(\tau
)+\int {\ d^{3}\sigma \vec{\pi}_{\perp }(\tau ,\vec{\sigma})\times \vec{B}%
(\tau ,\vec{\sigma})}\approx 0.  \label{IV4}
\end{eqnarray}

One can check that these constraints are still first class.

The Dirac Hamiltonian (\ref{III45}) is replaced by

\begin{equation}
{\hat H}^{^{\prime}}_D = M^{^{\prime}} - \vec \lambda (\tau )\cdot {\vec {%
{\cal H}}}_p(\tau ).  \label{IV5}
\end{equation}

See Subsection 2 of Appendix B for an attempt to find the constraints before
the restriction to the radiation gauge and on arbitrary spacelike
hypersurfaces. Since we do not know the Lagrangian, we must use Hamiltonian
methods\footnote{%
If we have the action $S=\int dt L$ of a non-singular system with
configuration variables $q^{\alpha}$ (we consider a finite-dimensional case
for the sake of simplicity) coupled to an external gravitational field, the
energy momentum tensor is defined as $T^{\mu\nu}(z)=-{\frac{2}{\sqrt{g(z)}}}
{\frac{{\delta S}}{{\delta g_{\mu\nu}(z)}}}{|}_{g=\eta}$. The canonical
momenta and the Hamiltonian are $p_{\alpha}={\frac{{\partial L}}{{\partial {%
\ \dot q}^{\alpha}}}}$ and $H={\frac{{\partial L}}{{\partial {\dot q}%
^{\alpha}} }} {\dot q}^{\alpha}-L$, respectively. Note that the momenta
depend on the gravitational field. Therefore, by using the notation $%
S_H=\int dt H=\int dt p_{\alpha}{\dot q}^{\alpha}-S$, we get $T^{\mu\nu}(z)={%
\frac{2}{\sqrt{g(z)}}} {\frac{{\delta S_H}}{{\delta g_{\mu\nu}(z)}}}{|}_{p}{|%
}_{g=\eta } -{\frac{2}{\sqrt{g(z)}}} \int dt [{\dot q}^{\alpha}-{\frac{{%
\partial H}}{{\partial p_{\alpha}}}}] {\frac{{\partial T^{\mu\nu}(z)}}{{%
\partial {\dot q}^{\alpha}}}} {|}_{g=\eta}\, \stackrel{ \circ }{=}\, \Big(
{\frac{2}{\sqrt{g(z)}}} {\frac{{\delta S_H}}{{\delta g_{\mu\nu}(z)}}}{|}_{p}%
\Big) {|}_{g=\eta }$ by using the first half of Hamilton equations. This
gives the Hamiltonian form of the energy momentum tensor. In the case of
singular systems, the Hamiltonian has to be replaced with the Dirac
Hamiltonian: the Hamiltonian form of energy momentum tensor is now dependent
explicitly on the Dirac multipliers.} for determining the energy momentum
tensor $T^{AB}$ and this requires the modified Dirac Hamiltonian on
arbitrary hypersurfaces, which replaces Eq.(\ref{III4}), as input.

\subsection{External and internal Poincar\'e Groups.}

The {\it external} realization of the Poincar\'e algebra, given in Eq.(\ref
{III28}), is

\begin{eqnarray}
J_{s}^{ij} &:&={\hat{x}}_{s}^{i}p_{s}^{j}-{\hat{x}}_{s}^{j}p_{s}^{i}+\delta
^{ir}\delta ^{js}{\bar{S}}^{rs},  \nonumber \\
J_{s}^{oi} &:&={\hat{x}}_{s}^{0}p_{s}^{i}-{\hat{x}}_{s}^{i}p_{s}^{0}-{\
\frac{{\delta ^{ir}{\bar{S}}^{rs}p_{s}^{s}}}{{\ p_{s}^{0}+\eta _{s}\sqrt{%
p_{s}^{2}}}}},  \label{IV6}
\end{eqnarray}

\noindent where ${\bar S}^r\equiv {\bar S}^r_s+{\bar S}^r_{\xi}=
\epsilon^{ruv}\Big( \sum_{i=1}^N \eta^u_i(\tau ) {\check \kappa}_i^v(\tau )
+ \int d^3\sigma\, \sigma^u ({\vec \pi}_{\perp}\times \vec B)^v(\tau ,\vec %
\sigma ) -{\frac{i}{2}} \sum_{i=1}^N \xi^u_i(\tau )\xi^v(\tau )\Big)$ as a
consequence of Eq.(\ref{III41}).

The {\it internal} unfaithful realization of the Poincar\'{e} algebra has
the generators [from now on we denote by $M$ the $M^{^{\prime }}$ of Eqs.(
\ref{IV4}), (\ref{IV5}); the Green function $\vec c(\vec \sigma )$ is
defined in Eq.(\ref{b15}) of Appendix B]

\begin{eqnarray}
{\cal P}_{(int)}^{\tau } &=&M=\sum_{i=1}^{N}\Big[\sqrt{m_{i}^{2}+(\check{%
\vec{\kappa}}_{i}(\tau )-Q_{i}{\check{\vec{A}}}_{\perp }(\tau ,\vec{\eta}%
_{i}(\tau )))^{2}}-  \nonumber \\
&+&i\frac{Q_{i}\vec{\xi}_{i}(\tau )\times \vec{\xi}_{i}(\tau )\cdot \check{%
\vec{B}}(\tau ,\vec{\eta}_{i}(\tau ))}{2\sqrt{m_{i}^{2}+\check{\vec{\kappa}}%
_{i}^{2}(\tau )}}-i\frac{Q_{i}\check{\vec{\kappa}}_{i}(\tau )\cdot \vec{\xi}%
_{i}(\tau )\,\vec{\xi}_{i}(\tau )\cdot \check{\vec{\pi}}_{\perp }(\tau ,\vec{%
\eta}_{i}(\tau ))}{(m_{i}+\sqrt{m_{i}^{2}+\check{\vec{\kappa}}_{i}^{2}(\tau )%
})\sqrt{m_{i}^{2}+\check{\vec{\kappa}}_{i}^{2}(\tau )}}\Big]+  \nonumber \\
&+&\sum_{i\neq j}\Big[\frac{Q_{i}Q_{j}}{4\pi \mid \vec{\eta}_{i}(\tau )-\vec{%
\eta}_{j}(\tau )\mid }-  \nonumber \\
&-&i\frac{Q_{i}Q_{j}\check{\vec{\kappa}}_{i}(\tau )\cdot \vec{\xi}_{i}(\tau
)\,\vec{\xi}_{i}(\tau )\cdot (\vec{\eta}_{i}(\tau )-\vec{\eta}_{j}(\tau ))}{%
4\pi \mid \vec{\eta}_{i}(\tau )-\vec{\eta}_{j}(\tau )\mid ^{3}(m_{i}+\sqrt{%
m_{i}^{2}+\check{\vec{\kappa}}_{i}^{2}(\tau )})\sqrt{m_{i}^{2}+\check{\vec{%
\kappa}}_{i}^{2}(\tau )}}\Big]+  \nonumber \\
&+&\int d^{3}\sigma {\frac{1}{2}}[\check{\vec{\pi}}_{\perp }^{2}+\check{\vec{%
B}}^{2}](\tau ,\vec{\sigma})=  \nonumber \\
&&{}  \nonumber \\
&=&\sum_{i=1}^{N}\Big[\sqrt{m_{i}^{2}+(\check{\vec{\kappa}}_{i}(\tau )-Q_{i}{%
\check{\vec{A}}}_{\perp }(\tau ,\vec{\eta}_{i}(\tau )))^{2}}-  \nonumber \\
&&+i\frac{Q_{i}\vec{\xi}_{i}(\tau )\times \vec{\xi}_{i}(\tau )\cdot \check{%
\vec{B}}(\tau ,\vec{\eta}_{i}(\tau ))}{2\sqrt{m_{i}^{2}+\check{\vec{\kappa}}%
_{i}^{2}(\tau )}}-i\frac{Q_{i}\check{\vec{\kappa}}_{i}(\tau )\cdot \vec{\xi}%
_{i}(\tau )\,\vec{\xi}_{i}(\tau )\cdot \check{\vec{\pi}}_{\perp }(\tau ,\vec{%
\eta}_{i}(\tau ))}{(m_{i}+\sqrt{m_{i}^{2}+\check{\vec{\kappa}}_{i}^{2}(\tau )%
})\sqrt{m_{i}^{2}+\check{\vec{\kappa}}_{i}^{2}(\tau )}}\Big]+  \nonumber \\
&+&\sum_{i\neq j}\Big[\frac{Q_{i}Q_{j}}{4\pi \mid \vec{\eta}_{i}(\tau )-\vec{%
\eta}_{j}(\tau )\mid }-i\frac{Q_{i}Q_{j}\check{\vec{\kappa}}_{i}(\tau )\cdot
\vec{\xi}_{i}(\tau )}{(m_{i}+\sqrt{m_{i}^{2}+\check{\vec{\kappa}}%
_{i}^{2}(\tau )})\sqrt{m_{i}^{2}+\check{\vec{\kappa}}_{i}^{2}(\tau )}}
\nonumber \\
&&\times {\vec{\xi}}_{i}(\tau )\cdot {\frac{{\partial }}{{\partial {\vec{\eta%
}}_{i}}}}\int d^{3}\sigma _{1}\,\vec{c}({\vec{\sigma}}_{1}-{\vec{\eta}}%
_{i}(\tau ))\cdot \vec{c}({\vec{\sigma}}_{1}-{\vec{\eta}}_{j}(\tau ))\Big]+
\nonumber \\
&+&\int d^{3}\sigma {\frac{1}{2}}[\check{\vec{\pi}}_{\perp }^{2}+\check{\vec{%
B}}^{2}](\tau ,\vec{\sigma}),  \nonumber \\
&&{}  \nonumber \\
&&{}  \nonumber \\
{\cal \vec{P}}_{(int)} &=&{\vec{{\cal H}}}_{p}=\sum_{i=1}^{N}{\check{\vec{%
\kappa}}}_{i}(\tau )+\int d^{3}\sigma \lbrack {\check{\vec{\pi}}}_{\perp
}\times {\check{\vec{B}}}](\tau ,\vec{\sigma})\approx 0,  \nonumber \\
&&{}  \nonumber \\
{\cal J}_{(int)}^{r} &=&{\bar{S}}^{r}=\varepsilon ^{rst}{\bar{S}}%
^{st}=\sum_{i=1}^{N}(\vec{\eta}_{i}(\tau )\times {\check{\vec{\kappa}}}%
_{i}(\tau ))^{r}-{\frac{i}{2}}\sum_{i=1}^{N}({\vec{\xi}}_{i}\times {\vec{\xi}%
}_{i})^{r}+  \nonumber \\
&+&\int d^{3}\sigma \,(\vec{\sigma}\times \,{[{\check{\vec{\pi}}}}_{\perp }{%
\times {\check{\vec{B}}]}}^{r}{(\tau ,\vec{\sigma})},  \nonumber \\
&&{}  \nonumber \\
{\cal K}_{(int)}^{r} &=&{\bar{S}}_{s}^{\bar{o}r}=-{\bar{S}}_{s}^{r\bar{o}%
}=-\sum_{i=1}^{N}\vec{\eta}_{i}(\tau )\sqrt{m_{i}^{2}+[{{\check{\vec{\kappa}}%
}}_{i}(\tau )-Q_{i}{\check{\vec{A}}}_{\perp }(\tau ,{\ \vec{\eta}}_{i}(\tau
))]{}^{2}}+  \nonumber \\
&+&\sum_{i=1}^{N}\Big[\sum_{j\not=i}^{1..N}Q_{i}Q_{j}[{\frac{1}{{\triangle _{%
{\vec{\eta}}_{j}}}}}{\frac{{\partial }}{{\partial \eta _{j}^{r}}}}c({\vec{%
\eta}}_{i}(\tau )-{\vec{\eta}}_{j}(\tau ))-\eta _{j}^{r}(\tau )c({\vec{\eta}}%
_{i}(\tau )-{\vec{\eta}}_{j}(\tau ))]+  \nonumber \\
&+Q_{i}&\int d^{3}\sigma {\check{\pi}}_{\perp }^{r}(\tau ,\vec{\sigma})c(%
\vec{\sigma}-{\ \vec{\eta}}_{i}(\tau ))\Big]-{\frac{1}{2}}\int d^{3}\sigma
\sigma ^{r}\,({{\check{\vec{\pi}}}}_{\perp }^{2}+{{\check{\vec{B}}}}%
^{2})(\tau ,\vec{\sigma})-  \nonumber \\
&+&\sum_{i=1}^{N}\Big[i\frac{Q_{i}\eta _{i}^{r}(\tau )\vec{\xi}_{i}(\tau
)\times \vec{\xi}_{i}(\tau )\cdot \check{\vec{B}}(\tau ,\vec{\eta}_{i}(\tau
))}{2\sqrt{m_{i}^{2}+\check{\vec{\kappa}}_{i}^{2}(\tau )}}-i\frac{Q_{i}\eta
_{i}^{r}(\tau )\check{\vec{\kappa}}_{i}(\tau )\cdot \vec{\xi}_{i}(\tau )\,%
\vec{\xi}_{i}(\tau )\cdot \check{\vec{\pi}}_{\perp }(\tau ,\vec{\eta}%
_{i}(\tau ))}{(m_{i}+\sqrt{m_{i}^{2}+\check{\vec{\kappa}}_{i}^{2}(\tau )})%
\sqrt{m_{i}^{2}+\check{\vec{\kappa}}_{i}^{2}(\tau )}}\Big]-  \nonumber \\
&-&i\sum_{i\neq j}\frac{Q_{i}Q_{j}\eta _{i}^{r}(\tau )\check{\vec{\kappa}}%
_{i}(\tau )\cdot \vec{\xi}_{i}(\tau )\,\vec{\xi}_{i}(\tau )\cdot (\vec{\eta}%
_{i}(\tau )-\vec{\eta}_{j}(\tau ))}{4\pi \mid \vec{\eta}_{i}(\tau )-\vec{\eta%
}_{j}(\tau )\mid ^{3}(m_{i}+\sqrt{m_{i}^{2}+\check{\vec{\kappa}}%
_{i}^{2}(\tau )})\sqrt{m_{i}^{2}+\check{\vec{\kappa}}_{i}^{2}(\tau )}}=
\nonumber \\
&&{}  \nonumber \\
&=&{\bar{S}}_{s}^{\bar{o}r}=-{\bar{S}}_{s}^{r\bar{o}}=-\sum_{i=1}^{N}\vec{%
\eta}_{i}(\tau )\sqrt{m_{i}^{2}+[{{\check{\vec{\kappa}}}}_{i}(\tau )-Q_{i}{%
\check{\vec{A}}}_{\perp }(\tau ,{\ \vec{\eta}}_{i}(\tau ))]{}^{2}}+
\nonumber \\
&+&\sum_{i=1}^{N}\Big[\sum_{j\not=i}^{1..N}Q_{i}Q_{j}[{\frac{1}{{\triangle _{%
{\vec{\eta}}_{j}}}}}{\frac{{\partial }}{{\partial \eta _{j}^{r}}}}c({\vec{%
\eta}}_{i}(\tau )-{\vec{\eta}}_{j}(\tau ))-\eta _{j}^{r}(\tau )c({\vec{\eta}}%
_{i}(\tau )-{\vec{\eta}}_{j}(\tau ))]+  \nonumber \\
&+Q_{i}&\int d^{3}\sigma {\check{\pi}}_{\perp }^{r}(\tau ,\vec{\sigma})c(%
\vec{\sigma}-{\ \vec{\eta}}_{i}(\tau ))\Big]-{\frac{1}{2}}\int d^{3}\sigma
\sigma ^{r}\,({{\check{\vec{\pi}}}}_{\perp }^{2}+{{\check{\vec{B}}}}%
^{2})(\tau ,\vec{\sigma})-  \nonumber \\
&+&\sum_{i=1}^{N}\Big[i\frac{Q_{i}\eta _{i}^{r}(\tau )\vec{\xi}_{i}(\tau
)\times \vec{\xi}_{i}(\tau )\cdot \check{\vec{B}}(\tau ,\vec{\eta}_{i}(\tau
))}{2\sqrt{m_{i}^{2}+\check{\vec{\kappa}}_{i}^{2}(\tau )}}-i\frac{Q_{i}\eta
_{i}^{r}(\tau )\check{\vec{\kappa}}_{i}(\tau )\cdot \vec{\xi}_{i}(\tau )\,%
\vec{\xi}_{i}(\tau )\cdot \check{\vec{\pi}}_{\perp }(\tau ,\vec{\eta}%
_{i}(\tau ))}{(m_{i}+\sqrt{m_{i}^{2}+\check{\vec{\kappa}}_{i}^{2}(\tau )})%
\sqrt{m_{i}^{2}+\check{\vec{\kappa}}_{i}^{2}(\tau )}}\Big]-  \nonumber \\
&-&i\sum_{i\not=j}^{1..N}{\frac{{Q_{i}Q_{j}\eta _{i}^{r}(\tau ){\check{\vec{%
\kappa}}}_{i}(\tau )\cdot {\vec{\xi}}_{i}(\tau )}}{{(m_{i}+\sqrt{m_{i}^{2}+{%
\check{\vec{\kappa}}}_{i}^{2}(\tau )})\sqrt{m_{i}^{2}+{\check{\vec{\kappa}}}%
_{i}^{2}(\tau )}}}}  \nonumber \\
&&{\vec{\xi}}_{i}(\tau )\cdot {\frac{{\partial }}{{\partial {\vec{\eta}}_{i}}%
}}\int d^{3}\sigma _{1}\,\vec{c}({\vec{\sigma}}_{1}-{\vec{\eta}}_{i}(\tau
))\cdot \vec{c}({\vec{\sigma}}_{1}-{\vec{\eta}}_{j}(\tau )).  \nonumber \\
&&  \label{IV7}
\end{eqnarray}

As said in Refs.\cite{ap,mate}, the natural gauge fixings for the rest-frame
conditions ${\vec {{\cal P}}}_{(int)}\approx 0$ are ${\vec {{\cal K}}}%
_{(int)}\approx 0$. These conditions identify the {\it internal} 3-center of
energy of M\o ller, which weakly coincides with the {\it internal} canonical
3-center of mass ${\vec q}_{+}$ due to ${\vec {{\cal P}}}_{(int)}\approx 0$.
Therefore the gauge fixings ${\vec {{\cal K}}}_{(int)}\approx 0$ are
equivalent to ${\vec q}_{+}\approx 0$: they put the {\it internal} 3-center
of mass in the origin $x^{\mu}_s(\tau )=z^{\mu}(\tau ,\vec 0)$ of the
3-coordinates in each Wigner hyperplane. The gauge fixings ${\vec q}
_{+}\approx 0$ imply $\vec \lambda (\tau )=0$.

\vfill\eject

\section{\protect\bigskip The Equations of Motion for N Charged
Positive-energy Spinning Particles Plus the Electromagnetic Field.}

The modified Dirac Hamiltonian (\ref{IV5}), (\ref{IV7}) in the rest-frame
instant form on Wigner hyperplanes and with $T_s\equiv \tau$ can be written
in the form

\begin{eqnarray}
{\hat H}_D&=& M -\vec{\lambda}(\tau )\cdot {\cal \vec{H }} _{p}(\tau )=
\nonumber \\
&=&\sum_{i=1}^{N}\Big[\sqrt{m_{i}^{2}+(\check{\vec{\kappa}} _{i}(\tau )-Q_{i}%
{\check{\vec{A}}}_{\perp }(\tau ,\vec{\eta}_{i}(\tau )))^{2} }+\frac{{\cal R}%
_{i}(\tau )}{2\sqrt{m_{i}^{2}+\check{\vec{\kappa}} _{i}^{2}(\tau )}}\Big]+
\nonumber \\
&+&\sum_{i\neq j}\frac{Q_{i}Q_{j}}{4\pi \mid \vec{\eta}_{i}(\tau )-\vec{\eta}%
_{j}(\tau )\mid }\Big[1-  \nonumber \\
&-&i\frac{\check{\vec{\kappa}}_{i}(\tau )\cdot \vec{\xi}_{i}(\tau )\, \vec{%
\xi}_{i}(\tau )\cdot (\vec{\eta}_{i}(\tau )-\vec{\eta}_{j}(\tau ))}{\mid
\vec{\eta}_{i}(\tau )-\vec{\eta}_{j}(\tau )\mid ^{2}(m_i+\sqrt{m_{i}^{2}+
\check{\vec{\kappa}}_{i}^{2}(\tau )})\sqrt{m_{i}^{2}+\check{\vec{\kappa}}
_{i}^{2}(\tau )}}\Big] +  \nonumber \\
&+&\int d^{3}\sigma {\frac{1}{2}}[\check{\vec{\pi}}_{\perp }^{2}+\check{\vec{
B}}^{2}](\tau ,\vec{\sigma})-\vec{\lambda}(\tau )\cdot {\cal \vec{H }}%
_{p}(\tau )=  \nonumber \\
&&{}  \nonumber \\
&&{}  \nonumber \\
&=&\sum_{i=1}^N \sqrt{m_{i}^{2}+{\cal R}_i(\tau )+(\check{\vec{\kappa}}
_{i}(\tau )-Q_{i}{\check{\vec{A}}}_{\perp }(\tau ,\vec{\eta}_{i}(\tau
)))^{2} } +  \nonumber \\
&+& \sum_{i\neq j}\frac{Q_{i}Q_{j}}{4\pi \mid \vec{\eta}_{i}(\tau )-\vec{%
\eta }_{j}(\tau )\mid }\Big[1-  \nonumber \\
&-&i\frac{\check{\vec{\kappa}}_{i}(\tau )\cdot \vec{\xi}_{i}(\tau )\, \vec{%
\xi}_{i}(\tau )\cdot (\vec{\eta}_{i}(\tau )-\vec{\eta}_{j}(\tau ))}{\mid
\vec{\eta}_{i}(\tau )-\vec{\eta}_{j}(\tau )\mid ^{2}(m_i+\sqrt{m_{i}^{2}+
\check{\vec{\kappa}}_{i}^{2}(\tau )})\sqrt{m_{i}^{2}+\check{\vec{\kappa}}
_{i}^{2}(\tau )}}\Big] +  \nonumber \\
&+&\int d^{3}\sigma {\frac{1}{2}}[\check{\vec{\pi}}_{\perp }^{2}+\check{\vec{
B}}^{2}](\tau ,\vec{\sigma})-\vec{\lambda}(\tau )\cdot {\cal \vec{H }}%
_{p}(\tau ),  \label{V1}
\end{eqnarray}

\noindent where we introduced the quantity

\begin{eqnarray}
{\cal R}_{i}(\tau ) &=&+iQ_{i}\vec{\xi}_{i}(\tau )\times \vec{\xi}_{i}(\tau
)\cdot \check{\vec{B}}(\tau ,\vec{\eta}_{i}(\tau ))-2i\frac{Q_{i}\check{\vec{%
\kappa}}_{i}(\tau )\cdot \vec{\xi}_{i}(\tau )\,\vec{\xi}_{i}(\tau )\cdot
\check{\vec{\pi}}_{\perp }(\tau ,\vec{\eta}_{i}(\tau ))}{m_{i}+\sqrt{%
m_{i}^{2}+\check{\vec{\kappa}}_{i}^{2}(\tau )}}=  \nonumber \\
&=&-2Q_{i}\Big[{\vec{\bar{S}}}_{i\xi }(\tau )\cdot {\check{\vec{B}}}(\tau ,{%
\vec{ \eta}}_{i}(\tau ))-{\frac{{{\check{\vec{\kappa}}}_{i}(\tau )\cdot {%
\check{\vec{\pi}}}_{\perp }(\tau ,{\vec{\eta}}_{i}(\tau ))\times {\vec{\bar{S%
}}}_{i\xi }(\tau )}}{{m_{i}+\sqrt{m_{i}^{2}+ {\check{\vec{\kappa}}}%
_{i}^{2}(\tau )}}}}\Big] =  \nonumber \\
&=&-2 Q_i {\vec {\bar S}}_{i\xi}(\tau ) \cdot \Big[ \check{\vec{B}}(\tau ,%
\vec{\eta}_{i}(\tau ))+ {\frac{{\ \check{\vec{\pi}}_{\perp }(\tau ,\vec{\eta}%
_{i}(\tau )) \times {\check {\vec \kappa}}_i(\tau )}}{{\ m_{i}+\sqrt{%
m_{i}^{2}+ {\check{\vec{\kappa}}}_{i}^{2}(\tau )} }}} \Big].  \label{V2}
\end{eqnarray}

Analogously the modified Coulomb potential may be written as

\begin{eqnarray}
&&\sum_{i\neq j}\frac{Q_{i}Q_{j}}{4\pi \mid \vec{\eta}_{i}(\tau )-\vec{\eta}%
_{j}(\tau )\mid }\Big[ 1 +  \nonumber \\
&&+ {\frac{{{\check {\vec \kappa}}_i(\tau )\cdot ({\vec \eta}_i(\tau )- {%
\vec \eta}_j(\tau ))\times {\vec {\bar S}}_{i\xi}(\tau )}}{{\mid \vec{\eta}%
_{i}(\tau )-\vec{\eta}_{j}(\tau )\mid ^{2}(m_i+\sqrt{m_{i}^{2}+\check{\vec{%
\kappa}}_{i}^{2}(\tau )})\sqrt{m_{i}^{2}+\check{\vec{\kappa}}_{i}^{2}(\tau )}
}}}\Big] .  \label{V3}
\end{eqnarray}

By using the Hamilton-Dirac equations of motion, $\dot{O}\,\stackrel{\circ }{%
=}\{O,{\hat{H}}_{D}\}$, we obtain (where '$\stackrel{\circ }{=}$' means
evaluated on the solutions of the equations of motion)

\begin{eqnarray}
\,\dot{\vec{\eta}}_{l}(\tau ) &\stackrel{\circ }{=}&\frac{\check{\vec{\kappa}%
}_{l}(\tau )-Q_{l}\check{\vec{A}}_{\perp }(\tau ,\vec{\eta}_{l}(\tau ))+%
\frac{1}{2}\frac{\partial {\cal R}_{l}(\tau )}{\partial \check{\vec{\kappa}}%
_{l}}}{\sqrt{m_{l}^{2}+{\cal R}_{l}+(\check{\vec{\kappa}}_{l}(\tau )-Q_{l}%
\check{\vec{A}}_{\perp }(\tau ,\vec{\eta}_{l}(\tau )))^{2}}}-\vec{\lambda}%
(\tau )-  \nonumber \\
&-&i\sum_{j\neq l}^{N}\frac{Q_{l}Q_{j}\vec{\xi}_{l}(\tau )\,\vec{\xi}%
_{l}(\tau )\cdot (\vec{\eta}_{l}(\tau )-\vec{\eta}_{j}(\tau ))}{4\pi \mid
\vec{\eta}_{l}(\tau )-\vec{\eta}_{j}(\tau )\mid ^{3}(m_{l}+\sqrt{m_{l}^{2}+%
\check{\vec{\kappa}}_{l}^{2}(\tau )})\sqrt{m_{l}^{2}+\check{\vec{\kappa}}%
_{l}^{2}(\tau )}}+  \nonumber \\
&+&i\sum_{j\neq l}^{N}\frac{Q_{l}Q_{j}\check{\vec{\kappa}}_{l}(\tau )\cdot
\vec{\xi}_{l}(\tau )\,\vec{\xi}_{l}(\tau )\cdot (\vec{\eta}_{l}(\tau )-\vec{%
\eta}_{j}(\tau ))\check{\vec{\kappa}}_{l}(\tau )}{4\pi \mid \vec{\eta}%
_{l}(\tau )-\vec{\eta}_{j}(\tau )\mid ^{3}(m_{l}+\sqrt{m_{l}^{2}+\check{\vec{%
\kappa}}_{l}^{2}(\tau )})(m_{l}^{2}+\check{\vec{\kappa}}_{l}^{2}(\tau
))^{3/2}}+  \nonumber \\
&+&i\sum_{j\neq l}^{N}\frac{Q_{l}Q_{j}\check{\vec{\kappa}}_{l}(\tau )\cdot
\vec{\xi}_{l}(\tau )\,\vec{\xi}_{l}(\tau )\cdot (\vec{\eta}_{l}(\tau )-\vec{%
\eta}_{j}(\tau ))\check{\vec{\kappa}}_{l}}{4\pi \mid \vec{\eta}_{l}(\tau )-%
\vec{\eta}_{j}(\tau )\mid ^{3}(m_{l}+\sqrt{m_{l}^{2}+\check{\vec{\kappa}}%
_{l}^{2}(\tau )})^{2}(m_{l}^{2}+\check{\vec{\kappa}}_{l}^{2}(\tau ))}=
\nonumber \\
&&{}  \nonumber \\
&=&\frac{\check{\vec{\kappa}}_{l}(\tau )}{\sqrt{m_{l}^{2}+{\cal R}_{l}(\tau
)+(\check{\vec{\kappa}}_{l}(\tau )-Q_{l}\check{\vec{A}}_{\perp }(\tau ,\vec{%
\eta}_{l}(\tau )))^{2}}}+  \nonumber \\
&+&\frac{-Q_{l}\check{\vec{A}}_{\perp }(\tau ,\vec{\eta}_{l}(\tau ))+\frac{1%
}{2}\frac{\partial {\cal R}_{l}(\tau )}{\partial \check{\vec{\kappa}}_{l}}}{%
\sqrt{m_{l}^{2}+\check{\vec{\kappa}}_{l}^{2}(\tau )}}-\vec{\lambda}(\tau )-
\nonumber \\
&-&i\sum_{j\neq l}^{N}\frac{Q_{l}Q_{j}\vec{\xi}_{l}(\tau )\,\vec{\xi}%
_{l}(\tau )\cdot (\vec{\eta}_{l}(\tau )-\vec{\eta}_{j}(\tau ))}{4\pi \mid
\vec{\eta}_{l}(\tau )-\vec{\eta}_{j}(\tau )\mid ^{3}(m_{l}+\sqrt{m_{l}^{2}+%
\check{\vec{\kappa}}_{l}^{2}(\tau )})\sqrt{m_{l}^{2}+\check{\vec{\kappa}}%
_{l}^{2}(\tau )}}+  \nonumber \\
&+&i\sum_{j\neq l}^{N}\frac{Q_{l}Q_{j}\check{\vec{\kappa}}_{l}(\tau )\cdot
\vec{\xi}_{l}(\tau )\,\vec{\xi}_{l}(\tau )\cdot (\vec{\eta}_{l}(\tau )-\vec{%
\eta}_{j}(\tau ))\check{\vec{\kappa}}_{l}(\tau )}{4\pi \mid \vec{\eta}%
_{l}(\tau )-\vec{\eta}_{j}(\tau )\mid ^{3}(m_{l}+\sqrt{m_{l}^{2}+\check{\vec{%
\kappa}}_{l}^{2}(\tau )})(m_{l}^{2}+\check{\vec{\kappa}}_{l}^{2}(\tau
))^{3/2}}+  \nonumber \\
&+&i\sum_{j\neq l}^{N}\frac{Q_{l}Q_{j}\check{\vec{\kappa}}_{l}(\tau )\cdot
\vec{\xi}_{l}(\tau )\,\vec{\xi}_{l}(\tau )\cdot (\vec{\eta}_{l}(\tau )-\vec{%
\eta}_{j}(\tau ))\check{\vec{\kappa}}_{l}(\tau )}{4\pi \mid \vec{\eta}%
_{l}(\tau )-\vec{\eta}_{j}(\tau )\mid ^{3}(m_{l}+\sqrt{m_{l}^{2}+\check{\vec{%
\kappa}}_{l}^{2}(\tau )})^{2}(m_{l}^{2}+\check{\vec{\kappa}}_{l}^{2}(\tau ))}%
,  \nonumber \\
&&{}  \nonumber \\
&\Downarrow &  \nonumber \\
&&{}  \nonumber \\
Q_{i}{\dot{\vec{\eta}}}_{i}(\tau ) &\stackrel{\circ }{=}&Q_{i}{\frac{{\ {%
\check{\vec{\kappa}}}_{i}(\tau )}}{\sqrt{m_{i}^{2}+{\check{\vec{\kappa}}}%
_{i}^{2}(\tau )}}},\qquad Q_{i}\sqrt{1-{\dot{\vec{\eta}}}_{i}^{2}(\tau )}\,%
\stackrel{\circ }{=}\,Q_{i}{\frac{{|{\check{\vec{\kappa}}}_{i}(\tau )|}}{%
\sqrt{m_{i}^{2}+{\check{\vec{\kappa}}}_{i}^{2}(\tau )}}},  \label{V4}
\end{eqnarray}

\noindent and

\begin{eqnarray}
\dot{\check{\vec{\kappa}}}_{l}(\tau ) &=&-\sum_{k\neq l}\frac{Q_{l}Q_{k}({%
\vec{\eta}}_{l}(\tau )-{\vec{\eta}}_{k}(\tau ))}{4\pi \mid \vec{\eta}%
_{l}(\tau )-\vec{\eta}_{k}(\tau )\mid ^{3}} -  \nonumber \\
&-&\sum_{j\neq l}i\frac{Q_{l}Q_{j}\check{\vec{\kappa}}_{l}(\tau )\cdot \vec{%
\xi}_{l}(\tau )\, \vec{\xi}_{l}(\tau )\cdot \Big[ |\vec{\eta}_{l}(\tau )-%
\vec{\eta}_{j}(\tau )|^{2}I -3(\vec{\eta}_{l}(\tau )-\vec{\eta}_{j}(\tau ))(%
\vec{\eta}_{l}(\tau )-\vec{\eta}_{j}(\tau ))\Big]}{4\pi \mid \vec{\eta}%
_{l}(\tau )-\vec{\eta}_{j}(\tau )\mid ^{5}(m_l+\sqrt{m_{l}^{2}+\check{\vec{%
\kappa}}_{l}^{2}(\tau )})\sqrt{m_{l}^{2}+\check{\vec{\kappa}}_{l}^{2}(\tau )}%
} +  \nonumber \\
&+&\frac{1}{\sqrt{m_{l}^{2}+{\cal R}_{l}(\tau )+(\check{\vec{\kappa}}%
_{l}(\tau )-Q_{l}\check{\vec{A}}_{\perp }(\tau ,\vec{\eta}_{l}(\tau )))^{2}}}
\nonumber \\
&& \Big[ (\check{\vec{\kappa}}_{l}(\tau )-Q_{l}\check{\vec{A}}_{\perp }(\tau
,\vec{\eta}_{l}(\tau ))_{u}Q_{l}{\frac{{\partial }}{{\partial {\vec{\eta}}%
_{l}}}}{\check{A}}_{\perp }^{u}(\tau ,\vec{\eta}_{i}(\tau ))+\frac{ 1}{2}%
\frac{\partial {\cal R}_{l}(\tau )}{\partial \vec{\eta}_{l}}\Big] =
\nonumber \\
&&{}  \nonumber \\
&=&-\sum_{k\neq l}\frac{Q_{l}Q_{k}({\vec{\eta}}_{l}(\tau )-{\vec{\eta}}%
_{k}(\tau ))}{4\pi \mid \vec{\eta}_{l}(\tau )-\vec{\eta}_{k}(\tau )\mid ^{3}}%
+\frac{Q_l\check{\vec{\kappa}}_{l}(\tau )_{u}{\frac{{\partial }}{\partial {\
{\vec{\eta}}_{l}}}\check{A}}_{\perp }^{u}(\tau ,\vec{\eta}_{i}(\tau ))}{%
\sqrt{m_{l}^{2}+\check{\vec{\kappa}}_{l}^{2}(\tau )}}+\frac{\frac{1}{2}\frac{%
\partial {\cal R}_{l}(\tau )}{\partial \vec{\eta}_{l}}}{\sqrt{m_{l}^{2}+%
\check{\vec{\kappa}}_{l}^{2}(\tau )}} -  \nonumber \\
&-&Q_l\sum_{j\neq l}i\frac{Q_{j}\check{\vec{\kappa}}_{l}\cdot \vec{\xi}_{l}%
\vec{\xi}_{l}\cdot \Big[ |\vec{\eta}_{l}(\tau )-\vec{\eta}_{j}(\tau
)|^{2}I-3(\vec{\eta}_{l}(\tau )-\vec{\eta}_{j}(\tau ))(\vec{\eta}_{l}(\tau )-%
\vec{\eta}_{j}(\tau ))\Big]}{4\pi \mid \vec{\eta}_{l}(\tau )-\vec{\eta}%
_{j}(\tau )\mid ^{5}(m_l+\sqrt{m_{l}^{2}+\check{\vec{\kappa}}_{l}^{2}(\tau )}%
) \sqrt{ m_{l}^{2}+\check{\vec{\kappa}}_{l}^{2}(\tau )}}.  \label{V5}
\end{eqnarray}

To these Hamilton-Dirac equations we must add the rest-frame condition

\begin{equation}
\sum_{i=1}^N{\check{\vec{\kappa}}}_i(\tau )+\int d^{3}\sigma \lbrack {\check{%
\vec{\pi}} }_{\perp }\times {\check{\vec{B}}}](\tau ,\vec{\sigma})\approx 0,
\label{V6}
\end{equation}

\noindent whose natural gauge fixing, implying $\vec{\lambda}(\tau )=0$, is $%
{\vec{{\cal K}}}_{(int)}\approx 0$, as mentionied in the previous Section.

The equations of motion for the Grassmann variables and the spins are

\begin{eqnarray}
{\dot{\vec{\xi}}_{l}}(\tau )&\stackrel{\circ }{=}&\frac{-i}{\sqrt{m_{l}^{2}+%
\check{\vec{\kappa}}_{l}^{2}(\tau )}}\frac{1}{2}\frac{\partial {\cal R}%
_{l}(\tau )}{\partial \vec{\xi}_{l}}-  \nonumber \\
&&-Q_l\sum_{j\neq l}\frac{Q_{j}[\check{\vec{\kappa}}_{l}\vec{\xi}_{l}\cdot (%
\vec{\eta}_{l}(\tau )-\vec{\eta}_{j}(\tau ))]-\check{\vec{\kappa}}_{l}\cdot
\vec{\xi}_{l}(\vec{\eta}_{l}(\tau )-\vec{\eta}_{j}(\tau ))}{4\pi \mid \vec{%
\eta}_{l}(\tau )-\vec{\eta}_{j}(\tau )\mid ^{3}(m_l+\sqrt{m_{l}^{2}+\check{%
\vec{\kappa}}_{l}^{2}(\tau )})\sqrt{m_{l}^{2}+\check{\vec{\kappa}}%
_{l}^{2}(\tau )}},  \nonumber \\
&&{}  \nonumber \\
{\dot {\bar S}}^r_{i\xi}(\tau ) &\stackrel{\circ }{=}& \Omega^{rs}_i(\tau
)\, {\bar S}^s_{i\xi}(\tau ),  \nonumber \\
{\dot {\bar S}}^r_{\xi}(\tau ) &=& \sum_{i=1}^N {\dot {\bar S}}%
^r_{i\xi}(\tau )\, \stackrel{\circ }{=}\, \sum_{i=1}^N \Omega^{rs}_i(\tau ) {%
\bar S}^s_{i\xi}(\tau ),  \nonumber \\
&&{}  \nonumber \\
\Omega^{rs}_i(\tau ) &=& {\frac{{Q_i}}{{\sqrt{m_i^2+{\check {\vec \kappa}}%
^2_i} }}} \Big[ \epsilon^{rst} {\check {\vec B}}_{\perp}(\tau ,{\vec \eta}%
_i(\tau ))+  \nonumber \\
&+&\sum_{j\neq i}\frac{Q_{j}}{(m_{i}+\sqrt{m_{i}+\vec{\kappa}_{i}^{2}})(%
\sqrt{m_{i}+\vec{\kappa}_{i}^{2}})}  \nonumber \\
&&\Big( {\check \kappa}_i^r(\tau ) ({\check \pi}^s_{\perp}(\tau ,{\vec \eta}%
_i(\tau ))+ {\frac{{Q_j}}{{4\pi \eta _{ij}^{3}}}} \eta^s_{ij}(\tau ))- {%
\check \kappa}_i^s(\tau ) ({\check \pi}^r_{\perp}(\tau ,{\vec \eta}_i(\tau
))+ {\frac{{Q_j}}{{4\pi \eta _{ij}^{3}}}} \eta^r_{ij}(\tau ))\Big) \Big].
\label{V7}
\end{eqnarray}

By using

\begin{equation}
\{{\check{A}}_{\perp }^{r}(\tau ,\vec{\sigma}),\check{\vec{\pi}}_{\perp
s}(\tau ,\vec{\eta}_{i}(\tau ))\}=\delta _{rs}\delta ^{3}(\vec{\sigma}-\vec{%
\eta}_{i}\left( \tau )\right) ,  \label{V8}
\end{equation}

\noindent the Hamilton-Dirac equations for the transverse fields are [$%
P^{rs}_{\perp}(\vec \sigma )=\delta^{rs}+{\frac{{\partial^r\partial^s}}{{%
\triangle}}}$ is the transverse projector]

\begin{eqnarray}
\dot{{\check{A}}}_{\perp r}(\tau ,\vec{\sigma})\,&\stackrel{\circ }{=}&-{\
\check{\pi}}_{\perp r}(\tau ,\vec{\sigma})-\sum_{i=1}^{N}P_{\perp rs}(\vec{%
\sigma}) \frac{iQ_i\check{\vec{\kappa}}_{i}(\tau )\cdot \vec{\xi}_{i}(\tau ){%
\xi}_{i}^{s}(\tau ) \delta^{3}(\vec{\sigma}-\vec{\eta}_{i}\left( \tau
)\right) }{(m_i+\sqrt{m_{i}^{2}+\check{\vec{\kappa}}_{i}^{2}(\tau )}) \sqrt{%
m_{i}^{2}+\check{\vec{\kappa}}_{i}^{2}(\tau )}}-  \nonumber \\
&-&[\vec{\lambda}(\tau )\cdot \vec{\partial}]{\check{A}}_{\perp r}(\tau ,%
\vec{\sigma}),  \nonumber \\
&&{}  \nonumber \\
\dot{{\check{\pi}}}_{\perp }^{r}(\tau ,\vec{\sigma})\, &\stackrel{\circ }{=}%
&\,\Delta {\check{A}}_{\perp }^{r}(\tau ,\vec{\sigma})-[\vec{\lambda}(\tau
)\cdot \vec{\partial}]{\check{\pi}}_{\perp }^{r}(\tau ,\vec{\sigma})
\nonumber \\
&+&\sum_{i}\Big[i\frac{Q_{i}\xi _{i}^{r}(\tau )\, \vec{\xi}_{i}(\tau )\cdot
\vec{\partial}\delta ^{3}(\vec{\sigma}-\vec{\eta}_{i}(\tau ))}{\sqrt{%
m_{i}^{2}+\check{\vec{\kappa}}_{i}^{2}(\tau )}}-Q_{i}P_{\perp }^{rs}(\vec{%
\sigma})\dot{\eta}_{i}^{s}(\tau )\delta ^{3}(\vec{\sigma}-\vec{\eta}%
_{i}(\tau ))\Big]=  \nonumber \\
&&{}  \nonumber \\
&=&-\frac{\partial ^{2}{\check{A}}_{\perp }^{r}}{\partial \tau ^{2}}%
-\sum_{i=1}^{N}i Q_i P_{\perp rs}(\vec{\sigma}) \Big[ \dot{\check{\vec{\kappa%
}}}_{l}(\tau )\cdot \Big( \frac{\bar{1}}{(m_i+\sqrt{m_{i}^{2}+\check{\vec{%
\kappa}}_{i}^{2}(\tau )})\sqrt{ m_{i}^{2}+\check{\vec{\kappa}}_{i}^{2}(\tau )%
}}-  \nonumber \\
&-&\frac{\check{\vec{\kappa}}_{i}(\tau )\, \check{\vec{\kappa}}_{i}(\tau )}{%
(m_i+\sqrt{m_{i}^{2}+\check{\vec{\kappa}}_{i}^{2}(\tau )})(\sqrt{m_{i}^{2}+%
\check{\vec{\kappa}}_{i}^{2}(\tau )})^{3}}-  \nonumber \\
&-&\frac{\check{\vec{\kappa}}_{i}(\tau )\, \check{\vec{\kappa}}_{i}(\tau )} {
(m_i+\sqrt{m_{i}^{2}+ \check{\vec{\kappa}}_{i}^{2}(\tau )})^{2}(\sqrt{
m_{i}^{2}+\check{\vec{\kappa}}_{i}^{2}(\tau )})^{2}}\Big) \cdot \vec{\xi}%
_{i}(\tau ){\xi}_{i}^{s}(\tau ) \delta ^{3}(\vec{\sigma}-\vec{\eta}%
_{i}\left( \tau )\right) +  \nonumber \\
&+&\frac{\check{\vec{\kappa}}_{i}(\tau )\cdot {\dot{\vec{\xi}}}_{i}(\tau ){%
\xi}_{i}^{s}(\tau )\delta^{3}(\vec{ \sigma}-\vec{\eta} _{i}\left( \tau
)\right) }{(m_i+\sqrt{m_{i}^{2}+\check{\vec{\kappa}}_{i}^{2}(\tau )} )\sqrt{%
m_{i}^{2}+\check{\vec{\kappa}} _{i}^{2}(\tau )}}+  \nonumber \\
&+&\frac{\check{\vec{ \kappa}}_{i}(\tau )\cdot {\vec{\xi}} _{i}(\tau ){\dot{%
\xi}}_{i}^{s}(\tau ) \delta ^{3}(\vec{\sigma}-\vec{\eta}_{i}\left( \tau
)\right) }{(m_i+\sqrt{m_{i}^{2}+ \check{\vec{\kappa}}_{i}^{2}(\tau )})\sqrt{%
m_{i}^{2}+\check{\vec{\kappa}} _{i}^{2}(\tau )}}-  \nonumber \\
&-&\frac{\check{\vec{\kappa}}_{i}(\tau )\cdot \vec{\xi}_{i}(\tau ){\ \xi}%
_{i}^{s}(\tau ) \dot{\vec{\eta}}_{l}(\tau )\cdot \vec{\partial}_{\sigma
}\delta ^{3}(\vec{\sigma}-\vec{\eta}_{i}\left( \tau )\right) }{(m_i+\sqrt{%
m_{i}^{2}+\check{\vec{\kappa}}_{i}^{2}(\tau )}) \sqrt{m_{i}^{2}+\check{\vec{%
\kappa}}_{i}^{2}(\tau )}}\Big] .  \label{V9}
\end{eqnarray}

When we eliminate $\dot{\check{\vec{\kappa}}}_{l}(\tau ),{\ \dot{\vec{\xi}}}%
_{i}(\tau )$ with Eqs.(\ref{V5}), (\ref{V7}), by virtue of the Grassmann
truncations due to $Q_{i}^{2}=0$ we arrive at the following wave equation
for the transverse potential

\begin{eqnarray}
&&\frac{\partial ^{2}{\check{A}}_{\perp }^{r}(\tau ,\vec{\sigma})}{\partial
\tau ^{2}}+\,\Delta {\check{A}}_{\perp }^{r}(\tau ,\vec{\sigma})-[\vec{%
\lambda}(\tau )\cdot \vec{\partial}]{\check{\pi}}_{\perp }^{r}(\tau ,\vec{%
\sigma})=  \nonumber \\
&=&\sum_{i}Q_{i}\Big[P_{\perp }^{rs}(\vec{\sigma})\dot{\eta}_{i}^{s}(\tau
)\delta ^{3}(\vec{\sigma}-\vec{\eta}_{i}(\tau ))-  \nonumber \\
&-&i\frac{\xi _{i}^{r}(\tau )\vec{\xi}_{i}(\tau )\cdot \vec{\partial}\delta
^{3}(\vec{\sigma}-\vec{\eta}_{i}(\tau ))}{\sqrt{m_{i}^{2}+\check{\vec{\kappa}%
}_{i}^{2}(\tau )}}+  \nonumber \\
&+&P_{\perp }^{rs}(\vec{\sigma})i\frac{\check{\vec{\kappa}}_{i}(\tau )\cdot
\vec{\xi}_{i}(\tau ){\ \xi }_{i}^{s}(\tau )\dot{\vec{\eta}}_{i}(\tau )\cdot
\vec{\partial}_{\sigma }\delta ^{3}(\vec{\sigma}-\vec{\eta}_{i}\left( \tau
)\right) }{(m_{i}+\sqrt{m_{i}^{2}+\check{\vec{\kappa}}_{i}^{2}(\tau )})\sqrt{%
m_{i}^{2}+\check{\vec{\kappa}}_{i}^{2}(\tau )}}\Big].  \label{V10}
\end{eqnarray}
Expressed in terms of particle velocities the right hand side becomes

\begin{eqnarray}
&=&\sum_{i}Q_i\Big[P_{\perp }^{rs}(\vec{\sigma})\dot{\eta}_{i}^{s}(\tau
)\delta ^{3}(\vec{\sigma}-\vec{\eta}_{i}(\tau ))-  \nonumber \\
&-&i\frac{\sqrt{(1-\dot{\vec{\eta}}_{i}^{2}}\xi _{i}^{r}(\tau )\vec{\xi}%
_{i}(\tau )\cdot \vec{\partial}\delta^{3}(\vec{\sigma} -\vec{\eta}_{i}(\tau
))}{m_{i}}+  \nonumber \\
&+& P_{\perp }^{rs}(\vec{\sigma}) i\frac{\sqrt{1-\dot{\vec{\eta}}%
_{i}^{2}(\tau )}\,\, \dot{\vec{\eta}} _{i}(\tau )\cdot \vec{ \xi}_{i}(\tau ){%
\xi}_{i}^{s}(\tau ) \dot{\vec{\eta}}_{i}(\tau )\cdot \vec{\partial}_{\sigma
}\delta ^{3}(\vec{\sigma}-\vec{\eta}_{i}\left( \tau )\right) }{m_{i}(\sqrt{1-%
\dot{\vec{\eta}}_{i}^{2}(\tau )}+1)} \Big].  \label{V11}
\end{eqnarray}

Let us now choose a gauge (the natural one is ${\vec {{\cal K}}}%
_{(int)}\approx 0$) implying $\vec \lambda (\tau )=0$. In this gauge we
obtain

\begin{eqnarray}
\Box A_{\perp }^{r}(\tau ,\vec{\sigma}) &=&(\frac{\partial ^{2}}{\partial
\tau ^{2}}-\vec{\partial}^{2})A_{\perp }^{r}(\tau ,\vec{\sigma})=  \nonumber
\\
&=&\sum_{i=1}^{N}Q_{i}\Big[\frac{i\dot{\vec{\eta}}_{i}(\tau )\cdot \vec{%
\partial}_{\sigma }\,\dot{\vec{\eta}}_{i}(\tau )\cdot \vec{\xi}_{i}(\tau )\,%
\vec{\xi}_{i}^{s}(\tau )\sqrt{1-\dot{\vec{\eta}}_{i}^{2}(\tau )}}{(1+\sqrt{1-%
\dot{\vec{\eta}}_{i}^{2}(\tau )})m_{i}}-  \nonumber \\
&&-\frac{i\xi _{i}^{s}(\tau )\vec{\xi}_{i}(\tau )\cdot \vec{\partial}%
_{\sigma }\sqrt{1-\dot{\vec{\eta}}_{i}^{2}(\tau )}}{m_{i}}+\dot{\eta}%
_{i}^{s}(\tau )\Big](\delta ^{sr}+\frac{\partial ^{s}\partial ^{r}}{\Delta }%
)\delta ^{3}(\vec{ \sigma}-\vec{\eta}_{i}(\tau ))=  \nonumber \\
&:&= j^r_{\perp}(\tau ,\vec \sigma )=P_{\perp }^{rs}(\vec{\sigma}%
)\sum_{i=1}^{N}Q_{i}{\bf V}_{i}^{s}(\tau )\delta ^{3}(\vec{\sigma}-\vec{\eta}%
_{i}(\tau )).  \label{V12}
\end{eqnarray}

\noindent We see that the effective current ${\vec{j}}_{\perp }(\tau ,\vec{%
\sigma})$ has the structure of a {\it pole-dipole}. It is convenient to
define an effective particle velocity operator (using boldface) ${\vec{{\bf V%
}}}_{i}$

\begin{eqnarray}
{\bf V}_{i}^{s}(\tau )&=&\dot{\eta}_{i}^{s}(\tau )-\frac{i\sqrt{1-\dot{\vec{%
\eta}} _{i}^{2}(\tau )}\xi_{i}^{s}(\tau ) \vec{\xi}_{i}(\tau )\cdot \vec{%
\partial}_{\sigma }} {m_{i}}+\frac{i\sqrt{1-\dot{\vec{\eta}} _{i}^{2}(\tau )}%
\dot{\vec{\eta}}_{i}(\tau )\cdot \vec{\xi}_{i}(\tau ){\xi} _{i}^{s}(\tau )%
\dot{\vec{\eta}}_{i}(\tau )\cdot \vec{\partial} _{\sigma }}{m_i(1+\sqrt{1-%
\dot{\vec{\eta}}_{i}^{2}(\tau )})},  \nonumber \\
&&{}  \nonumber \\
&&\Downarrow  \nonumber \\
&&{}  \nonumber \\
Q_i{\vec {{\bf V}}}_i(\tau )&\stackrel{\circ }{=} & {\frac{{Q_i}}{{\ \sqrt{%
m_i^2+{\check {\vec \kappa}}^2_i(\tau )} }}} \Big[ {\check {\vec \kappa}}%
_i(\tau )- {\frac{{|{\check {\vec \kappa}}_i(\tau )|}}{{m_i}}} {\vec {\bar S}%
}_{i\xi}(\tau )\times {\vec \partial}_{\sigma} +  \nonumber \\
&+& {\frac{{|{\check {\vec \kappa}}_i(\tau )|\,\, {\vec {\bar S}}%
_{i\xi}(\tau )\times {\check {\vec \kappa}}_i(\tau )\,\, {\check {\vec \kappa%
}}_i(\tau )\cdot {\vec \partial}_{\sigma}}}{{m_i \sqrt{m_i^2+{\check {\vec %
\kappa}}^2_i(\tau )} ( |{\check {\vec \kappa}}_i(\tau )|+ \sqrt{m_i^2+{%
\check {\vec \kappa}}^2_i(\tau )}) }}} \Big].  \label{V13}
\end{eqnarray}

\noindent In the last line the operators ${\vec {{\bf V}}}_i$ have been
reexpressed in terms of the momenta by using Eqs.(\ref{V4}).

We point out that defining $\vec{\beta}_{i}(\tau )=\dot{\vec{\eta}}_{i}(\tau
)={\frac{{d{\vec{\eta}}_{i}(\tau )}}{{d\tau }}}={\frac{1}{c}}{\frac{{d{\vec{
\eta}}_{i}^{^{\prime }}(t)}}{{dt}}}$ \footnote{$\tau =ct$, ${\vec{\eta}}
^{^{\prime}}_{i}(t)={\vec{\eta}}_{i}(\tau )$; though using everywhere $c=1$
, we have momentarily re-introduced it.} and $\vec{\beta}_{i}^{(h)}=d^{h}%
\vec{ \beta}_{i}/d\tau ^{h}$ and writing the particle equations of motion as
(no sum over $i$)

\begin{equation}
\frac{d}{d\tau }(m_{i}\frac{\vec{\beta}_{i}(\tau )}{\sqrt{1-\vec{\beta}%
_{i}^{2}(\tau )}})\,=\frac{m_{i}}{\sqrt{1-\vec{\beta}_{i}^{2}(\tau )}}[\vec{%
\beta}_{i}^{(1)}+\vec{\beta}_{i}\frac{\vec{\beta}_{i}^{(1)}\cdot \vec{\beta}%
_{i}}{1-\vec{\beta}_{i}^{2}(\tau )}]\stackrel{\circ }{=}{Q}_{i}\vec{F}_{i},
\label{V14}
\end{equation}

\noindent we obtain

\begin{equation}
m_{i}\frac{\vec{\beta}_{i}^{(1)}\cdot \vec{\beta}_{i}}{(1-\vec{\beta}%
_{i}^{2}(\tau ))^{3/2}}\stackrel{\circ }{=}{Q}_{i}\vec{\beta}_{i}\cdot \vec{F%
}_{i},  \label{V15}
\end{equation}

\noindent so that

\begin{equation}
\vec{\beta}_{i}^{(1)}\stackrel{\circ }{=}\frac{\sqrt{1-\vec{\beta}%
_{i}^{2}(\tau )}}{m_{i}}Q_{i}(\vec{F}_{i}-\vec{\beta}_{i}\vec{\beta}%
_{i}\cdot \vec{F}_{i}).  \label{V16}
\end{equation}

Thus in general we will have for every $h\geq 1$

\begin{equation}
\vec{\beta}_{i}^{(h)}\stackrel{\circ }{=}{Q}_{i}\vec{G}_{i},  \label{V17}
\end{equation}

\noindent so that using the Grassmann property of the charges

\begin{equation}
Q_{i}\vec{\beta}_{i}^{(h)}\stackrel{\circ }{=}{0},\quad \quad h\geq 1.
\label{V18}
\end{equation}

This will lead to important simplifications later, allowing us to drop
acceleration dependent terms in the force.

Due to the projector $P_{\perp }^{rs}(\vec{\sigma})$ required by the
rest-frame radiation gauge, the sources of the transverse (Wigner spin 1)
vector potential becomes non-local and one has a system of
integrodifferential equations (like the equations generated by
Fokker-Tetrode-type actions) for which it is not known how to define an
initial value problem.

Let us end this Section with some comments on the equations of motion of the
spin. Eq.(\ref{V12}) shows that, besides the standard term for scalar
particles \cite{lu6}, the particle current contains also a dipole term $%
P_{\perp}^{rs}(\vec \sigma )\sum_{i=1}^N {\frac{{|{\check {\vec \kappa}}%
_i(\tau )|}}{{m_i}}} \Big({-\frac{{\ ({\vec {\bar S}}_{i\xi} \times \vec %
\partial )^s}}{{\sqrt{m_i^2+{\check {\vec \kappa}} ^2_i} }}}+ {\frac{{\ {%
\vec {\bar S}}_{i\xi}(\tau )\times {\check {\vec \kappa}}_i(\tau )\,\, {%
\check {\vec \kappa}}_i(\tau )\cdot \vec \partial}}{{\sqrt{m_i^2+{\check {%
\vec \kappa}}^2_i(\tau )} ( |{\check {\vec \kappa}}_i(\tau )|+ \sqrt{m_i^2+{%
\check {\vec \kappa}}^2_i(\tau )}) }}} \Big) \delta^3(\vec \sigma -{\vec \eta%
}_i(\tau ))$ in accord with the fact that the spinning particle has a
pole-dipole structure \cite{g1} according to Papapetrou's classification\cite
{g2} (see Refs.\cite{g3} for other pole-dipole models and Ref.\cite{g4} for
their influence on the energy momentum tensor of action-at-a-distance
models).

For N=1 (${\vec \eta}_i \mapsto \vec \eta$, $m_i \mapsto m$,...) we have
\footnote{%
See Eq.(\ref{III42}) for the definition of $\Sigma_{\mu}={\frac{1}{2}}
\epsilon_{\mu\nu\rho\sigma}P^{\nu}S^{\rho\sigma}$ with ${\dot P}^{\mu}\not=
0 $.}:

i) ${\bar S}_{\xi}^{rs}=\epsilon^{rst}{\bar S}_{\xi}^t=-i \xi^r\xi^s$, ${%
\bar S}^{rs}_s\approx \int d^3\sigma [(\sigma^r-\eta^r)({\vec \pi}_{\perp}
\times \vec B)^s-(\sigma^s-\eta^s)({\vec \pi}_{\perp}\times \vec B)^r](\tau ,%
\vec \sigma )$;

ii) $\Sigma_{s \mu}={\frac{1}{2}}\epsilon_{\mu\nu\rho\sigma}p_s ^{\nu}{%
\tilde S}_{\xi}^{\rho\sigma}=({\frac{1}{2}}\epsilon_{\mu\nu ij}\delta^{ir}
\delta^{js}-\epsilon_{\mu\nu oi}{\frac{{\delta^{ir}p^s_s}}{{\
p^o_s+\epsilon_s}}}) p^{\nu}_s {\bar S}^{rs}_{\xi}$.

The Bargmann-Michel-Telegdi equation\cite{bmt} for the spin part of the
Pauli-Lubanski 4-vector of Eq.(\ref{III42})

\begin{equation}
{\dot \Sigma}_{\mu}\, \stackrel{ \circ }{=}\, {\frac{e}{m}} F_{\mu\nu}
\Sigma^{\nu},  \label{V19}
\end{equation}

\noindent for a spinning particle in an external electromagnetic field is
replaced by the following equation in the canonical realization ( \ref{III28}%
) of the {\it external} Poincar\'e group in the rest-frame instant form for
the isolated system of a positive-energy spinning particle plus the
electromagnetic field (${\dot p}_s^{\mu}=0$)

\begin{eqnarray}
{\dot{\Sigma}}_{s\mu }(\tau ) &=&({\frac{1}{2}}\epsilon _{\mu \nu ij}\delta
^{ir}\delta ^{js}-\epsilon _{\mu \nu oi}{\frac{{\delta ^{ir}p_{s}^{s}}}{{%
p_{s}^{o}+\epsilon _{s}}}})p_{s}^{\nu }\epsilon ^{rst}{\dot{\bar{S}}}_{\xi
}^{t}(\tau )\,\stackrel{\circ }{=}  \nonumber \\
&\stackrel{\circ }{=}\,& ({\frac{1}{2}}\epsilon _{\mu \nu ij}\delta
^{ir}\delta^{js}-\epsilon _{\mu \nu oi}{\frac{{\delta ^{ir}p_{s}^{s}}}{{%
p_{s}^{o}+\epsilon _{s}}}})p_{s}^{\nu } \epsilon^{rst} \Omega^{tn}(\tau ) {%
\bar S}^n_{\xi}(\tau ).  \label{V20}
\end{eqnarray}

\noindent In the last line we used Eqs.(\ref{V7}) for the spin.

\vfill\eject

\section{The Semi-Classical Lienard-Wiechert Solution, the Second Class
Constraints and the Reduced Phase Space.}

In this Section we study the retarded, advanced and symmetric
Lienard-Wiechert solutions of Eqs.(\ref{V12}). After an equal time
development of the delay, we show that, by using the particle equations of
motion, the higher order accelerations disappear by virtue of $Q_{i}^{2}=0$.
Therefore, at the semi-classical level there is a {\it unique}
Lienard-Wiechert solution. Then we obtain the phase space expression of the
semi-classical solution. This allows to eliminate the radiation field
degrees of freedom by adding second class constraints and to arrive at a
reduced phase space containing only particle degrees of freedom. The result
is a canonical basis of this reduced phase space.

To simplify the notation from now on we shall denote the Dirac observables ${%
\check {\vec \kappa}}_i$, ${\check {\vec A}}_{\perp}$,... as ${\vec \kappa}%
_i $, ${\vec A}_{\perp}$,...

\subsection{The Lienard-Wiechert Solutions, the Equal Time Expansion of the
Delay and the Semi-Classical Approximation.}

The symmetric solution of Eq.(\ref{V12}) is

\begin{eqnarray}
A_{\perp S}^{r}(\tau ,\vec{\sigma}) &=&\frac{1}{2}[A_{\perp +}^{r}+A_{\perp
-}^{r}](\tau ,\vec{\sigma})=  \nonumber \\
&=&{\frac{1}{2}}\sum_{i=1}^{N}{\frac{{Q_{i}}}{{2\pi }}}\int d\tau
_{1}d^{3}\sigma _{1}P_{\perp }^{rs}(\vec{\sigma} )\,{\bf V}_{i}^{s}(\tau
_{1}) [\theta (\tau -\tau _{1})+\theta (\tau -\tau _{1})]  \nonumber \\
&&\delta \lbrack (\tau -\tau _{1})^{2}-(\vec{\sigma}-{\vec{\sigma}}%
_{1})^{2}]\delta ^{3}({\vec{\sigma}}_{1}-{\vec{\eta}}_{i}(\tau _{1}))=
\nonumber \\
&&{}  \nonumber \\
&=&\sum_{i=1}^{N}\frac{Q_{i}}{2\pi c}\int dt_{1} P_{\perp }^{rs}(\vec{\sigma}%
){\bf V}_{i}^{s}(\tau _{1})\delta \lbrack (t-t_{1})^{2}-\frac{1}{c^{2}}(\vec{%
\sigma}-\vec{\eta}_{i}(ct_{1}))^{2}]:=  \nonumber \\
&:&=\sum_{i=1}^{N}Q_{i}A_{\perp \,Si}^{r}(\tau ,\vec{\sigma}),  \label{VI1}
\end{eqnarray}

\noindent in which we have put $\tau =ct$, $\vec{\beta}_{i}(\tau )=\dot{\vec{%
\eta}} _{i}(\tau )\,={\frac{1}{c}}{\frac{{d{\vec{\eta}}_{i}^{^{\prime }}(t)}%
}{{dt}}} $ and ${\vec{A}}_{\perp +}={\vec{A}}_{\perp RET}$ (${\vec{A}}%
_{\perp -}={\vec{A}}_{\perp ADV}$) for the retarded (advanced) solution. \
The equation for $t_{1}$ is $c^{2}(t-t_{1})^{2}=(\vec{\sigma}-\vec{\eta}%
_{i}(ct_{1}))^{2}$ with the two solutions being

\begin{eqnarray}
t_{i+}(\tau ,\vec \sigma )&=&{\frac{1}{c}} \tau _{i+}(\tau ,\vec{\sigma}) =t-%
\frac{1}{c}r_{i+}(\tau_{i+}(\tau , \vec{\sigma}),\vec{\sigma})={\frac{{\tau}%
}{c}} -T_{i+}(\tau ,\vec \sigma ),  \nonumber \\
t_{i-}(\tau ,\vec \sigma )&=&{\frac{1}{c}} \tau _{i-}(\tau ,\vec{\sigma}) =t
+\frac{1}{c}r_{i-}(\tau _{i-}(\tau , \vec{\sigma}),\vec{\sigma})={\frac{{\tau%
}}{c}} +T_{i-}(\tau ,\vec \sigma ),  \label{VI2}
\end{eqnarray}

\noindent for the retarded and for the advanced case respectively. The light
cone delta function is

\begin{eqnarray}
&&\delta \lbrack (\tau -\tau _{1})^{2}-(\vec{\sigma}-{\vec{\eta}}_{i}(\tau
_{1}))^{2}]={\frac{1}{{c^{2}}}}\delta \lbrack (t-t_{1})^{2}-\frac{1}{c^{2}}(%
\vec{\sigma}-\vec{\eta}_{i}(ct_{1}))^{2}]=  \nonumber \\
&=&\frac{\delta \lbrack \tau _{1-}-\tau _{i+}(\tau ,\vec{\sigma})]}{2|\tau
-\tau _{1}-\vec{\beta}_{i}(\tau _{1})\cdot (\vec{\sigma}-\vec{\eta}_{i}(\tau
_{1}))|}+\frac{\delta \lbrack \tau _{1-}-\tau _{i-}(\tau ,\vec{\sigma})]}{%
2|\tau -\tau _{1}-\vec{\beta}_{i}(\tau _{1})\cdot (\vec{\sigma}-\vec{\eta}%
_{i}(\tau _{1}))|}.  \label{VI3}
\end{eqnarray}

The relative space location between the field point and the retarded or
advanced particle position is

\begin{equation}
\vec{\sigma}-\vec{\eta}_{i}(\tau _{i\pm }(\tau ,\vec{\sigma}))=\vec{r}_{i\pm
}(\tau _{i\pm }(\tau ,\vec{\sigma}),\vec{\sigma})=r_{i\pm }(\tau _{i\pm
}(\tau ,\vec{\sigma}),\vec{\sigma})\hat{r}_{i\pm }(\tau _{i\pm }(\tau ,\vec{%
\sigma}),\vec{\sigma}),  \label{VI4}
\end{equation}

\noindent and its length is related to the time interval by

\begin{eqnarray}
&&r_{i\pm }(\tau _{i\pm }(\tau ,\vec{\sigma}),\vec{\sigma})=|\,\vec{\sigma}-{%
\vec{\eta}}_{i}(\tau _{i\pm }(\tau ,\vec{\sigma}))|=cT_{i\pm }(\tau ,\vec{%
\sigma})=|\tau -\tau _{i\pm }(\tau ,\vec{\sigma})|,  \nonumber \\
&&{}  \nonumber \\
&\Rightarrow &\quad \tau -\tau _{i\pm }(\tau ,\vec{\sigma})=\pm cT_{i\pm
}(\tau ,\vec{\sigma})=\pm r_{i\pm }(\tau _{i\pm }(\tau ,\vec{\sigma}),\vec{%
\sigma}).  \label{VI5}
\end{eqnarray}

The effective spatial interval is defined by

\begin{equation}
\rho _{i\pm }(\tau _{i\pm }(\tau ,\vec{\sigma}),\vec{\sigma})=r_{i\pm }(\tau
_{i\pm }(\tau ,\vec{\sigma}),\vec{\sigma})[1\mp \vec{\beta}_{i}(\tau _{i\pm
}(\tau ,\vec{\sigma}))\cdot \hat{r}_{i\pm }(\tau _{i\pm }(\tau ,\vec{\sigma}%
),\vec{\sigma})].  \label{VI6}
\end{equation}
In terms of these variables, the retarded, advanced and time symmetric
solutions are

\begin{eqnarray}
A_{\perp \pm }^{r}(\tau ,\vec{\sigma}) &=&P_{\perp}^{rs}(\vec{\sigma}%
)\sum_{i=1}^{N}{\frac{{Q_{i}}}{{\ 4\pi }}\ {\bf V}_{i}^{s}(\tau _{i\pm
}(\tau ,\vec{\sigma}))\frac{{1}}{{\rho _{i\pm }(\tau _{i\pm }(\tau ,\vec{%
\sigma }),\vec{\sigma})}}},  \nonumber \\
A_{\perp S}^{r}(\tau ,\vec{\sigma}) &=&\sum_{i=1}^{N}Q_{i}A_{\perp
\,Si}^{r}(\tau ,\vec{\sigma})=  \nonumber \\
&=& P_{\perp}^{rs}(\vec{\sigma}) \sum_{i=1}^{N}\frac{Q_{i}}{8\pi }[{\bf V}%
_{i}^{s}(\tau _{i+}(\tau ,\vec{ \sigma}))\frac{{1}}{\rho _{i+}(\tau
_{i+}(\tau ,\vec{\sigma}),\vec{\sigma})}  \nonumber \\
&&+{\ }{\bf V}_{i}^{s}(\tau _{i-}(\tau ,\vec{\sigma})) \frac{{1}}{\rho
_{i-}(\tau _{i-}(\tau ,\vec{\sigma}),\vec{\sigma })}].  \label{VI7}
\end{eqnarray}

We use the Smart-Wintner expansion \cite{sw,tet3,tet4}

\begin{eqnarray}
f(\tau _{i\pm }) &=&f(\tau \mp cT_{i\pm }(\tau _{i\pm }((\tau ,\vec{\sigma}),%
\vec{\sigma})))=f(\tau -[\pm r_{i\pm }(\tau _{i\pm }(\tau ,\vec{\sigma}),%
\vec{\sigma})])=  \nonumber \\
&=&f(\tau )+\sum_{k=1}^{\infty }\frac{(-)^{k}}{k!}\frac{d^{k-1}}{d\tau ^{k-1}%
}\left[ \frac{df(\tau )}{d\tau }\left( \pm r_{i}(\tau ,\vec{\sigma})\right)
^{k}\right] =  \nonumber \\
&&{}  \nonumber \\
&=&\sum_{k=0}^{\infty }\frac{(-)^{k}}{k!}\frac{d^{k}}{d\tau ^{k}}\left[
f(\tau )\left( \pm r_{i}(\tau ,\vec{\sigma})\right) ^{k}[1\mp \vec{\beta}%
_{i}(\tau )\cdot \hat{r}_{i}(\tau ,\vec{\sigma})]\right] ,  \label{VI8}
\end{eqnarray}

\noindent where

\begin{eqnarray}
{\vec{r}}_{i}(\tau ,\vec{\sigma}) &=&r_{i}(\tau ,\vec{\sigma}){\hat{r}}%
_{i}(\tau ,\vec{\sigma})=\vec{\sigma}-{\vec{\eta}}_{i}(\tau )={\vec{r}}%
_{i\pm }(\tau _{\pm }(\tau ,\vec{\sigma}),\vec{\sigma}){|}_{\tau _{i\pm
}(\tau ,\vec{\sigma})=\tau },  \nonumber \\
&&{}  \nonumber \\
f(\tau _{i\pm }) &\mapsto& P_{\perp}^{rs}(\vec{\sigma}){\bf V}_{i}^{s}(\tau
_{i\pm }) \frac{1}{\rho _{i\pm }(\tau _{i\pm })}=P_{\perp}^{rs}(\vec{\sigma})%
{\bf V}_{i}^{s}(\tau _{i\pm })\frac{1}{r_{i\pm }(\tau _{i\pm })[1\mp \vec{%
\beta}_{i}(\tau _{i\pm })\cdot \hat{r}_{i\pm }(\tau ,\vec{\sigma})]}.
\label{VI9}
\end{eqnarray}

\noindent The last line in Eq.(\ref{VI8}) is identical to the previous one
since ${\frac{{dr_{i}(\tau ,\vec{\sigma})}}{{d\tau }}}=-{\vec{\beta}}%
_{i}(\tau )\cdot {\hat{r}}_{i}(\tau ,\vec{\sigma})$. \ Hence we get

\begin{equation}
A_{\perp \pm }^{r}(\tau ,\vec{\sigma})=P_{\perp}^{rs}(\vec{\sigma})
\sum_{i=1}^{N}\frac{Q_{i}}{4\pi }\sum_{k=0}^{\infty }\frac{(\mp )^{k}}{k!}%
\frac{d^{k}}{d\tau ^{k}}[{\bf V}_{i}^{s}(\tau )r_{i}^{k-1}(\tau ,\vec{\sigma}%
)],\quad A_{\perp S}^{r}={\frac{1}{2}}(A_{\perp +}^{r}+A_{\perp -}^{r}).
\label{VI10}
\end{equation}

In order to evaluate the above derivatives we need the Leibnitz formula for
the $k$th derivative of the product $f(\tau )g(\tau )$

\begin{equation}
\frac{d^{k}}{d\tau ^{k}}(fg)=\sum_{m=0}^{k}{\frac{{k!}}{{m!(k-m)!}}}\frac{%
d^{m}f}{d\tau ^{m}}\,\,\frac{d^{k-m}g}{d\tau ^{k-m}},  \label{VI11}
\end{equation}
Thus

\begin{eqnarray}
\sum_{k=0}^{\infty }\frac{(\mp )^{k}}{k!}\frac{d^{k}}{d\tau ^{k}}[{\bf V}%
_{i}^{s}r_{i}^{k-1}] &=&\sum_{k=0}^{\infty }\frac{(\mp )^{k}}{k!}%
\sum_{m=0}^{k}\frac{k!}{m!(k-m)!}\frac{d^{k-m}{\bf V}_i^{s}}{d\tau ^{k-m}}%
\frac{d^{m}r_i^{k-1}}{d\tau ^{m}}=  \nonumber \\
\sum_{m=0}^{\infty }\sum_{k=m}^{\infty }\frac{(\mp )^{k}}{m!(k-m)!}\frac{%
d^{k-m}{\bf V}_i^{s}}{d\tau ^{k-m}}\frac{d^{m}r_i^{k-1}}{d\tau ^{m}}%
&=&\sum_{m=0}^{\infty }\sum_{h=0}^{\infty }\frac{(\mp )^{h+m}}{m!h!}\frac{%
d^{h}{\bf V}_i^{s}}{d\tau ^{h}}\frac{d^{m}r_i^{h+m-1}}{d\tau ^{m}}.
\label{VI12}
\end{eqnarray}

Using the notation ${\bf V}_i^{(h)s}=\frac{d^{h}{\bf V}_i^{s}}{d\tau ^{h}}$,
$\,$we obtain the following expression for the vector potential

\begin{equation}
A_{\perp \pm }^{r}(\tau ,\vec{\sigma})=P_{\perp}^{rs}(\vec{\sigma})
\sum_{i=1}^{N}\frac{Q_{i}}{4\pi } \sum_{h=0}^{\infty }\frac{(\mp )^{h}}{h!}%
{\bf V}_{i}^{(h)s}(\tau ) \phi _{i\pm ,h}(\tau ,\vec{\sigma}),  \label{VI13}
\end{equation}

\noindent in which

\begin{equation}
\phi _{i\pm ,h}(\tau ,\vec{\sigma})=\sum_{m=0}^{\infty }\frac{(\mp )^{m}}{m!}%
\frac{d^{m}r_{i}^{h+m-1}(\tau ,\vec{\sigma})}{d\tau ^{m}}=\sum_{m=0}^{\infty
}\frac{(\mp )^{m}}{m!}\frac{d^{m}}{d\tau ^{m}}[\sqrt{(\vec{\sigma}-\vec{\eta}%
_{i}(\tau ))^{2}}]^{m+h-1}.  \label{VI14}
\end{equation}

In order to display the result of the evaluation of the derivative we use
the formula

\begin{equation}
\frac{d^{m}}{d\tau ^{m}}R(f(\tau ))=\sum_{n=0}^{m}\sum_{n_{1}n_{2..}}\frac{m!%
}{n_{1}!n_{2}!..}\frac{d^{n}R(f(\tau ))}{df^{n}}|_{f=f(\tau )}(\frac{1}{1!}%
\frac{df(\tau )}{d\tau })^{n_{1}}(\frac{1}{2!}\frac{d^{2}f(\tau )}{d\tau ^{2}%
})^{n_{2}}...  \label{VI15}
\end{equation}

\noindent (with the summations restricted so that $\sum_{r}n_{r}=n,\,\,%
\sum_{r}rn_{r}=m $ ) to obtain

\begin{eqnarray}
\phi _{i\pm ,h}(\tau ,\vec{\sigma}) &=&\sum_{m=0}^{\infty }\frac{(\mp )^{m}}{%
m!}\sum_{n=0}^{m}\sum_{n_{1}n_{2..}}\frac{m!}{n_{1}!n_{2}!..}  \nonumber \\
&&\frac{\partial ^{n}r_{i}^{m+h-1}(\tau ,\vec{\sigma})}{\partial \vec{r}%
_{i}^{n}}\circ \left( \frac{-\vec{\beta}_{i}(\tau )}{1!}\right)
^{n_{1}}\left( \frac{-\vec{\beta}_{i}^{(1)}(\tau )}{2!}\right) ^{n_{2}}...
\label{VI16}
\end{eqnarray}

In this expression the symbol $\circ $ represents a scalar product between
the tensors to the left and to the right with the summation $\sum_{r}n_{r}=n$
indicating how the indices would be matched. Changing the $m$ summation
index to $k=m-n$ we obtain

\begin{eqnarray}
\phi _{i\pm ,h}(\tau ,\vec{\sigma}) &=&\sum_{n=0}^{\infty
}\sum_{k=0}^{\infty }\sum_{n_{1}n_{2..}}\frac{(\mp )^{k+n}(-)^{\Sigma
_{r}n_{r}=n}}{n_{1}!n_{2}!..}  \nonumber \\
&&\frac{\partial ^{n}r_{i}^{k+n+h-1}(\tau ,\vec{\sigma})}{\partial \vec{r}%
_{i}^{n}}\circ \left( \frac{\vec{\beta}_{i}(\tau )}{1!}\right)
^{n_{1}}\left( \frac{\vec{\beta}_{i}^{(1)}(\tau )}{2!}\right) ^{n_{2}}...
\label{VI17}
\end{eqnarray}

\noindent (in this latter summation $\sum_{r}n_{r}=n,\,\,\sum_{r}rn_{r}=n+k.$%
)

Now we can take advantage of the Grassmann charges to significantly simplify
the above multi-summations. \ As we have seen above, with a semi-classical $%
Q_{i}$ there are {\it no accelerations} on shell [$Q_{i}\vec{\beta}_{i}^{(h)}%
\stackrel{\circ }{=}{0}$ from Eq.(\ref{V18})] in the equations of motion of
the particle `$i$', since both the Coulomb potential and the
Lienard-Wiechert Lorentz force on particle `$i$' produced by the other
particles, i.e. $Q_{i}[{\vec{E}}_{\perp }(\tau ,{\vec{\eta}}_{i}(\tau ))+{%
\vec{\beta}}_{i}(\tau )\times \vec{B}(\tau ,{\vec{\eta}}_{i}(\tau ))]$, are
proportional to $Q_{i}$. Therefore, the full set of Hamilton equations (\ref
{V5}), (\ref{V6}) for both fields and particles imply that at the
semi-classical level we have a natural {\it order reduction} of the final
particle equation of motion in the Lienard-Wiechert sector (only second
order differential equations). One effect of this truncation is the
elimination of multi-particle forces; all the interactions will be {\it %
pairwise}, in both the Lagrangian and Hamiltonian formalisms. This was to be
expected since the rest-frame instant form is an equal-time description of
the $N$ particle system: (acceleration-independent) 3-body,.. N-body forces
appear as soon as we go to a description with no concept of equal time, like
in the standard approach with $N$ first class constraints \cite{manybody}.

Thus the only contributing indices are $n_{2}=n_{3}=..=0,\,n_{1}=n$ and our
expression for the transverse vector potentials simplify to

\begin{eqnarray}
A_{\perp \pm }^{r}(\tau ,\vec{\sigma}) &\stackrel{\circ }{=}&P_{\perp}^{rs}(%
\vec{\sigma})\,\sum_{i=1}^{N} \frac{Q_{i}}{4\pi }{\bf V}_{i}^{s}(\tau ) \phi
_{i\pm ,0}(\tau ,\vec{\sigma})\stackrel{\circ }{=}  \nonumber \\
\stackrel{\circ }{=}\, &&P_{\perp}^{rs}(\vec{\sigma}) \sum_{i=1}^{N}\frac{%
Q_{i}}{4\pi }{\bf V}_{i}^{s}(\tau )\sum_{n=0}^{\infty }\frac{(\pm )^{n}}{n!}%
\frac{\partial ^{n}r_{i}^{n-1}(\tau ,\vec{\sigma})}{\partial \vec{r}_{i}^{n}}%
\circ \left( \frac{\vec{\beta}_{i}(\tau )}{1!}\right) ^{n},  \nonumber \\
A_{\perp S}^{r}(\tau ,\vec{\sigma}) &=&{\frac{1}{2}}(A_{\perp
+}^{r}+A_{\perp -}^{r})(\tau ,\vec{\sigma}).  \label{VI18}
\end{eqnarray}

Since $r_{i}=\sqrt{{\vec{r}}_{i}^{2}}$, we see that for odd $n=2m+1$ we get

\begin{equation}
{\frac{{\partial^{2m+1}}}{{\partial {\vec r}_i^{2m+1}}}} (\sqrt{{\vec r}_i^2}%
)^{2m} ={\frac{{\partial^{2m+1}}}{{\partial {\vec r}_i^{2m+1}}}} ({\vec r}%
_i^2 )^m =0,  \label{VI19}
\end{equation}

\noindent and this implies the equality of the retarded, advanced and
symmetric Lienard-Wiechert potentials on-shell

\begin{eqnarray}
A_{\perp S}^{r}(\tau ,\vec{\sigma}) &\stackrel{\circ }{=}&A_{\perp \pm
}^{r}(\tau ,\vec{\sigma})\stackrel{\circ }{=}\,P_{\perp }^{rs}(\vec{\sigma}%
)\sum_{i=1,i\neq u}^{N}\frac{Q_{i}}{4\pi }V_{i}^{s}(\tau )  \nonumber \\
&&\times \sum_{m=0}^{\infty }\frac{1}{(2m)!}\left( \vec{\beta}_{i}(\tau
)\cdot \frac{\partial ^{2m}}{\partial \vec{r}_{i}^{2m}}\right)
r_{i}^{2m-1}(\tau ,\vec{\sigma}).  \label{VI20}
\end{eqnarray}
Therefore, at the semi-classical level there is only one Lienard-Wiechert
sector with a uniquely determined standard action-at-a-distance interaction.

\subsection{The Phase Space Expression of the Semi-Classical
Lienard-Wiechert Solution.}

We use a tensor notation to write the transverse symmetric vector potential
above as

\begin{equation}
\vec{A}_{\perp S}(\tau ,\vec{\sigma})\stackrel{\circ }{=}\,{\vec {\vec P}}%
_{\perp}(\vec{\sigma}) \cdot \sum_{i=1}^{N}{\ \frac{Q_{i}}{4\pi }}{\bf \vec{V%
}}_{i}(\tau ) \sum_{m=0}^{\infty }{\frac{\dot{\vec{\eta}}_{ij_{1}}(\tau )..%
\dot{\vec{\eta}}_{ij_{2m}}(\tau )}{(2m)!}}{\ \frac{\partial ^{2m}|\vec{\sigma%
}-\vec{\eta}_{i}(\tau )|^{2m-1}}{\partial \sigma _{j_{1}}..\partial \sigma
_{j_{2m}}}}.  \label{VI21}
\end{equation}

Using the definition of the Coulomb projection operator

\begin{equation}
P_{\perp hk}(\vec{\sigma})F(\vec{\sigma})=\delta _{hk}F(\vec{\sigma})- {%
\frac{1}{4\pi }}\int d^{3}\sigma ^{\prime }{\frac{\partial ^{2}}{\partial
\sigma _{h}\partial \sigma _{k}}}{\frac{1}{|{\vec{\sigma}}^{\prime }-\vec{%
\sigma}|}}F({\vec{\sigma}}^{\prime }),  \label{VI22}
\end{equation}

\noindent and compactifying the notation still further we obtain

\begin{eqnarray}
&&\vec{A}_{\perp S}(\tau ,\vec{\sigma})\,\stackrel{\circ }{=}\,\sum_{i=1}^{N}%
{\frac{Q_{i}}{4\pi }}\sum_{m=0}^{\infty }{\frac{1}{(2m)!}}\Big[{\bf \vec{V}}%
_{i}(\tau )(\dot{\vec{\eta}}_{i}(\tau )\cdot \vec{\partial}_{\sigma
})^{2m})\,|\vec{\sigma}-\vec{\eta}_{i}(\tau )|^{2m-1}-  \nonumber \\
&&-{\frac{1}{4\pi }}\int d^{3}\sigma ^{\prime }[\vec{\partial}_{\sigma }(%
{\bf \vec{V}}_{i}(\tau )\cdot \vec{\partial}_{\sigma }){\frac{1}{|{\vec{%
\sigma}}^{\prime }-\vec{\sigma}|}}](\dot{\vec{\eta}}_{i}(\tau )\cdot \vec{%
\partial}_{\sigma ^{\prime }})^{2m}\,|{\vec{\sigma}}^{\prime }-\vec{\eta}%
_{i}(\tau )|^{2m-1}\Big].  \label{VI23}
\end{eqnarray}
Integration by parts and changing from ${\frac{\partial }{\partial {\vec{%
\sigma}}^{\prime }}}$ to ${\frac{\partial }{\partial \vec{\sigma}}}$ and
translation gives

\begin{eqnarray}
\vec{A}_{\perp S}(\tau ,\vec{\sigma})\,\stackrel{\circ }{=} &&\sum_{i=1}^{N}{%
\frac{Q_{i}}{4\pi }}\sum_{m=0}^{\infty }{\frac{1}{(2m)!}}\Big[{\bf \vec{V}}%
_{i}(\tau )(\dot{\vec{\eta}}_{i}(\tau )\cdot \vec{\partial}_{\sigma
})^{2m})\,|\vec{\sigma}-\vec{\eta}_{i}(\tau )|^{2m-1}-  \nonumber \\
&&-{\frac{1}{4\pi }}\int d^{3}\sigma ^{\prime }\Big(\vec{\partial}_{\sigma
}( {\bf \vec{V}}_{i}(\tau )\cdot \vec{\partial}_{\sigma })(\dot{\vec{\eta}}
_{i}(\tau )\cdot \vec{\partial}_{\sigma })^{2m}\,{\frac{1}{|{\vec{\sigma}}%
^{\prime }-(\vec{\sigma}-\vec{\eta}_{i}(\tau ))|}}\Big)\,{\sigma ^{\prime }}%
^{2m-1}\Big].  \nonumber \\
&&  \label{VI24}
\end{eqnarray}

We remind ourselves that, in addition to a convective part, the generalized
current include magnetic and electric parts

\begin{equation}
{\bf \vec{V}}_{i}(\tau )=\dot{\vec{\eta}}_{i}(\tau )- \frac{i\sqrt{1-\dot{%
\vec{\eta}} _{i}^{2}(\tau )}\xi_{i}^{s}(\tau ) \vec{\xi}_{i}(\tau )\cdot
\vec{\partial}_{\sigma }} {m_{i}}+\frac{i\sqrt{1-\dot{\vec{\eta}}
_{i}^{2}(\tau )}\dot{\vec{\eta}}_{i}(\tau )\cdot \vec{\xi}_{i}(\tau ){\xi}
_{i}^{s}(\tau )\dot{\vec{\eta}}_{i}(\tau )\cdot \vec{\partial} _{\sigma }}{%
m_i(1+\sqrt{1-\dot{\vec{\eta}}_{i}^{2}(\tau )})},  \label{VI25}
\end{equation}

\noindent in which the derivatives act on $\vec \sigma$. Due to the delta
function in Eq.(\ref{V12}), this derivative can be replaced with minus the
derivative on ${\vec \eta}_i(\tau )$.

The integral in Eq.(\ref{VI24}) is finite, and thus we can view it as the ${%
\Lambda \rightarrow \infty }$ limit of an integral with a cutoff $\Lambda $
and take the derivatives out. The integral is thus of the form

\begin{equation}
-{\frac{1}{4\pi }}{\vec{\partial}}_{\sigma }({\bf \vec{V}}_{i}(\tau )\cdot
\vec{\partial}_{\sigma })(\dot{\vec{\eta}}_{i}(\tau )\cdot \vec{\partial}%
_{\sigma })^{2m}\int d^{3}\sigma ^{\prime }{\frac{{\sigma ^{\prime }}^{2m-1}%
}{|{\vec{\sigma}}^{\prime }-(\vec{\sigma}-\vec{\eta}_{i}(\tau ))|}},
\label{VI26}
\end{equation}

\noindent and

\begin{eqnarray}
{\frac{1}{4\pi }}\int_{\Lambda }d^{3}\sigma ^{\prime }{\frac{{\sigma
^{\prime }}^{2m-1}}{|\vec{\sigma}^{\prime }-(\vec{\sigma}-\vec{\eta}_{i})|}}
&=&{{\frac{1}{2}}}\int_{0}^{\Lambda }d\sigma ^{\prime }{\sigma ^{\prime }}%
^{2m+1}\int_{-1}^{1}{\frac{dz}{\sqrt{{\vec{\sigma ^{\prime }}}^{2}+(\vec{%
\sigma}-\vec{\eta}_{i})^{2}-2\sigma ^{\prime }|\vec{\sigma}-\vec{\eta}_{i}|z}%
}}  \nonumber \\
&=&{\frac{1}{2}}\int_{0}^{\Lambda }d\sigma ^{\prime }{\sigma ^{\prime }}%
^{2m+1}{\frac{-1}{\sigma ^{\prime }|\vec{\sigma}-\vec{\eta}_{i}|}}\sqrt{{\
\vec{\sigma ^{\prime }}}^{2}+(\vec{\sigma}-\vec{\eta}_{i})^{2}-2\sigma |\vec{%
\sigma}-\vec{\eta}_{i}|}  \nonumber \\
&=&-{\frac{1}{2|\vec{\sigma}-\vec{\eta}_{i}|}}\int_{0}^{\Lambda }d\sigma
^{\prime }{\sigma ^{\prime }}^{2m}(|{\vec{\sigma}}^{\prime }-|\vec{\sigma}-%
\vec{\eta}_{i}||-|{\vec{\sigma}}^{\prime }+|\vec{\sigma}-\vec{\eta}_{i}||)
\nonumber \\
&=&{\frac{\Lambda ^{2m+1}}{2m+1}}-{\frac{|\vec{\sigma}-\vec{\eta}_{i}|^{2m+1}%
}{(2m+1)(2m+2)}}.  \label{VI27}
\end{eqnarray}

Note that the $\Lambda $ cutoff will get killed by the $\sigma $
derivatives. Thus, we obtain

\begin{eqnarray}
\vec{A}_{\perp S}(\tau ,\vec{\sigma})\,=\, &&\sum_{i=1}^{N}{\frac{Q_{i}}{%
4\pi }}\sum_{m=0}^{\infty }\Big[{\frac{1}{(2m)!}}{\bf \vec{V}}_{i}(\tau )(%
\dot{\vec{\eta}}_{i}(\tau )\cdot {\vec{\partial}}_{\sigma })^{2m}\,|\vec{%
\sigma}-\vec{\eta}_{i}(\tau )|^{2m-1}-  \nonumber \\
&&-{\frac{1}{(2m+2)!}}{\vec{\partial}}_{\sigma }({\bf \vec{V}}_{i}(\tau
)\cdot {\vec{\partial}}_{\sigma })(\dot{\vec{\eta}}_{i}(\tau )\cdot {\vec{%
\partial}}_{\sigma })^{2m\,}|\vec{\sigma}-\vec{\eta}_{i}(\tau )|^{2m+1}\Big]%
:=  \nonumber \\
&:&=\sum_{i=1}^{N}Q_{i}\vec{A}_{\perp Si}(\vec{\sigma}-\vec{\eta}_{i}(\tau ),%
{\dot{\vec{\eta}}}_{i}(\tau )):=\vec{A}_{\perp 1}(\vec{\sigma},\tau )+\vec{A}%
_{\perp 2}(\vec{\sigma},\tau )+\vec{A}_{\perp 3}(\vec{\sigma},\tau ).
\label{VI28}
\end{eqnarray}

\noindent We have introduced the notations

\begin{eqnarray}
\vec{A}_{\perp 1}(\tau ,\vec{\sigma}) &=&\sum_{i=1}^{N}{\frac{Q_{i}}{4\pi }}%
\sum_{m=0}^{\infty }{\frac{1}{(2m)!}}{\bf \vec{V}}_{i}(\tau )(\dot{\vec{\eta}%
}_{i}(\tau )\cdot {\vec{\partial}}_{\sigma })^{2m}\,|\vec{\sigma}-\vec{\eta}%
_{i}(\tau )|^{2m-1},  \nonumber \\
&&{}  \nonumber \\
\vec{A}_{\perp 2}(\tau ,\vec{\sigma}) &=&(-)\sum_{i=1}^{N}{\frac{Q_{i}}{4\pi
}}\Big[(1+\frac{i\dot{\vec{\eta}}_{i}(\tau )\cdot \vec{\xi}_{i}(\tau )\vec{%
\xi}_{i}(\tau )\cdot {\vec{\partial}}_{\sigma }\sqrt{1-\dot{\vec{\eta}}%
_{i}^{2}(\tau )}}{m_{i}(1+\sqrt{1-\dot{\vec{\eta}}_{i}^{2}(\tau )})}%
\sum_{m=0}^{\infty }\Big[{\frac{1}{(2m+2)!}}  \nonumber \\
&&{\vec{\partial}}_{\sigma }(\dot{\vec{\eta}}_{i}(\tau )\cdot {\vec{\partial}%
}_{\sigma })^{2m+1\,}|\vec{\sigma}-\vec{\eta}_{i}(\tau )|^{2m+1}\Big],
\label{VI29}
\end{eqnarray}

\noindent and

\begin{eqnarray}
\vec{A}_{\perp 3}(\tau ,\vec{\sigma}) &=&-\sum_{i=1}^{N}{\frac{Q_{i}}{4\pi }}
\frac{i\vec{\xi}_{i}(\tau )\cdot \vec{\partial}_{i}\sqrt{1-\dot{\vec{\eta}}%
_{i}^{2}(\tau )}}{m_{i}} \sum_{m=0}^{\infty }[{\frac{1}{(2m+2)!}}{\ \vec{%
\partial}}_{\sigma }(\vec{\xi}_{i}(\tau )\cdot {\vec{\partial}}_{\sigma })
\nonumber \\
&& (\dot{\vec{\eta}}_{i}(\tau )\cdot {\vec{\partial}}_{\sigma })^{2m\,}|\vec{%
\sigma}-\vec{\eta}_{i}(\tau )|^{2m+1}].  \label{VI30}
\end{eqnarray}

Note that since

\begin{equation}
(\vec{\xi}_{i}(\tau )\cdot {\vec{\partial}}_{\sigma })(\vec{\xi}_{i}(\tau
)\cdot {\vec{\partial}}_{i})F(|\vec{\sigma}-\vec{\eta}_{i}(\tau )|)=0,
\label{VI31}
\end{equation}

\noindent the $\vec{A}_{\perp 3}(\tau ,\vec{\sigma})$ term is zero.

Using the result that $(\dot{\vec{\eta}}_{i}\cdot {\vec{\partial}}_{\eta
})^{2m}|\vec{\eta}|^{2m-1}=[(2m-1)!!]^{2}{\frac{1}{|\vec{\eta}|}}[\dot{\vec{%
\eta}}_{i}^{2}-(\dot{\vec{\eta}}_{i}\cdot {\frac{{\vec{\eta}}}{{|\vec{\eta}|}%
}})^{2}]^{m}$ [$\,\,\, {\vec{\partial}}_{\eta }=\partial /\partial \vec{\eta}
$], we get

\begin{equation}
\vec{A}_{\perp 1}(\tau ,\vec{\sigma})=\sum_{i=1}^{N}{\frac{Q_{i}}{4\pi }}%
\sum_{m=0}^{\infty }\Big[{\frac{[(2m-1)!!]^{2}}{(2m)!|\vec{\sigma}-\vec{\eta}%
_{i}|}}{\bf \vec{V}}_{i}(\tau )(\dot{\vec{\eta}}_{i}^{2}-(\dot{\vec{\eta}}%
_{i}\cdot \frac{\vec{\sigma}-\vec{\eta}_{i}}{|\vec{\sigma}-\vec{\eta}_{i}|}%
)^{2})^{m}\Big].  \label{VI32}
\end{equation}

We point out that all three currents contribute to $\vec{A}_{\perp 1}$,
while only the convective and spin-electric parts contribute to $\vec{A}%
_{\perp 2}$. \ Using

\begin{eqnarray}
\frac{\lbrack (2m-1)!!]^{2}}{(2m)!} &=&\frac{(2m)!}{(m!)^{2}2^{2m}}=\frac{%
(m-1/2)!}{\sqrt{\pi }m!}=\frac{\sqrt{\pi }(-)^{m-1}(m-1/2)}{(1/2-m)!m!}=
\nonumber \\
&=&\frac{\sqrt{\pi }(-)^{m}}{(-1/2-m)!m!}=(-)^{m}%
{-1/2 \choose m}%
\label{VI33}
\end{eqnarray}

\noindent we find that

\begin{eqnarray}
\vec{A}_{\perp 1}(\tau ,\vec{\sigma}) &=&\sum_{i=1}^{N}{\frac{Q_{i}}{4\pi }%
{\bf \vec{V}}_{i}(\tau )\frac{1}{\,|\vec{\sigma}-\vec{\eta}_{i}|}}%
\sum_{m=0}^{\infty }(-)^{m}%
{-1/2 \choose m}%
\left( \dot{\vec{\eta}}_{i}^{2}-(\dot{\vec{\eta}}_{i}\cdot \frac{\vec{\sigma}%
-\vec{\eta}_{i}}{|\vec{\sigma}-\vec{\eta}_{i}|})^{2}\right) ^{m}=  \nonumber
\\
&=&\sum_{i=1}^{N}{\frac{Q_{i}}{4\pi }{\bf \vec{V}}_{i}(\tau )\frac{1}{|\vec{%
\sigma}-\vec{\eta}_{i}|}}\frac{1}{\sqrt{1-\dot{\vec{\eta}}_{i}^{2}+(\dot{%
\vec{\eta}}_{i}\cdot \frac{\vec{\sigma}-\vec{\eta}_{i}}{|\vec{\sigma}-\vec{%
\eta}_{i}|})^{2}}}.  \label{VI34}
\end{eqnarray}

In terms of canonically conjugate variables this becomes

\begin{eqnarray}
\vec{A}_{\perp 1}(\tau ,\vec{\sigma}) &=&\sum_{i=1}^{N}{\frac{Q_{i}}{4\pi }%
\Big[}\vec{\kappa}_{i}-i\vec{\xi}_{i}(\tau )\vec{\xi}_{i}(\tau )\cdot \vec{%
\partial}_{\sigma }{+\frac{i\vec{\kappa}_{i}(\tau )\cdot \vec{\partial}%
_{\sigma }\vec{\kappa}_{i}(\tau )\cdot \vec{\xi}_{i}(\tau )\vec{\xi}%
_{i}(\tau )}{\sqrt{m_{i}^{2}+\vec{\kappa}_{i}^{2}}(m_i+\sqrt{m_{i}^{2}+\vec{%
\kappa}_{i}^{2}})}]}  \nonumber \\
&& {\frac{1}{|\vec{\sigma}-\vec{\eta}_{i}|}}\frac{1}{\sqrt{m_{i}^{2}+(\vec{%
\kappa}_{i}\cdot {\frac{{\vec{\sigma}-\vec{\eta}_{i}}}{{|\vec{\sigma}-\vec{%
\eta}_{i}|}}})^{2}}}.  \label{VI35}
\end{eqnarray}

For $\vec{A}_{\perp 2}(\tau ,\vec{\sigma})$, we see from Eq.(\ref{VI25})
that we need an expression for $(\dot{\vec{\eta}}_{i}(\tau )\cdot {\vec{%
\partial}}_{\eta })^{2m+1}|\vec{\eta}|^{2m+1}$ and $\vec{\xi}_{i}(\tau
)\cdot {\vec{\partial}}_{\eta }(\dot{\vec{\eta}}_{i}(\tau )\cdot {\vec{%
\partial}}_{\eta })^{2m}|\vec{\eta}|^{2m+1}$. One can show by an induction
procedure that

\begin{equation}
(\dot{\vec{\eta}}_{i}(\tau )\cdot {\vec{\partial}}_{\sigma })^{2m+1}|\vec{%
\sigma}-\vec{\eta}_{i}|^{2m+1}=[(2m+1)!!]^{2}\sum_{l=0}^{m}(-)^{l}%
{m \choose l}%
(\dot{\vec{\eta}}_{i}^{2}(\tau ))^{m-l}\frac{(\dot{\vec{\eta}}_{i}(\tau
)\cdot \frac{\vec{\sigma}-\vec{\eta}_{i}}{|\vec{\sigma}-\vec{\eta}_{i}|}%
)^{2l+1}}{2l+1},  \label{VI36}
\end{equation}

\noindent so that we get

\begin{eqnarray}
\vec{A}_{\perp 2}(\tau ,\vec{\sigma}) &=&-)\sum_{i=1}^{N}{\frac{Q_{i}}{4\pi }%
}(1-\frac{i\dot{\vec{\eta}} _{i}(\tau )\cdot \vec{\xi}_{i}(\tau )\vec{\xi}%
_{i}(\tau )\cdot {\vec{\partial}}_{\sigma }\sqrt{1-\dot{\vec{\eta}}%
_{i}^{2}(\tau )}}{m_i(1+\sqrt{1-\dot{\vec{\eta}}_{i}^{2}(\tau )})})\vec{%
\partial }_{\sigma }  \nonumber \\
&& \sum_{m=0}^{\infty }\frac{[(2m+1)!!]^{2}}{(2m+2)!}\left( \dot{\vec{\eta}}%
_{i}^{2}(\tau )\right) ^{m}\sum_{l=0}^{m}(-)^{l}%
{m \choose l}%
(\dot{\vec{\eta}}_{i}^{2}(\tau ))^{-l}\frac{(\dot{\vec{\eta}}_{i}(\tau
)\cdot \frac{\vec{\sigma}-\vec{\eta}_{i}}{|\vec{\sigma}-\vec{\eta}_{i}|}%
)^{2l+1}}{2l+1}.  \label{VI37}
\end{eqnarray}

Since we have

\begin{equation}
\frac{\lbrack (2m+1)!!]^{2}}{(2m+2)!}=(-)^{m+1}%
{-1/2 \choose m+1}%
\label{VI38}
\end{equation}

\noindent we obtain (where ${\bf I}$ is the unit dyad)

\begin{eqnarray}
\vec{A}_{\perp 2}(\tau ,\vec{\sigma}) &=&-\sum_{i=1}^{N}{\frac{Q_{i}}{4\pi }}%
\Big(1+\frac{i{\vec{\kappa}_{i}}(\tau )\cdot \vec{\xi}_{i}(\tau )\vec{\xi}%
_{i}(\tau )\cdot {\vec{\partial}}_{\sigma }}{\sqrt{m_{i}^{2}+{\vec{\kappa}%
_{i}}^{2}}(m_{i}+\sqrt{m_{i}^{2}+{\vec{\kappa}_{i}}^{2}})}\Big)\frac{\vec{%
\kappa}_{i}}{|\vec{\sigma}-\vec{\eta}_{i}|}\cdot \Big({\bf I}-\frac{\vec{%
\sigma}-\vec{\eta}_{i}}{|\vec{\sigma}-\vec{\eta}_{i}|}\frac{\vec{\sigma}-%
\vec{\eta}_{i}}{|\vec{\sigma}-\vec{\eta}_{i}|}\Big)  \nonumber \\
&&\Big({\frac{\sqrt{m_{i}^{2}+{\vec{\kappa}_{i}}^{2}}}{\sqrt{m_{i}^{2}+(\vec{%
\kappa}_{i}\cdot {\frac{{\vec{\sigma}-\vec{\eta}_{i}}}{{|\vec{\sigma}-\vec{%
\eta}_{i}|}}})^{2}}}}-1\Big){\frac{\sqrt{m_{i}^{2}+{\vec{\kappa}_{i}}^{2}}}{%
\vec{\kappa}_{i}^{2}-(\vec{\kappa}_{i}\cdot {\frac{{\vec{\sigma}-\vec{\eta}%
_{i}}}{{|\vec{\sigma}-\vec{\eta}_{i}|}}})^{2}}},  \label{VI39}
\end{eqnarray}

After combining some terms in $\vec{A}_{\perp 2}(\tau ,\vec{\sigma})$, our
{\it exact} result for the vector potential is

\begin{eqnarray}
\vec{A}_{\perp S}(\tau ,\vec{\sigma}) &=&\sum_{i=1}^{N}{\frac{Q_{i}}{4\pi }%
\Big(\Big[\vec{\kappa}_{i}-i\vec{\xi}_{i}(\tau )\vec{\xi}_{i}(\tau )\cdot
\vec{\partial}_{\sigma }{+\frac{i\vec{\kappa}_{i}(\tau )\cdot \vec{\partial}%
_{\sigma }\vec{\kappa}_{i}(\tau )\cdot \vec{\xi}_{i}(\tau )\vec{\xi}%
_{i}(\tau )}{\sqrt{m_{i}^{2}+\vec{\kappa}_{i}^{2}}(m_i+\sqrt{m_{i}^{2}+\vec{%
\kappa}_{i}^{2}})}}\Big]}  \nonumber \\
&&{\frac{1}{|\vec{\sigma}-\vec{\eta}_{i}|}}\frac{1}{\sqrt{m_{i}^{2}+(\vec{%
\kappa}_{i}\cdot {\frac{{\vec{\sigma}-\vec{\eta}_{i}}}{{|\vec{\sigma}-\vec{%
\eta}_{i}|}}})^{2}}}-(1+\frac{i{\vec{\kappa}_{i}}(\tau )\cdot \vec{\xi}%
_{i}(\tau )\vec{\xi}_{i}(\tau )\cdot {\vec{\partial}}_{\sigma }}{\sqrt{%
m_{i}^{2}+{\vec{\kappa}_{i}}^{2}}(m_{i}+\sqrt{m_{i}^{2}+{\vec{\kappa}_{i}}%
^{2}})})\frac{\vec{\kappa}_{i}}{|\vec{\sigma}-\vec{\eta}_{i}|}  \nonumber \\
&&\cdot ({\bf I}-\frac{\vec{\sigma}-\vec{\eta}_{i}}{|\vec{\sigma}-\vec{\eta}%
_{i}|}\frac{\vec{\sigma}-\vec{\eta}_{i}}{|\vec{\sigma}-\vec{\eta}_{i}|}){%
\frac{\sqrt{m_{i}^{2}+{\vec{\kappa}_{i}}^{2}}}{\sqrt{m_{i}^{2}+(\vec{\kappa}%
_{i}\cdot {\frac{{\vec{\sigma}-\vec{\eta}_{i}}}{{|\vec{\sigma}-\vec{\eta}%
_{i}|}}})^{2}}(\sqrt{m_{i}^{2}+{\vec{\kappa}_{i}}^{2}}+\sqrt{m_{i}^{2}+(\vec{%
\kappa}_{i}\cdot {\frac{{\vec{\sigma}-\vec{\eta}_{i}}}{{|\vec{\sigma}-\vec{%
\eta}_{i}|}}})^{2}})}} \Big) =  \nonumber \\
&&{}  \nonumber \\
&=&\sum_{i=1}^{N}{\frac{Q_{i}}{4\pi }\Big(\vec{\kappa}_{i}\cdot ({\bf I}}+{%
\frac{\vec{\sigma}-\vec{\eta}_{i}}{|\vec{\sigma}-\vec{\eta}_{i}|}\frac{\vec{%
\sigma}-\vec{\eta}_{i}}{|\vec{\sigma}-\vec{\eta}_{i}|})\frac{\sqrt{m_{i}^{2}+%
{\vec{\kappa}_{i}}^{2}}}{|\vec{\sigma}-\vec{\eta}_{i}|\sqrt{m_{i}^{2}+(\vec{%
\kappa}_{i}\cdot {\frac{{\vec{\sigma}-\vec{\eta}_{i}}}{{|\vec{\sigma}-\vec{%
\eta}_{i}|}}})^{2}}}}-  \nonumber \\
&&{-i\vec{\xi}_{i}(\tau )\vec{\xi}_{i}(\tau )\cdot \vec{\partial}_{\sigma }%
\frac{1}{|\vec{\sigma}-\vec{\eta}_{i}|\sqrt{m_{i}^{2}+(\vec{\kappa}_{i}\cdot
{\frac{{\vec{\sigma}-\vec{\eta}_{i}}}{{|\vec{\sigma}-\vec{\eta}_{i}|}}})^{2}}%
}} +  \nonumber \\
&&+{{\frac{i\vec{\kappa}_{i}(\tau )\cdot \vec{\partial}_{\sigma }\vec{\kappa}%
_{i}(\tau )\cdot \vec{\xi}_{i}(\tau )\vec{\xi}_{i}(\tau )}{\sqrt{m_{i}^{2}+%
\vec{\kappa}_{i}^{2}}(\sqrt{m_{i}^{2}+\vec{\kappa}_{i}^{2}}+m_{i})}}}\frac{1%
}{|\vec{\sigma}-\vec{\eta}_{i}|\sqrt{m_{i}^{2}+(\vec{\kappa}_{i}\cdot {\frac{%
{\vec{\sigma}-\vec{\eta}_{i}}}{{|\vec{\sigma}-\vec{\eta}_{i}|}}})^{2}}} -
\nonumber \\
&&-\frac{i{\vec{\kappa}_{i}}(\tau )\cdot \vec{\xi}_{i}(\tau )\vec{\xi}%
_{i}(\tau )\cdot {\vec{\partial}}_{\sigma }}{(m_{i}+\sqrt{m_{i}^{2}+{\vec{%
\kappa}_{i}}^{2}})}\vec{\kappa}_{i}\cdot ({\bf I}-\frac{\vec{\sigma}-\vec{%
\eta}_{i}}{|\vec{\sigma}-\vec{\eta}_{i}|}\frac{\vec{\sigma}-\vec{\eta}_{i}}{|%
\vec{\sigma}-\vec{\eta}_{i}|})\frac{1}{|\vec{\sigma}-\vec{\eta}_{i}|\sqrt{%
m_{i}^{2}+(\vec{\kappa}_{i}\cdot {\frac{{\vec{\sigma}-\vec{\eta}_{i}}}{{|%
\vec{\sigma}-\vec{\eta}_{i}|}}})^{2}}}\Big).  \nonumber \\
&&  \label{VI40}
\end{eqnarray}

The corresponding $\vec{E}_{\perp S}(\tau ,\vec{\sigma})$ and $\vec{B}%
_{\perp S}(\tau ,\vec{\sigma})$ fields have the series forms

\begin{eqnarray}
\vec{E}_{\perp S}(\tau ,\vec{\sigma})\,=-\, &&\sum_{i=1}^{N}{\frac{Q_{i}}{%
4\pi }}\sum_{m=0}^{\infty }\Big[{\frac{1}{(2m)!}}{\bf \vec{V}}_{i}(\tau )(%
\dot{\vec{\eta}}_{i}(\tau )\cdot {\vec{\partial}}_{\sigma })^{2m+1}\,|\vec{%
\sigma}-\vec{\eta}_{i}(\tau )|^{2m-1}-  \nonumber \\
&&-{\frac{1}{(2m+2)!}}{\vec{\partial}}_{\sigma }({\bf \vec{U}}_{i}(\tau
)\cdot {\vec{\partial}}_{\sigma })(\dot{\vec{\eta}}_{i}(\tau )\cdot {\vec{%
\partial}}_{\sigma })^{2m+1\,}|\vec{\sigma}-\vec{\eta}_{i}(\tau )|^{2m+1}%
\Big] ,  \label{VI41}
\end{eqnarray}

\noindent where

\begin{equation}
{\bf \vec{U}}_{i}(\tau )=\dot{\vec{\eta}}_{i}(\tau )-\frac{i \sqrt{1-\dot{%
\vec{\eta}}_{i}^{2}(\tau )}\dot{\vec{\eta}}_{i}(\tau )\cdot \vec{\xi}%
_{i}(\tau )\vec{\xi}_{i}(\tau )\, \dot{\vec{\eta}}_{i}(\tau )\cdot \vec{%
\partial}_{\sigma }}{m_i(1+\sqrt{1-\dot{\vec{\eta}}_{i}^{2}(\tau )})},
\label{VI42}
\end{equation}

\noindent and

\begin{equation}
\vec{B}_{S}(\tau ,\vec{\sigma})\,=\,-\sum_{i=1}^{N}{\frac{Q_{i}}{4\pi }}%
\sum_{m=0}^{\infty }{\frac{1}{(2m)!}}{\bf \vec{V}}_{i}(\tau )\times \vec{%
\partial}_{\sigma }(\dot{\vec{\eta}}_{i}(\tau )\cdot {\vec{\partial}}%
_{\sigma })^{2m}\,|\vec{\sigma}-\vec{\eta}_{i}(\tau )|^{2m-1}.  \label{VI43}
\end{equation}

\noindent The corresponding {\it exact} forms in terms of the canonical
variables are

\begin{eqnarray}
\vec{B}_{\perp S}(\tau ,\vec{\sigma}) &=&-\sum_{j=1}^{N}{\frac{Q_{j}}{4\pi }%
\Big[}\vec{\kappa}_{j}-i\vec{\xi}_{j}\vec{\xi}_{j}\cdot \vec{\partial}%
_{\sigma }{+\frac{i\vec{\kappa}_{j}\cdot \vec{\partial}_{\sigma }\vec{\kappa}%
_{j}(\tau )\cdot \vec{\xi}_{j}(\tau )\vec{\xi}_{j}(\tau )}{\sqrt{m_{j}^{2}+%
\vec{\kappa}_{j}^{2}}(m_j+\sqrt{m_{j}^{2}+\vec{\kappa}_{j}^{2}})}}\Big]
\nonumber \\
&& \vec{\partial}_{\sigma }\Big[{\frac{1}{|\vec{\sigma}-\vec{\eta}_{j}|}}%
\frac{1}{\sqrt{m_{j}^{2}+(\vec{\kappa}_{j}\cdot {\frac{{\vec{\sigma}-\vec{%
\eta}_{j}}}{{|\vec{\sigma}-\vec{\eta}_{j}|}}})^{2}}}\Big],  \label{VI44}
\end{eqnarray}

\noindent and

\begin{eqnarray}
\vec{\pi}_{\perp S}(\tau ,\vec{\sigma}) &=& {\vec E}_{\perp S}(\tau ,\vec %
\sigma ) =-\sum_{j=1}^{N}{\frac{Q_{j}}{4\pi } \Big[({\bf I+}\frac{i\vec{%
\kappa}_{j}(\tau )\cdot \vec{\xi}_{j}(\tau ) \vec{\xi}_{j}(\tau )\vec{%
\partial}_{\sigma }}{\sqrt{m_{j}^{2}+\vec{\kappa}_{j}^{2}}(m_j+\sqrt{%
m_{j}^{2}+\vec{\kappa}_{j}^{2}})})\cdot {\frac{\vec{\sigma}-\vec{\eta}_{j}}{|%
\vec{\sigma}-\vec{\eta}_{j}|^{3}} }}  \nonumber \\
&&(\frac{m_{j}^{2}\sqrt{ m_{j}^{2}+\vec{\kappa}_{j}^{2}}}{[m_{j}^{2}+(\vec{%
\kappa}_{j}\cdot {\frac{{\vec{\sigma}-\vec{\eta}_{j}}}{{|\vec{\sigma}-\vec{%
\eta}_{j}|}}})^{2}]^{3/2}}-1)-  \nonumber \\
&&-{i\vec{\xi}_{j}\vec{\xi}_{j}\cdot \vec{\partial}_{\sigma }\vec{\kappa}%
_{j}\cdot \vec{\partial}}_{\sigma }{\frac{1}{|\vec{\sigma}-\vec{\eta}_{j}|}}%
\frac{1}{\sqrt{m_{j}^{2}+(\vec{\kappa}_{j}\cdot {\frac{{\vec{\sigma}-\vec{%
\eta}_{j}}}{{|\vec{\sigma}-\vec{\eta}_{j}|}}})^{2}}}\Big] .  \label{VI45}
\end{eqnarray}

For the evaluation of the field energy we need ${\vec{E}}_{\perp S}^{2}+{%
\vec{B}}_{S}^{2}$, while for the field 3-momentum we need ${\vec{E}}_{\perp
S}\times {\vec{B}}_{S}$. \ Since the exact forms cannot be integrated to
closed forms, we need the series forms for the fields. From Appendix D we
get [${\vec{\eta}}_{ij}(\tau )=\eta _{ij}(\tau ){\hat{\eta}}_{ij}(\tau )={%
\vec{\eta}}_{i}(\tau )-{\vec{\eta}}_{j}(\tau )$; ${\vec{\partial}}%
_{ij}=\partial /\partial {\vec{\eta}}_{ij}$]

\begin{eqnarray}
&&U(\tau )\, :=\, {\frac{1}{2}}\int d^{3}\sigma ({\vec{E}}_{\perp S}^{2}+{%
\vec{B}} _{S}^{2})(\tau ,\vec{\sigma}):= \, \sum_{i<j}^{1..N}{\frac{{%
Q_{i}Q_{j}}}{{4\pi }}}h_{1}(\dot{\vec{\eta}_{i}},\dot{\vec{\eta}_{j}},\vec{%
\eta}_{ij})=  \nonumber \\
&&  \nonumber \\
&=&\sum_{i<j}^{1..N}{\frac{Q_{i}Q_{j}}{4\pi }}\sum_{m=0}^{\infty
}\sum_{n=0}^{\infty }\Big[{\bf \vec{V}}_{i}\cdot {\bf \vec{V}}_{j}{\frac{(%
\dot{\vec{\eta}_{i}}\cdot \vec{\partial}_{ij})^{2m+1}(\dot{\vec{\eta}_{j}}%
\cdot \vec{\partial}_{ij})^{2n+1}\eta _{ij}^{2n+2m+1}}{(2n+2m+2)!}}-
\nonumber \\
&-&{\frac{({\bf \vec{V}}_{i}\cdot \vec{\partial}_{ij})({\bf \vec{U}}%
_{j}\cdot \vec{\partial}_{ij})(\dot{\vec{\eta}_{i}}\cdot \vec{\partial}%
_{ij})^{2m+1}(\dot{\vec{\eta}_{j}}\cdot \vec{\partial}_{ij})^{2n+1}\eta
^{2n+2m+3}}{(2n+2m+4)!}}+  \nonumber \\
&&-{\frac{({\bf \vec{U}}_{i}\cdot \vec{\partial}_{ij})({\bf \vec{V}}%
_{j}\cdot \vec{\partial}_{ij})(\dot{\vec{\eta}_{i}}\cdot \vec{\partial}%
_{ij})^{2m+1}(\dot{\vec{\eta}_{j}}\cdot \vec{\partial}_{ij})^{2n+1}\eta
^{2n+2m+3}}{(2n+2m+4)!}}+  \nonumber \\
&&+{\frac{({\bf \vec{U}}_{i}\cdot \vec{\partial}_{ij})({\bf \vec{U}}%
_{j}\cdot \vec{\partial}_{ij})(\dot{\vec{\eta}_{i}}\cdot \vec{\partial}%
_{ij})^{2m+1}(\dot{\vec{\eta}_{j}}\cdot \vec{\partial}_{ij})^{2n+1}\eta
^{2n+2m+3}}{(2n+2m+4)!}}+  \nonumber \\
&+&{\bf \vec{V}}_{i}\cdot {\bf \vec{V}}_{j}{\frac{(\dot{\vec{\eta}_{i}}\cdot
\vec{\partial}_{ij})^{2m}(\dot{\vec{\eta}_{j}}\cdot \vec{\partial}_{ij})^{2n}%
\vec{\partial}_{ij}^{2}\eta _{ij}^{2n+2m+1}}{(2n+2m+2)!}}-  \nonumber \\
&-&{\frac{({\bf \vec{V}}_{i}\cdot \vec{\partial}_{ij})({\bf \vec{V}}%
_{j}\cdot \vec{\partial}_{ij})(\dot{\vec{\eta}_{i}}\cdot \vec{\partial}%
_{ij})^{2m}(\dot{\vec{\eta}_{j}}\cdot \vec{\partial}_{ij})^{2n}\eta
_{ij}^{2n+2m+1}}{(2n+2m+2)!}}\Big],  \label{VI46}
\end{eqnarray}

\noindent and

\begin{eqnarray}
&&\int d^{3}\sigma ({\vec{E}}_{\perp S}\times {\vec{B}}_{S})(\tau ,\vec{%
\sigma})\, :=\, \sum_{i<j}^{1..N}Q_{i}Q_{j}\vec{h}_{1}(\dot{\vec{\eta}_{i}},%
\dot{\vec{\eta}_{j}},\vec{\eta}_{ij})=  \nonumber \\
&&  \nonumber \\
&=&\sum_{i<j}^{1..N}{\frac{Q_{i}Q_{j}}{4\pi }}\Big[\sum_{m=0}^{\infty
}\sum_{n=0}^{\infty }\Big({\vec{\partial}}_{ij}[{\bf \vec{V}}_{i}\cdot {\bf
\vec{V}}_{j}{\frac{(\dot{\vec{\eta}_{i}}\cdot \vec{\partial}_{ij})^{2m+1}(%
\dot{\vec{\eta}_{j}}\cdot \vec{\partial}_{ij})^{2n}\eta _{ij}^{2n+2m+1}}{%
(2n+2m+2)!}}-  \nonumber \\
&-&{\frac{({\bf \vec{U}}_{i}\cdot \vec{\partial}_{ij})({\bf \vec{V}}%
_{j}\cdot \vec{\partial}_{ij})(\dot{\vec{\eta}_{i}}\cdot \vec{\partial}%
_{ij})^{2m+1}(\dot{\vec{\eta}_{j}}\cdot \vec{\partial}_{ij})^{2n}\eta
_{ij}^{2n+2m+3}}{(2n+2m+4)!}}]  \nonumber \\
&&-\frac{{\bf \vec{V}}_{j}({\bf \vec{V}}_{i}-{\bf \vec{U}}_{i})\cdot \vec{%
\partial}_{ij}(\dot{\vec{\eta}_{i}}\cdot \vec{\partial}_{ij})^{2m+1}(\dot{%
\vec{\eta}_{j}}\cdot \vec{\partial}_{ij})^{2n}\eta _{ij}^{2n+2m+1}}{%
(2n+2m+2)!}\Big)  \nonumber \\
&&+(i \longleftrightarrow j)\Big].  \label{VI47}
\end{eqnarray}

\subsection{Final Dirac brackets and their Darboux basis}

Till now we have worked in the reduced phase space of $N$ positive-energy
charged spinning particles plus the transverse electromagnetic field. This
is a well defined isolated system with a global Darboux basis [ $\vec{\eta}%
_{i},\vec{\kappa}_{i},\vec{\xi}_{i},\vec{A}_{\perp }(\tau ,\vec{\sigma}),%
\vec{\pi}_{\perp }(\tau ,\vec{\sigma})$] and a well defined physical
Hamiltonian, the invariant mass $M={\cal P}_{(int)}^{\tau }$ of Eqs.(\ref
{IV7}), (\ref{V1}). All possible configurations of motion take place in this
reduced phase space. \ The space of solutions of Hamilton's equations is a
symplectic space, since there is a definition of Poisson brackets on the
space of solutions. In Ref.\cite{ap} we found that we could select a subset
of solutions of the equations of motion which is still a symplectic
manifold. The method we used was to add by hand a set of second class
constraints {\it compatible with the equations of motion}: this amounts to
the selection of a symplectic submanifold of the symplectic manifold of
solutions. We follow the same method here including the pseudo-classical
spin variables $\vec{\xi}_{i}$.

\bigskip As in the case of \ scalar particle, the Grassmann truncated
semiclassical Lienard Wiechert solution $\vec{A}_{\perp S}$ given in Eq.(\ref
{VI40}) for the vector potential, with $\vec{\pi}_{\perp S}={\vec{E}}_{\perp
S}=-\frac{\partial }{\partial \tau }\vec{A}_{\perp S}$ for the canonical
conjugate field momentum given by Eq.(\ref{VI45}), provides us such a set of
second class constraints \ by way of

\begin{eqnarray}
\vec{\chi}_{1}(\tau ,\vec{\sigma}) &=&\vec{A}_{\perp }(\tau ,\vec{\sigma}%
)-\sum_{i=1}^{N}Q_{i}\vec{A}_{\perp Si}(\vec{\sigma}-\vec{\eta}_{i}(\tau ),{%
\vec{\kappa}}_{i}(\tau ),\vec{\xi}_{i}(\tau ))\approx 0,  \nonumber \\
\vec{\chi}_{2}(\tau ,\vec{\sigma}) &=&\vec{\pi}_{\perp }(\tau ,\vec{\sigma}%
)-\sum_{i=1}^{N}Q_{i}\vec{\pi}_{\perp Si}(\vec{\sigma}-\vec{\eta}_{i}(\tau ),%
{\vec{\kappa}}_{i}(\tau ),\vec{\xi}_{i}(\tau ))\approx 0.  \label{VI48}
\end{eqnarray}

These constraints allow us to eliminate the canonical degrees of freedom of
the radiation field and to get the symmetric semi-classical Lienard-Wiechert
reduced phase space, in which there are only particle degrees of freedom. In
analogy to what occurred in Ref.\cite{ap} for spinless particle the
independent variables $\vec{\eta}_{i},\vec{\kappa}_{i}$ and now $\vec{\xi}%
_{i}$ as well will no longer be canonical when one imposes these constraints
by way of modified Dirac brackets.

To determine the effects of these constraints in the construction of Dirac
brackets we must compute the 6x6 matrix of brackets

\begin{equation}
\bigg({%
{\{\vec{\chi}_{1},\vec{\chi}_{1}\} \atop \{\vec{\chi}_{2},\vec{\chi}_{1}\}}%
}{%
{\{\vec{\chi}_{1},\vec{\chi}_{2}\} \atop \{\vec{\chi}_{2},\vec{\chi}_{2}\}}%
}\bigg ).  \label{VI49}
\end{equation}
It turns out that this matrix bracket is relatively simple, due to the
Grassmann charges. \ Consider, for example the case of two particles. The
particle or Lienard-Wiechert parts of the matrix bracket vanish since $%
Q_{1}^{2}=0=Q_{2}^{2}$ and cross terms vanish because they involve Poisson
brackets of particle one variables with particle two variables. Thus the
only part of the 6x6 matrix bracket that contributes is from the field
variables. It has the form

\begin{equation}
\{\vec{\chi}_{1}(\tau ,\vec{\sigma}_{1}),\vec{\chi}_{2}(\tau ,\vec{\sigma}%
_{2})\}=({\bf I}-{\frac{\vec{\partial}\vec{\partial}}{\vec{\partial}^{2}}}%
)\delta ^{3}(\vec{\sigma}_{1}-\vec{\sigma}_{2}),  \label{VI50}
\end{equation}

\noindent and since

\begin{equation}
\{\vec{\chi}_{1},\vec{\chi}_{1}\}=0=\{\vec{\chi}_{2},\vec{\chi}_{2}\},
\label{VI51}
\end{equation}

\noindent only the 3x3 off diagonal portion contributes.

In order to have a well defined Dirac bracket we need to use a modified form
of the Poisson bracket in which the inverse of the matrix of constraint
Poisson brackets is used. Calling this matrix $C$, we define $\tilde{C}^{-1}$
so that $C\tilde{C}^{-1}=({\bf I}-{\frac{\vec{\partial}\vec{ \partial}}{\vec{%
\partial}^{2}}})\delta ^{3}(\vec{\sigma}_{1}-\vec{\sigma}_{2})$. But the
transverse form of the delta function allows us to use the idempotent
property of the projector to show that the inverse of $C$ in this sense is
just $C$ itself. In that case for two functions $f(\vec{\kappa}_{i},\vec{\eta%
}_{i}),g(\vec{\kappa}_{i},\vec{\eta}_{i})$ of the particle variables the
Dirac bracket becomes

\begin{eqnarray}
\{f,g\}^{\ast } &=&\{f,g\}-  \nonumber \\
&&-Big[\int d^{3}\sigma \{f,-\sum_{i}Q_{i}\vec{A}_{\perp Si}(\vec{\sigma}-%
\vec{ \eta}_{i}(\tau ),{\vec{\kappa}}_{i}{(\tau )},\vec{\xi}_{i}(\tau
))\}\cdot  \nonumber \\
&&\cdot \{-\sum_{j}Q_{j}\vec{\pi}_{\perp Sj}(\vec{\sigma}-\vec{\eta}%
_{j}(\tau ),{\vec{\kappa}}_{j}(\tau ),\vec{\xi}_{i}(\tau )),g\}-  \nonumber
\\
&&-\{f,-\sum_{j}Q_{j}\vec{\pi}_{\perp Sj}(\vec{\sigma}-\vec{\eta}_{j}(\tau
), {\vec{\kappa}}_{j}(\tau ),\vec{\xi}_{i}(\tau ))\}\cdot  \nonumber \\
&&\cdot \{-\sum_{i}Q_{i}\vec{A }_{\perp Si}(\vec{\sigma}-\vec{\eta}_{i}(\tau
),{\vec{\kappa}}_{i}(\tau ), \vec{\xi}_{i}(\tau )),g\}\Big].  \label{VI52}
\end{eqnarray}

This is the bracket for the new reduced symplectic manifold containing only
particles. To find the new canonical basis for the particles with respect to
these Dirac brackets, we define the following scalar function

\begin{equation}
{\cal K}=\sum_{i=1}^{N-1}\sum_{j=i+1}^{N}Q_{i}Q_{j}{\cal K}_{ij}(\vec{\xi}%
_{i},\vec{\xi}_{j};\vec{\kappa}_{i},\vec{\kappa}_{j};\vec{\eta}_{i}-\vec{\eta%
}_{j}),  \label{VI53}
\end{equation}

\noindent in which there appear the following functions

\begin{eqnarray}
{\cal K}_{ij}=\int d^{3}\vec{\sigma}[\vec{A}_{\perp Si}(\vec{\sigma}-\vec{%
\eta}_{i},{\vec{\kappa}}_{i},\vec{\xi}_{i})\cdot \vec{\pi}_{\perp Sj}(\vec{%
\sigma}-\vec{\eta}_{j},{\vec{\kappa}}_{j},\vec{\xi}_{j}) &&  \nonumber \\
-\vec{A}_{\perp Sj}(\vec{\sigma}-\vec{\eta}_{j},{\vec{\kappa}}_{j},\vec{\xi}%
_{j})\cdot \vec{\pi}_{\perp Si}(\vec{\sigma}-\vec{\eta}_{i},{\vec{\kappa}}%
_{i},\vec{\xi}_{i})] &=&{\cal K}_{ij}(\vec{\xi}_{i},\vec{\xi}_{j};\vec{\kappa%
}_{i},\vec{\kappa}_{j};\vec{\eta}_{i}-\vec{\eta}_{j})=-{\cal K}_{ji}.
\nonumber \\
&&  \label{VI54}
\end{eqnarray}

In Ref.\cite{ap} we found that the old canonical variables $\vec{\eta}_{i}$
and $\vec{\kappa}_{i}$ $\ $\ need to be modified to

\begin{eqnarray}
\widetilde{\vec{\eta}_{i}} &=&\vec{\eta}_{i}+\frac{1}{2}\sum_{j\neq
i}Q_{i}Q_{j}\vec{\partial}_{\kappa _{i}}{\cal K}_{ij}=\vec{\eta}%
_{i}+\sum_{j\neq i}\frac{1}{2}Q_{i}Q_{j}\{\vec{\eta}_{i},{\cal K}_{ij}\},
\nonumber \\
\widetilde{\vec{\kappa}_{i}} &=&\vec{\kappa}_{i}-\frac{1}{2}\sum_{j\neq
i}Q_{i}Q_{j}\vec{\partial}_{\eta _{i}}{\cal K}_{ij}=\vec{\kappa}%
_{i}+\sum_{j\neq i}\frac{1}{2}Q_{i}Q_{j}\{\vec{\kappa}_{i},{\cal K}_{ij}\},
\label{VI55}
\end{eqnarray}

\noindent so that they satisfy

\begin{eqnarray}
\{\widetilde{\vec{\eta}_{k}},\widetilde{\vec{\eta}_{l}}\}^{\ast } &=&0,
\nonumber \\
\{ \widetilde{\vec{\kappa}_{k}},\widetilde{\vec{\kappa}_{l}}\}^{\ast } &=&0,
\nonumber \\
\{\widetilde{\vec{\eta}_{i}},\widetilde{\vec{\kappa}_{j}}\}^{\ast } &=&{\bf I%
}\delta _{ij}.  \label{VI56}
\end{eqnarray}

In analogy to the above modifications we assume that

\begin{equation}
\widetilde{\vec{\xi}_{i}}=\vec{\xi}_{i}+\frac{1}{2}\sum_{j\neq i}Q_{i}Q_{j}\{%
\vec{\xi}_{i},{\cal K}_{ij}\}.  \label{VI57}
\end{equation}
We now show that this definition, together with the previous ones, are
sufficient to guarantee that the new variables are canonical. To this end we
let both

\begin{eqnarray}
\widetilde{\vec{\mu}_{i}} &=&\vec{\mu}_{i}+\frac{1}{2}\sum_{j\neq
i}Q_{i}Q_{j}\{\vec{\mu}_{i},{\cal K}_{ij}\} ,  \nonumber \\
\widetilde{\vec{\nu}_{i}} &=&\vec{\nu}_{i}+\frac{1}{2}\sum_{j\neq
i}Q_{i}Q_{j}\{\vec{\nu}_{i},{\cal K}_{ij}\} ,  \label{VI58}
\end{eqnarray}

\noindent represent any of the new candidate canonical variables. Then we get

\begin{eqnarray}
\{\widetilde{\vec{\mu}_{i}},\widetilde{\vec{\nu}_{j}}\}^{\ast } &=&\{%
\widetilde{\vec{\mu}_{i}},\widetilde{\vec{\nu}_{j}}\}  \nonumber \\
&&-\Big[\int d^{3}\sigma \{\widetilde{\vec{\mu}_{i}},-\sum_{k}Q_{k}\vec{A}%
_{\perp Sk}(\vec{\sigma}-\vec{\eta}_{k}(\tau ),{\vec{\kappa}}_{k}{(\tau )},%
\vec{\xi}_{k}(\tau ))\}\cdot  \nonumber \\
&&\cdot \{-\sum_{l}Q_{l}\vec{\pi}_{\perp Sl}(\vec{\sigma}-\vec{\eta}%
_{l}(\tau ),{\vec{\kappa}}_{l}(\tau ),\vec{\xi}_{l}(\tau )),\widetilde{\vec{%
\nu}_{j}}\}+  \nonumber \\
&&+[\int d^{3}\sigma \{\widetilde{\vec{\mu}_{i}},-\sum_{k}Q_{k}\vec{\pi}%
_{\perp Sk}(\vec{\sigma}-\vec{\eta}_{k}(\tau ),{\vec{\kappa}}_{k}{(\tau )},%
\vec{\xi}_{k}(\tau ))\}\cdot  \nonumber \\
&&\cdot \{-\sum_{l}Q_{l}\vec{A}_{\perp Sl}(\vec{\sigma}-\vec{\eta}_{l}(\tau
),{\vec{\kappa}}_{l}(\tau ),\vec{\xi}_{l}(\tau )),\widetilde{\vec{\nu}_{j}}\}%
\Big].  \label{VI59}
\end{eqnarray}

Since we have

\begin{eqnarray}
\{\widetilde{\vec{\mu}_{i}},\widetilde{\vec{\nu}_{j}}\} &=&\{\vec{\mu}_{i},%
\vec{\nu}_{j}\}+\frac{1}{2}\{\vec{\mu}_{i},\sum_{k\neq j}Q_{j}Q_{k}\{\vec{\nu%
}_{j},{\cal K}_{jk}\}\}-\frac{1}{2}\{\vec{\nu}_{j},\sum_{k\neq i}Q_{i}Q_{k}\{%
\vec{\mu}_{i},{\cal K}_{ik}\}\} +  \nonumber \\
&&+\frac{1}{4}\{\sum_{k\neq i}Q_{i}Q_{k}\{\vec{\mu}_{i},{\cal K}%
_{ik}\},\sum_{l\neq j}Q_{j}Q_{l}\{\vec{\nu}_{j},{\cal K}_{jl}\}\} ,
\label{VI60}
\end{eqnarray}

\noindent we see from the definition of ${\cal K}_{ij}$ that the last
bracket would require that either $i=j,i=l,$ or $j=l$. \ Due to the Grassman
charges this forces the last bracket to be zero. \ For similar reasons we
can replace the new canonical variables by the old ones in the last two
lines of Eq.(\ref{VI59}) and with the sums truncating so that

\begin{eqnarray}
\{\widetilde{\vec{\mu}_{i}},\widetilde{\vec{\nu}_{j}}\} &=&\{\vec{\mu}_{i},%
\vec{\nu}_{j}\}+\frac{1}{2}\{\vec{\mu}_{i},\sum_{k\neq j}Q_{j}Q_{k}\{\vec{\nu%
}_{j},{\cal K}_{jk}\}\}-\frac{1}{2}\{\vec{\nu}_{j},\sum_{k\neq i}Q_{i}Q_{k}\{%
\vec{\mu}_{i},{\cal K}_{ik}\}\}  \nonumber \\
&&-{Q}_{i}Q_{j}\int d^{3}\sigma \Big[\{\vec{\mu}_{i},\vec{A}_{\perp Si}(\vec{%
\sigma}-\vec{\eta}_{i},{\vec{\kappa}}_{i},\vec{\xi}_{i})\}\cdot \{\vec{\pi}%
_{\perp Sj}(\vec{\sigma}-\vec{\eta}_{j},{\vec{\kappa}}_{j},\vec{\xi}_{j}),%
\vec{\nu}_{j}\}  \nonumber \\
&&-\{\vec{\mu}_{i},\vec{\pi}_{\perp Si}(\vec{\sigma}-\vec{\eta}_{i},{\vec{%
\kappa}}_{i},\vec{\xi}_{i})\}\cdot \{\vec{A}_{\perp Sj}(\vec{\sigma}-\vec{%
\eta}_{j},{\vec{\kappa}}_{j},\vec{\xi}_{j}),\vec{\nu}_{j}\}\Big].
\label{VI61}
\end{eqnarray}

Using the definition of the ${\cal K}_{jk}$, we find that the second and
third terms on the right hand side become

\begin{eqnarray}
&&\frac{1}{2}\{\vec{\mu}_{i},\sum_{k\neq j}Q_{j}Q_{k}\{\vec{\nu}_{j},{\cal K}%
_{jk}\}\}-\frac{1}{2}\{\vec{\nu}_{j},\sum_{k\neq i}Q_{i}Q_{k}\{\vec{\mu}_{i},%
{\cal K}_{ik}\}\} =  \nonumber \\
&=&\sum_{k\neq j}Q_{i}Q_{k}\frac{1}{2}\Big(\delta _{ij}\int d^{3}\sigma %
\Big[ \{\vec{\mu}_{i},\{\vec{\nu}_{i},\vec{A}_{\perp
\mathop{\rm Si}%
}(\vec{\sigma}-\vec{\eta}_{i},{\vec{\kappa}}_{i},\vec{\xi}_{i})\}\}\cdot
\vec{\pi}_{\perp k}(\vec{\sigma}-\vec{\eta}_{k},{\vec{\kappa}}_{k},\vec{\xi}%
_{k}) -  \nonumber \\
&&-\{\vec{\mu}_{i},\{\vec{\nu}_{i},\vec{\pi}_{\perp
\mathop{\rm Si}%
}(\vec{\sigma}-\vec{\eta}_{i},{\vec{\kappa}}_{i},\vec{\xi}_{i})\}\}\cdot
\vec{A}_{\perp k}(\vec{\sigma}-\vec{\eta}_{k},{\vec{\kappa}}_{k},\vec{\xi}%
_{k})\Big] -  \nonumber \\
&&-\delta _{ij}\int d^{3}\sigma \Big[ \{\vec{\nu}_{i},\{\vec{\mu}_{i},\vec{A}%
_{\perp
\mathop{\rm Si}%
}(\vec{\sigma}-\vec{\eta}_{i},{\vec{\kappa}}_{i},\vec{\xi}_{i})\}\}\cdot
\vec{\pi}_{\perp k}(\vec{\sigma}-\vec{\eta}_{k},{\vec{\kappa}}_{k},\vec{\xi}%
_{k}) -  \nonumber \\
&&-\{\vec{\nu}_{i},\{\vec{\mu}_{i},\vec{\pi}_{\perp
\mathop{\rm Si}%
}(\vec{\sigma}-\vec{\eta}_{i},{\vec{\kappa}}_{i},\vec{\xi}_{i})\}\}\cdot
\vec{A}_{\perp k}(\vec{\sigma}-\vec{\eta}_{k},{\vec{\kappa}}_{k},\vec{\xi}%
_{k})\Big]\Big)+  \nonumber \\
&&+Q_{j}Q_{i}\frac{1}{2}\Big(\int d^{3}\sigma \Big[ \{\vec{\mu}_{i},\vec{\pi}%
_{\perp i}(\vec{\sigma}-\vec{\eta}_{k},{\vec{\kappa}}_{k},\vec{\xi}%
_{k})\}\cdot \{\vec{\nu}_{j},\vec{A}_{\perp Sj}(\vec{\sigma}-\vec{\eta}_{j},{%
\vec{\kappa}}_{j},\vec{\xi}_{j})\} -  \nonumber \\
&&-\{\vec{\mu}_{i},\vec{A}_{\perp i}(\vec{\sigma}-\vec{\eta}_{i},{\vec{\kappa%
}}_{i},\vec{\xi}_{i})\}\cdot \{\vec{\nu}_{j},\vec{\pi}_{\perp Sj}(\vec{%
\sigma }-\vec{\eta}_{j},{\vec{\kappa}}_{j},\vec{\xi}_{j})\}\Big]\Big) -
\nonumber \\
&&-Q_{j}Q_{i}\frac{1}{2}\Big(\int d^{3}\sigma \Big[ \{\vec{\nu}_{j},\vec{\pi}%
_{\perp j}(\vec{\sigma}-\vec{\eta}_{j},{\vec{\kappa}}_{j},\vec{\xi}%
_{j})\}\cdot \{\vec{\mu}_{i},\vec{A}_{\perp
\mathop{\rm Si}%
}(\vec{\sigma}-\vec{\eta}_{i},{\vec{\kappa}}_{i},\vec{\xi}_{i})\} -
\nonumber \\
&&-\{\vec{\nu}_{j},\vec{A}_{\perp j}(\vec{\sigma}-\vec{\eta}_{j},{\vec{\kappa%
}}_{j},\vec{\xi}_{j})\}\cdot \{\vec{\mu}_{i},\vec{\pi}_{\perp
\mathop{\rm Si}%
}(\vec{\sigma}-\vec{\eta}_{i},{\vec{\kappa}}_{i},\vec{\xi}_{i})\}\Big] \Big).
\label{VI62}
\end{eqnarray}

Combining the terms proportional to $\delta _{ij}$ by using the Jacobi
identity and the fact that $\{\vec{\nu}_{i},\vec{\mu}_{i}\}$ has a zero
bracket with both $\vec{\pi}_{\perp
\mathop{\rm Si}%
}$ and $\vec{A}_{\perp
\mathop{\rm Si}%
}$, implies that these terms vanish, while the remaining terms combine to
exactly cancel the last two lines of Eq.(\ref{VI61}). \ Thus we have that

\begin{equation}
\{\widetilde{\vec{\mu}_{i}},\widetilde{\vec{\nu}_{j}}\}^{\ast }=\{\vec{\mu}%
_{i},\vec{\nu}_{j}\} ,  \label{VI63}
\end{equation}

\noindent and hence the new dynamical variables have the same Dirac brackets
as the Poisson brackets of the old dynamical variables, that is, they are
canonical. Thus, not only do we have same brackets as in Eq.(\ref{VI56}) but
also

\begin{eqnarray}
\{\widetilde{\vec{\xi}_{i}},\widetilde{\vec{\xi}_{j}}\}^{\ast } &=&\{\vec{\xi%
}_{i},\vec{\xi}_{j}\}=-i{\bf I}\delta _{ij}  \nonumber \\
\{\widetilde{\vec{\eta}_{i}},\widetilde{\vec{\xi}_{j}}\}^{\ast } &=&\{\vec{%
\eta}_{i},\vec{\xi}_{j}\}=0  \nonumber \\
\{\widetilde{\kappa _{i}},\widetilde{\vec{\xi}_{j}}\}^{\ast } &=&\{\vec{%
\kappa}_{i},\vec{\xi}_{j}\}=0.  \label{VI64}
\end{eqnarray}
\ \

As in Ref.\cite{ap} the rest frame condition

\begin{eqnarray}
&&{\vec{{\cal H}}}_{p}={\vec{{\cal P}}}_{(int)}=\sum_{i-1}^{N}\vec{\kappa}%
_{i}+\int d^{3}\sigma \lbrack {\vec{\pi}}_{\perp }\times {\vec{B}}](\tau ,%
\vec{\sigma})=  \nonumber \\
&=&\sum_{i-1}^{N}\vec{\kappa}_{i}+\sum_{i<j}Q_{i}Q_{j}\int d^{3}\sigma \Big[%
\vec{\pi}_{\perp Si}(\vec{\sigma}-\vec{\eta}_{i},{\vec{\kappa}}_{i})\times ({%
\vec{\partial}}_{\sigma }\times \vec{A}_{\perp Sj}(\vec{\sigma}-\vec{\eta}%
_{j},{\vec{\kappa}}_{j}))+  \nonumber \\
&&+\vec{\pi}_{\perp Sj}(\vec{\sigma}-\vec{\eta}_{j},{\vec{\kappa}}%
_{j})\times ({\vec{\partial}}_{\sigma }\times \vec{A}_{\perp Si}(\vec{\sigma}%
-\vec{\eta}_{i},{\vec{\kappa}}_{i}))\Big]\approx 0.  \label{VI65}
\end{eqnarray}

\noindent can be rewritten in these new canonical variables as

\begin{equation}
{\vec{{\cal H}}}_{p}={\vec{{\cal P}}}_{(int)}=\widetilde{\vec{\kappa}_{+}}%
=\sum_{i=1}^{N}\widetilde{\vec{\kappa}_{i}}=0.  \label{VI66}
\end{equation}
This is due to the fact that the steps of expanding the cross products,
integrating by parts and using the transverse gauge condition make no
additonal reference to the spin dependence of the fields. \

The previous results allow to get ${\cal \vec{P}}_{(int)}$ in terms of the
final canonical variables. For the internal angular momentum in terms of
these variables we get

\begin{eqnarray}
{\cal J}_{(int)}^{r} &=&\varepsilon ^{rst}{\bar{S}}_{s}^{st}=\sum_{i=1}^{N}(%
\vec{\eta}_{i}(\tau )\times {\vec{\kappa}}_{i}(\tau ))^{r}-{\frac{i}{2}}%
\sum_{i=1}^{N}({\vec{\xi}}_{i}\times {\vec{\xi}}_{i})^{r}+\int d^{3}\sigma
\,(\vec{\sigma}\times \,{[\vec{\pi}}_{\perp }{\times }\vec{B}{{]}(\tau ,\vec{%
\sigma})})^{r}=  \nonumber \\
&=&\sum_{i=1}^{N}[(\widetilde{\vec{\eta}_{i}}-\vec{\alpha}_{i})\times (%
\widetilde{\vec{\kappa}_{i}}-\vec{\beta}_{i})]^{r}-{\frac{i}{2}}%
\sum_{i=1}^{N}[(\widetilde{\vec{\xi}_{i}}-\vec{\gamma}_{i})\times (%
\widetilde{\vec{\xi}_{i}}-\vec{\gamma}_{i})]^{r} +  \nonumber \\
&+&\frac{1}{2}\sum_{i=1}^{N}\sum_{j\neq i}^{N}Q_{i}Q_{j}\int d^{3}\sigma \,(%
\vec{\sigma}\times \,{[{\vec{\pi}}_{\perp Si}\times (\vec{\partial}}_{\sigma
}\times {\vec{A}}_{\perp Sj}){{]}+i\leftrightarrow j)}.  \label{VI67}
\end{eqnarray}

In Ref.\cite{ap} we showed that $-\vec{\alpha}_{i}\times \widetilde{\vec{%
\kappa}_{i}}-\widetilde{\vec{\eta}_{i}}\times \vec{\beta}_{i}$ exactly
compensates the last expression for the field portion ($\vec{\alpha}%
_{i}\times \vec{\beta}_{i}$ turned out to be zero due to Grassman
truncation). \ The arguement we used depended explicitly on the fact that ${{%
\vec{\pi}}_{\perp Si},\vec{A}}_{\perp Sj}$ have expected transformation
properties. \ In the case here of spinning particles, we need in addition
the terms $-\vec{\gamma}_{i}\times \widetilde{\vec{\xi}_{i}}-\widetilde{\vec{%
\xi}_{i}}\times \vec{\gamma}_{i}$. Using the forms for $\vec{\alpha}_{i}$ , $%
\vec{\beta}_{i}$, and $\vec{\gamma}_{i}$ given in Eqs.(\ref{VI55}), (\ref
{VI57}), together with the expression for ${\cal K}_{ij}$, and expanding the
cross products in the integral, using the transverse nature of the field
together with vanishing surface terms, we find that

\begin{eqnarray}
{\cal J}_{(int)}^{r} &=&\sum_{i=1}^{N}(\widetilde{\vec{\eta}_{i}}\times
\widetilde{\vec{\kappa}_{i}})^{r}-{\frac{i}{2}}\sum_{i=1}^{N}(\widetilde{%
\vec{\xi}_{i}}\times \widetilde{\vec{\xi}_{i}})^{r} +  \nonumber \\
&&+\frac{1}{2}\sum_{i=1}^{N}\sum_{j\neq i}^{N}Q_{i}Q_{j}\Big[(\vec{\eta}%
_{i}\times \vec{\partial}_{\eta _{i}}+\vec{\kappa}_{i}\times \vec{\partial}%
_{\kappa _{i}}+\vec{\xi}_{i}\times \vec{\partial}_{\xi _{i}})^{r}  \nonumber
\\
&&\int d^{3}\sigma ({\vec{A}_{\perp Si}\cdot {\vec{\pi}}_{\perp Sj}-\vec{A}%
_{\perp Sj}\cdot {\vec{\pi}}_{\perp Si}})\Big] -  \nonumber \\
&&-\frac{1}{2}\sum_{i=1}^{N}\sum_{j\neq i}^{N}Q_{i}Q_{j}\int d^{3}\sigma %
\Big[ \vec{A}_{\perp Si}\times {\vec{\pi}}_{\perp Sj}-  \nonumber \\
&-&\vec{\sigma}\times (\vec{\partial}_{\sigma }A^{k}{_{\perp Si})\pi _{\perp
Sj}^{k}+\vec{A}_{\perp Sj}\times {\vec{\pi}}_{\perp Si}+\vec{\sigma}\times (}
A^{k}{_{\perp Si}{\vec{\partial}}_{\sigma }\pi _{\perp Sj}^{k})\Big]} .
\label{VI68}
\end{eqnarray}

\ In order to see how the cancellation between the last two lines works, we
first spell out the new vector and scalar dependence of the vector potential
and its canonical momentum. \ Their general forms are [from Eq.(\ref{VI40})
and Eq.(\ref{VI45})]

\begin{eqnarray}
{\vec{A}_{\perp Si}} &=&\frac{1}{4\pi \rho _{i}}\Big[\vec{\kappa}%
_{i}f_{i}(\kappa _{i}^{2},\vec{\kappa}_{i}\cdot \hat{\rho}_{i})+\hat{\rho}%
_{i}g_{i}(\kappa _{i}^{2},\vec{\kappa}_{i}\cdot \hat{\rho}_{i})+\vec{\kappa}%
_{i}\cdot \vec{\xi}_{i}\vec{\xi}_{i}\sum_{n}h_{n}(\rho _{i})l_{ni}(\kappa
_{i}^{2},\vec{\kappa}_{i}\cdot \hat{\rho}_{i}) +  \nonumber \\
&&+\frac{\vec{\kappa}_{i}\vec{\kappa}_{i}\cdot \vec{\xi}_{i}\vec{\xi}%
_{i}\cdot \hat{\rho}_{i}}{\rho _{i}}j_{i}(\kappa _{i}^{2},\vec{\kappa}%
_{i}\cdot \hat{\rho}_{i})+\frac{\hat{\rho}_{i}\vec{\kappa}_{i}\cdot \vec{\xi}%
_{i}\vec{\xi}_{i}\cdot \hat{\rho}_{i}}{\rho _{i}}m_{i}(\kappa _{i}^{2},\vec{%
\kappa}_{i}\cdot \hat{\rho}_{i})\Big],  \nonumber \\
{{\vec{\pi}}_{\perp Si}} &=&\frac{1}{4\pi \rho _{i}^{2}}\Big[\hat{\rho}%
_{i}c_{i}(\kappa _{i}^{2},\vec{\kappa}_{i}\cdot \hat{\rho}_{i})+  \nonumber
\\
&+&\vec{\xi}_{i}\vec{\xi}_{i}\cdot \hat{\rho}_{i}\sum_{n}e_{n}(\rho
_{i})d_{ni}(\kappa _{i}^{2},\vec{\kappa}_{i}\cdot \hat{\rho}_{i})+\vec{\xi}%
_{i}\vec{\xi}_{i}\cdot \vec{\kappa}_{i}\sum_{n}a_{n}(\rho _{i})b_{ni}(\kappa
_{i}^{2},\vec{\kappa}_{i}\cdot \hat{\rho}_{i})\Big],  \label{VI69}
\end{eqnarray}

\noindent where

\begin{eqnarray}
\hat{\rho}_{i} &:&=\frac{(\vec{\sigma}-\vec{\eta}_{i})}{|\vec{\sigma}-\vec{%
\eta}_{i}|};\quad \rho _{i}:=|\vec{\sigma}-\vec{\eta}_{i}|:=|\vec{\rho}%
_{i}|;\quad \vec{\partial}_{\sigma }\hat{\rho}_{i}=\frac{1}{\rho _{i}}({\bf %
I-}\hat{\rho}_{i}\hat{\rho}_{i}).  \nonumber \\
&&  \label{VI70}
\end{eqnarray}

Now we use

\begin{equation}
\vec{\eta}_{i}\times \vec{\partial}_{\eta _{i}}=-\vec{\sigma}\times \vec{%
\partial }_{\sigma }+\vec{\rho}_{i}\times \vec{\partial}_{\sigma },
\label{VI71}
\end{equation}

\noindent and several facts (using generic functions $h$):

\begin{eqnarray}
\vec{\rho}_{i}\times \vec{\partial}_{\sigma }h(\rho _{i}) &=&0,  \nonumber \\
(\vec{\rho}_{i}\times \vec{\partial}_{\sigma }+\vec{\kappa}_{i}\times \vec{%
\partial}_{\kappa _{i}})h_{i}(\kappa _{i}^{2},\vec{\kappa}_{i}\cdot \hat{\rho%
}_{i}) &=&0,  \nonumber \\
(\vec{\rho}_{i}\times \vec{\partial}_{\sigma }+\vec{\kappa}_{i}\times \vec{%
\partial}_{\kappa _{i}})\vec{\kappa}_{i}\cdot &=&\vec{\kappa}_{i}{\bf \times
},  \nonumber \\
(\vec{\rho}_{i}\times \vec{\partial}_{\sigma }+\vec{\kappa}_{i}\times \vec{%
\partial}_{\kappa _{i}})\hat{\rho}_{i}\cdot &=&\hat{\rho}_{i}\times ,
\nonumber \\
(\vec{\rho}_{i}\times \vec{\partial}_{\sigma }+\vec{\kappa}_{i}\times \vec{%
\partial}_{\kappa _{i}}+\vec{\xi}_{i}\times \vec{\partial}_{\xi _{i}})(\vec{%
\kappa}_{i}\cdot \vec{\xi}_{i})\vec{\xi}_{i}\cdot &=&\vec{\xi}_{i}\times ,
\nonumber \\
(\vec{\rho}_{i}\times \vec{\partial}_{\sigma }+\vec{\kappa}_{i}\times \vec{%
\partial}_{\kappa _{i}}+\vec{\xi}_{i}\times \vec{\partial}_{\xi _{i}})\vec{%
\kappa}_{i}\cdot \vec{\xi}_{i}\vec{\xi}_{i}\cdot \hat{\rho}_{i}\vec{\kappa}%
_{i}\cdot &=&\vec{\kappa}_{i}\cdot \vec{\xi}_{i}\vec{\xi}_{i}\cdot \hat{\rho}%
_{i}\vec{\kappa}\times ,  \nonumber \\
(\vec{\rho}_{i}\times \vec{\partial}_{\sigma }+\vec{\kappa}_{i}\times \vec{%
\partial}_{\kappa _{i}}+\vec{\xi}_{i}\times \vec{\partial}_{\xi _{i}})\hat{%
\rho}_{i}\vec{\kappa}_{i}\cdot \vec{\xi}_{i}\vec{\xi}_{i}\cdot \hat{\rho}_{i}%
\hat{\rho}_{i}\cdot &=&\hat{\rho}_{i}\vec{\kappa}_{i}\cdot \vec{\xi}_{i}\vec{%
\xi}_{i}\cdot \hat{\rho}_{i}\hat{\rho}_{i}\times ,  \nonumber \\
(\vec{\rho}_{i}\times \vec{\partial}_{\sigma }+\vec{\kappa}_{i}\times \vec{%
\partial}_{\kappa _{i}}+\vec{\xi}_{i}\times \vec{\partial}_{\xi _{i}})(-)%
\vec{\xi}_{i}\vec{\xi}_{i}\cdot \hat{\rho}_{i}\vec{\xi}_{i}\cdot &=&(-)\vec{%
\xi}_{i}\vec{\xi}_{i}\cdot \hat{\rho}_{i}\vec{\xi}_{i}\times .  \label{VI72}
\end{eqnarray}

\noindent where the $\times $ indicates vector cross product. \ Thus the
effect of the bracket $(\vec{\rho}_{i}\times \vec{\partial}_{\sigma }+\vec{%
\kappa}_{i}\times \vec{\partial}_{\kappa _{i}}+\vec{\xi}_{i}\times \vec{%
\partial}_{\xi _{i}})$ is to turn scalar product into cross product, \ which
together with the first part of Eq.(\ref{VI68}) leads to the cancellation
referred to above yielding

\begin{equation}
{\cal \vec{J}}_{(int)}=\sum_{i=1}^{N}(\widetilde{\vec{\eta}_{i}}\times
\widetilde{\vec{\kappa}_{i}})-{\frac{i}{2}}\sum_{i=1}^{N}(\widetilde{\vec{\xi%
}_{i}}\times \widetilde{\vec{\xi}_{i}}).  \label{VI73}
\end{equation}

Thus the total angular momentum of fields plus particles reduces to an
expression involving just the new canonical particle variables when the
fields are eliminated in pairs by using the modified Dirac brackets.

\vfill\eject

\section{The Semi-Classical Hamiltonian with the Darwin and Spin-Dependent
Potentials.}

In this Section we study the Hamiltonian $M$ of Eq.(\ref{V1}), which
contains both scalar (inside and outside the square roots) and vector
(inside the square roots) direct interparticle potentials. By using the
semi-classical property $Q^2_i=0$, these potentials can be reexpressed in
terms of a unique scalar potential outside the square roots. This
semi-classical potential contains the Coulomb potential and, moreover,
semi-classical relativistic Darwin and spin-dependent potentials. The
determination of these potentials will be done firstly by using the old
variables (not canonical with respect to the Dirac brackets (\ref{VI52}))
and then in the final canonical variables.

\subsection{The Semi-Classical Hamiltonian in the Old Variables.}

If we rewrite the Hamiltonian $M$ (\ref{V1}) in the following form

\begin{eqnarray}
M &=&\sum_{i=1}^{N}\Big[\sqrt{m_{i}^{2}+({\vec{\kappa}}_{i}(\tau )-Q_{i}{\ {%
\vec{A}}}_{\perp }(\tau ,\vec{\eta}_{i}(\tau )))^{2}}+\frac{{\cal R} _{i}}{2%
\sqrt{m_{i}^{2}+{\vec{\kappa}}_{i}(\tau )^{2}}}\Big] +  \nonumber \\
&+&\sum_{i\neq j}\frac{Q_{i}Q_{j}}{4\pi \mid \vec{\eta}_{i}(\tau )-\vec{\eta}
_{j}(\tau )\mid }\Big[1-i\frac{{\vec{\kappa}}_{i}\cdot \vec{\xi}_{i}\vec{ \xi%
}_{i}\cdot (\vec{\eta}_{i}(\tau )-\vec{\eta}_{j}(\tau ))}{\mid \vec{\eta}
_{i}(\tau )-\vec{\eta}_{j}(\tau )\mid ^{2}(m_i+\sqrt{m_{i}^{2}+{\vec{\kappa }%
}_{i}^{2}})\sqrt{m_{i}^{2}+{\vec{\kappa}}_{i}^{2}}}\Big]+  \nonumber \\
&+&\int d^{3}\sigma {\frac{1}{2}}[{\vec{\pi}}_{\perp }^{2}+{\vec{ B}}%
^{2}](\tau ,\vec{\sigma}) =  \nonumber \\
&&{}  \nonumber \\
&:=&\sum_{i=1}^{N}\sqrt{m_{i}^{2}+{\vec{\kappa}}_{i}(\tau )^{2}}
+\sum_{i\neq j}\frac{Q_{i}Q_{j}}{4\pi \mid \vec{\eta}_{i}(\tau )-\vec{\eta}%
_{j}(\tau )\mid }+V_{DS}(\tau ),  \label{VII1}
\end{eqnarray}

\noindent with the scalar potential $V_{DS}(\tau )$ containing the
generalized Darwin and spin-dependent interactions. This potential can be
decomposed in three terms

\begin{equation}
V_{DS}(\tau )\,:=\, U_{1}(\tau )+{V}_{DSO}(\tau )+U(\tau ),  \label{VII2}
\end{equation}

\noindent where $U(\tau )$, given in Eq.(\ref{VI46}), contains the
contribution coming from the radiation field energy evaluated on the
semi-classical Lienard-Wiechert solution.

The potential $U_1(\tau )$ has the form

\begin{equation}
U_{1}(\tau )=\sum_{i=1}^{N}\frac{{\cal R}_{i}-2Q_{i}{\vec{\kappa}} _{i}(\tau
)\cdot {{\vec{A}}}_{\perp }(\tau ,\vec{\eta}_{i}(\tau ))}{2 \sqrt{m_{i}^{2}+{%
\vec{\kappa}}_{i}(\tau )^{2}}},  \label{VII3}
\end{equation}

\noindent containing the Grassman truncated vector potential, which was in
the kinetic terms under the square roots, and the spin-magnetic field and
spin-electric field interactions present in Eq.(\ref{V2})

\begin{eqnarray}
{\cal R}_{i}&=&iQ_{i}\vec{\xi}_{i}\times \vec{\xi}_{i}\cdot {\vec{B}} (\tau ,%
\vec{\eta}_{i})-2i\frac{Q_{i}{\vec{\kappa}}_{i}\cdot \vec{\xi}_{i}\vec{\xi}%
_{i}\cdot {\vec{\pi}}_{\perp }(\tau ,\vec{\eta}_{i})}{ m_i+\sqrt{m_{i}^{2}+{%
\vec{\kappa}}_{i}^{2}}}=  \nonumber \\
&=& -2 Q_i {\vec {\bar S}}_{i\xi}(\tau ) \cdot \Big[ {\vec{B}}(\tau ,\vec{%
\eta}_{i}(\tau ))+ {\frac{{\ {\vec{\pi}}_{\perp }(\tau ,\vec{\eta}_{i}(\tau
)) \times {\ {\vec \kappa}}_i(\tau )}}{{\ m_{i}+\sqrt{m_{i}^{2}+ {{\vec{%
\kappa}}}_{i}^{2}(\tau )} }}} \Big].  \label{VII4}
\end{eqnarray}

Finally, the potential $V_{DSO}(\tau )$ come from the interaction of the
spin with the Coulomb electric field (after quantization it will contain
important Darwin and spin orbit terms)

\begin{eqnarray}
{V}_{DSO}(\tau )&=&-i\sum_{i\neq j}\frac{Q_{i}Q_{j}}{4\pi \mid \vec{\eta}%
_{i}(\tau )-\vec{\eta}_{j}(\tau )\mid ^{3}}\frac{{\vec{\kappa}} _{i}\cdot
\vec{\xi}_{i}\vec{\xi}_{i}\cdot (\vec{\eta}_{i}(\tau )-\vec{\eta} _{j}(\tau
))}{(m_i+\sqrt{m_{i}^{2}+{\vec{\kappa}}_{i}^{2}})\sqrt{ m_{i}^{2}+{\vec{%
\kappa}}_{i}^{2}}}=  \nonumber \\
&=&\sum_{i\neq j}\frac{Q_{i}Q_{j}}{4\pi \mid \vec{\eta}_{i}(\tau )-\vec{\eta}%
_{j}(\tau )\mid ^{3}}\frac{{\vec{\kappa}} _{i}(\tau )\cdot (\vec{\eta}%
_{i}(\tau )-\vec{\eta} _{j}(\tau ))\times {\vec {\bar S}}_{i\xi}(\tau )} {%
(m_i+\sqrt{m_{i}^{2}+{\vec{\kappa}}_{i}^{2}})\sqrt{ m_{i}^{2}+{\vec{\kappa}}%
_{i}^{2}}}.  \label{VII5}
\end{eqnarray}

\noindent The rest-frame conditions are given by the vanishing of the {\it %
internal} 3-momentum [the last line is Eq.(\ref{VI66})]

\begin{equation}
{\vec{{\cal P}}}_{(int)}={\vec{{\cal H}}}_{p}(\tau )={\vec{\kappa}}_{+}(\tau
)+\int d^{3}\sigma \lbrack {\vec{\pi}}_{\perp S}\times {\vec{B}}_{S}](\tau ,%
\vec{\sigma}) = \sum_{i=1}^N {\tilde {\vec \kappa}}_{+} \approx 0,
\label{VII6}
\end{equation}

\noindent with the term coming from radiation field 3-momentum given in Eq.(%
\ref{VI47}).

Using the closed forms for the vector potential and the electric and
magnetic fields given in Eqs. (\ref{VI40}), (\ref{VI45}) and (\ref{VI44}),
respectively, we can evaluate both the spin-independent and spin-dependent
parts of the scalar potential $U_{1}(\tau )$ of Eq.(\ref{VII3}).

While the last term in $U_1(\tau )$ gives

\begin{eqnarray}
&&-\sum_{i=1}^{N}\frac{Q_{i}\vec{\kappa}_{i}(\tau )\cdot {\vec{A}}_{\perp
}(\tau ,\vec{\eta}_{i}(\tau ))}{\sqrt{m_{i}^{2}+\check{\vec{\kappa}}%
_{i}(\tau )^{2}}} =  \nonumber \\
&=&-\sum_{i\neq j}^{N}{\frac{Q_{j}Q_{i}}{4\pi \sqrt{m_{i}^{2}+\vec{\kappa}%
_{i}{}^{2}}}\Big(\Big[}\vec{\kappa}_{i}{+\frac{i\vec{\kappa}_{j}\cdot \vec{%
\xi}_{j}\vec{\kappa}_{i}\cdot \vec{\xi}_{j}\vec{\partial}_{ij}}{\sqrt{%
m_{j}^{2}+\vec{\kappa}_{j}^{2}}(m_j+\sqrt{m_{j}^{2}+\vec{\kappa}_{j}^{2}})}%
\Big]\cdot \lbrack }\vec{\kappa}_{j}+\hat{\eta}_{ij}\vec{\kappa}_{j}\cdot
\hat{\eta}_{ij}]  \nonumber \\
&& {\frac{1}{|\vec{\eta}_{i}-\vec{\eta}_{j}|(\sqrt{m_{j}^{2}+{\vec{\kappa}%
_{j}}^{2}}+\sqrt{m_{j}^{2}+(\vec{\kappa}_{j}\cdot {\hat{\eta}}_{ij})^{2}})}-%
\vec{\kappa}_{i}\cdot \vec{\xi}_{j}\vec{\xi}_{j}\cdot \vec{\partial}_{ij}}%
\frac{1}{|\vec{\eta}_{i}-\vec{\eta}_{j}|\sqrt{m_{j}^{2}+(\vec{\kappa}%
_{j}\cdot {\hat{\eta}}_{ij})^{2}}}\Big),  \nonumber \\
&&  \label{VII7}
\end{eqnarray}

\noindent the magnetic and electric terms in ${\cal R}_i(\tau )$ give

\begin{eqnarray}
&&+\sum_{i=1}^{N}\frac{iQ_{i}\vec{\xi}_{i}\times \vec{\xi}_{i}\cdot \check{%
\vec{B}}(\tau ,\vec{\eta}_{i})}{2\sqrt{m_{i}^{2}+\check{\vec{\kappa}}%
_{i}(\tau )^{2}}} =  \nonumber \\
&=&-\sum_{i\neq j}^{N}{\frac{iQ_{i}Q_{j}\vec{\xi}_{i}\times \vec{\xi}%
_{i}\cdot \{{[}\vec{\kappa}_{j}-i\vec{\xi}_{j}\vec{\xi}_{j}\cdot \vec{%
\partial}_{ij}{+\frac{i\vec{\kappa}_{j}\cdot \vec{\partial}\vec{\kappa}%
_{j}(\tau )\cdot \vec{\xi}_{j}(\tau )\vec{\xi}_{j}(\tau )}{\sqrt{m_{j}^{2}+%
\vec{\kappa}_{j}^{2}}(\sqrt{m_{j}^{2}+\vec{\kappa}_{j}^{2}}+m_{j})}]}}{8\pi
\sqrt{m_{i}^{2}+\vec{\kappa}_{i}{}^{2}}}}  \nonumber \\
&&\times \vec{\partial}_{ij}[{\frac{1}{|\vec{\eta}_{i}-\vec{\eta}_{j}|}}%
\frac{1}{\sqrt{m_{j}^{2}+(\vec{\kappa}_{j}\cdot {\hat{\eta}}_{ij})^{2}}}]\},
\label{VII8}
\end{eqnarray}

\noindent and

\begin{eqnarray}
&&-i\sum_{i=1}^{N}\frac{Q_{i}\check{\vec{\kappa}}_{i}\cdot \vec{\xi}_{i}\vec{%
\xi}_{i}\cdot \check{\vec{\pi}}_{\perp }(\tau ,\vec{\eta}_{i})}{\sqrt{%
m_{i}^{2}+\check{\vec{\kappa}}_{i}{}^{2}}(m_i+\sqrt{m_{i}^{2}+\check{\vec{%
\kappa}}_{i}^{2}})} =  \nonumber \\
&=&i\sum_{i\neq j}^{N}{\frac{Q_{i}Q_{j}\vec{\kappa}_{i}\cdot \vec{\xi}_{i}}{%
\sqrt{m_{i}^{2}+\vec{\kappa}_{i}{}^{2}}(m_i+\sqrt{m_{i}^{2}+\vec{\kappa}%
_{i}^{2}})}\Big(\Big[}\vec{\xi}_{i}{+\frac{i\vec{\kappa}_{j}\cdot \vec{\xi}%
_{j}\vec{\xi}_{i}\cdot \vec{\xi}_{j}\vec{\partial}_{ij}}{\sqrt{m_{j}^{2}+%
\vec{\kappa}_{j}^{2}}(m_j+\sqrt{m_{j}^{2}+\vec{\kappa}_{j}^{2}})}\Big]\cdot }%
\frac{\hat{\eta}_{ij}}{\eta _{ij}^{2}}  \nonumber \\
&& {(}\frac{m_{j}^{2}\sqrt{m_{j}^{2}+\vec{\kappa}_{j}^{2}}}{[m_{j}^{2}+(\vec{%
\kappa}_{j}\cdot \hat{\eta}_{ij})^{2}]^{3/2}}-1)-{i\vec{\xi} }_{i}\cdot {%
\vec{\xi}_{j}\vec{\xi}_{j}\cdot \vec{\partial}_{ij}\vec{\kappa}_{j}\cdot
\vec{\partial}}_{ij}\frac{1}{\eta _{ij}}\frac{1}{\sqrt{m_{j}^{2}+(\vec{\kappa%
}_{j}\cdot \hat{\eta}_{ij})^{2}}}\Big) ,  \label{VII9}
\end{eqnarray}

\noindent respectively.

In Appendix D there is a comparison between these terms for the potential $%
U_{1}(\tau )$ with the terms appearing in the potential $U(\tau )$ of Eq.(%
\ref{VI47}). Introducing the differential operators ${\bf {\vec{k}}}_{i},
{\bf {\vec{l}}}_{i},{\bf {\vec{f}}}_{i}$ and ${\bf {\vec{g}}}_{i}$ defined by

\begin{eqnarray}
{\bf {\vec{k}}} &:&=\vec{\kappa}_{i}+{\bf {\vec{f}}}_{i},\,\,\,\,\,{\bf {%
\vec{f}}}_{i}:=-i\vec{\xi}_{i}\vec{\xi}_{i}\cdot \vec{\partial}_{ij}+\frac{i%
\vec{\kappa}_{i}\cdot \vec{\partial}_{ij}\vec{\kappa}_{i}\cdot \vec{\xi}_{i}%
\vec{\xi}_{i}}{(m_{i}+\sqrt{m_{i}^{2}+\vec{\kappa}_{i}{}^{2}})\sqrt{%
m_{i}^{2}+\vec{\kappa}_{i}{}^{2}}},  \nonumber \\
{\bf {\vec{l}}}_{i} &:&=\vec{\kappa}_{i}+{\bf {\vec{g}}}_{i},\,\,\,{\bf {%
\vec{g}}}_{i}:=\frac{i\vec{\kappa}_{i}\cdot \vec{\partial}_{ij}\vec{\kappa}%
_{i}\cdot \vec{\xi}_{i}\vec{\xi}_{i}}{(m_{i}+\sqrt{m_{i}^{2}+\vec{\kappa}%
_{i}{}^{2}})\sqrt{m_{i}^{2}+\vec{\kappa}_{i}{}^{2}}},  \label{VII10}
\end{eqnarray}

\noindent and combining all terms, we find that the potential $V_{DS}(\tau )$
can be rewritten in the form

\begin{equation}
V_{DS}(\tau )=V_{DSO}(\tau )+U_{1}(\tau )+U(\tau )=V_{LDS}+U_{HDS},
\label{VII11}
\end{equation}

\noindent where

\begin{eqnarray}
V_{LDS} &:&=\sum_{i\neq j}\frac{Q_{i}Q_{j}}{4\pi }\Big[-\frac{\vec{\kappa}%
_{i}\cdot \vec{\kappa}_{j}+\vec{\kappa}_{i}\cdot {\hat{\eta}}_{ij}\vec{\kappa%
}_{j}\cdot {\hat{\eta}}_{ij}}{4\sqrt{m_{i}^{2}+\vec{\kappa}_{i}{}^{2}}\sqrt{%
m_{j}^{2}+\vec{\kappa}_{j}{}^{2}}}\frac{1}{\eta _{ij}}-i\frac{\vec{\kappa}%
_{i}\cdot \vec{\xi}_{i}\vec{\xi}_{i}\cdot \vec{\eta}_{ij}}{\eta
_{ij}^{3}(m_i+\sqrt{m_{i}^{2}+\vec{\kappa}_{i}^{2}})\sqrt{m_{i}^{2}+\vec{%
\kappa}_{i}^{2}}} -  \nonumber \\
&&-i{\frac{\vec{\xi}_{i}\times \vec{\xi}_{i}\cdot ({\bf \vec{k}}_{j}\times
\vec{\partial}_{ij})}{2\sqrt{m_{i}^{2}+\vec{\kappa}_{i}{}^{2}}}}\frac{1}{%
\eta _{ij}\sqrt{m_{j}^{2}+(\vec{\kappa}_{j}\cdot {\hat{\eta}}_{ij})^{2}}}+i{%
\frac{\vec{\kappa}_{i}\cdot \vec{\xi}_{i}}{\sqrt{m_{i}^{2}+\vec{\kappa}%
_{i}{}^{2}}(m_i+\sqrt{m_{i}^{2}+\vec{\kappa}_{i}^{2}})}}  \nonumber \\
&& {\Big([}\vec{\xi}_{i}{+\frac{i\vec{\kappa}_{j}\cdot \vec{\xi}_{j}\vec{\xi}%
_{i}\cdot \vec{\xi}_{j}\vec{\partial}_{ij}}{\sqrt{m_{j}^{2}+\vec{\kappa}%
_{j}^{2}}(\sqrt{m_{j}^{2}+\vec{\kappa}_{j}^{2}}+m_{j})}]\cdot }\frac{\hat{%
\eta}_{ij}}{\eta _{ij}^{2}}{(}\frac{m_{j}^{2}\sqrt{m_{j}^{2}+\vec{\kappa}%
_{j}^{2}}}{[m_{j}^{2}+(\vec{\kappa}_{j}\cdot \hat{\eta}_{ij})^{2}]^{3/2}}-1)
-  \nonumber \\
&&-{i\vec{\xi}}_{i}\cdot {\vec{\xi}_{j}\vec{\xi}_{j}\cdot \vec{\partial}_{ij}%
\vec{\kappa}_{j}\cdot \vec{\partial}}_{ij}\frac{1}{\eta _{ij}}\frac{1}{\sqrt{%
m_{j}^{2}+(\vec{\kappa}_{j}\cdot \hat{\eta}_{ij})^{2}}}\Big)+  \nonumber \\
&+&\frac{{\bf {\vec{f}}_{i}\cdot (\vec{\kappa}_{j}+{\vec{f}}_{j})}}{\sqrt{%
m_{j}^{2}+\vec{\kappa}_{j}{}^{2}}}\frac{1}{\eta _{ij}\sqrt{m_{i}^{2}+(\vec{%
\kappa}_{i}\cdot {\hat{\eta}}_{ij})^{2}}}+\frac{{\bf {\vec{f}}_{j}\cdot (%
\vec{\kappa}_{i}+{\vec{f}}_{i})}}{\sqrt{m_{i}^{2}+\vec{\kappa}_{i}{}^{2}}}%
\frac{1}{\eta _{ij}\sqrt{m_{j}^{2}+(\vec{\kappa}_{j}\cdot {\hat{\eta}}%
_{ij})^{2}}}-  \nonumber \\
&&-\frac{\vec{\kappa}_{i}\cdot {\bf {\vec{f}}}_{j}+\vec{\kappa}_{j}\cdot
{\bf {\vec{f}}}_{i}+{\bf {\vec{f}}}_{i}\cdot {\bf {\vec{f}}}_{j}}{2\sqrt{%
m_{i}^{2}+\vec{\kappa}_{i}{}^{2}}\sqrt{m_{j}^{2}+\vec{\kappa}_{j}{}^{2}}}%
\frac{1}{\eta _{ij}} +  \nonumber \\
&+&\Big(\vec{\kappa}_{i}\cdot {\bf {\vec{f}}}_{j}+\vec{\kappa}_{j}\cdot {\bf
{\vec{f}}}_{i}+{\bf {\vec{f}}}_{i}\cdot {\bf {\vec{f}}}_{j}-({\bf {\vec{f}}}%
_{j}\vec{\kappa}_{i}\cdot {\hat{\eta}}_{ij}+{\bf {\vec{f}}}_{i}\vec{\kappa}%
_{j}\cdot {\hat{\eta}}_{ij}+{\bf {\vec{f}}}_{j}{\bf {\vec{f}}}_{i}\cdot {%
\hat{\eta}}_{ij})\cdot {\hat{\eta}}_{ij}\Big)  \nonumber \\
&&\Big( \frac{1}{2\sqrt{m_{i}^{2}+\vec{\kappa}_{i}{}^{2}}\sqrt{m_{j}^{2}+%
\vec{\kappa}_{j}{}^{2}}\eta _{ij}}-\frac{2}{\eta _{ij}\sqrt{m_{j}^{2}+\vec{%
\kappa}_{j}{}^{2}}{\large (}\sqrt{m_{i}^{2}+\vec{\kappa}_{i}{}^{2}}+\sqrt{%
m_{i}^{2}+(\vec{\kappa}_{i}{}\cdot \hat{\eta}_{ij})^{2}}{\large )}}\Big)-
\nonumber \\
&&-2\Big((\vec{\kappa}_{i}\cdot \vec{\kappa}_{i})({\bf {\vec{f}}}_{j}\cdot
\vec{\kappa}_{i}-({\bf {\vec{f}}}_{j}\cdot {\hat{\eta}}_{ij})(\vec{\kappa}%
_{i}\cdot {\hat{\eta}}_{ij}))-{\bf {\vec{f}}}_{j}\cdot [ (\vec{\kappa}%
_{i}\cdot {\hat{\eta}}_{ij})(\vec{\kappa}_{i}-{\hat{\eta}}_{ij}(\vec{\kappa}%
_{i}\cdot {\hat{\eta}}_{ij}))](\vec{\kappa}_{i}\cdot {\hat{\eta}}_{ij}) +
\nonumber \\
&&+({\bf {\vec{f}}}_{i}\cdot \vec{\kappa}_{i})(\vec{\kappa}_{j}\cdot \vec{%
\kappa}_{i}-(\vec{\kappa}_{j}\cdot {\hat{\eta}}_{ij})(\vec{\kappa}_{i}\cdot {%
\hat{\eta}}_{ij}))-\vec{\kappa}_{j}\cdot [ ({\bf {\vec{f}}}_{i}\cdot {\hat{%
\eta}}_{ij})(\vec{\kappa}_{i}-{\hat{\eta}}_{ij}(\vec{\kappa}_{i}\cdot {\hat{%
\eta}}_{ij}))](\vec{\kappa}_{i}\cdot {\hat{\eta}}_{ij}) +  \nonumber \\
&&+({\bf {\vec{f}}}_{i}\cdot \vec{\kappa}_{i})({\bf {\vec{f}}}_{j}\cdot \vec{%
\kappa}_{i}-({\bf {\vec{f}}}_{j}\cdot {\hat{\eta}}_{ij})(\vec{\kappa}%
_{i}\cdot {\hat{\eta}}_{ij}))-{\bf {\vec{f}}}_{j}\cdot [ ({\bf {\vec{f}}}%
_{i}\cdot {\hat{\eta}}_{ij})(\vec{\kappa}_{i}-{\hat{\eta}}_{ij}(\vec{\kappa}%
_{i}\cdot {\hat{\eta}}_{ij}))](\vec{\kappa}_{i}\cdot {\hat{\eta}}_{ij})\Big)
\nonumber \\
&& \frac{1}{\eta _{ij}\sqrt{m_{j}^{2}+\vec{\kappa}_{j}{}^{2}}\sqrt{%
m_{i}^{2}+(\vec{\kappa}_{i}{}\cdot \hat{\eta}_{ij})^{2}}}\frac{1}{(\sqrt{%
m_{i}^{2}+\vec{\kappa}_{i}{}^{2}}+\sqrt{m_{i}^{2}+(\vec{\kappa}_{i}{}\cdot
\hat{\eta}_{ij})^{2}})^{2}}) +  \nonumber \\
&&+[\vec{\kappa}_{i}\cdot {\bf {\vec{g}}}_{j}-\vec{\kappa}_{i}\cdot ({\bf {%
\vec{g}}}_{j}\cdot ({\hat{\eta}}_{ij}{\hat{\eta}}_{ij}))]  \nonumber \\
&&\frac{2}{\eta _{ij}\sqrt{m_{i}^{2}+{\vec{\kappa}_{i}}^{2}}\sqrt{m_{j}^{2}+(%
\vec{\kappa}_{j}\cdot {\hat{\eta}}_{ij})^{2}}}({\frac{\sqrt{m_{j}^{2}+\vec{%
\kappa}_{j}^{2}}}{\sqrt{m_{j}^{2}+{\vec{\kappa}_{j}}^{2}}+\sqrt{m_{j}^{2}+(%
\vec{\kappa}_{j}\cdot {\hat{\eta}}_{ij})^{2}}})}\Big].  \nonumber \\
&&{}  \label{VII12}
\end{eqnarray}

The term $V_{LDS}$ above includes all of the terms in Appendix D that are in
closed form and are of order $1/c^{2}$ and higher while the term $U_{HDS}$
below includes all of the terms that involve the double infinite series and
are of order $1/c^{4}$ and higher.

\begin{eqnarray}
&&U_{HDS}:=\sum_{i\neq j}{\frac{Q_{i}Q_{j}}{8\pi }}\sum_{m=0}^{\infty
}\sum_{n=0}^{\infty }\Big[\frac{{\bf {\vec{k}}_{i}}}{\sqrt{m_{i}^{2}+\vec{%
\kappa}_{i}{}^{2}}}_{i}\cdot \frac{{\bf {\vec{k}}_{j}}}{\sqrt{m_{j}^{2}+\vec{%
\kappa}_{j}{}^{2}}}  \nonumber \\
&& (\frac{\vec{\kappa}_{i}}{\sqrt{m_{i}^{2}+\vec{\kappa}_{i}{}^{2}}}\cdot
\vec{\partial}_{ij})^{2m+1}(\frac{\vec{\kappa}_{j}}{\sqrt{m_{j}^{2}+\vec{%
\kappa}_{j}{}^{2}}}\cdot \vec{\partial}_{ij})^{2n+1}\frac{\eta
_{ij}^{2n+2m+1}}{(2n+2m+2)!}-  \nonumber \\
&-&(\frac{{\bf {\vec{k}}_{i}}}{\sqrt{m_{i}^{2}+\vec{\kappa}_{i}{}^{2}}}\cdot
\vec{\partial}_{ij})(\frac{{\bf {\vec{l}}_{j}}}{\sqrt{m_{j}^{2}+\vec{\kappa}%
_{j}{}^{2}}}\cdot \vec{\partial}_{ij})(\frac{\vec{\kappa}_{i}}{\sqrt{%
m_{i}^{2}+\vec{\kappa}_{i}{}^{2}}}\cdot \vec{\partial}_{ij})^{2m+1}
\nonumber \\
&&(\frac{\vec{\kappa}_{j}}{\sqrt{m_{j}^{2}+\vec{\kappa}_{j}{}^{2}}}\cdot
\vec{\partial}_{ij})^{2n+1}\frac{\eta _{ij}^{2n+2m+3}}{(2n+2m+4)!}-
\nonumber \\
&&-(\frac{{\bf {\vec{l}}_{i}}}{\sqrt{m_{i}^{2}+\vec{\kappa}_{i}{}^{2}}}\cdot
\vec{\partial}_{ij})(\frac{{\bf {\vec{k}}_{j}}}{\sqrt{m_{j}^{2}+\vec{\kappa}%
_{j}{}^{2}}}\cdot \vec{\partial}_{ij})(\frac{\vec{\kappa}_{i}}{\sqrt{%
m_{i}^{2}+\vec{\kappa}_{i}{}^{2}}}\cdot \vec{\partial}_{ij})^{2m+1}
\nonumber \\
&&(\frac{\vec{\kappa}_{j}}{\sqrt{m_{j}^{2}+\vec{\kappa}_{j}{}^{2}}}\cdot
\vec{\partial}_{ij})^{2n+1}\frac{\eta _{ij}^{2n+2m+3}}{(2n+2m+4)!}+
\nonumber \\
&&+(\frac{{\bf {\vec{l}}_{i}}}{\sqrt{m_{i}^{2}+\vec{\kappa}_{i}{}^{2}}}\cdot
\vec{\partial}_{ij})(\frac{{\bf {\vec{l}}_{j}}}{\sqrt{m_{j}^{2}+\vec{\kappa}%
_{j}{}^{2}}}\cdot \vec{\partial}_{ij})(\frac{\vec{\kappa}_{i}}{\sqrt{%
m_{i}^{2}+\vec{\kappa}_{i}{}^{2}}}\cdot \vec{\partial}_{ij})^{2m+1}
\nonumber \\
&&(\frac{\vec{\kappa}_{j}}{\sqrt{m_{j}^{2}+\vec{\kappa}_{j}{}^{2}}}\cdot
\vec{\partial}_{ij})^{2n+1}\frac{\eta _{ij}^{2n+2m+3}}{(2n+2m+4)!}+
\nonumber \\
&&+\frac{{\bf {\vec{k}}_{i}}}{\sqrt{m_{i}^{2}+\vec{\kappa}_{i}{}^{2}}}\cdot
\frac{{\bf {\vec{k}}_{j}}}{\sqrt{m_{j}^{2}+\vec{\kappa}_{j}{}^{2}}}(\frac{%
\vec{\kappa}_{i}}{\sqrt{m_{i}^{2}+\vec{\kappa}_{i}{}^{2}}}\cdot \vec{\partial%
}_{ij})^{2m+2}(\frac{\vec{\kappa}_{j}}{\sqrt{m_{j}^{2}+\vec{\kappa}_{j}{}^{2}%
}}\cdot \vec{\partial}_{ij})^{2n+2}\frac{\eta _{ij}^{2n+2m+3}}{(2n+2m+4)!} -
\nonumber \\
&&-(\frac{{\bf {\vec{k}}_{i}}}{\sqrt{m_{i}^{2}+\vec{\kappa}_{i}{}^{2}}}\cdot
\vec{\partial}_{ij})(\frac{{\bf {\vec{k}}_{j}}}{\sqrt{m_{j}^{2}+\vec{\kappa}%
_{j}{}^{2}}}\cdot \vec{\partial}_{ij})(\frac{\vec{\kappa}_{i}}{\sqrt{%
m_{i}^{2}+\vec{\kappa}_{i}{}^{2}}}\cdot \vec{\partial}_{ij})^{2m+2}
\nonumber \\
&&(\frac{\vec{\kappa}_{j}}{\sqrt{m_{j}^{2}+\vec{\kappa}_{j}{}^{2}}}\cdot
\vec{\partial}_{ij})^{2n+2}\frac{\eta _{ij}^{2n+2m+5}}{(2n+2m+6)!}\Big].
\label{VII13}
\end{eqnarray}

As discussed in Appendix D this complex formula for the lower and higher
order Darwin and spin-dependent terms has been checked to reduce to the
corresponding spinless results given in Eq.(6.14) in Ref.\cite{ap} when $%
{\bf {\vec{f}}}_{i}={\bf {\vec{g}}}_{i}=0$.

\subsection{The Semi-Classical Hamiltonian in the Final Canonical Variables.}

We now reexpress our final result in terms of the final canonical variables (%
\ref{VI56}), (\ref{VI64}) forming a Darboux basis for the Dirac brackets.
These are the {\it physical} particle variable in this reduced phase space.
By using Eqs.(\ref{VI55}) we get

\begin{eqnarray}
\sqrt{m_{i}^{2}+\vec{\kappa}_{i}{}^{2}} &=&\sqrt{m_{i}^{2}+\widetilde{\vec{%
\kappa}_{i}}^{2}+\widetilde{\vec{\kappa}_{i}}\cdot \sum_{j\neq i}Q_{i}Q_{j}%
\vec{\partial}_{\eta _{i}}{\cal K}_{ij}}=  \nonumber \\
&=&\sqrt{m_{i}^{2}+\widetilde{\vec{\kappa}_{i}}^{2}}+\frac{\widetilde{\vec{%
\kappa}_{i}}\cdot \sum_{j\neq i}Q_{i}Q_{j}\vec{\partial}_{\tilde{\eta}_{i}}%
\widetilde{{\cal K}}_{ij}}{2\sqrt{m_{i}^{2}+\widetilde{\vec{\kappa}_{i}}^{2}}%
},  \label{VII14}
\end{eqnarray}

\noindent with

\begin{eqnarray}
&&Q_{i}Q_{j}{\cal K}_{ij}(\vec{\kappa}_{i,}\vec{\kappa}_{j,}\vec{\eta}_{i}-%
\vec{\eta}_{j};\vec{\xi}_{i},\vec{\xi}_{j}) = Q_{i}Q_{j}\widetilde{{\cal K}}%
_{ij}(\widetilde{\vec{\kappa}_{i}},\widetilde{\vec{\kappa}_{j}},\widetilde{%
\vec{\eta}_{i}}-\widetilde{\vec{\eta}_{j}};\widetilde{\vec{\xi}_{i}},%
\widetilde{\vec{\xi}_{j}}),  \nonumber \\
&&{}  \nonumber \\
&&\widetilde{\vec{\kappa}_{i}}\cdot \vec \partial _{\tilde{\eta}_{i}}%
\widetilde{{\cal K}}_{ij} =\int d^{3}\sigma \Big[ \Big(\widetilde{\vec{\kappa%
}_{i}}\cdot \vec{\partial}_{\tilde{\eta}_{i}}\vec{A}_{\perp Si}(\widetilde{%
\vec{\kappa}_{i}},\widetilde{\vec{\kappa}_{j}},\widetilde{\vec{\eta}_{i}}-%
\widetilde{\vec{\eta}_{j}};\widetilde{\vec{\xi}_{i}},\widetilde{\vec{\xi}_{j}%
})\Big)\cdot \vec{\pi}_{\perp Si}(\widetilde{\vec{\kappa}_{i}},\widetilde{%
\vec{\kappa}_{j}},\widetilde{\vec{\eta}_{i}}-\widetilde{\vec{\eta}_{j}};%
\widetilde{\vec{\xi}_{i}},\widetilde{\vec{\xi}_{j}})-  \nonumber \\
&&-\vec{A}_{\perp Si}(\widetilde{\vec{\kappa}_{i}},\widetilde{\vec{\kappa}%
_{j}},\widetilde{\vec{\eta}_{i}}-\widetilde{\vec{\eta}_{j}};\widetilde{\vec{%
\xi}_{i}},\widetilde{\vec{\xi}_{j}})\cdot \Big(\widetilde{\vec{\kappa}_{i}}%
\cdot \vec{\partial}_{\tilde{\eta}_{i}}\vec{\pi}_{\perp Si}(\widetilde{\vec{%
\kappa}_{i}},\widetilde{\vec{\kappa}_{j}},\widetilde{\vec{\eta}_{i}}-%
\widetilde{\vec{\eta}_{j}};\widetilde{\vec{\xi}_{i}},\widetilde{\vec{\xi}_{j}%
})\Big)\Big],  \label{VII15}
\end{eqnarray}

But as we have seen in Eqs.(\ref{VI65}) and (\ref{VI66}), we can replace $%
\sum_{i<j}{\vec{\partial}}_{\eta _{i}}Q_{i}Q_{j}{\cal K}_{ij}$ by $-\int
d^{3}\sigma \lbrack {\vec{\pi}}_{\perp S}\times {\vec{B}}_{S}](\tau ,\vec{%
\sigma}).$ Using the integral in Eq.(\ref{VI47}), written in terms of
canonical variables, we obtain that the old kinetic term can be rewritten as
the new final kinetic term plus an extra scalar potential $U_{HDS}^{^{\prime
}}$ ($\eta _{ij}=|{\vec{\eta}}_{ij}|=|{\vec{\eta}}_{i}-{\vec{\eta}}_{j}|$)

\begin{eqnarray}
&&\sum_{i=1}^{N}\sqrt{m_{i}^{2}+\vec{\kappa}_{i}(\tau )^{2}}=\sqrt{%
m_{i}^{2}+ \widetilde{\vec{\kappa}_{i}}(\tau )^{2}}+U_{HDS}^{\prime }(\tau ),
\nonumber \\
&&  \nonumber \\
&&{}  \nonumber \\
&&U_{HDS}^{\prime }(\tau )=-\sum_{i<j}{\frac{Q_{i}Q_{j}}{8\pi }}%
\sum_{m=0}^{\infty }\sum_{n=0}^{\infty }\Big[{\frac{\widetilde{{\bf \vec{k}}}%
_{i}}{\sqrt{m_{i}^{2}+{\widetilde{\vec{\kappa}}_{i}}^{2}}}}\cdot {\frac{%
\widetilde{{\bf \vec{k}}}_{j}}{\sqrt{m_{j}^{2}+{\widetilde{\vec{\kappa}}_{j}}%
^{2}}}}\frac{1}{(2n+2m+2)!}  \nonumber \\
&& \Big(({\frac{\widetilde{\vec{\kappa}}_{i}}{\sqrt{m_{i}^{2}+{\widetilde{%
\vec{\kappa}}_{i}}^{2}}}}\cdot \vec{\partial _{ij}})^{2m+2}({\frac{%
\widetilde{\vec{\kappa}}_{j}}{\sqrt{m_{j}^{2}+{\widetilde{\vec{\kappa}}_{j}}%
^{2}}}}\cdot \vec{\partial _{ij}})^{2n}+  \nonumber \\
&&+2({\frac{\widetilde{\vec{\kappa}}_{i}}{\sqrt{m_{i}^{2}+{\widetilde{\vec{%
\kappa}}_{i}}^{2}}}}\cdot \vec{\partial _{ij}})^{2m+1}({\frac{\widetilde{%
\vec{\kappa}}_{j}}{\sqrt{m_{j}^{2}+{\vec{\kappa}_{j}}^{2}}}}\cdot \vec{%
\partial _{ij}})^{2n+1}  \nonumber \\
&&({\frac{\widetilde{\vec{\kappa}}_{i}}{\sqrt{m_{i}^{2}+{\widetilde{\vec{%
\kappa}}_{i}}^{2}}}}\cdot \vec{\partial _{ij}})^{2m}({\frac{\widetilde{\vec{%
\kappa}}_{j}}{\sqrt{m_{j}^{2}+{\vec{\kappa}_{j}}^{2}}}}\cdot \vec{\partial
_{ij}})^{2n+2}\Big)\tilde{\eta}_{ij}^{2n+2m+1}-  \nonumber \\
&&-\frac{1}{(2n+2m+4)!}\Big[({\frac{\widetilde{{\bf \vec{k}}}_{i}}{\sqrt{%
m_{i}^{2}+{\vec{\kappa}_{i}}^{2}}}}\cdot \vec{\partial _{ij}})({\frac{%
\widetilde{{\bf \vec{l}}}_{j}}{\sqrt{m_{j}^{2}+{\widetilde{\vec{\kappa}}_{j}}%
^{2}}}}\cdot \vec{\partial _{ij}})  \nonumber \\
&& \Big(({\frac{\widetilde{\vec{\kappa}}_{i}}{\sqrt{m_{i}^{2}+{\vec{\kappa}%
_{i}}^{2}}}}\cdot \vec{\partial _{ij}})^{2m+2}({\frac{\widetilde{\vec{\kappa}%
}_{j}}{\sqrt{m_{j}^{2}+{\widetilde{\vec{\kappa}}_{j}}^{2}}}}\cdot \vec{%
\partial _{ij}})^{2n} +  \nonumber \\
&&+({\frac{\widetilde{\vec{\kappa}}_{i}}{\sqrt{m_{i}^{2}+{\widetilde{\vec{%
\kappa}}_{i}}^{2}}}}\cdot \vec{\partial _{ij}})^{2m+1}({\frac{\widetilde{%
\vec{\kappa}}_{j}}{\sqrt{m_{j}^{2}+{\vec{\kappa}_{j}}^{2}}}}\cdot \vec{%
\partial _{ij}})^{2n+1}\Big)\tilde{\eta}_{ij}^{2n+2m+3}+  \nonumber \\
&&+({\frac{\widetilde{{\bf \vec{l}}}_{i}}{\sqrt{m_{i}^{2}+{\vec{\kappa}_{i}}%
^{2}}}}\cdot \vec{\partial _{ij}})({\frac{\widetilde{{\bf \vec{k}}}_{j}}{%
\sqrt{m_{j}^{2}+{\widetilde{\vec{\kappa}}_{j}}^{2}}}}\cdot \vec{\partial
_{ij}})\Big(({\frac{\widetilde{\vec{\kappa}}_{i}}{\sqrt{m_{i}^{2}+{%
\widetilde{\vec{\kappa}}_{i}}^{2}}}}\cdot \vec{\partial _{ij}})^{2m+1}({%
\frac{\widetilde{\vec{\kappa}}_{j}}{\sqrt{m_{j}^{2}+{\vec{\kappa}_{j}}^{2}}}}%
\cdot \vec{\partial _{ij}} )^{2n+1} +  \nonumber \\
&+&({\frac{\widetilde{\vec{\kappa}}_{i}}{\sqrt{m_{i}^{2}+{\widetilde{\vec{%
\kappa}}_{i}}^{2}}}}\cdot \vec{\partial _{ij}})^{2m}({\frac{\widetilde{\vec{%
\kappa}}_{j}}{\sqrt{m_{j}^{2}+{\vec{\kappa}_{j}}^{2}}}}\cdot \vec{\partial
_{ij}})^{2n+2}\Big)\tilde{\eta}_{ij}^{2n+2m+3}\Big] -  \nonumber \\
&&-\frac{1}{(2n+2m+2)!}\Big({\frac{\widetilde{\vec{\kappa}}_{i}}{\sqrt{%
m_{i}^{2}+{\widetilde{\vec{\kappa}}_{i}}^{2}}}}\cdot {\frac{\widetilde{{\bf
\vec{k}}}_{j}}{\sqrt{m_{j}^{2}+{\widetilde{\vec{\kappa}}_{j}}^{2}}}}
\nonumber \\
&& {\lbrack }\frac{\widetilde{{\bf \vec{k}}}_{i}-\widetilde{{\bf \vec{l}}}%
_{i}}{\sqrt{m_{i}^{2}+{\widetilde{\vec{\kappa}}_{i}}^{2}}}\cdot \vec{%
\partial _{ij}}({\frac{\widetilde{\vec{\kappa}}_{i}}{\sqrt{m_{i}^{2}+{\vec{%
\kappa}_{i}}^{2}}}}\cdot \vec{\partial _{ij}})^{2m+1}({\frac{\widetilde{\vec{%
\kappa}}_{j}}{\sqrt{m_{j}^{2}+{\widetilde{\vec{\kappa}}_{j}}^{2}}}}\cdot
\vec{\partial _{ij}})^{2n}+  \nonumber \\
&&+\frac{\widetilde{{\bf \vec{k}}}_{j}-\widetilde{{\bf \vec{l}}}_{j}}{\sqrt{%
m_{j}^{2}+{\widetilde{\vec{\kappa}}_{j}}^{2}}}\cdot \vec{\partial _{ij}}({%
\frac{\widetilde{\vec{\kappa}}_{i}}{\sqrt{m_{i}^{2}+{\vec{\kappa}_{i}}^{2}}}}%
\cdot \vec{\partial _{ij}})^{2m}({\frac{\widetilde{\vec{\kappa}}_{j}}{\sqrt{%
m_{j}^{2}+{\widetilde{\vec{\kappa}}_{j}}^{2}}}}\cdot \vec{\partial _{ij}}
)^{2n+1}]\Big)\tilde{\eta}_{ij}^{2n+2m+1}\Big].  \label{VII16}
\end{eqnarray}

There is some cancellations between the terms in $U_{HDS}^{^{\prime}}(\tau )$
and those in $V_{HDS}(\tau )$ when expressed in the final variables.

The final result for the semi-classical Hamiltonian is

\begin{eqnarray}
M &=&{\cal P}_{(int)}^{\tau }=\sum_{i=1}^{N}\sqrt{m_{i}^{2}+\widetilde{\vec{%
\kappa}_{i}}^{2}}+\sum_{i=1}^{N}\frac{\widetilde{\vec{\kappa}_{i}}\cdot
\sum_{j\neq i}Q_{i}Q_{j}\vec{\partial}_{\tilde{\eta}_{i}}{\cal K}_{ij}}{2%
\sqrt{m_{i}^{2}+\widetilde{\vec{\kappa}_{i}}^{2}}}+\sum_{i\neq j}\frac{%
Q_{i}Q_{j}}{\widetilde{\vec{\eta}_{i}}-\widetilde{\vec{\eta}_{j}}\mid }%
+V_{DS}(\tau ) =  \nonumber \\
&=&\sum_{i=1}^{N}\sqrt{m_{i}^{2}+\widetilde{\vec{\kappa}_{i}}^{2}}%
+\sum_{i\neq j}\frac{Q_{i}Q_{j}}{4\pi \mid \widetilde{\vec{\eta}_{i}}-%
\widetilde{\vec{\eta}_{j}}\mid }+\tilde{V}_{DS}(\tau
),\,\,\,\,\,\,\,\,\,\,\,\,\,\,\,[\tilde{V}_{DS}\,\,=V_{DS}+U_{HDS}^{\prime }]
\label{VII17}
\end{eqnarray}

\noindent where

\begin{equation}
\tilde{V}_{DS}=V_{LDS}+U_{HDS}+U_{HDS}^{\prime }:=V_{LDS}+V_{HDS},
\label{VII18}
\end{equation}

\noindent with

\begin{eqnarray}
&&V_{HDS}:=-\sum_{i<j}{\frac{Q_{i}Q_{j}}{8\pi }}\sum_{m=0}^{\infty
}\sum_{n=0}^{\infty }\Big[{\frac{\widetilde{{\bf \vec{k}}}_{i}}{\sqrt{%
m_{i}^{2}+{\widetilde{\vec{\kappa}}_{i}}^{2}}}}\cdot {\frac{\widetilde{{\bf
\vec{k}}}_{j}}{\sqrt{m_{j}^{2}+{\widetilde{\vec{\kappa}}_{j}}^{2}}}}\frac{1}{%
(2n+2m+2)!}  \nonumber \\
&& \Big(({\frac{\widetilde{\vec{\kappa}}_{i}}{\sqrt{m_{i}^{2}+{\widetilde{%
\vec{\kappa}}_{i}}^{2}}}}\cdot \vec{\partial _{ij}})^{2m+2}({\frac{%
\widetilde{\vec{\kappa}}_{j}}{\sqrt{m_{j}^{2}+{\widetilde{\vec{\kappa}}_{j}}%
^{2}}}}\cdot \vec{\partial _{ij}})^{2n}+  \nonumber \\
&+&({\frac{\widetilde{\vec{\kappa}}_{i}}{\sqrt{m_{i}^{2}+{\widetilde{\vec{%
\kappa}}_{i}}^{2}}}}\cdot \vec{\partial _{ij}})^{2m}({\frac{\widetilde{\vec{%
\kappa}}_{j}}{\sqrt{m_{j}^{2}+{\widetilde{\vec{\kappa}}_{j}}^{2}}}}\cdot
\vec{\partial _{ij}})^{2n+2}\Big)\tilde{\eta}_{ij}^{2n+2m+1}-  \nonumber \\
&&-\frac{1}{(2n+2m+4)!}\Big[({\frac{\widetilde{{\bf \vec{k}}}_{i}}{\sqrt{%
m_{i}^{2}+{\widetilde{\vec{\kappa}}_{i}}^{2}}}}\cdot \vec{\partial _{ij}})({%
\frac{\widetilde{{\bf \vec{l}}}_{j}}{\sqrt{m_{j}^{2}+{\widetilde{\vec{\kappa}%
}_{j}}^{2}}}}\cdot \vec{\partial _{ij}})  \nonumber \\
&& \Big(({\frac{\widetilde{\vec{\kappa}}_{i}}{\sqrt{m_{i}^{2}+{\widetilde{%
\vec{\kappa}}_{i}}^{2}}}}\cdot \vec{\partial _{ij}})^{2m+2}({\frac{%
\widetilde{\vec{\kappa}}_{j}}{\sqrt{m_{j}^{2}+{\widetilde{\vec{\kappa}}_{j}}%
^{2}}}}\cdot \vec{\partial _{ij}})^{2n} -  \nonumber \\
&&-({\frac{\widetilde{\vec{\kappa}}_{i}}{\sqrt{m_{i}^{2}+{\widetilde{\vec{%
\kappa}}_{i}}^{2}}}}\cdot \vec{\partial _{ij}})^{2m+1}({\frac{\widetilde{%
\vec{\kappa}}_{j}}{\sqrt{m_{j}^{2}+{\widetilde{\vec{\kappa}}_{j}}^{2}}}}%
\cdot \vec{\partial _{ij}})^{2n+1}\Big)\tilde{\eta}_{ij}^{2n+2m+3} +
\nonumber \\
&&+({\frac{\widetilde{{\bf \vec{l}}}_{i}}{\sqrt{m_{i}^{2}+{\widetilde{\vec{%
\kappa}}_{i}}^{2}}}}\cdot \vec{\partial _{ij}})({\frac{\widetilde{{\bf \vec{k%
}}}_{j}}{\sqrt{m_{j}^{2}+{\widetilde{\vec{\kappa}}_{j}}^{2}}}}\cdot \vec{%
\partial _{ij}})\Big(({\frac{\widetilde{\vec{\kappa}}_{i}}{\sqrt{m_{i}^{2}+{%
\widetilde{\vec{\kappa}}_{i}}^{2}}}}\cdot \vec{\partial _{ij}})^{2m}(\frac{%
\widetilde{\vec{\kappa}}_{j}}{\sqrt{m_{j}^{2}+{\widetilde{\vec{\kappa}}_{j}}%
^{2}}}\cdot \vec{\partial _{ij}})^{2n+2} -  \nonumber \\
&&-({\frac{\widetilde{\vec{\kappa}}_{i}}{\sqrt{m_{i}^{2}+{\widetilde{\vec{%
\kappa}}_{i}}^{2}}}}\cdot \vec{\partial _{ij}})^{2m+1}(\frac{\widetilde{\vec{%
\kappa}}_{j}}{\sqrt{m_{j}^{2}+{\widetilde{\vec{\kappa}}_{j}}^{2}}}\cdot \vec{%
\partial _{ij}})^{2n+1}\Big)\tilde{\eta}_{ij}^{2n+2m+3}+  \nonumber \\
&+&2(\frac{{\bf {\vec{l}}_{i}}}{\sqrt{m_{i}^{2}+\vec{\kappa}_{i}{}^{2}}}%
\cdot \vec{\partial}_{ij})(\frac{{\bf {\vec{l}}}_{j}}{\sqrt{m_{j}^{2}+{%
\widetilde{\vec{\kappa}}_{j}}^{2}}}\cdot \vec{\partial}_{ij})  \nonumber \\
&& (\frac{\widetilde{\vec{\kappa}}_{i}}{\sqrt{m_{i}^{2}+{\widetilde{\vec{%
\kappa}}_{i}}^{2}}}\cdot \vec{\partial}_{ij})^{2m+1}(\frac{\widetilde{\vec{%
\kappa}}_{j}}{\sqrt{m_{j}^{2}+{\widetilde{\vec{\kappa}}_{j}}^{2}}}\cdot \vec{%
\partial}_{ij})^{2n+1}\tilde{\eta}_{ij}^{2n+2m+3}\Big] -  \nonumber \\
&&-\frac{1}{(2n+2m+2)!}\Big({\frac{\widetilde{\vec{\kappa}}_{i}}{\sqrt{%
m_{i}^{2}+{\widetilde{\vec{\kappa}}_{i}}^{2}}}}\cdot {\frac{\widetilde{{\bf
\vec{k}}}_{j}}{\sqrt{m_{j}^{2}+{\widetilde{\vec{\kappa}}_{j}}^{2}}}}
\nonumber \\
&&[\frac{\widetilde{{\bf \vec{k}}}_{i}-\widetilde{{\bf \vec{l}}}_{i}}{\sqrt{%
m_{i}^{2}+{\widetilde{\vec{\kappa}}_{i}}^{2}}}\cdot \vec{\partial _{ij}}(%
\frac{\widetilde{\vec{\kappa}}_{i}}{\sqrt{m_{i}^{2}+{\widetilde{\vec{\kappa}}%
_{i}}^{2}}}\cdot \vec{\partial _{ij}})^{2m+1}({\frac{\widetilde{\vec{\kappa}}%
_{j}}{\sqrt{m_{j}^{2}+{\widetilde{\vec{\kappa}}_{j}}^{2}}}}\cdot \vec{%
\partial _{ij}})^{2n}]+  \nonumber \\
&&+{\frac{\widetilde{{\bf \vec{k}}}_{i}}{\sqrt{m_{i}^{2}+{\widetilde{\vec{%
\kappa}}_{i}}^{2}}}}\cdot {\frac{\widetilde{\vec{\kappa}}_{j}}{\sqrt{
m_{j}^{2}+{\widetilde{\vec{\kappa}}_{j}}^{2}}}}  \nonumber \\
&&[\frac{\widetilde{{\bf \vec{k} }}_{j}-\widetilde{{\bf \vec{l}}}_{j}}{\sqrt{%
m_{j}^{2}+{\widetilde{\vec{\kappa}}_{j}}^{2}}}\cdot \vec{\partial _{ij}}({%
\frac{\widetilde{\vec{\kappa}}_{i}}{\sqrt{m_{i}^{2}+{\widetilde{\vec{\kappa}}%
_{i}}^{2}}}}\cdot \vec{\partial _{ij}})^{2m}({\frac{\widetilde{\vec{\kappa}}%
_{j}}{\sqrt{m_{j}^{2}+{\widetilde{\vec{\kappa}}_{j}}^{2}}}}\cdot \vec{%
\partial _{ij}})^{2n+1}]\Big)\tilde{\eta}_{ij}^{2n+2m+1}-  \nonumber \\
&&-\frac{{\bf {\vec{k}}_{i}}}{\sqrt{m_{i}^{2}+{\widetilde{\vec{\kappa}}_{i}}%
^{2}}}\cdot \frac{{\bf {\vec{k}}_{j}}}{\sqrt{m_{j}^{2}+{\widetilde{\vec{%
\kappa}}_{j}}^{2}}}(\frac{\widetilde{\vec{\kappa}}_{i}}{\sqrt{m_{i}^{2}+{%
\widetilde{\vec{\kappa}}_{i}}^{2}}}\cdot \vec{\partial}_{ij})^{2m+2}(\frac{%
\widetilde{\vec{\kappa}}_{j}}{\sqrt{m_{j}^{2}+{\widetilde{\vec{\kappa}}_{j}}%
^{2}}}\cdot \vec{\partial}_{ij})^{2n+2}\frac{\eta _{ij}^{2n+2m+3}}{(2n+2m+4)!%
} +  \nonumber \\
&&+(\frac{{\bf {\vec{k}}_{i}}}{\sqrt{m_{i}^{2}+{\widetilde{\vec{\kappa}}_{i}}%
^{2}}}\cdot \vec{\partial}_{ij})(\frac{{\bf {\vec{k}}_{j}}}{\sqrt{m_{j}^{2}+{%
\widetilde{\vec{\kappa}}_{j}}^{2}}}\cdot \vec{\partial}_{ij})  \nonumber \\
&& (\frac{\widetilde{\vec{\kappa}}_{i}}{\sqrt{m_{i}^{2}+{\widetilde{\vec{%
\kappa}}_{i}}^{2}}}\cdot \vec{\partial}_{ij})^{2m+2}(\frac{\widetilde{\vec{%
\kappa}}_{j}}{\sqrt{m_{j}^{2}+{\widetilde{\vec{\kappa}}_{j}}^{2}}}\cdot \vec{%
\partial}_{ij})^{2n+2}\frac{\eta _{ij}^{2n+2m+5}}{(2n+2m+6)!}\Big].
\label{VII19}
\end{eqnarray}

The term $V_{LDS}$ is defined as in Eq.(\ref{VII12}) but with all the old
non-canonical variables replaced with the new canonical ones. \ Due to
Grassmann truncation the expressions are equivalent. As we shall show in the
next Section, it reduces in lowest order to the pseudo-classical version of
the Breit results (of order $1/c^{2}$). The term $V_{HDS}$ involves the
double infinite series and is of order $1/c^{4}$ and higher. In general
there is no closed form for it (see Section VI of Ref.\cite{ap} for related
consideration for spinless charged particles), but in the case of the two
body problem the double infinite series can be summed to closed form in the
rest system.

\vfill\eject

\section{The Semi-Classical 2-Body Problem: Hydrogen Atom, Muonium and
Positronium.}

\subsection{The Semi-Classical 2-Body Problem.}

\ \ In this Section we specialize to the two body problem, which includes
the pseudo-classical analogues of \ the hydrogen atom, muonium and
positronium.

${}$

${}$

1) The rest frame form of the energy in the unequal mass case ({\it muonium}%
) is ($\tilde{\eta}:=|\widetilde{\vec{\eta}_{1}}-\widetilde{\vec{\eta}_{2}}|$
and $\widetilde{\vec{\kappa}}:=\widetilde{\vec{\kappa}}_{1}=-\widetilde{\vec{%
\kappa}}_{2}$)

\begin{equation}
M=\sqrt{m_{1}^{2}+\widetilde{\vec{\kappa}}^{2}}+\sqrt{m_{2}^{2}+\widetilde{%
\vec{\kappa}}^{2}}+\frac{Q_{1}Q_{2}}{4\pi \widetilde{\eta }}+\tilde{V}%
_{DS}(\tau ),  \label{VIII1}
\end{equation}

\noindent where $\tilde{V}_{DS}=V_{LDS}+V_{HDS}.$ \ From Eq.(\ref{VII12}) we
obtain for arbitrary masses

\begin{eqnarray}
V_{LDS} &:&=\frac{Q_{1}Q_{2}}{4\pi }\Big[\frac{\widetilde{\vec{\kappa}}^{2}+(%
\widetilde{\vec{\kappa}}\cdot {\hat{\eta})}^{2}}{2\sqrt{m_{1}^{2}+\widetilde{%
\vec{\kappa}}{}^{2}}\sqrt{m_{2}^{2}+\widetilde{\vec{\kappa}}{}^{2}}}\frac{1}{%
\tilde{\eta}}-  \nonumber \\
&&-i\frac{\widetilde{\vec{\kappa}}\cdot \vec{\xi}_{1}\vec{\xi}_{1}\cdot \vec{%
\eta}}{\tilde{\eta}^{3}(m_1+\sqrt{m_{1}^{2}+\widetilde{\vec{\kappa}}^{2}})%
\sqrt{m_{1}^{2}+\widetilde{\vec{\kappa}}^{2}}}-i\frac{\widetilde{\vec{\kappa}%
}\cdot \vec{\xi}_{2}\vec{\xi}_{2}\cdot \vec{\eta}}{\tilde{\eta}^{3}(m_2+%
\sqrt{m_{2}^{2}+\widetilde{\vec{\kappa}}^{2}})\sqrt{m_{2}^{2}+\widetilde{%
\vec{\kappa}}^{2}}}-  \nonumber \\
&&-i{\frac{\vec{\xi}_{1}\times \vec{\xi}_{1}\cdot (\widetilde{{\bf \vec{k}}}%
_{2}\times \vec{\partial}_{\eta })}{2\sqrt{m_{1}^{2}+\widetilde{\vec{\kappa}}%
{}^{2}}}}\frac{1}{\tilde{\eta}\sqrt{m_{2}^{2}+(\widetilde{\vec{\kappa}}\cdot
{\hat{\eta}})^{2}}}+i{\frac{\vec{\xi}_{2}\times \vec{\xi}_{2}\cdot (%
\widetilde{{\bf \vec{k}}}_{1}\times \vec{\partial}_{\eta })}{2\sqrt{%
m_{2}^{2}+\widetilde{\vec{\kappa}}{}^{2}}}}\frac{1}{\tilde{\eta}\sqrt{%
m_{1}^{2}+(\widetilde{\vec{\kappa}}\cdot {\hat{\eta}})^{2}}} +  \nonumber \\
&&+i{\frac{\widetilde{\vec{\kappa}}\cdot \vec{\xi}_{1}}{\sqrt{m_{1}^{2}+%
\widetilde{\vec{\kappa}}{}^{2}}(m_1+\sqrt{m_{1}^{2}+\widetilde{\vec{\kappa}}%
^{2}})}}  \nonumber \\
&&\Big([\vec{\xi}_{1}-{\frac{i\widetilde{\vec{\kappa}}\cdot \vec{\xi}_{2}
\vec{\xi}_{1}\cdot \vec{\xi}_{2}\vec{\partial}_{\eta }}{\sqrt{m_{2}^{2}+%
\widetilde{\vec{\kappa}}^{2}}(m_2+\sqrt{m_{2}^{2}+\widetilde{\vec{\kappa}}%
^{2}})}]\cdot }\frac{\hat{\eta}}{\tilde{\eta}^{2}}{(}\frac{m_{2}^{2}\sqrt{%
m_{2}^{2}+\widetilde{\vec{\kappa}}^{2}}}{[m_{2}^{2}+(\widetilde{\vec{\kappa}}%
\cdot \hat{\eta})^{2}]^{3/2}}-1) +  \nonumber \\
&&+{i\vec{\xi}}_{1}\cdot {\vec{\xi}_{2}\vec{\xi}_{2}\cdot \vec{\partial}%
_{\eta }\widetilde{\vec{\kappa}}\cdot \vec{\partial}}_{\eta }\frac{1}{\tilde{%
\eta}}\frac{1}{\sqrt{m_{2}^{2}+(\widetilde{\vec{\kappa}}\cdot \hat{\eta})^{2}%
}}\Big)+  \nonumber \\
&&+i{\frac{\widetilde{\vec{\kappa}}\cdot \vec{\xi}_{2}}{\sqrt{m_{2}^{2}+%
\widetilde{\vec{\kappa}}{}^{2}}(m_2+\sqrt{m_{2}^{2}+\widetilde{\vec{\kappa}}%
^{2}})}}  \nonumber \\
&&\Big([\vec{\xi}_{2}-{\frac{i\widetilde{\vec{\kappa}}\cdot \vec{\xi}_{1}
\vec{\xi}_{2}\cdot \vec{\xi}_{1}\vec{\partial}_{\eta }}{\sqrt{m_{1}^{2}+%
\widetilde{\vec{\kappa}}^{2}}(m_1+\sqrt{m_{1}^{2}+\widetilde{\vec{\kappa}}%
^{2}})}]\cdot }\frac{\hat{\eta}}{\tilde{\eta}^{2}}{(}\frac{m_{1}^{2}\sqrt{%
m_{1}^{2}+\widetilde{\vec{\kappa}}^{2}}}{[m_{1}^{2}+(\widetilde{\vec{\kappa}}%
\cdot \hat{\eta})^{2}]^{3/2}}-1) +  \nonumber \\
&&+{i\vec{\xi}}_{2}\cdot {\vec{\xi}_{1}\vec{\xi}_{1}\cdot \vec{\partial}%
_{\eta }\widetilde{\vec{\kappa}}\cdot \vec{\partial}}_{\eta }\frac{1}{\tilde{%
\eta}}\frac{1}{\sqrt{m_{2}^{2}+(\widetilde{\vec{\kappa}}\cdot \hat{\eta})^{2}%
}}\Big)+  \nonumber \\
&+&\frac{{\bf {\vec{f}}_{1}\cdot }\widetilde{{\bf \vec{k}}}_{2}}{\sqrt{%
m_{2}^{2}+\widetilde{\vec{\kappa}}{}^{2}}}\frac{1}{\tilde{\eta}\sqrt{%
m_{1}^{2}+(\widetilde{\vec{\kappa}}\cdot {\hat{\eta}})^{2}}}+\frac{{\bf {%
\vec{f}}_{2}\cdot }\widetilde{{\bf \vec{k}}}_{1}}{\sqrt{m_{1}^{2}+\widetilde{%
\vec{\kappa}}{}^{2}}}\frac{1}{\tilde{\eta}\sqrt{m_{2}^{2}+(\widetilde{\vec{%
\kappa}}\cdot {\hat{\eta}})^{2}}}-  \nonumber \\
&-&\frac{\widetilde{\vec{\kappa}}\cdot {\bf {\vec{f}}}_{2}-\widetilde{\vec{%
\kappa}}\cdot {\bf {\vec{f}}}_{1}+{\bf {\vec{f}}}_{1}\cdot {\bf {\vec{f}}}%
_{2}}{2\sqrt{m_{1}^{2}+\widetilde{\vec{\kappa}}{}^{2}}\sqrt{m_{2}^{2}+%
\widetilde{\vec{\kappa}}{}^{2}}}\frac{1}{\tilde{\eta}} +  \nonumber \\
&+&\Big(\widetilde{\vec{\kappa}}\cdot {\bf {\vec{f}}}_{2}-\widetilde{\vec{%
\kappa} }\cdot {\bf {\vec{f}}}_{1}+{\bf {\vec{f}}}_{1}\cdot {\bf {\vec{f}}}%
_{2}-( {\bf {\vec{f}}}_{2}\vec{\kappa}\cdot {\hat{\eta}}-{\bf {\vec{f}}}_{1}%
\widetilde{\vec{\kappa}}_{j}\cdot {\hat{\eta}}+{\bf {\vec{f}}}_{2}{\bf {\vec{%
f}}}_{1}\cdot {\hat{\eta}})\cdot {\hat{\eta}}\Big)  \nonumber \\
&&\Big( \frac{1}{2\sqrt{m_{1}^{2}+\widetilde{\vec{\kappa}}{}^{2}}\sqrt{%
m_{2}^{2}+\widetilde{\vec{\kappa}}{}^{2}}\tilde{\eta}}-\frac{2}{\tilde{\eta}%
\sqrt{m_{2}^{2}+\widetilde{\vec{\kappa}}{}^{2}}{\large (}\sqrt{m_{1}^{2}+%
\widetilde{\vec{\kappa}}{}^{2}}+\sqrt{m_{1}^{2}+(\widetilde{\vec{\kappa}}%
{}\cdot \hat{\eta})^{2}}{\large )}}\Big)-  \nonumber \\
&&-2\Big(\widetilde{\vec{\kappa}}^{2}({\bf {\vec{f}}}_{2}\cdot \widetilde{%
\vec{\kappa}}-({\bf {\vec{f}}}_{2}\cdot {\hat{\eta}})(\widetilde{\vec{\kappa}%
}_{i}\cdot {\hat{\eta}}))-{\bf {\vec{f}}}_{2}\cdot [ (\widetilde{\vec{\kappa}%
}_{i}\cdot {\hat{\eta}})(\widetilde{\vec{\kappa}}-{\hat{\eta}}(\widetilde{%
\vec{\kappa}}\cdot {\hat{\eta}}))](\widetilde{\vec{\kappa}}\cdot {\hat{\eta}}%
) -  \nonumber \\
&&-\widetilde{\vec{\kappa}}^{2}({\bf {\vec{f}}}_{1}\cdot \widetilde{\vec{%
\kappa}}-({\bf {\vec{f}}}_{1}\cdot {\hat{\eta}})(\widetilde{\vec{\kappa}}%
\cdot {\hat{\eta}}))-{\bf {\vec{f}}}_{1}\cdot [ (\widetilde{\vec{\kappa}}%
\cdot {\hat{\eta}})(\widetilde{\vec{\kappa}}-{\hat{\eta}}(\widetilde{\vec{%
\kappa}}\cdot {\hat{\eta}}))](\widetilde{\vec{\kappa}}\cdot {\hat{\eta}})-
\nonumber \\
&&-({\bf {\vec{f}}}_{1}\cdot \widetilde{\vec{\kappa}})(\widetilde{\vec{\kappa%
}}\cdot \widetilde{\vec{\kappa}}-(\widetilde{\vec{\kappa}}\cdot {\hat{\eta}}%
)(\widetilde{\vec{\kappa}}\cdot {\hat{\eta}}))-\widetilde{\vec{\kappa}}\cdot
[ ({\bf {\vec{f}}}_{1}\cdot {\hat{\eta}})(\widetilde{\vec{\kappa}}-{\hat{\eta%
}}(\widetilde{\vec{\kappa}}\cdot {\hat{\eta}}))](\widetilde{\vec{\kappa}}%
\cdot {\hat{\eta}}) +  \nonumber \\
&&+({\bf {\vec{f}}}_{2}\cdot \widetilde{\vec{\kappa}})(\widetilde{\vec{\kappa%
}}\cdot \widetilde{\vec{\kappa}}-(\widetilde{\vec{\kappa}}\cdot {\hat{\eta}}%
)(\widetilde{\vec{\kappa}}\cdot {\hat{\eta}}))-\widetilde{\vec{\kappa}}\cdot
[ ({\bf {\vec{f}}}_{2}\cdot {\hat{\eta}})(\widetilde{\vec{\kappa}}-{\hat{\eta%
}}(\widetilde{\vec{\kappa}}\cdot {\hat{\eta}}))](\widetilde{\vec{\kappa}}%
\cdot {\hat{\eta}}) +  \nonumber \\
&&+({\bf {\vec{f}}}_{1}\cdot \widetilde{\vec{\kappa}})({\bf {\vec{f}}}%
_{2}\cdot \vec{\kappa}-({\bf {\vec{f}}}_{2}\cdot {\hat{\eta}})(\widetilde{%
\vec{\kappa}}\cdot {\hat{\eta}}))-{\bf {\vec{f}}}_{2}\cdot [ ({\bf {\vec{f}}}%
_{1}\cdot {\hat{\eta}})(\widetilde{\vec{\kappa}}-{\hat{\eta}}(\widetilde{%
\vec{\kappa}}\cdot {\hat{\eta}}))](\widetilde{\vec{\kappa}}\cdot {\hat{\eta}}%
)\Big)  \nonumber \\
&& \frac{1}{\tilde{\eta}\sqrt{m_{2}^{2}+\widetilde{\vec{\kappa}}{}^{2}}\sqrt{%
m_{1}^{2}+(\widetilde{\vec{\kappa}}{}\cdot \hat{\eta})^{2}}}\frac{1}{(\sqrt{%
m_{1}^{2}+\widetilde{\vec{\kappa}}{}^{2}}+\sqrt{m_{1}^{2}+(\widetilde{\vec{%
\kappa}}{}\cdot \hat{\eta})^{2}})^{2}} +  \nonumber \\
&&+[\widetilde{\vec{\kappa}}\cdot {\bf {\vec{g}}}_{1}-\widetilde{\vec{\kappa}%
}\cdot ({\bf {\vec{g}}}_{1}\cdot ({\hat{\eta}}{\hat{\eta}}))]\frac{2}{\tilde{%
\eta}\sqrt{m_{1}^{2}+{\widetilde{\vec{\kappa}}}^{2}}\sqrt{m_{2}^{2}+(%
\widetilde{\vec{\kappa}}\cdot {\hat{\eta}})^{2}}}({\frac{\sqrt{m_{2}^{2}+%
\widetilde{\vec{\kappa}}^{2}}}{\sqrt{m_{2}^{2}+{\widetilde{\vec{\kappa}}}^{2}%
}+\sqrt{m_{2}^{2}+(\widetilde{\vec{\kappa}}\cdot {\hat{\eta}})^{2}}})} -
\nonumber \\
&&-[\widetilde{\vec{\kappa}}\cdot {\bf {\vec{g}}}_{2}-\widetilde{\vec{\kappa}%
}\cdot ({\bf {\vec{g}}}_{2}\cdot ({\hat{\eta}}{\hat{\eta}}))]\frac{2}{\tilde{%
\eta}\sqrt{m_{2}^{2}+{\widetilde{\vec{\kappa}}}^{2}}\sqrt{m_{1}^{2}+(%
\widetilde{\vec{\kappa}}\cdot {\hat{\eta}})^{2}}}({\frac{\sqrt{m_{1}^{2}+%
\widetilde{\vec{\kappa}}^{2}}}{\sqrt{m_{1}^{2}+{\widetilde{\vec{\kappa}}}^{2}%
}+\sqrt{m_{1}^{2}+(\widetilde{\vec{\kappa}}\cdot {\hat{\eta}})^{2}}})} \Big].
\nonumber \\
&&  \label{VIII2}
\end{eqnarray}

In Appendix E we obtain the result

\begin{eqnarray}
&&V_{HDS}=-{\frac{Q_{1}Q_{2}}{8\pi (m_{1}^{2}-m_{2}^{2})}}\Big[\widetilde{%
{\bf \vec{k}}}_{1}\cdot \widetilde{{\bf \vec{k}}}_{2}\frac{1}{\tilde{\eta}}
\nonumber \\
&&\Big(\sqrt{{\frac{m_{2}^{2}++{\widetilde{\vec{\kappa}}}^{2}}{m_{1}^{2}+{%
\widetilde{\vec{\kappa}}}^{2}}}}+\sqrt{{\frac{m_{1}^{2}+{\widetilde{\vec{%
\kappa}}}^{2}}{m_{2}^{2}+{\widetilde{\vec{\kappa}}}^{2}}}}\Big)\Big(\sqrt{{%
\frac{m_{2}^{2}+{\widetilde{\vec{\kappa}}}^{2}}{m_{2}^{2}+(\widetilde{\vec{%
\kappa}}\cdot \hat{\eta})^{2}}}}-\sqrt{{\frac{m_{1}^{2}+{\widetilde{\vec{%
\kappa}}}^{2}}{m_{1}^{2}+(\widetilde{\vec{\kappa}}\cdot \hat{\eta})^{2}}}}%
\Big)+  \nonumber \\
&&+\Big[-\Big((\sqrt{{\frac{m_{2}^{2}+{\widetilde{\vec{\kappa}}}^{2}}{%
m_{1}^{2}+{\widetilde{\vec{\kappa}}}^{2}}}}-{1})\widetilde{{\bf \vec{k}}}%
_{1}\cdot \widetilde{{\bf \vec{l}}}_{2}+(\sqrt{{\frac{m_{1}^{2}+{\widetilde{%
\vec{\kappa}}}^{2}}{m_{2}^{2}+{\widetilde{\vec{\kappa}}}^{2}}}}-{1})%
\widetilde{{\bf \vec{k}}}_{2}\cdot \widetilde{{\bf \vec{l}}}_{1}+\widetilde{%
{\bf \vec{l}}}_{1}\cdot \widetilde{{\bf \vec{l}}}_{2}\Big)  \nonumber \\
&&\frac{1}{2}\Big(\frac{\sqrt{m_{2}^{2}+{\widetilde{\vec{\kappa}}}^{2}}-%
\sqrt{m_{2}^{2}+(\widetilde{\vec{\kappa}}\cdot \hat{\eta})^{2}}}{\sqrt{%
m_{2}^{2}+{\widetilde{\vec{\kappa}}}^{2}}+\sqrt{m_{2}^{2}+(\widetilde{\vec{%
\kappa}}\cdot \hat{\eta})^{2}}}-\frac{\sqrt{m_{1}^{2}+{\widetilde{\vec{\kappa%
}}}^{2}}-\sqrt{m_{1}^{2}+(\widetilde{\vec{\kappa}}\cdot \hat{\eta})^{2}}}{%
\sqrt{m_{1}^{2}+{\widetilde{\vec{\kappa}}}^{2}}+\sqrt{m_{1}^{2}+(\widetilde{%
\vec{\kappa}}\cdot \hat{\eta})^{2}}}\Big)+  \nonumber \\
&&+\Big((\sqrt{{\frac{m_{2}^{2}+{\widetilde{\vec{\kappa}}}^{2}}{m_{1}^{2}+{%
\widetilde{\vec{\kappa}}}^{2}}}}-{1})\frac{\widetilde{{\bf \vec{k}}}_{1}%
\widetilde{{\bf \vec{l}}}_{2}\cdot \cdot {\hat{\eta}\hat{\eta}}\widetilde{%
\vec{\kappa}}^{2}}{2}+(\sqrt{{\frac{m_{1}^{2}+{\widetilde{\vec{\kappa}}}^{2}%
}{m_{2}^{2}+{\widetilde{\vec{\kappa}}}^{2}}}}-{1})\frac{\widetilde{{\bf \vec{%
k}}}_{2}\cdot (\widetilde{{\bf \vec{l}}}_{1}\cdot {\hat{\eta}\hat{\eta})}%
\widetilde{\vec{\kappa}}^{2}}{2}+\frac{\widetilde{{\bf \vec{l}}}_{1}\cdot (%
\widetilde{{\bf \vec{l}}}_{2}\cdot {\hat{\eta}\hat{\eta})}\widetilde{\vec{%
\kappa}}^{2}}{2}\Big)  \nonumber \\
&&\Big(\frac{1}{(\sqrt{m_{2}^{2}+{\widetilde{\vec{\kappa}}}^{2}}+\sqrt{%
m_{2}^{2}+(\widetilde{\vec{\kappa}}\cdot \hat{\eta})^{2}})^{2}}-\frac{1}{(%
\sqrt{m_{1}^{2}+{\widetilde{\vec{\kappa}}}^{2}}+\sqrt{m_{1}^{2}+(\widetilde{%
\vec{\kappa}}\cdot \hat{\eta})^{2}})^{2}}\Big)-  \nonumber \\
&&-\Big((\sqrt{{\frac{m_{2}^{2}+{\widetilde{\vec{\kappa}}}^{2}}{m_{1}^{2}+{%
\widetilde{\vec{\kappa}}}^{2}}}}-{1})[\widetilde{{\bf \vec{k}}}_{1}\cdot {%
\widetilde{\vec{\kappa}}}\widetilde{{\bf \vec{l}}}_{2}\cdot \widetilde{\vec{%
\kappa}}-\widetilde{{\bf \vec{k}}}_{1}\cdot (\widetilde{{\bf \vec{l}}}%
_{2}\cdot (\widetilde{\vec{\kappa}}{\hat{\eta}}+{\hat{\eta}}\widetilde{\vec{%
\kappa}}))(\widetilde{\vec{\kappa}}\cdot {\hat{\eta}})]+  \nonumber \\
&&+(\sqrt{{\frac{m_{1}^{2}+{\widetilde{\vec{\kappa}}}^{2}}{m_{2}^{2}+{%
\widetilde{\vec{\kappa}}}^{2}}}}-{1})[\widetilde{{\bf \vec{k}}}_{2}\cdot {%
\widetilde{\vec{\kappa}}}\widetilde{{\bf \vec{l}}}_{1}\cdot \widetilde{\vec{%
\kappa}}-\widetilde{{\bf \vec{k}}}_{2}\cdot (\widetilde{{\bf \vec{l}}}%
_{1}\cdot (\widetilde{\vec{\kappa}}{\hat{\eta}}+{\hat{\eta}}\widetilde{\vec{%
\kappa}}))(\widetilde{\vec{\kappa}}\cdot {\hat{\eta}})]+  \nonumber \\
&&+[\widetilde{{\bf \vec{l}}}_{2}\cdot {\widetilde{\vec{\kappa}}}\widetilde{%
{\bf \vec{l}}}_{1}\cdot \widetilde{\vec{\kappa}}-\widetilde{{\bf \vec{l}}}%
_{2}\cdot (\widetilde{{\bf \vec{l}}}_{1}\cdot (\widetilde{\vec{\kappa}}{\hat{%
\eta}}+{\hat{\eta}}\widetilde{\vec{\kappa}}))(\widetilde{\vec{\kappa}}\cdot {%
\hat{\eta}})]\Big)  \nonumber \\
&&\Big(\sqrt{{\frac{m_{2}^{2}+{\widetilde{\vec{\kappa}}}^{2}}{m_{2}^{2}+(%
\widetilde{\vec{\kappa}}\cdot \hat{\eta})^{2}}}}\frac{1}{(\sqrt{m_{2}^{2}+{%
\widetilde{\vec{\kappa}}}^{2}}+\sqrt{m_{2}^{2}+(\widetilde{\vec{\kappa}}%
\cdot \hat{\eta})^{2}})^{2}}-  \nonumber \\
&-&\sqrt{{\frac{m_{1}^{2}+{\widetilde{\vec{\kappa}}}^{2}}{m_{1}^{2}+(%
\widetilde{\vec{\kappa}}\cdot \hat{\eta})^{2}}}}\frac{1}{(\sqrt{m_{1}^{2}+{%
\widetilde{\vec{\kappa}}}^{2}}+\sqrt{m_{1}^{2}+(\widetilde{\vec{\kappa}}%
\cdot \hat{\eta})^{2}})^{2}}\Big)-  \nonumber \\
&&-\Big((\sqrt{{\frac{m_{2}^{2}+{\widetilde{\vec{\kappa}}}^{2}}{m_{1}^{2}+{%
\widetilde{\vec{\kappa}}}^{2}}}}-{1})\widetilde{{\bf \vec{k}}}_{1}\cdot (%
\widetilde{{\bf \vec{l}}}_{2}\cdot {\hat{\eta}\hat{\eta})}(\widetilde{\vec{%
\kappa}}\cdot {\hat{\eta}})^{2}+  \nonumber \\
&+&(\sqrt{{\frac{m_{2}^{2}+{\widetilde{\vec{\kappa}}}^{2}}{m_{1}^{2}+{%
\widetilde{\vec{\kappa}}}^{2}}}}-{1})\widetilde{{\bf \vec{k}}}_{1}\cdot (%
\widetilde{{\bf \vec{l}}}_{2}\cdot {\hat{\eta}\hat{\eta})}(\widetilde{\vec{%
\kappa}}\cdot {\hat{\eta}})^{2}+\widetilde{{\bf \vec{l}}}_{1}\cdot (%
\widetilde{{\bf \vec{l}}}_{2}\cdot {\hat{\eta}\hat{\eta})}(\widetilde{\vec{%
\kappa}}\cdot {\hat{\eta}})^{2}\Big)  \nonumber \\
&&\Big((\frac{1}{(\sqrt{m_{2}^{2}+{\widetilde{\vec{\kappa}}}^{2}}+\sqrt{%
m_{2}^{2}+(\widetilde{\vec{\kappa}}\cdot \hat{\eta})^{2}})^{2}})(2+\sqrt{{%
\frac{m_{2}^{2}+(\widetilde{\vec{\kappa}}\cdot \hat{\eta})^{2}}{m_{2}^{2}+{%
\widetilde{\vec{\kappa}}}^{2}}}})-  \nonumber \\
&&-(\frac{1}{(\sqrt{m_{1}^{2}+{\widetilde{\vec{\kappa}}}^{2}}+\sqrt{%
m_{1}^{2}+(\widetilde{\vec{\kappa}}\cdot \hat{\eta})^{2}})^{2}})(2+\sqrt{{%
\frac{m_{1}^{2}+(\widetilde{\vec{\kappa}}\cdot \hat{\eta})^{2}}{m_{1}^{2}+{%
\widetilde{\vec{\kappa}}}^{2}}}})\Big)\Big]\frac{1}{\tilde{\eta}}-  \nonumber
\\
&&-\Big(\widetilde{\vec{\kappa}}\cdot \widetilde{{\bf \vec{k}}}_{2}\frac{(%
\widetilde{{\bf \vec{k}}}_{1}-\widetilde{{\bf \vec{l}}}_{1})\cdot \widetilde{%
\vec{\kappa}}-(\widetilde{{\bf \vec{k}}}_{1}-\widetilde{{\bf \vec{l}}}%
_{1})\cdot {\hat{\eta}}\widetilde{\vec{\kappa}}\cdot {\hat{\eta}}}{m_{1}^{2}+%
{\widetilde{\vec{\kappa}}}{}^{2}}-\widetilde{\vec{\kappa}}\cdot \widetilde{%
{\bf \vec{k}}}_{1}\frac{(\widetilde{{\bf \vec{k}}}_{2}-\widetilde{{\bf \vec{l%
}}_{2}})\cdot \widetilde{\vec{\kappa}}-(\widetilde{{\bf \vec{k}}}_{2}-%
\widetilde{{\bf \vec{l}}}_{2})\cdot {\hat{\eta}}\widetilde{\vec{\kappa}}%
\cdot {\hat{\eta}}}{m_{2}^{2}+{\widetilde{\vec{\kappa}}}{}^{2}}\Big)
\nonumber \\
&&\Big(\sqrt{{\frac{1}{m_{2}^{2}+(\widetilde{\vec{\kappa}}\cdot \hat{\eta}%
)^{2}}}}\frac{\sqrt{m_{1}^{2}+{\widetilde{\vec{\kappa}}}{}^{2}}\sqrt{%
m_{2}^{2}+{\widetilde{\vec{\kappa}}}{}^{2}}}{\sqrt{m_{2}^{2}+{\widetilde{%
\vec{\kappa}}}^{2}}+\sqrt{m_{2}^{2}+(\widetilde{\vec{\kappa}}\cdot \hat{\eta}%
)^{2}}}-  \nonumber \\
&-&\sqrt{{\frac{1}{m_{1}^{2}+(\widetilde{\vec{\kappa}}\cdot \hat{\eta})^{2}}}%
}\frac{\sqrt{m_{1}^{2}+{\widetilde{\vec{\kappa}}}{}^{2}}\sqrt{m_{2}^{2}+{%
\widetilde{\vec{\kappa}}}{}^{2}}}{\sqrt{m_{1}^{2}+{\widetilde{\vec{\kappa}}}%
^{2}}+\sqrt{m_{1}^{2}+(\widetilde{\vec{\kappa}}\cdot \hat{\eta})^{2}}}\Big)%
\frac{1}{\tilde{\eta}}  \nonumber \\
&&-\widetilde{{\bf \vec{k}}}_{1}\cdot \widetilde{{\bf \vec{k}}}_{2}\frac{1}{2%
\sqrt{m_{1}^{2}+{\widetilde{\vec{\kappa}}}{}^{2}}\sqrt{m_{2}^{2}+{\widetilde{%
\vec{\kappa}}}{}^{2}}\tilde{\eta}}\Big((\sqrt{m_{2}^{2}+{\widetilde{\vec{%
\kappa}}}^{2}}-\sqrt{m_{2}^{2}+(\widetilde{\vec{\kappa}}\cdot \hat{\eta})^{2}%
})^{2}(2+\sqrt{{\frac{m_{2}^{2}+(\widetilde{\vec{\kappa}}\cdot \hat{\eta}%
)^{2}}{m_{2}^{2}+{\widetilde{\vec{\kappa}}}^{2}}}})-  \nonumber \\
&&-(\sqrt{m_{1}^{2}+{\widetilde{\vec{\kappa}}}^{2}}-\sqrt{m_{1}^{2}+(%
\widetilde{\vec{\kappa}}\cdot \hat{\eta})^{2}})^{2}(2+\sqrt{{\frac{%
m_{1}^{2}+(\widetilde{\vec{\kappa}}\cdot \hat{\eta})^{2}}{m_{1}^{2}+{%
\widetilde{\vec{\kappa}}}^{2}}}})\Big)-  \nonumber \\
&&-\Big[-\widetilde{{\bf \vec{k}}}_{1}\cdot \widetilde{{\bf \vec{k}}}_{2}%
\Big(\frac{m_{2}^{2}+{\widetilde{\vec{\kappa}}}^{2}}{2}(\frac{\sqrt{%
m_{2}^{2}+{\widetilde{\vec{\kappa}}}^{2}}-\sqrt{m_{2}^{2}+(\widetilde{\vec{%
\kappa}}\cdot \hat{\eta})^{2}}}{\sqrt{m_{2}^{2}+{\widetilde{\vec{\kappa}}}%
^{2}}+\sqrt{m_{2}^{2}+(\widetilde{\vec{\kappa}}\cdot \hat{\eta})^{2}}})-
\nonumber \\
&-&\frac{m_{1}^{2}+{\widetilde{\vec{\kappa}}}^{2}}{2}(\frac{\sqrt{m_{1}^{2}+{%
\widetilde{\vec{\kappa}}}^{2}}-\sqrt{m_{1}^{2}+(\widetilde{\vec{\kappa}}%
\cdot \hat{\eta})^{2}}}{\sqrt{m_{1}^{2}+{\widetilde{\vec{\kappa}}}^{2}}+%
\sqrt{m_{1}^{2}+(\widetilde{\vec{\kappa}}\cdot \hat{\eta})^{2}}})\Big)+
\nonumber \\
&&+\Big(\widetilde{{\bf \vec{k}}}_{1}\cdot {\widetilde{\vec{\kappa}}}%
\widetilde{{\bf \vec{k}}}_{2}\cdot \widetilde{\vec{\kappa}}-\widetilde{{\bf
\vec{k}}}_{1}\cdot (\widetilde{{\bf \vec{k}}}_{2}\cdot (\widetilde{\vec{%
\kappa}}{\hat{\eta}}+{\hat{\eta}}\widetilde{\vec{\kappa}}))(\widetilde{\vec{%
\kappa}}\cdot {\hat{\eta}})-\widetilde{{\bf \vec{k}}}_{1}\cdot (\widetilde{%
{\bf \vec{k}}}_{2}\cdot {\hat{\eta}\hat{\eta})}\widetilde{\vec{\kappa}}^{2}%
\Big)  \nonumber \\
&&\Big(\frac{(\sqrt{m_{2}^{2}+{\widetilde{\vec{\kappa}}}{}^{2}})^{3}}{\sqrt{%
m_{2}^{2}+(\widetilde{\vec{\kappa}}\cdot \hat{\eta})^{2}}}\frac{1}{(\sqrt{%
m_{2}^{2}+{\widetilde{\vec{\kappa}}}^{2}}+\sqrt{m_{2}^{2}+(\widetilde{\vec{%
\kappa}}\cdot \hat{\eta})^{2}})^{2}}-  \nonumber \\
&-&\frac{(\sqrt{m_{1}^{2}+{\widetilde{\vec{\kappa}}}{}^{2}})^{3}}{\sqrt{%
m_{1}^{2}+(\widetilde{\vec{\kappa}}\cdot \hat{\eta})^{2}}}\frac{1}{(\sqrt{%
m_{1}^{2}+{\widetilde{\vec{\kappa}}}^{2}}+\sqrt{m_{1}^{2}+(\widetilde{\vec{%
\kappa}}\cdot \hat{\eta})^{2}})^{2}}\Big)-  \nonumber \\
&&-\widetilde{{\bf \vec{k}}}_{1}\widetilde{{\bf \vec{k}}}_{2}\cdot \cdot {%
\hat{\eta}\hat{\eta}}(\widetilde{\vec{\kappa}}\cdot {\hat{\eta}})^{2}\Big(%
\frac{\sqrt{m_{2}^{2}+{\widetilde{\vec{\kappa}}}{}^{2}}(2\sqrt{m_{2}^{2}+{%
\widetilde{\vec{\kappa}}}{}^{2}}+\sqrt{m_{2}^{2}+(\widetilde{\vec{\kappa}}%
\cdot \hat{\eta})^{2}}}{(\sqrt{m_{2}^{2}+{\widetilde{\vec{\kappa}}}^{2}}+%
\sqrt{m_{2}^{2}+(\widetilde{\vec{\kappa}}\cdot \hat{\eta})^{2}})^{2}}-
\nonumber \\
&&-\frac{\sqrt{m_{1}^{2}+{\widetilde{\vec{\kappa}}}{}^{2}}(2\sqrt{m_{1}^{2}+{%
\widetilde{\vec{\kappa}}}{}^{2}}+\sqrt{m_{1}^{2}+(\widetilde{\vec{\kappa}}%
\cdot \hat{\eta})^{2}}}{(\sqrt{m_{1}^{2}+{\widetilde{\vec{\kappa}}}^{2}}+%
\sqrt{m_{1}^{2}+(\widetilde{\vec{\kappa}}\cdot \hat{\eta})^{2}})^{2}}\Big)%
\Big]\frac{1}{\sqrt{m_{1}^{2}+{\widetilde{\vec{\kappa}}}{}^{2}}\sqrt{%
m_{2}^{2}+{\widetilde{\vec{\kappa}}}{}^{2}}\tilde{\eta}}{\huge ]}\Big].
\label{VIII3}
\end{eqnarray}

2) For the pseudo-classical version of {\it positronium} ($m_{1}=m_{2}$)

\begin{equation}
M=2\sqrt{m^{2}+\widetilde{\vec{\kappa}}^{2}}+\frac{Q_{1}Q_{2}}{4\pi
\widetilde{\eta }}+V_{LDS}+V_{HDS},  \label{VIII4}
\end{equation}

\noindent where now

\begin{eqnarray}
V_{LDS} &:&=\frac{Q_{1}Q_{2}}{4\pi }\Big[\frac{\widetilde{\vec{\kappa}}^{2}+(%
\widetilde{\vec{\kappa}}\cdot {\hat{\eta})}^{2}}{2(m^{2}+\widetilde{\vec{%
\kappa}}{}^{2})^{2}}\frac{1}{\tilde{\eta}}-i\frac{\widetilde{\vec{\kappa}}%
\cdot \vec{\xi}_{1}\vec{\xi}_{1}\cdot \vec{\eta}+\widetilde{\vec{\kappa}}%
\cdot \vec{\xi}_{2}\vec{\xi}_{2}\cdot \vec{\eta}}{\tilde{\eta}^{3}(m+\sqrt{%
m^{2}+\widetilde{\vec{\kappa}}^{2}})\sqrt{m^{2}+\widetilde{\vec{\kappa}}^{2}}%
}-  \nonumber \\
&&-i{\frac{\vec{\xi}_{1}\times \vec{\xi}_{1}\cdot (\widetilde{{\bf \vec{k}}}%
_{2}\times \vec{\partial}_{\eta })-\vec{\xi}_{2}\times \vec{\xi}_{2}\cdot (%
\widetilde{{\bf \vec{k}}}_{1}\times \vec{\partial}_{\eta })}{2\sqrt{m^{2}+%
\widetilde{\vec{\kappa}}{}^{2}}}}\frac{1}{\tilde{\eta}\sqrt{m^{2}+(%
\widetilde{\vec{\kappa}}\cdot {\hat{\eta}})^{2}}}+  \nonumber \\
&&+i{\frac{\widetilde{\vec{\kappa}}\cdot \vec{\xi}_{1}}{\sqrt{m^{2}+%
\widetilde{\vec{\kappa}}{}^{2}}(m+\sqrt{m^{2}+\widetilde{\vec{\kappa}}^{2}})}%
}  \nonumber \\
&&\cdot \Big([\vec{\xi}_{1}-{\frac{i\widetilde{\vec{\kappa}}\cdot \vec{\xi}%
_{2}\vec{\xi}_{1}\cdot \vec{\xi}_{2}\vec{\partial}_{\eta }}{\sqrt{m^{2}+%
\widetilde{\vec{\kappa}}^{2}}(m+\sqrt{m^{2}+\widetilde{\vec{\kappa}}^{2}})}%
]\cdot }\frac{\hat{\eta}}{\tilde{\eta}^{2}}{(}\frac{m^{2}\sqrt{m^{2}+%
\widetilde{\vec{\kappa}}^{2}}}{[m^{2}+(\widetilde{\vec{\kappa}}\cdot \hat{%
\eta})^{2}]^{3/2}}-1)+  \nonumber \\
&&+{i\vec{\xi}}_{1}\cdot {\vec{\xi}_{2}\vec{\xi}_{2}\cdot \vec{\partial}%
_{\eta }\widetilde{\vec{\kappa}}\cdot \vec{\partial}}_{\eta }\frac{1}{\tilde{%
\eta}}\frac{1}{\sqrt{m^{2}+(\widetilde{\vec{\kappa}}\cdot \hat{\eta})^{2}}}%
\Big)+  \nonumber \\
&&+i{\frac{\widetilde{\vec{\kappa}}\cdot \vec{\xi}_{2}}{\sqrt{m^{2}+%
\widetilde{\vec{\kappa}}{}^{2}}(m+\sqrt{m^{2}+\widetilde{\vec{\kappa}}^{2}})}%
}  \nonumber \\
&&\cdot \Big([\vec{\xi}_{2}-{\frac{i\widetilde{\vec{\kappa}}\cdot \vec{\xi}%
_{1}\vec{\xi}_{2}\cdot \vec{\xi}_{1}\vec{\partial}_{\eta }}{\sqrt{m^{2}+%
\widetilde{\vec{\kappa}}^{2}}(m+\sqrt{m^{2}+\widetilde{\vec{\kappa}}^{2}})}%
]\cdot }\frac{\hat{\eta}}{\tilde{\eta}^{2}}{(}\frac{m^{2}\sqrt{m^{2}+%
\widetilde{\vec{\kappa}}^{2}}}{[m^{2}+(\widetilde{\vec{\kappa}}\cdot \hat{%
\eta})^{2}]^{3/2}}-1)+  \nonumber \\
&&+{i\vec{\xi}}_{2}\cdot {\vec{\xi}_{1}\vec{\xi}_{1}\cdot \vec{\partial}%
_{\eta }\widetilde{\vec{\kappa}}\cdot \vec{\partial}}_{\eta }\frac{1}{\tilde{%
\eta}}\frac{1}{\sqrt{m^{2}+(\widetilde{\vec{\kappa}}\cdot \hat{\eta})^{2}}}%
\Big)+  \nonumber \\
&+&\frac{{\bf {\vec{f}}_{1}\cdot }\widetilde{{\bf \vec{k}}}_{2}}{\sqrt{m^{2}+%
\widetilde{\vec{\kappa}}{}^{2}}}\frac{1}{\tilde{\eta}\sqrt{m^{2}+(\widetilde{%
\vec{\kappa}}\cdot {\hat{\eta}})^{2}}}+\frac{{\bf {\vec{f}}_{2}\cdot }%
\widetilde{{\bf \vec{k}}}_{1}}{\sqrt{m^{2}+\widetilde{\vec{\kappa}}{}^{2}}}%
\frac{1}{\tilde{\eta}\sqrt{m^{2}+(\widetilde{\vec{\kappa}}\cdot {\hat{\eta}}%
)^{2}}}-  \nonumber \\
&-&\frac{\widetilde{\vec{\kappa}}\cdot {\bf {\vec{f}}}_{2}-\widetilde{\vec{%
\kappa}}\cdot {\bf {\vec{f}}}_{1}+{\bf {\vec{f}}}_{1}\cdot {\bf {\vec{f}}}%
_{2}}{2\sqrt{m^{2}+\widetilde{\vec{\kappa}}{}^{2}}\sqrt{m^{2}+\widetilde{%
\vec{\kappa}}{}^{2}}}\frac{1}{\tilde{\eta}}+  \nonumber \\
&+&[\widetilde{\vec{\kappa}}\cdot {\bf {\vec{f}}}_{2}-\widetilde{\vec{\kappa}%
}\cdot {\bf {\vec{f}}}_{1}+{\bf {\vec{f}}}_{1}\cdot {\bf {\vec{f}}}_{2}-(%
{\bf {\vec{f}}}_{2}\vec{\kappa}\cdot {\hat{\eta}}-{\bf {\vec{f}}}_{1}%
\widetilde{\vec{\kappa}}_{j}\cdot {\hat{\eta}}+{\bf {\vec{f}}}_{2}{\bf {\vec{%
f}}}_{1}\cdot {\hat{\eta}})\cdot {\hat{\eta}}]  \nonumber \\
&&\Big(\frac{1}{2(m^{2}+\widetilde{\vec{\kappa}}{}^{2})\tilde{\eta}}-\frac{2%
}{\tilde{\eta}\sqrt{m^{2}+\widetilde{\vec{\kappa}}_{j}{}^{2}}\Big(\sqrt{%
m^{2}+\widetilde{\vec{\kappa}}{}^{2}}+\sqrt{m^{2}+(\widetilde{\vec{\kappa}}%
{}\cdot \hat{\eta})^{2}}\Big)}\Big)-  \nonumber \\
&&-2\Big(\widetilde{\vec{\kappa}}^{2}({\bf {\vec{f}}}_{2}\cdot \widetilde{%
\vec{\kappa}}-({\bf {\vec{f}}}_{2}\cdot {\hat{\eta}})(\widetilde{\vec{\kappa}%
}_{i}\cdot {\hat{\eta}}))-{\bf {\vec{f}}}_{2}\cdot \lbrack (\widetilde{\vec{%
\kappa}}_{i}\cdot {\hat{\eta}})(\widetilde{\vec{\kappa}}-{\hat{\eta}}(%
\widetilde{\vec{\kappa}}\cdot {\hat{\eta}}))](\widetilde{\vec{\kappa}}\cdot {%
\hat{\eta}})-  \nonumber \\
&&-\widetilde{\vec{\kappa}}^{2}({\bf {\vec{f}}}_{1}\cdot \widetilde{\vec{%
\kappa}}-({\bf {\vec{f}}}_{1}\cdot {\hat{\eta}})(\widetilde{\vec{\kappa}}%
\cdot {\hat{\eta}}))-{\bf {\vec{f}}}_{1}\cdot \lbrack (\widetilde{\vec{\kappa%
}}\cdot {\hat{\eta}})(\widetilde{\vec{\kappa}}-{\hat{\eta}}(\widetilde{\vec{%
\kappa}}\cdot {\hat{\eta}}))](\widetilde{\vec{\kappa}}\cdot {\hat{\eta}})-
\nonumber \\
&&-({\bf {\vec{f}}}_{1}\cdot \widetilde{\vec{\kappa}})(\widetilde{\vec{\kappa%
}}\cdot \widetilde{\vec{\kappa}}-(\widetilde{\vec{\kappa}}\cdot {\hat{\eta}}%
)(\widetilde{\vec{\kappa}}\cdot {\hat{\eta}}))-\widetilde{\vec{\kappa}}\cdot
\lbrack ({\bf {\vec{f}}}_{1}\cdot {\hat{\eta}})(\widetilde{\vec{\kappa}}-{%
\hat{\eta}}(\widetilde{\vec{\kappa}}\cdot {\hat{\eta}}))](\widetilde{\vec{%
\kappa}}\cdot {\hat{\eta}})+  \nonumber \\
&&+({\bf {\vec{f}}}_{2}\cdot \widetilde{\vec{\kappa}})(\widetilde{\vec{\kappa%
}}\cdot \widetilde{\vec{\kappa}}-(\widetilde{\vec{\kappa}}\cdot {\hat{\eta}}%
)(\widetilde{\vec{\kappa}}\cdot {\hat{\eta}}))-\widetilde{\vec{\kappa}}\cdot
\lbrack ({\bf {\vec{f}}}_{2}\cdot {\hat{\eta}})(\widetilde{\vec{\kappa}}-{%
\hat{\eta}}(\widetilde{\vec{\kappa}}\cdot {\hat{\eta}}))](\widetilde{\vec{%
\kappa}}\cdot {\hat{\eta}})+  \nonumber \\
&&+({\bf {\vec{f}}}_{1}\cdot \widetilde{\vec{\kappa}})({\bf {\vec{f}}}%
_{2}\cdot \vec{\kappa}-({\bf {\vec{f}}}_{2}\cdot {\hat{\eta}})(\widetilde{%
\vec{\kappa}}\cdot {\hat{\eta}}))-{\bf {\vec{f}}}_{2}\cdot \lbrack ({\bf {%
\vec{f}}}_{1}\cdot {\hat{\eta}})(\widetilde{\vec{\kappa}}-{\hat{\eta}}(%
\widetilde{\vec{\kappa}}\cdot {\hat{\eta}}))](\widetilde{\vec{\kappa}}\cdot {%
\hat{\eta}})\Big)  \nonumber \\
&&\frac{1}{\tilde{\eta}\sqrt{m^{2}+\widetilde{\vec{\kappa}}{}^{2}}\sqrt{%
m^{2}+(\widetilde{\vec{\kappa}}{}\cdot \hat{\eta})^{2}}}\frac{1}{(\sqrt{%
m^{2}+\widetilde{\vec{\kappa}}{}^{2}}+\sqrt{m^{2}+(\widetilde{\vec{\kappa}}%
{}\cdot \hat{\eta})^{2}})^{2}}+  \nonumber \\
&&+\Big([\widetilde{\vec{\kappa}}\cdot {\bf {\vec{g}}}_{1}-\widetilde{\vec{%
\kappa}}\cdot ({\bf {\vec{g}}}_{1}\cdot ({\hat{\eta}}{\hat{\eta}}))]-[%
\widetilde{\vec{\kappa}}\cdot {\bf {\vec{g}}}_{2}-\widetilde{\vec{\kappa}}%
\cdot ({\bf {\vec{g}}}_{2}\cdot ({\hat{\eta}}{\hat{\eta}}))]\Big)  \nonumber
\\
&&\frac{2}{\tilde{\eta}\sqrt{m^{2}+{\widetilde{\vec{\kappa}}}^{2}}\sqrt{%
m^{2}+(\widetilde{\vec{\kappa}}\cdot {\hat{\eta}})^{2}}}({\frac{\sqrt{m^{2}+%
\widetilde{\vec{\kappa}}^{2}}}{\sqrt{m^{2}+{\widetilde{\vec{\kappa}}}^{2}}+%
\sqrt{m^{2}+(\widetilde{\vec{\kappa}}\cdot {\hat{\eta}})^{2}}})}\Big].
\label{VIII5}
\end{eqnarray}

\noindent while

\begin{eqnarray}
V_{HDS} &=&-{\frac{Q_{1}Q_{2}}{8\pi }}\Big[\widetilde{{\bf \vec{k}}}%
_{1}\cdot \widetilde{{\bf \vec{k}}}_{2}\frac{1}{\tilde{\eta}}(\frac{1}{%
m^{2}+(\widetilde{\vec{\kappa}}\cdot \hat{\eta})^{2}}-\frac{1}{m^{2}+{%
\widetilde{\vec{\kappa}}}^{2}})+  \nonumber \\
&&+\Big(-\widetilde{{\bf \vec{l}}}_{1}\cdot \widetilde{{\bf \vec{l}}}_{2}%
\frac{1}{2}\frac{\sqrt{m^{2}+{\widetilde{\vec{\kappa}}}^{2}}-\sqrt{m^{2}+(%
\widetilde{\vec{\kappa}}\cdot \hat{\eta})^{2}}}{\sqrt{m^{2}+{\widetilde{\vec{%
\kappa}}}^{2}}+\sqrt{m^{2}+(\widetilde{\vec{\kappa}}\cdot \hat{\eta})^{2}}}%
\frac{1}{\sqrt{m^{2}+{\widetilde{\vec{\kappa}}}^{2}}\sqrt{m^{2}+(\widetilde{%
\vec{\kappa}}\cdot \hat{\eta})^{2}}}+  \nonumber \\
&&+\frac{\widetilde{{\bf \vec{l}}}_{1}\cdot (\widetilde{{\bf \vec{l}}}%
_{2}\cdot {\hat{\eta}\hat{\eta})}\widetilde{\vec{\kappa}}^{2}}{2}(\frac{1}{(%
\sqrt{m^{2}+{\widetilde{\vec{\kappa}}}^{2}}+\sqrt{m^{2}+(\widetilde{\vec{%
\kappa}}\cdot \hat{\eta})^{2}})^{2}})\frac{1}{\sqrt{m^{2}+{\widetilde{\vec{%
\kappa}}}^{2}}\sqrt{m^{2}+(\widetilde{\vec{\kappa}}\cdot \hat{\eta})^{2}}}-
\nonumber \\
&&-\Big(\widetilde{{\bf \vec{l}}}_{2}\cdot {\widetilde{\vec{\kappa}}}%
\widetilde{{\bf \vec{l}}}_{1}\cdot \widetilde{\vec{\kappa}}-\widetilde{{\bf
\vec{l}}}_{2}\cdot (\widetilde{{\bf \vec{l}}}_{1}\cdot (\widetilde{\vec{%
\kappa}}{\hat{\eta}}+{\hat{\eta}}\widetilde{\vec{\kappa}})(\widetilde{\vec{%
\kappa}}\cdot {\hat{\eta}}))\Big)  \nonumber \\
&&[\sqrt{{\frac{m^{2}+{\widetilde{\vec{\kappa}}}^{2}}{m^{2}+(\widetilde{\vec{%
\kappa}}\cdot \hat{\eta})^{2}}}}\frac{1}{(\sqrt{m^{2}+{\widetilde{\vec{\kappa%
}}}^{2}}+\sqrt{m^{2}+(\widetilde{\vec{\kappa}}\cdot \hat{\eta})^{2}})^{2}}
\nonumber \\
&&\Big(\frac{1}{\sqrt{m^{2}+{\widetilde{\vec{\kappa}}}^{2}}\sqrt{m^{2}+(%
\widetilde{\vec{\kappa}}\cdot \hat{\eta})^{2}}}-\frac{1}{2}(\frac{1}{m^{2}+(%
\widetilde{\vec{\kappa}}\cdot \hat{\eta})^{2}}-\frac{1}{m^{2}+{\widetilde{%
\vec{\kappa}}}^{2}}\Big)]-  \nonumber \\
&&-\widetilde{{\bf \vec{l}}}_{1}\widetilde{{\bf \vec{l}}}_{2}\cdot \cdot {%
\hat{\eta}\hat{\eta}}(\widetilde{\vec{\kappa}}\cdot {\hat{\eta}})^{2}(\frac{1%
}{(\sqrt{m^{2}+{\widetilde{\vec{\kappa}}}^{2}}+\sqrt{m^{2}+(\widetilde{\vec{%
\kappa}}\cdot \hat{\eta})^{2}})^{2}})(\frac{2}{\sqrt{m^{2}+{\widetilde{\vec{%
\kappa}}}^{2}}\sqrt{m^{2}+(\widetilde{\vec{\kappa}}\cdot \hat{\eta})^{2}}}-
\nonumber \\
&&-\frac{1}{2}\sqrt{{\frac{m^{2}+(\widetilde{\vec{\kappa}}\cdot \hat{\eta}%
)^{2}}{m^{2}+{\widetilde{\vec{\kappa}}}^{2}}}}(\frac{1}{m^{2}+(\widetilde{%
\vec{\kappa}}\cdot \hat{\eta})^{2}}-\frac{1}{m^{2}+{\widetilde{\vec{\kappa}}}%
^{2}}))\Big)\frac{1}{\tilde{\eta}}+  \nonumber \\
&&+\Big(\widetilde{\vec{\kappa}}\cdot \widetilde{{\bf \vec{k}}}_{2}(%
\widetilde{{\bf \vec{l}}}_{1}\cdot \widetilde{\vec{\kappa}}-(\widetilde{{\bf
\vec{k}}}_{1}-\widetilde{{\bf \vec{l}}}_{1})\cdot {\hat{\eta}}\widetilde{%
\vec{\kappa}}\cdot {\hat{\eta})}-\widetilde{\vec{\kappa}}\cdot \widetilde{%
{\bf \vec{k}}}_{1}(\widetilde{{\bf \vec{l}}_{2}}\cdot \widetilde{\vec{\kappa}%
}-(\widetilde{{\bf \vec{k}}}_{2}-\widetilde{{\bf \vec{l}}}_{2})\cdot {\hat{%
\eta}}\widetilde{\vec{\kappa}}\cdot {\hat{\eta})}\Big]  \nonumber \\
&&\Big(\sqrt{{\frac{1}{m^{2}+(\widetilde{\vec{\kappa}}\cdot \hat{\eta})^{2}}}%
}\frac{1}{\sqrt{m^{2}+{\widetilde{\vec{\kappa}}}^{2}}+\sqrt{m^{2}+(%
\widetilde{\vec{\kappa}}\cdot \hat{\eta})^{2}}}  \nonumber \\
&&(\frac{1}{m^{2}+(\widetilde{\vec{\kappa}}\cdot \hat{\eta})^{2}}+\frac{1}{%
\sqrt{m^{2}+{\widetilde{\vec{\kappa}}}^{2}}\sqrt{m^{2}+(\widetilde{\vec{%
\kappa}}\cdot \hat{\eta})^{2}}})\Big)\frac{1}{2\tilde{\eta}}+  \nonumber \\
&&+\frac{\widetilde{{\bf \vec{k}}}_{1}\cdot \widetilde{{\bf \vec{k}}}_{2}}{%
4(m^{2}+{\widetilde{\vec{\kappa}}}{}^{2})}\Big((\sqrt{m^{2}+{\widetilde{\vec{%
\kappa}}}^{2}}-\sqrt{m^{2}+(\widetilde{\vec{\kappa}}\cdot \hat{\eta})^{2}}%
)^{2}\sqrt{{\frac{m^{2}+(\widetilde{\vec{\kappa}}\cdot \hat{\eta})^{2}}{%
m^{2}+{\widetilde{\vec{\kappa}}}^{2}}}})  \nonumber \\
&&(\frac{1}{m^{2}+(\widetilde{\vec{\kappa}}\cdot \hat{\eta})^{2}}-\frac{1}{%
m^{2}+{\widetilde{\vec{\kappa}}}^{2}})\frac{1}{\tilde{\eta}}-  \nonumber \\
&&-4(\sqrt{m^{2}+{\widetilde{\vec{\kappa}}}^{2}}-\sqrt{m^{2}+(\widetilde{%
\vec{\kappa}}\cdot \hat{\eta})^{2}})(\frac{1}{\sqrt{m^{2}+{\widetilde{\vec{%
\kappa}}}^{2}}}-\frac{1}{\sqrt{m^{2}+(\widetilde{\vec{\kappa}}\cdot \hat{\eta%
})^{2}}})\Big)-  \nonumber \\
&&-\Big(\widetilde{{\bf \vec{k}}}_{1}\cdot \widetilde{{\bf \vec{k}}}_{2}\Big(%
\frac{1}{2(\sqrt{m_{2}^{2}+{\widetilde{\vec{\kappa}}}^{2}}+\sqrt{m_{2}^{2}+(%
\widetilde{\vec{\kappa}}\cdot \hat{\eta})^{2}})}  \nonumber \\
&&(2\sqrt{m^{2}+{\widetilde{\vec{\kappa}}}^{2}}-\sqrt{m^{2}+(\widetilde{\vec{%
\kappa}}\cdot \hat{\eta})^{2}}-\frac{m^{2}+{\widetilde{\vec{\kappa}}}^{2}}{%
\sqrt{m^{2}+(\widetilde{\vec{\kappa}}\cdot \hat{\eta})^{2}}})\Big)-
\nonumber \\
&&-\Big(\widetilde{{\bf \vec{k}}}_{1}\cdot {\widetilde{\vec{\kappa}}}%
\widetilde{{\bf \vec{k}}}_{2}\cdot \widetilde{\vec{\kappa}}-\widetilde{{\bf
\vec{k}}}_{1}\cdot (\widetilde{{\bf \vec{k}}}_{2}\cdot (\widetilde{\vec{%
\kappa}}{\hat{\eta}}+{\hat{\eta}}\widetilde{\vec{\kappa}}))(\widetilde{\vec{%
\kappa}}\cdot {\hat{\eta}})-\widetilde{{\bf \vec{k}}}_{1}\widetilde{{\bf
\vec{k}}}_{2}\cdot \cdot {\hat{\eta}\hat{\eta}}\widetilde{\vec{\kappa}}^{2}%
\Big)  \nonumber \\
&&(\frac{1}{(\sqrt{m^{2}+{\widetilde{\vec{\kappa}}}^{2}}+\sqrt{m^{2}+(%
\widetilde{\vec{\kappa}}\cdot \hat{\eta})^{2}})^{2}})  \nonumber \\
&&\Big(3\sqrt{m^{2}+{\widetilde{\vec{\kappa}}}^{2}}-\frac{m^{2}+{\widetilde{%
\vec{\kappa}}}^{2}}{\sqrt{m^{2}+(\widetilde{\vec{\kappa}}\cdot \hat{\eta}%
)^{2}}}-\frac{(\sqrt{m^{2}+{\widetilde{\vec{\kappa}}}^{2}})^{3}}{2(m^{2}+(%
\widetilde{\vec{\kappa}}\cdot \hat{\eta})^{2})}\Big)+  \nonumber \\
&&+\widetilde{{\bf \vec{k}}}_{1}\cdot (\widetilde{{\bf \vec{k}}}_{2}\cdot {%
\hat{\eta}\hat{\eta})}(\widetilde{\vec{\kappa}}\cdot {\hat{\eta}})^{2}[\frac{%
\sqrt{m^{2}+{\widetilde{\vec{\kappa}}}{}^{2}}}{(\sqrt{m^{2}+{\widetilde{\vec{%
\kappa}}}^{2}}+\sqrt{m^{2}+(\widetilde{\vec{\kappa}}\cdot \hat{\eta})^{2}}%
)^{2}}  \nonumber \\
&&(\frac{2}{\sqrt{m^{2}+{\widetilde{\vec{\kappa}}}^{2}}}-\frac{1}{\sqrt{%
m^{2}+(\widetilde{\vec{\kappa}}\cdot \hat{\eta})^{2}}}+\frac{\sqrt{m^{2}+(%
\widetilde{\vec{\kappa}}\cdot \hat{\eta})^{2}}}{2(m^{2}+{\widetilde{\vec{%
\kappa}}}^{2})})]\Big)\frac{1}{(m^{2}+{\widetilde{\vec{\kappa}}}^{2})\tilde{%
\eta}}\Big].  \label{VIII6}
\end{eqnarray}

3) For the pseudo-classical version of {\it hydrogen-like atoms} ($%
m_{2}\rightarrow \infty )$ the active part of the rest energy is just

\begin{equation}
M=\sqrt{m_{1}^{2}+\widetilde{\vec{\kappa}}^{2}}+\frac{Q_{1}Q_{2}}{4\pi
\widetilde{\eta }}+V_{LDS},  \label{VIII7}
\end{equation}

\noindent since $V_{HDS}\rightarrow 0$ in this limit. \ In this case $%
V_{LDS} $ simplifies dramatically to just

\begin{equation}
V_{LDS}=\frac{Q_{1}Q_{2}}{4\pi }{\large [}-i\frac{\widetilde{\vec{\kappa}}%
\cdot \vec{\xi}_{1}\vec{\xi}_{1}\cdot \vec{\eta}}{\tilde{\eta}^{3}(m_1+\sqrt{%
m_{1}^{2}+\widetilde{\vec{\kappa}}^{2}})\sqrt{m_{1}^{2}+\widetilde{\vec{%
\kappa}}^{2}}}].  \label{VIII8}
\end{equation}

In the remaining part of this section we check the lowest order portion for
comparison with the known {\it Breit results}. At lowest order ($O(1/c^{2})$%
) Eq.( \ref{VII12})

\begin{eqnarray}
V_{LDS}(\tau ) &=&\sum_{i\neq j}\frac{Q_{i}Q_{j}}{4\pi }\Big[-\frac{\vec{%
\kappa}_{i}\cdot \vec{\kappa}_{j}+\vec{\kappa}_{i}\cdot {\hat{\eta}}_{ij}%
\vec{\kappa}_{j}\cdot {\hat{\eta}}_{ij}}{4m_{i}m_{j}}\frac{1}{\eta _{ij}}-i%
\frac{\vec{\kappa}_{i}\cdot \vec{\xi}_{i}\vec{\xi}_{1}\cdot \vec{\eta}_{ij}}{%
2\eta _{ij}^{3}m_{i}^{2}} -  \nonumber \\
&&-i\vec{\xi}_{i}\times \vec{\xi}_{i}\cdot ({\vec{\kappa}}_{j}\times \vec{%
\partial}_{ij})\frac{1}{2\eta _{ij}m_{i}m_{j}}+\frac{{\bf {\vec{f}}_{i}\cdot
{\vec{f}}_{j}}}{2m_{i}m_{j}}\frac{1}{\eta _{ij}}+  \nonumber \\
&+&\Big(\vec{\kappa}_{i}\cdot {\bf {\vec{f}}}_{j}+\vec{\kappa}_{j}\cdot \hat{%
\vec{f}}_{i}+{\bf {\vec{f}}}_{i}\cdot {\bf {\vec{f}}}_{j}-({\bf {\vec{f}}}%
_{j}\vec{\kappa}_{i}\cdot {\hat{\eta}}_{ij}+{\bf {\vec{f}}}_{i}\vec{\kappa}%
_{j}\cdot {\hat{\eta}}_{ij}+{\bf {\vec{f}}}_{j}{\bf {\vec{f}}}_{i}\cdot {%
\hat{\eta}}_{ij})\cdot {\hat{\eta}}_{ij}\Big)[\frac{-1}{2m_{i}m_{j}\eta _{ij}%
}] +  \nonumber \\
&&+\Big(\vec{\kappa}_{i}\cdot {\bf {\vec{g}}}_{j}-\vec{\kappa}_{i}\cdot (%
\hat{\vec{g}}_{j}\cdot ({\hat{\eta}}_{ij}{\hat{\eta}}_{ij}))\Big)\frac{1}{%
\eta _{ij}m_{i}m_{j}}\Big].  \label{VIII9}
\end{eqnarray}

To lowest order ${\bf {\vec{f}}}_{i}=-i\vec{\xi}_{i}\vec{\xi}_{i}\cdot \vec{%
\partial}_{ij}=i\vec{\xi}_{i}\cdot \vec{\partial}_{ij}\vec{\xi}_{i},\,\,\hat{%
\vec{g}}_{i}=0$ so

\begin{eqnarray}
V_{LDS}(\tau ) &=&\sum_{i\neq j}\frac{Q_{i}Q_{j}}{4\pi }\Big[-\frac{\vec{%
\kappa}_{i}\cdot \vec{\kappa}_{j}+\vec{\kappa}_{i}\cdot {\hat{\eta}}_{ij}%
\vec{\kappa}_{j}\cdot {\hat{\eta}}_{ij}}{4m_{i}m_{j}}\frac{1}{\eta _{ij}}-i%
\frac{\vec{\kappa}_{i}\cdot \vec{\xi}_{i}\vec{\xi}_{i}\cdot \vec{\eta}_{ij}}{%
2\eta _{ij}^{3}m_{i}^{2}}-  \nonumber \\
&&-i\vec{\xi}_{i}\times \vec{\xi}_{i}\cdot ({\vec{\kappa}}_{j}\times \vec{%
\partial}_{ij})\frac{1}{2\eta _{ij}m_{i}m_{j}}+\vec{\xi}_{i}\times \vec{\xi}%
_{i}\cdot (\vec{\xi}_{2}\cdot \vec{\partial}_{ij}\vec{\xi}_{j}\times \vec{%
\partial}_{ij})\frac{1}{2\eta _{ij}m_{i}m_{j}}+  \nonumber \\
&&+\frac{(\vec{\xi}_{i}\cdot \vec{\partial}_{ij})(\vec{\xi}_{i}\cdot \vec{\xi%
}_{j})(\vec{\xi}_{2}\cdot \vec{\partial}_{ij})}{2m_{i}m_{j}}\frac{1}{\eta
_{ij}} +  \nonumber \\
&&+[-2i\vec{\kappa}_{i}\cdot \vec{\xi}_{2}\vec{\xi}_{2}\cdot \vec{\partial}%
_{ij}+(\vec{\xi}_{i}\cdot \vec{\partial}_{ij})(\vec{\xi}_{i}\cdot \vec{\xi}%
_{j})\vec{\xi}_{2}\cdot \vec{\partial}_{ij} -  \nonumber \\
&&-(2i\vec{\xi}_{2}\cdot \vec{\partial}_{ij}\vec{\xi}_{2}\cdot {\hat{\eta}}%
_{ij}\vec{\kappa}_{i}\cdot {\hat{\eta}}_{ij}+\vec{\xi}_{2}\cdot \vec{\partial%
}_{ij}\vec{\xi}_{2}\cdot {\hat{\eta}}_{ij}\vec{\xi}_{i}\cdot \vec{\partial}%
_{ij}\vec{\xi}_{i}\cdot {\hat{\eta}}_{ij})][\frac{-1}{2m_{i}m_{j}\eta _{ij}}]%
\Big].  \label{VIII10}
\end{eqnarray}

Including the derivative terms and canceling like terms we obtain our lowest
order pseudo-classical expression, giving the semi-relativistic (order $%
1/c^{2}$) potential energy terms for the pseudo-classical analogue of {\it %
hydrogen-like systems} (either mass $\rightarrow \infty $), {\it %
positronium-like systems} ($m_{1}=m_{2}$), or general unequal mass systems (%
{\it muonium -like}):

\begin{eqnarray}
V_{LDS}(\tau ) &=&\sum_{i\neq j}\frac{Q_{i}Q_{j}}{4\pi }\Big[-\frac{\vec{%
\kappa}_{i}\cdot \vec{\kappa}_{j}+\vec{\kappa}_{i}\cdot {\hat{\eta}}_{ij}%
\vec{\kappa}_{j}\cdot {\hat{\eta}}_{ij}}{4m_{i}m_{j}}\frac{1}{\eta _{ij}}-i%
\frac{\vec{\kappa}_{i}\cdot \vec{\xi}_{i}\vec{\xi}_{i}\cdot \vec{\eta}_{ij}}{%
2\eta _{ij}^{3}m_{i}^{2}}-  \nonumber \\
&&-i\frac{\vec{\xi}_{i}\times \vec{\xi}_{i}\cdot (\vec{\eta}_{ij}\times {%
\vec{\kappa}}_{j})}{2\eta _{ij}^{3}m_{i}m_{j}}+\frac{\vec{\xi}_{i}\times
\vec{\xi}_{i}\cdot \vec{\xi}_{2}\times \vec{\xi}_{2}}{2m_{i}m_{j}}(\frac{1}{%
\eta _{ij}^{3}}+\frac{4\pi \delta (\vec{\eta}_{ij})}{3})-  \nonumber \\
&&-\frac{3\vec{\xi}_{i}\times \vec{\xi}_{i}\cdot \vec{\xi}_{2}\times \hat{%
\eta}_{ij}(\vec{\xi}_{j}\cdot \hat{\eta}_{ij})}{2\eta _{ij}^{3}m_{i}m_{j}}%
\Big].  \label{VIII11}
\end{eqnarray}

\subsection{Quantization of the Lowest Order Potential.}

Next we examine the quantum version of this interaction for comparison with
the standard results of the {\it reduction of the Bethe-Salpeter equation}.
\ In the manifestly Lorentz covariant formulation with the two signs of the
energy the 5 Grassmann variables $\xi ^{\mu }$and $\xi _{5}$ are quantized
to $\gamma _{5}\gamma ^{\mu }/\sqrt{2}$and $\gamma _{5}/\sqrt{2},$so that
the Dirac constraint $p_{\mu }\xi ^{\mu }-m\xi _{5}\approx 0$ becomes the
wave operator $\gamma _{5}[p_{\mu }\gamma ^{\mu }-m]$. But now with the
positive-energy spinning particles we have only 3 Grassmann variables $\xi
^{r}$ and no spinor wave equation (the spinors satisfy only the Klein-Gordon
equation ). Therefore we have a different quantization (corresponding to the
pseudo-classical spin $\vec{S}=(-i/2)\vec{\xi}\times \vec{\xi}$ going into $%
\vec{\sigma}/2$) holding only on the positive-energy branch of the mass
spectrum. The quantization is of Eq.(\ref{VI64}), $\{\xi _{i}^{h},\xi
_{i}^{k}\}=-i\delta ^{hk}$, is

\begin{equation}
\xi ^{r}\mapsto \frac{\sigma _{r}}{\sqrt{2}},  \label{VIII12}
\end{equation}

\noindent and is compatible with the Pauli matrix algebra of

\begin{equation}
\sigma ^{h}\sigma ^{k}+\sigma ^{k}\sigma ^{h}=2\delta ^{hk}.  \label{VIII13}
\end{equation}

With this conversion rules and with the algebra of the Pauli matrices we
obtain

\begin{eqnarray}
\vec{A}\cdot \vec{\xi}_{i}\vec{B}\cdot \vec{\xi}_{i} &=&\frac{1}{2}\vec{A}
\cdot \vec{B}+\frac{i}{2}\vec{A}\times \vec{B}\cdot \vec{\sigma},  \nonumber
\\
\vec{\xi}_{i}\times \vec{\xi}_{i} &=&+i\vec{\sigma}_{i},  \nonumber \\
(\vec{A}\cdot \vec{\xi}_{i})(\vec{B}\times \vec{\xi}_{i}) &=&-\frac{1}{2}(%
\vec{A}\times \vec{B})-\frac{i}{2}\vec{A}(\vec{B}\cdot \vec{\sigma}_{i})+%
\frac{i}{2}(\vec{A}\cdot \vec{B})\vec{\sigma}_{i},  \label{VIII14}
\end{eqnarray}

\noindent giving

\begin{eqnarray}
V_{LDS}(\tau ) &=&\sum_{i\neq j}\frac{Q_{i}Q_{j}}{4\pi }\Big[\Big(-\,\frac{%
\vec{\kappa}_{i}\cdot \vec{\kappa}_{j}+\vec{\kappa}_{i}\cdot {\hat{\eta}}%
_{ij}\vec{\kappa}_{j}\cdot {\hat{\eta}}_{ij}}{4m_{i}m_{j}\eta _{ij}}-i\frac{%
\vec{\kappa}_{i}\cdot \vec{\eta}_{ij}}{4\eta _{ij}^{3}m_{i}^{2}}\Big)%
_{ordered} -  \nonumber \\
&&-\frac{\vec{\eta}_{ij}\times \vec{\kappa}_{i}\cdot \vec{\sigma}_{i}}{4\eta
_{ij}^{3}m_{i}^{2}}+\frac{\vec{\eta}_{ij}\times \vec{\kappa}_{j}\cdot \vec{%
\sigma}_{i}}{2\eta _{ij}^{3}m_{i}m_{j}} -  \nonumber \\
&&-\frac{\vec{\sigma}_{i}\cdot \vec{\sigma}_{j}}{2m_{i}m_{j}}(\frac{1}{\eta
_{ij}^{3}}+\frac{4\pi \delta (\vec{\eta}_{ij})}{3})+\frac{3(\vec{\sigma}%
_{i}\cdot \vec{\sigma}_{j}-\vec{\sigma}_{i}\cdot \hat{\eta}_{ij}\vec{\sigma}%
_{j}\cdot \hat{\eta}_{ij})}{4m_{i}m_{j}\eta _{ij}^{3}}\Big].  \label{VIII15}
\end{eqnarray}

\noindent

\noindent\

For two particles in the CM system the above reduces to

\begin{eqnarray}
V_{LDS}(\tau ) &=&\frac{Q_{1}Q_{2}}{4\pi }\Big[\Big(\frac{\vec{\kappa}^{2}+(%
\vec{\kappa}\cdot {\hat{\eta}})^{2}}{2m_{1}m_{2}\eta }\frac{1}{\eta }-i\frac{%
\vec{\kappa}\cdot \vec{\eta}}{4\eta ^{3}m_{1}^{2}}-i\frac{\vec{\kappa}\cdot
\vec{\eta}}{4\eta ^{3}m_{2}^{2}}\Big)_{ordered}+  \nonumber \\
&-&\frac{\vec{L}\cdot \vec{\sigma}_{1}}{4\eta ^{3}m_{1}^{2}}-\frac{\vec{L}%
\cdot \vec{\sigma}_{2}}{4\eta ^{3}m_{2}^{2}}-\frac{\vec{L}\cdot (\vec{\sigma}%
_{1}+\vec{\sigma}_{2})}{2\eta ^{3}m_{1}m_{2}}+  \nonumber \\
&+&(-\frac{2\pi }{3}\delta (\vec{\eta})\vec{\sigma}_{1}\cdot \vec{\sigma}%
_{2}+\frac{1}{4}\frac{\vec{\sigma}_{1}\cdot \vec{\sigma}_{2}}{\eta ^{3}}-%
\frac{3}{4}\frac{\vec{\sigma}_{1}\cdot \hat{\eta}\vec{\sigma}_{2}\cdot \hat{%
\eta}}{\eta ^{3}})\frac{1}{m_{1}m_{2}}\Big].  \label{VIII16}
\end{eqnarray}
These terms, derived from the order $1/c^{2}$ \ relativistic corrections
beyond Coulomb potential, are valid for hydrogen-like systems (either mass $%
\rightarrow \infty $), equal mass (positronium-like systems) or general
unequal mass (muonium-like systems). This expression is the same as derived
from the Bethe-Salpeter equation and Breit equation (see Refs.\cite
{bethe,das,breit,stroscio} ) and by quantization of Wheeler-Feynman dynamics
for particles with pseudo-classical spin \cite{alstine}.

\vfill\eject

\section{Conclusions.}

In this paper we have shown that the effective lowest order Hamiltonian used
for the theory of relativistic bound states with spin 1/2 constituents of
arbitrary mass (muonium-, hydrogen- and positronium-like systems) may be
derived as a {\it semi-classical} result implied by the Lienard-Wiechert
solution in presence of positive-energy spinning particles in the rest-frame
instant form of dynamics.

The result is non trivial in two respects:

1) At least to us it was not evident that the semi-classical treatment of
the delay with Grassmann-valued electric charges with $Q_{i}^{2}=0$ and $%
Q_{i}Q_{j}\not=0$ (regularizing the Coulomb energy and producing a unique
semi-classical Lienard-Wiechert solution) could produce {exactly} a result
derived from quantum field theory through the reduction of the
Bethe-Salpeter equation.

2) It confirms the validity of the concept of spinning particle, with its
semi-classical description of spin with Grassmann variables, as a
semi-classical simulation of fermions and the relevance of the
Foldy-Wouthuysen transformation in determining the effective
action-at-a-distance interparticle potential.

Moreover, we obtain the full relativistic structure of the semi-classical
approximation without any $1/c^2$ expansions. This could help in evaluating
relativistic recoil effects.

We have already a semi-classical formulation for scalar quarks in terms of
the Dirac observables in the non-Abelian radiation gauge in the rest-frame
instant form of dynamics\cite{lu4}. This has been obtained in the framework
of weighted Sobolev spaces, where there is no Gribov ambiguity, in the case
of a trivial SU(3)-principal bundle over each spacelike hypersurface. Having
solved the Gauss laws, there is already a {\it $1/r$ Yang-Mills
transverse-potential-dependent potential} in the final Hamiltonian. It acts
between any pair of color charge densities, $\rho_a(\tau ,{\vec \sigma}_1 )$
and $\rho_b(\tau ,{\vec \sigma}_2)$, [either quarks or localized charged
Yang-Mills field (classical background of gluons)] and consists of three
terms:

i) a Coulomb potential between $\rho_a(\tau ,{\vec \sigma}_1 )$ and $%
\rho_b(\tau ,{\vec \sigma}_2)$;

ii) a potential in which $\rho _{a}(\tau ,{\vec{\sigma}}_{1})$ has a Coulomb
interaction with an arbitrary {\it color-carrying center} located at ${\vec{%
\sigma}}_{3}$, while $\rho _{b}(\tau ,{\vec{\sigma}}_{2})$ interacts with
the same center through a {\it Wilson line along the flat geodesics} ${\vec{%
\sigma}}_{2}-{\vec{\sigma}}_{3}$ (plus the case $1\leftrightarrow2$): then
one integrates over the location of the center;

iii) an interaction with {\it two centers} ${\vec \sigma}_3$ and ${\vec %
\sigma}_4$ (over whose location one integrates): $\rho_a(\tau,{\vec \sigma}%
_1)$ interacts with ${\vec \sigma}_3$ through a Wilson line, the same
happens between $\rho_b(\tau ,{\vec \sigma}_2)$ and ${\vec \sigma}_4$, and,
finally, the two centers have a mutual Coulomb interaction.

As shown in Ref.\cite{lu4} in the case of mesons (quark-antiquark system)
the {\it semiclassical approximation of Grassmann-valued color charges}
regularizes the potential and produces a {\it semiclassical asymptotic
freedom}. However, due to the presence of the Wilson lines, it is not
possible to check whether there is confinement. Neither baryons (3-quark
system) nor the introduction of the spin along the lines of this paper have
been studied till now.

The challenge now is to see whether with the technology of this paper it is
possible to treat the non-Abelian case of the quark model, by using a
perturbative treatment based on the iterative Lienard-Wiechert-type
solutions developed in Ref.\cite{drec} adapted to this non-Abelian radiation
gauge. Namely, whether it is possible to get a non-Abelian Darwin potential
and to check if it is a confining potential.

\vfill\eject

\appendix

\section{Explicit Forms of the F and G Functions of Section II.}

To find the functions $F=F_{2}-F_{1}$, $G=G_{2}-G_{1}$ of Eqs.(\ref{II18})
we must preliminarly determine the functions $\Phi _{n}$, $\Psi _{n}$ of
Eqs. (\ref{II12}), (\ref{II15}). In order to calculate $\Phi _{n}$ and $\Psi
_{n}$ from the definitions (\ref{II12}),(\ref{II15}), we observe that, since
Eqs.(\ref{II6}), (\ref{II7}) imply

\begin{eqnarray}
S &=&S_{free}+{\cal O}(Q),  \nonumber \\
S_{free} &=&2i(\vec{p}\cdot \vec{\xi})\xi _{5}\theta (p),  \label{a1}
\end{eqnarray}

\noindent then, Eq.(\ref{II12}) implies that for every $n$ we have

\begin{eqnarray}
2i\xi _{0}\Phi _{2n}=\{\Phi _{2n-1}\xi _{5}\xi _{0},S_{free}\} &\Rightarrow
&\Phi _{2n}=-i(\vec{p}\cdot \vec{\xi})\theta (p)\Phi _{2n-1},  \nonumber \\
2i\xi _{5}\xi _{0}\Phi _{2n+1}=\{\Phi _{2n}\xi _{0},S_{free}\} &\Rightarrow
&\Phi _{2n+1}=-\{\Phi _{2n},(\vec{p}\cdot \vec{\xi})\theta (p)\} .
\label{a2}
\end{eqnarray}

\noindent Analogously for $n\geq 2$ Eq.(\ref{II15}) implies

\begin{eqnarray}
2i\xi _{0}\Psi _{2n}=\{\Psi _{2n-1}\xi _{5}\xi _{0},S_{free}\} &\Rightarrow
&\Psi _{2n}=-i(\vec{p}\cdot \vec{\xi})\theta (p)\Psi _{2n-1},  \nonumber \\
2i\xi _{5}\xi _{0}\Psi _{2n+1}=\{\Psi _{2n}\xi _{0},S_{free}\} &\Rightarrow
&\Psi _{2n+1}=-\{\Psi _{2n},(\vec{p}\cdot \vec{\xi})\theta (p)\} .
\label{a3}
\end{eqnarray}

From Eqs.(\ref{II13}), (\ref{II15}) we get at order $Q$ [$\{QA_{0}\xi
_{0},S\}=2iQ\Phi _{1}\xi _{5}\xi _{0}$, $\{p_{0}\xi _{0},S\}=2iQ\Psi _{1}\xi
_{5}\xi _{0}$]

\begin{eqnarray}
\Phi _{1} &=&\theta (p)\xi ^{k}\frac{\partial A_{0}}{\partial x^{k}}+(\vec{p}
\cdot \vec{\xi})\frac{\partial \theta (p)}{\partial p^{u}}\frac{\partial
A_{0}}{\partial x^{u}},  \nonumber \\
&&  \nonumber \\
\Psi _{1} &=&-\theta (p)\vec{\xi}\cdot \frac{\partial \vec{A}}{\partial x^o}%
-2 \frac{d\theta (s)}{ds}{|}_{s=p^2}(\vec{p}\cdot \vec{\xi})\vec{p}\cdot
\frac{\partial \vec{A}}{\partial x^o}+\frac{d\theta (s)}{ds}{|}_{s=p^2}(\vec{%
p}\cdot \vec{\xi})\left( -i\xi ^{h}\xi ^{k}\frac{\partial F_{hk}}{\partial
x^o}\right).  \label{a4}
\end{eqnarray}

Then, from $\{ \{ Q A_o \xi_o, S \} , S \} =-4Q \Phi_2 \xi_5\xi_o$, $\{ \{
p_o\xi_o ,S \} , S \} =-4 Q \Psi_2 \xi_5\xi_o$ we obtain

\begin{eqnarray}
\Phi _{2} &=&-i\theta ^{2}(p)(\vec{p}\cdot \vec{\xi})\vec{\xi}\cdot \vec %
\partial A_{0}\equiv f_{2,uv}\xi ^{u}\xi ^{v},  \nonumber \\
&&  \nonumber \\
\Psi _{2} &=&+i\theta ^{2}(p)(\vec{p}\cdot \vec{\xi})\vec{\xi}\cdot \frac{
\partial \vec{A}}{\partial x^o}\equiv g_{2,uv}\xi ^{u}\xi ^{v},  \label{a5}
\end{eqnarray}

\noindent where

\begin{eqnarray}
f_{2,uv} &:=&-i\theta ^{2}(p)p^{u}\frac{\partial A_{0}}{\partial x^{v}},
\nonumber \\
&&  \nonumber \\
g_{2,uv} &:=&+i\theta^{2}(p)p^{u} \frac{\partial A^{v}}{\partial x^o}.
\label{a6}
\end{eqnarray}

\subsection{Sum of Even Terms (Diagonal Terms).}

We prove the following lemma:

{\bf Lemma 1 }\newtheorem{lemma}{Lemma}
\begin{lemma}
For $n\geq 1$:
\begin{equation}
\Phi_{2n}=T_{(2n)}(p)(\vec{p}\cdot\xi)\vec{\xi}\cdot \vec \partial A_0(x),
\label{a7}
\end{equation}

\noindent with

\begin{equation}
T_{(2n)}(p)=-i\frac{[p^2\theta^2(p)]^n}{p^2}.
  \label{a8}
\end{equation}
\end{lemma}

{\bf Proof}: We first prove that, for $n\geq 1$ we have:

\begin{equation}
\Phi _{2n}=f_{2n,uv}\xi ^{u}\xi ^{v}.  \label{a9}
\end{equation}

This equation is true for $\Phi _{2}$; by assuming that it is true for $%
\Phi_{2n-2}$, then by using Eq.(\ref{a2}) we obtain:

\begin{eqnarray}
\Phi _{2n-1} &=&-\theta (p)p^{k}f_{2n-2,uv}\{\xi ^{u}\xi ^{v},\xi
^{k}\}-\{f_{2n-2,uv},p^{k}\theta (p)\}\xi ^{u}\xi ^{v}\xi ^{k}=  \nonumber \\
&=&-i\theta (p)\left[ p^{u}f_{2n-2,uv}\xi ^{v}-\xi ^{u}f_{2n-2,uv}p^{v} %
\right] +h_{uvk}\xi ^{u}\xi ^{v}\xi ^{k},  \label{a10}
\end{eqnarray}

\noindent where

\begin{equation}
h_{uvk}=\{f_{2n-2,uv},p^{k}\theta (p)\} .  \label{a11}
\end{equation}

Using again Eq.(\ref{a2}) and observing that $\xi ^{u}\xi ^{v}\xi
^{k}\xi^{h}\equiv 0$, we get

\begin{eqnarray}
\Phi _{2n} &=&-i\theta (p)(\vec{p}\cdot \vec{\xi})\Phi _{2n-1}=  \nonumber \\
&=&-\theta ^{2}(p)(\vec{p}\cdot \vec{\xi})\left[ p^{u}f_{2n-2,uv}\xi
^{v}-\xi ^{u}f_{2n-2,uv}p^{v}\right] .  \label{a12}
\end{eqnarray}

Thus Eq.(\ref{a9}) is true with:

\begin{equation}
f_{2n,uv}=-\theta ^{2}(p)p^{u}\left[ p^{k}f_{2n-2,kv}-f_{2n-2,vk}p^{k}\right]%
,  \label{a13}
\end{equation}

\noindent and thus by induction it is true for all $n\geq 1$.

To complete the proof, we observe that the lemma is true for $\Phi _{2}$
because:

\begin{equation}
f_{2,uv}=T_{(2)}(p)p^{u}\frac{\partial A_{0}}{\partial x^{v}},  \label{a14}
\end{equation}

\noindent with

\begin{equation}
T_{(2)}(p)=-i\theta ^{2}(p).  \label{a15}
\end{equation}

Again by induction, since we have

\begin{equation}
f_{2n-2,uv}=T_{(2n-2)}(p)p^{u}\frac{\partial A_{0}}{\partial x^{v}},
\label{a16}
\end{equation}

\noindent we can substitute this formula in Eq.(\ref{a9}) and, using Eq.(\ref
{a13} ), obtain

\begin{eqnarray}
\Phi _{2n} &=&-\theta ^{2}(p)p^{u}\left[ p^{k}f_{2n-2,kv}-f_{2n-2,vk}p^{k}%
\right] \xi ^{u}\xi ^{v}=  \nonumber \\
&=&\theta ^{2}(p)p^{2}T_{(2n-2)}(p)(\vec{p}\cdot \xi )\vec{\xi}\cdot \vec %
\partial A_{0}(x).  \label{a17}
\end{eqnarray}

Finally, by using Eq.(\ref{a15}) we get

\begin{equation}
T_{(2n)}(p)=\theta ^{2}(p)p^{2}T_{(2n-2)}(p)=-i\frac{[p^{2}\theta
^{2}(p)]^{n}}{p^{2}}.  \label{a18}
\end{equation}

{\bf Q.E.D.}

In the same way we can to prove the following lemma:

{\bf Lemma 2 }%
\begin{lemma}
For $n\ge 1$:
\begin{equation}
\Psi_{2n}=-T_{(2n)}(p)(\vec{p}\cdot\xi)\vec{\xi}\cdot \frac{\partial\vec{A}%
(x)}{\partial x^o},
  \label{a19}
\end{equation}

\noindent with:

\begin{equation}
T_{(2n)}(p)=-i\frac{[p^2\theta^2(p)]^n}{p^2}.
  \label{a20}
\end{equation}
\end{lemma}

The proof is the same of lemma 1, with $\Phi_{2n}$ replaced by ($n\geq 1$)

\begin{equation}
\Psi _{2n}=g_{2n,uv}\xi ^{u}\xi ^{v},  \label{a21}
\end{equation}

\noindent and $f_{2n,uv}$ by

\begin{equation}
g_{2n,uv}=-T_{(2n)}(p)p^{u}\frac{\partial A^{v}(x)}{\partial x^o}.
\label{a22}
\end{equation}

Eq.(\ref{a10}) is replaced by

\begin{equation}
\Psi _{2n-1}=-i\theta (p)\left[ p^{u}g_{2n-2,uv}\xi ^{v}-\xi
^{u}g_{2n-2,uv}p^{v}\right] +k_{uvk}\xi ^{u}\xi ^{v}\xi ^{k},  \label{a23}
\end{equation}

\noindent with

\begin{equation}
k_{uvk}=\{g_{2n-2,uv},p^{k}\theta (p)\} .  \label{a24}
\end{equation}

Using the results of the two lemmas we find:

\begin{equation}
\Psi _{2n}-\Phi _{2n}=-T_{(2n)}(p)(\vec{p}\cdot \xi )\vec{\xi}\cdot \left[
\frac{\partial \vec{A}(x)}{\partial x^{0}}+\vec{\partial}A_{0}(x)\right]
=+T_{(2n)}(p)(\vec{p}\cdot \xi )\vec{\xi}\cdot \vec{E}(x).  \label{a25}
\end{equation}

Then, we have the explicit form of the $F$ function given in Eqs.(\ref{II14},%
\ref{II17},\ref{II18})

\begin{eqnarray}
F &=&F_{2}-F_{1}=  \nonumber \\
&=&A_{0}(x)\xi _{0}+(\vec{p}\cdot \xi )\vec{\xi}\cdot \vec{E}%
(x)\sum_{k=1}^{\infty }\frac{(2i)^{2k}}{2k!}T_{(2k)}(p)=  \nonumber \\
&=&A_{0}(x)\xi _{0}-\frac{i}{p^{2}}(\vec{p}\cdot \xi )\vec{\xi}\cdot \vec{E}%
(x)\sum_{k=1}^{\infty }(-)^{k}\frac{2^{2k}p^{2k}\theta ^{2k}(p)}{2k!}=
\nonumber \\
&=&A_{0}(x)\xi _{0}-\frac{i}{p^{2}}(\vec{p}\cdot \xi )\vec{\xi}\cdot \vec{E}%
(x)\left[ \cos (2p\theta (p))-1\right] =  \nonumber \\
&=&A_{0}(x)\xi _{0}-i(\vec{p}\cdot \xi )(\vec{\xi}\cdot \vec{E}(x))\frac{m-%
\sqrt{m^{2}+{\vec{p}}^{2}}}{{\vec{p}}^{2}\sqrt{m^{2}+{\vec{p}}^{2}}}
\nonumber \\
&=&A_{0}(x)\xi _{0}-i(\vec{p}\cdot \xi )(\vec{\xi}\cdot \vec{E}(x))\frac{1}{%
(m+\sqrt{m^{2}+{\vec{p}}^{2}})\sqrt{m^{2}+{\vec{p}}^{2}}}.  \label{a26}
\end{eqnarray}

\subsection{Sum of Odd Terms (Skew-Diagonal Terms)}

We calculate first $G_1$. In Eq.(\ref{a10}) for $\Phi_{2n-1}$

\begin{equation}
\Phi_{2n-1}= -i\theta(p)\left[ p^uf_{2n-2,uv}\xi^v-\xi^uf_{2n-2,uv}p^v\right]
+ h_{uvk}\xi^u\xi^v\xi^k,  \label{a27}
\end{equation}

\noindent we can substitute the expression for $f_{2n,uv}$

\begin{equation}
f_{2n,uv}=T_{(2n)}(p)p^u\frac{\partial A_0}{\partial x^v}.  \label{a28}
\end{equation}

Since we have

\begin{equation}
h_{uvk}=\{f_{2n-2,uv},p^k\theta(p)\} = h_{ukv},  \label{a29}
\end{equation}

\noindent then, by symmetry considerations, we get

\begin{equation}
h_{uvk}\xi^u\xi^v\xi^k=0.  \label{a30}
\end{equation}

We have

\begin{eqnarray}
\Phi_{2n-1}&=&-i\theta(p)\left[ -i\frac{(p\theta(p))^{2n-2}}{p^2}\right]
\left( (\vec{p}\cdot\vec \partial A_0)(\vec{p}\cdot\vec{\xi})-p^2(\vec{\xi}
\cdot\vec \partial A_0) \right)=  \nonumber \\
&=& -\frac{(p\theta(p))^{2n-1})}{p^3} \left( (\vec{p}\cdot\vec \partial A_0)(%
\vec{p} \cdot\vec{\xi})-p^2(\vec{\xi}\cdot\vec \partial A_0) \right) .
\label{a31}
\end{eqnarray}

Then we obtain

\begin{eqnarray}
G_1&=&2i\phi_1+ \left( (\vec{p}\cdot\vec \partial A_0)(\vec{p}\cdot\vec{\xi}%
)-p^2( \vec{\xi}\cdot\vec \partial A_0) \right) \sum_{k=1}^{\infty}(-)\frac{%
i^{2k+1}}{2k+1! }\frac{(2p\theta(p))^{2n-1}}{p^3} =  \nonumber \\
&=&\frac{2ip\theta(p)}{p}(\vec{\xi}\cdot\vec \partial A_0)+ 2i(\vec{p}\cdot%
\vec{\xi })\frac{\partial\theta (p)}{\partial p^u}\frac{\partial A_0}{%
\partial x^u}+  \nonumber \\
&+&i(\vec{\xi}\cdot\vec \partial A_0) \sum_{k=1}^{\infty}\frac{(-)^k}{2k+1!}
\frac{ (2p\theta(p))^{2n-1}}{p}+  \nonumber \\
&-&i(\vec{p}\cdot\vec{\xi})(\vec{p}\cdot\vec \partial A_0)
\sum_{k=1}^{\infty}\frac{ (-)^k}{2k+1!} \frac{(2p\theta(p))^{2n-1}}{p^3}=
\nonumber \\
&=&i(\vec{\xi}\cdot\vec \partial A_0) \sum_{k=0}^{\infty}\frac{(-)^k}{2k+1!}
\frac{ (2p\theta(p))^{2n-1}}{p}+  \nonumber \\
&-&i(\vec{p}\cdot\vec{\xi})(\vec{p}\cdot\vec \partial A_0)
\sum_{k=0}^{\infty}\frac{ (-)^k}{2k+1!} \frac{(2p\theta(p))^{2n-1}}{p^3}+
\nonumber \\
&+&2i(\vec{p}\cdot\vec{\xi})\frac{\partial\theta (p)}{\partial p^u}\frac{%
\partial A_0}{\partial x^u}+ i\frac{2p\theta(p)}{p^3} (\vec{p}\cdot\vec{\xi}
)(\vec{p}\cdot\vec \partial A_0)=  \nonumber \\
&=& +i\frac{(\vec{\xi}\cdot\vec \partial A_0)}{p}\sin[2p\theta(p)] -i\frac{(%
\vec{p} \cdot\vec{\xi})(\vec{p}\cdot\vec \partial A_0)}{p^3} \sin[2p\theta(p)%
]+  \nonumber \\
&+& 2i(\vec{p}\cdot\vec{\xi})\frac{\partial\theta (p)}{\partial p^u}\frac{
\partial A_0}{\partial x^u}+ i\frac{2p\theta(p)}{p^3} (\vec{p}\cdot\vec{\xi}
)(\vec{p}\cdot\vec \partial A_0).  \label{a32}
\end{eqnarray}

Now we calculate $G_2$. We have:

\begin{equation}
\Psi_{2n-1}= -i\theta(p)\left[ p^ug_{2n-2,uv}\xi^v-\xi^ug_{2n-2,uv}p^v\right]
+k_{uvk}\xi^u\xi^v\xi^k,  \label{a33}
\end{equation}

\noindent and we can substitute the expression for $g_{2n,uv}$ into it

\begin{equation}
g_{2n,uv}=-T_{(2n)}(p)p^u\frac{\partial A^v}{\partial x^o}.  \label{a34}
\end{equation}

With

\begin{equation}
k_{uvk}=\{g_{2n-2,uv},p^k\theta(p)\} ,  \label{a35}
\end{equation}

\noindent we have in this case

\begin{eqnarray}
k_{uvk}\xi ^{u}\xi ^{v}\xi ^{k} &=&-T_{(2n-2)}(p)p^{u}\left\{ \frac{\partial
A^{v}}{\partial x^{0}},p^{k}\theta (p)\right\} \xi ^{u}\xi ^{v}\xi ^{k}=
\nonumber \\
&=&-T_{(2n-2)}(p)p^{u}\theta (p)\frac{\partial }{\partial x^{0}}\frac{%
\partial A^{v}}{\partial x^{k}}\xi ^{u}\xi ^{v}\xi ^{k}=  \nonumber \\
&=&\frac{-i}{2}\frac{(p\theta (p))^{2n-1}}{p^{3}}(\vec{p}\cdot \vec{\xi})%
\frac{\partial F_{hk}}{\partial x^{0}}\xi ^{h}\xi ^{k}.  \label{a36}
\end{eqnarray}

We get

\begin{eqnarray}
\Psi_{2n-1}&=&+i\theta(p)\left[ -i\frac{(p\theta(p))^{2n-2}}{p^2}\right]
\left( \vec{p}\cdot \frac{\partial\vec{A}}{\partial x^o} (\vec{p}\cdot\vec{%
\xi} )-p^2\vec{\xi}\cdot \frac{\partial\vec{A}}{\partial x^o} \right)+
\nonumber \\
&-&\frac{i}{2}\frac{(p\theta(p))^{2n-1}}{p^3}(\vec{p}\cdot\vec{\xi}) \frac{%
\partial F_{hk}}{\partial x^o}\xi^h\xi^k =  \nonumber \\
&=& +\frac{(p\theta(p))^{2n-1})}{p^3} \left( \vec{p}\cdot \frac{\partial\vec{
A}}{\partial x^o} (\vec{p}\cdot\vec{\xi})-p^2\vec{\xi}\cdot \frac{\partial%
\vec{ A}}{\partial x^o} \right)+  \nonumber \\
&-&\frac{i}{2}\frac{(p\theta(p))^{2n-1}}{p^3}(\vec{p}\cdot\vec{\xi}) \frac{
\partial F_{hk}}{\partial x^o}\xi^h\xi^k.  \label{a37}
\end{eqnarray}

Then we have:

\begin{eqnarray}
G_2&=&2i\Psi_1+  \nonumber \\
&+& \left[\left(\vec{p}\cdot \frac{\partial\vec{A}}{\partial x^o} (\vec{p}%
\cdot \vec{\xi})-p^2\vec{\xi}\cdot \frac{\partial\vec{A}}{\partial x^o}
\right)- \frac{i}{2}(\vec{p}\cdot\vec{\xi}) \frac{\partial F_{hk}}{\partial
x^o} \xi^h\xi^k \right] \sum_{k=1}^{\infty} i\frac{(-)^k}{p^3}\frac{
(2p\theta(p))^{2k+1}}{2k+1!}=  \nonumber \\
&=& -2i\theta(p)\vec{\xi}\cdot\frac{\partial\vec{A}}{\partial x^o}-4i\frac{
d\theta(s)}{ds}{|}_{s=p^2} (\vec{p}\cdot\vec{\xi})\vec{p}\cdot\frac{\partial%
\vec{A}} { \partial t}+\frac{d\theta(s)}{ds}{|}_{s=p^2} (\vec{p}\cdot\vec{\xi%
})\left( +2\xi^h\xi^k \frac{\partial F_{hk}}{\partial x^o}\right)+  \nonumber
\\
&-&\vec{\xi}\cdot \frac{\partial\vec{A}}{\partial x^o} \sum_{k=1}^{\infty} i
\frac{(-)^k}{p}\frac{(2p\theta(p))^{2k+1}}{2k+1!}+  \nonumber \\
&+& \left(\vec{p}\cdot \frac{\partial\vec{A}}{\partial x^o} (\vec{p}\cdot%
\vec{ \xi}) -\frac{i}{2}(\vec{p}\cdot\vec{\xi}) \frac{\partial F_{hk}}{%
\partial x^o} \xi^h\xi^k \right) \sum_{k=1}^{\infty} i\frac{(-)^k}{p^3}\frac{
(2p\theta(p))^{2k+1}}{2k+1!}  \nonumber \\
&=& -\vec{\xi}\cdot \frac{\partial\vec{A}}{\partial x^o} \sum_{k=0}^{\infty}
i \frac{(-)^k}{p}\frac{(2p\theta(p))^{2k+1}}{2k+1!}+  \nonumber \\
&+& \left(\vec{p}\cdot \frac{\partial\vec{A}}{\partial x^o} (\vec{p}\cdot%
\vec{ \xi}) -\frac{i}{2}(\vec{p}\cdot\vec{\xi}) \frac{\partial F_{hk}}{%
\partial x^o} \xi^h\xi^k \right) \sum_{k=0}^{\infty} i\frac{(-)^k}{p^3}\frac{
(2p\theta(p))^{2k+1}}{2k+1!}+  \nonumber \\
&-&i\frac{2p\theta(p)}{p}\left(\vec{p}\cdot \frac{\partial\vec{A}}{\partial
x^o } (\vec{p}\cdot\vec{\xi}) -\frac{i}{2}(\vec{p}\cdot\vec{\xi}) \frac{%
\partial F_{hk}}{\partial x^o}\xi^h\xi^k \right)+  \nonumber \\
&-&4i\frac{d\theta(s)}{ds}{|}_{s=p^2} (\vec{p}\cdot\vec{\xi})\vec{p}\cdot%
\frac{\partial \vec{A}} {\partial x^o}+\frac{d\theta(s)}{ds}{|}_{s=p^2} (%
\vec{p}\cdot\vec{\xi})\left( +2\xi^h\xi^k\frac{\partial F_{hk}}{\partial x^o}%
\right) =  \nonumber \\
&=&-\vec{\xi}\cdot \frac{\partial\vec{A}}{\partial x^o}\frac{i}{p}\sin[
2p\theta(p)] +  \nonumber \\
&+&\left(\vec{p}\cdot \frac{\partial\vec{A}}{\partial x^o} (\vec{p}\cdot\vec{
\xi}) -\frac{i}{2}(\vec{p}\cdot\vec{\xi}) \frac{\partial F_{hk}}{\partial x^o%
} \xi^h\xi^k \right) \frac{i}{p^3}\sin[2p\theta(p)]+  \nonumber \\
&-&i\frac{2p\theta(p)}{p}\left(\vec{p}\cdot \frac{\partial\vec{A}}{\partial
x^o } (\vec{p}\cdot\vec{\xi}) -\frac{i}{2}(\vec{p}\cdot\vec{\xi}) \frac{%
\partial F_{hk}}{\partial x^o}\xi^h\xi^k \right)+  \nonumber \\
&-&4i\frac{d\theta(s)}{ds}{|}_{s=p^2} (\vec{p}\cdot\vec{\xi})\vec{p}\cdot%
\frac{\partial \vec{A}} {\partial x^o}+\frac{d\theta(s)}{ds}{|}_{s=p^2} (%
\vec{p}\cdot\vec{\xi})\left( +2\xi^h\xi^k\frac{\partial F_{hk}}{\partial x^o}%
\right) .  \label{a38}
\end{eqnarray}

The complete expression for the function $G$ is

\begin{eqnarray}
G&=&G_2-G_1= -(\vec{\xi}\cdot \vec{E})\frac{i}{p}\sin[2p\theta(p)] +
\nonumber \\
&+&\left((\vec{p}\cdot \vec{E}) (\vec{p}\cdot\vec{\xi}) -\frac{i}{2}(\vec{p}
\cdot\vec{\xi}) \frac{\partial F_{hk}}{\partial x^o}\xi^h\xi^k \right) \frac{%
i }{p^3}\sin[2p\theta(p)]+  \nonumber \\
&+&i\frac{2p\theta(p)}{p}\left((\vec{p}\cdot \vec{E}) (\vec{p}\cdot\vec{\xi}
) -\frac{i}{2}(\vec{p}\cdot\vec{\xi}) \frac{\partial F_{hk}}{\partial x^o}%
\xi^h\xi^k \right)+  \nonumber \\
&-&4i\frac{d\theta(s)}{ds}{|}_{s=p^2} (\vec{p}\cdot\vec{\xi})(\vec{p}\cdot
\vec{E})+ \frac{d\theta(s)}{ds}{|}_{s=p^2} (\vec{p}\cdot\vec{\xi})\left(
+2\xi^h\xi^k\frac{ \partial F_{hk}}{\partial x^o}\right) .  \label{a39}
\end{eqnarray}

\vfill\eject

\section{Spacelike hypersurfaces.}

\subsection{Review of Their Properties.}

Let us first review some preliminary results from Refs.\cite{lu1,lu6} needed
in the description of physical systems on spacelike hypersurfaces.

Let $\lbrace \Sigma_{\tau}\rbrace$ be a one-parameter family of spacelike
hypersurfaces foliating Minkowski spacetime $M^4$ and giving a 3+1
decomposition of it. At fixed $\tau$, let $z^{\mu}(\tau ,\vec \sigma )$ be
the coordinates of the points on $\Sigma _{\tau }$ in $M^4$, $\lbrace \vec %
\sigma \rbrace$ a system of coordinates on $\Sigma_{\tau}$. If $\sigma^{%
\check A}=(\sigma^{\tau}=\tau ;\vec \sigma =\lbrace \sigma^{\check r%
}\rbrace) $ [the notation ${\check A}=(\tau , {\check r})$ with ${\check r}%
=1,2,3$ will be used; note that ${\check A}= \tau$ and ${\check A}={\check r}%
=1,2,3$ are Lorentz-scalar indices] and $\partial_{\check A}=\partial
/\partial \sigma^{\check A}$, one can define the vierbeins

\begin{equation}
z^{\mu}_{\check A}(\tau ,\vec \sigma )=\partial_{\check A}z^{\mu}(\tau ,\vec %
\sigma ),\quad\quad \partial_{\check B}z^{\mu}_{\check A}-\partial_{\check A%
}z^{\mu}_{\check B}=0,  \label{b1}
\end{equation}

\noindent so that the metric on $\Sigma_{\tau}$ is

\begin{eqnarray}
&&g_{{\check A}{\check B}}(\tau ,\vec \sigma )=z^{\mu}_{\check A}(\tau ,\vec %
\sigma )\eta_{\mu\nu}z^{\nu}_{\check B}(\tau ,\vec \sigma ),\quad\quad
g_{\tau\tau}(\tau ,\vec \sigma ) > 0,  \nonumber \\
&&g(\tau ,\vec \sigma )=-det\, ||\, g_{{\check A}{\check B}}(\tau ,\vec %
\sigma )\, || ={(det\, ||\, z^{\mu}_{\check A}(\tau ,\vec \sigma )\, ||)}^2,
\nonumber \\
&&\gamma (\tau ,\vec \sigma )=-det\, ||\, g_{{\check r}{\check s}}(\tau ,%
\vec \sigma )\, ||.  \label{b2}
\end{eqnarray}

If $\gamma^{{\check r}{\check s}}(\tau ,\vec \sigma )$ is the inverse of the
3-metric $g_{{\check r}{\check s}}(\tau ,\vec \sigma )$ [$\gamma^{{\check r}
{\check u}}(\tau ,\vec \sigma )g_{{\check u}{\check s}}(\tau ,\vec \sigma
)=\delta^{\check r}_{\check s}$], the inverse $g^{{\check A}{\check B}}
(\tau ,\vec \sigma )$ of $g_{{\check A}{\check B}}(\tau ,\vec \sigma )$ [$g^{%
{\check A}{\check C}}(\tau ,\vec \sigma )g_{{\check c}{\check b}}(\tau ,
\vec \sigma )=\delta^{\check A}_{\check B}$] is given by

\begin{eqnarray}
&&g^{\tau\tau}(\tau ,\vec \sigma )={\frac{{\gamma (\tau ,\vec \sigma )}}{{%
g(\tau ,\vec \sigma )}}},  \nonumber \\
&&g^{\tau {\check r}}(\tau ,\vec \sigma )=-[{\frac{{\gamma}}{g}} g_{\tau {%
\check u}}\gamma^{{\check u}{\check r}}](\tau ,\vec \sigma ),  \nonumber \\
&&g^{{\check r}{\check s}}(\tau ,\vec \sigma )=\gamma^{{\check r}{\check s}}
(\tau ,\vec \sigma )+[{\frac{{\gamma}}{g}}g_{\tau {\check u}}g_{\tau {\check %
v}} \gamma^{{\check u}{\check r}}\gamma^{{\check v}{\check s}}](\tau ,\vec %
\sigma ),  \label{b3}
\end{eqnarray}

\noindent so that $1=g^{\tau {\check C}}(\tau ,\vec \sigma )g_{{\check C}%
\tau} (\tau ,\vec \sigma )$ is equivalent to

\begin{equation}
{\frac{{g(\tau ,\vec \sigma )}}{{\gamma (\tau ,\vec \sigma )}}}=g_{\tau\tau}
(\tau ,\vec \sigma )-\gamma^{{\check r}{\check s}}(\tau ,\vec \sigma )
g_{\tau {\check r}}(\tau ,\vec \sigma )g_{\tau {\check s}}(\tau ,\vec \sigma
).  \label{b4}
\end{equation}

We have

\begin{equation}
z^{\mu}_{\tau}(\tau ,\vec \sigma )=(\sqrt{{\frac{g}{{\gamma}}} }l^{\mu}+
g_{\tau {\check r}}\gamma^{{\check r}{\check s}}z^{\mu}_{\check s})(\tau ,
\vec \sigma ),  \label{b5}
\end{equation}

\noindent and

\begin{eqnarray}
\eta^{\mu\nu}&=&z^{\mu}_{\check A}(\tau ,\vec \sigma )g^{{\check A}{\check B}%
} (\tau ,\vec \sigma )z^{\nu}_{\check B}(\tau ,\vec \sigma )=  \nonumber \\
&=&(l^{\mu}l^{\nu}+z^{\mu}_{\check r}\gamma^{{\check r}{\check s}} z^{\nu}_{%
\check s})(\tau ,\vec \sigma ),  \label{b6}
\end{eqnarray}

\noindent where

\begin{eqnarray}
l^{\mu}(\tau ,\vec \sigma )&=&({\frac{1}{\sqrt{\gamma} }}\epsilon^{\mu}{}_{%
\alpha \beta\gamma}z^{\alpha}_{\check 1}z^{\beta}_{\check 2}z^{\gamma}_{%
\check 3}) (\tau ,\vec \sigma ),  \nonumber \\
&&l^2(\tau ,\vec \sigma )=1,\quad\quad l_{\mu}(\tau ,\vec \sigma )z^{\mu} _{%
\check r}(\tau ,\vec \sigma )=0,  \label{b7}
\end{eqnarray}

\noindent is the unit (future pointing) normal to $\Sigma_{\tau}$ at $%
z^{\mu}(\tau ,\vec \sigma )$.

For the volume element in Minkowski spacetime we have

\begin{eqnarray}
d^4z&=&z^{\mu}_{\tau}(\tau ,\vec \sigma )d\tau d^3\Sigma_{\mu}=d\tau
[z^{\mu} _{\tau}(\tau ,\vec \sigma )l_{\mu}(\tau ,\vec \sigma )]\sqrt{\gamma
(\tau ,\vec \sigma )}d^3\sigma=  \nonumber \\
&=&\sqrt{g(\tau ,\vec \sigma )} d\tau d^3\sigma.  \label{b8}
\end{eqnarray}

Let us remark that according to the geometrical approach of Ref.\cite{kuchar}%
,one can use Eq.(\ref{III5}) in the form $z^{\mu}_{\tau}(\tau ,\vec \sigma
)=N(\tau , \vec \sigma )l^{\mu}(\tau ,\vec \sigma )+N^{\check r}(\tau ,\vec %
\sigma ) z^{\mu}_{\check r}(\tau ,\vec \sigma )$, where $N=\sqrt{g/\gamma}=%
\sqrt{g _{\tau\tau}-\gamma^{{\check r}{\check s}}g_{\tau{\check r}}g_{\tau{%
\check s}}}$ and $N^{\check r}=g_{\tau \check s}\gamma^{\check s\check r}$
are the standard lapse and shift functions, so that $g_{\tau \tau}=N^2+ g_{%
\check r\check s}N^{\check r}N^{\check s}, g_{\tau \check r}= g_{\check r%
\check s}N^{\check s}, g^{\tau \tau}=N^{-2}, g^{\tau \check r}=-N^{\check r%
}/N^2, g^{\check r\check s}=\gamma^{\check r\check s}+{\frac{{N^{\check r}N^{%
\check s}}}{{N^2}}}$, ${\frac{{\partial}}{{\partial z^{\mu}_{\tau}}}}%
=l_{\mu}\, {\frac{{\partial}}{{\partial N}}}+z_{{\check s}\mu}\gamma^{{%
\check s}{\check r}} {\frac{{\partial}}{{\partial N^{\check r}}}}$, $d^4z=N%
\sqrt{\gamma}d\tau d^3\sigma$.

The rest frame form of a timelike fourvector $p^{\mu}$ is $\stackrel{\circ}{p%
}{}^{\mu}=\eta \sqrt{p^2} (1;\vec 0)= \eta^{\mu o}\eta \sqrt{p^2}$, $%
\stackrel{\circ}{p}{}^2=p^2$, where $\eta =sign\, p^o$. The standard Wigner
boost transforming $\stackrel{\circ}{p}{}^{\mu}$ into $p^{\mu}$ is

\begin{eqnarray}
L^{\mu}{}_{\nu}(p,\stackrel{\circ}{p})&=&\epsilon^{\mu}_{\nu}(u(p))=
\nonumber \\
&=&\eta^{\mu}_{\nu}+2{\frac{{p^{\mu}{\stackrel{\circ}{p}}_{\nu}}}{{p^2}}}- {%
\frac{{(p^{\mu}+{\stackrel{\circ}{p}}^{\mu})(p_{\nu}+{\stackrel{\circ}{p}}%
_{\nu})} }{{p\cdot \stackrel{\circ}{p} +p^2} }}=  \nonumber \\
&=&\eta^{\mu}_{\nu}+2u^{\mu}(p)u_{\nu}(\stackrel{\circ}{p})-{\frac{{%
(u^{\mu}(p)+ u^{\mu}(\stackrel{\circ}{p}))(u_{\nu}(p)+u_{\nu}(\stackrel{\circ%
}{p}))} }{{1+u^o(p)} }},  \nonumber \\
&&{}  \nonumber \\
\nu =0 &&\epsilon^{\mu}_o(u(p))=u^{\mu}(p)=p^{\mu}/\eta \sqrt{p^2},
\nonumber \\
\nu =r &&\epsilon^{\mu}_r(u(p))=(-u_r(p); \delta^i_r-{\frac{{u^i(p)u_r(p)}}{{%
1+u^o(p)} }}).  \label{b9}
\end{eqnarray}

The inverse of $L^{\mu}{}_{\nu}(p,\stackrel{\circ}{p})$ is $L^{\mu}{}_{\nu} (%
\stackrel{\circ}{p},p)$, the standard boost to the rest frame, defined by

\begin{equation}
L^{\mu}{}_{\nu}(\stackrel{\circ}{p},p)=L_{\nu}{}^{\mu}(p,\stackrel{\circ}{p}%
)= L^{\mu}{}_{\nu}(p,\stackrel{\circ}{p}){|}_{\vec p\rightarrow -\vec p}.
\label{b10}
\end{equation}

Therefore, we can define the following vierbeins \footnote{%
The $\epsilon^{\mu}_r(u(p)) $'s are also called polarization vectors; the
indices r, s will be used for A=1,2,3 and $\bar o$ for A=0.}

\begin{eqnarray}
&&\epsilon^{\mu}_A(u(p))=L^{\mu}{}_A(p,\stackrel{\circ}{p}),  \nonumber \\
&&\epsilon^A_{\mu}(u(p))=L^A{}_{\mu}(\stackrel{\circ}{p},p)=\eta^{AB}\eta
_{\mu\nu}\epsilon^{\nu}_B(u(p)),  \nonumber \\
&&{}  \nonumber \\
&&\epsilon^{\bar o}_{\mu}(u(p))=\eta_{\mu\nu}\epsilon^{\nu}_o(u(p))=u_{%
\mu}(p),  \nonumber \\
&&\epsilon^r_{\mu}(u(p))=-\delta^{rs}\eta_{\mu\nu}\epsilon^{\nu}_r(u(p))=
(\delta^{rs}u_s(p);\delta^r_j-\delta^{rs}\delta_{jh}{\frac{{u^h(p)u_s(p)}}{{%
1+u^o(p)} }}),  \nonumber \\
&&\epsilon^A_o(u(p))=u_A(p),  \label{b11}
\end{eqnarray}

\noindent which satisfy

\begin{eqnarray}
&&\epsilon^A_{\mu}(u(p))\epsilon^{\nu}_A(u(p))=\eta^{\mu}_{\nu},  \nonumber
\\
&&\epsilon^A_{\mu}(u(p))\epsilon^{\mu}_B(u(p))=\eta^A_B,  \nonumber \\
&&\eta^{\mu\nu}=\epsilon^{\mu}_A(u(p))\eta^{AB}\epsilon^{\nu}_B(u(p))=u^{%
\mu} (p)u^{\nu}(p)-\sum_{r=1}^3\epsilon^{\mu}_r(u(p))\epsilon^{\nu}_r(u(p)),
\nonumber \\
&&\eta_{AB}=\epsilon^{\mu}_A(u(p))\eta_{\mu\nu}\epsilon^{\nu}_B(u(p)),
\nonumber \\
&&p_{\alpha}{\frac{{\partial}}{{\partial p_{\alpha}} }}\epsilon^{%
\mu}_A(u(p))= p_{\alpha}{\frac{{\partial}}{{\partial p_{\alpha}} }}%
\epsilon^A_{\mu}(u(p)) =0.  \label{b12}
\end{eqnarray}

The Wigner rotation corresponding to the Lorentz transformation $\Lambda$ is

\begin{eqnarray}
R^{\mu}{}_{\nu}(\Lambda ,p)&=&{[L(\stackrel{\circ}{p},p)\Lambda^{-1}L(%
\Lambda p,\stackrel{\circ}{p})]}^{\mu}{}_{\nu}=\left(
\begin{array}{cc}
1 & 0 \\
0 & R^i{}_j(\Lambda ,p)
\end{array}
\right) ,  \nonumber \\
{} && {}  \nonumber \\
R^i{}_j(\Lambda ,p)&=&{(\Lambda^{-1})}^i{}_j-{\frac{{(\Lambda^{-1})^i{}_op_{%
\beta} (\Lambda^{-1})^{\beta}{}_j}}{{p_{\rho}(\Lambda^{-1})^{\rho}{}_o+\eta
\sqrt{p^2}} }}-  \nonumber \\
&-&{\frac{{p^i}}{{p^o+\eta \sqrt{p^2}} }}[(\Lambda^{-1})^o{}_j- {\frac{{%
((\Lambda^{-1})^o {}_o-1)p_{\beta}(\Lambda^{-1})^{\beta}{}_j}}{{%
p_{\rho}(\Lambda^{-1})^{\rho} {}_o+\eta \sqrt{p^2}} }}].  \label{b13}
\end{eqnarray}

The polarization vectors transform under the Poincar\'e transformations $%
(a,\Lambda )$ in the following way

\begin{equation}
\epsilon^{\mu}_r(u(\Lambda p))=(R^{-1})_r{}^s\, \Lambda^{\mu}{}_{\nu}\,
\epsilon^{\nu}_s(u(p)).  \label{b14}
\end{equation}

\subsection{General Form of the Constraints of Section IV on Arbitrary
Spacelike Hypersurfaces.}

Let us try to undo the canonical reduction of Section III and to find the
general form of the modified constraints (\ref{IV4}) outside the radiation
gauge and outside the Wigner hyperplanes.

Since Eq.(\ref{III34}) implies

\begin{eqnarray}
\pi^r_{\perp}(\tau ,\vec \sigma ) &=& \pi^r(\tau ,\vec \sigma )-{\frac{{%
\partial^r}}{{\triangle_{\sigma}}}} \Big[ \Gamma (\tau ,\vec \sigma )-
\sum_{i=1}^N Q_i \delta^3(\vec \sigma -{\vec \eta}_i(\tau ))\Big] \approx
\nonumber \\
&\approx & \pi^r(\tau ,\vec \sigma )+{\frac{{\partial^r}}{{%
\triangle_{\sigma} }}} \sum_{i=1}^N Q_i \delta^3(\vec \sigma -{\vec \eta}%
_i(\tau )) = \pi^r(\tau ,\vec \sigma )+\sum_{i=1}^NQ_i c^r(\tau ,\vec \sigma
),  \nonumber \\
&&{}  \nonumber \\
&&\triangle =-{\vec \partial}^2,\qquad \triangle c(\vec \sigma )= \delta^3(%
\vec \sigma ),  \nonumber \\
&&{\frac{1}{{\triangle}}} \delta^3(\vec \sigma ) = c(\vec \sigma ) ={\frac{{%
-1}}{{4\pi |\vec \sigma |}}},  \nonumber \\
&&c^r(\vec \sigma ) =\partial^r c(\vec \sigma ) ={\frac{{\partial^r}}{{%
\triangle}}} \delta^3(\vec \sigma ) = {\frac{{\sigma^r}}{{4\pi |\vec \sigma
|^3}}},  \nonumber \\
&&\vec \partial \cdot \vec c(\vec \sigma ) =-\delta^3(\vec \sigma ),
\label{b15}
\end{eqnarray}

\noindent and since Eq.(\ref{III38}) allows to write the Coulomb potential
in the following form

\begin{eqnarray}
V_{ij} &=& {\frac{{Q_iQ_j}}{{4\pi |{\vec \eta}_i(\tau )-{\vec \eta}_j(\tau
)| }}}= - Q_iQ_j \int d^3\sigma {\frac{{\vec \partial}}{{\triangle_{\sigma}}}%
} \delta^3(\vec \sigma -{\vec \eta}_i(\tau )) \cdot {\frac{{\vec \partial}}{{%
\triangle_{\sigma}}}} \delta^3(\vec \sigma -{\vec \eta}_j(\tau ))=  \nonumber
\\
&=& -Q_iQ_j \int d^3\sigma \, \vec c(\vec \sigma -{\vec \eta}_i(\tau ))\cdot
\vec c(\vec \sigma -{\vec \eta}_j(\tau )),  \label{b16}
\end{eqnarray}

\noindent the general form of the modified constraint ${\cal H}
^{^{\prime}}(\tau )\approx 0$ on the Wigner hyperplane without the
restriction to the radiation gauge is

\begin{eqnarray}
{\cal H}^{^{\prime }} &=&\epsilon _{s}-\Big(\sum_{i=1}^{N}\sqrt{%
m_{i}^{2}-iQ_{i}\xi _{i}^{r}(\tau )\xi _{i}^{s}(\tau )F_{rs}(\tau ,{\vec{\eta%
}}_{i}(\tau ))+(\check{\vec{\kappa}}_{i}(\tau )-Q_{i}{\check{\vec{A}}}%
_{\perp }(\tau ,\vec{\eta}_{i}(\tau )))^{2}}-  \nonumber \\
&-&i\sum_{i=1}^{N}\frac{Q_{i}\check{\vec{\kappa}}_{i}(\tau )\cdot \vec{\xi}%
_{i}(\tau )\,\vec{\xi}_{i}(\tau )\cdot \check{\vec{\pi}}_{\perp }(\tau ,\vec{%
\eta}_{i}(\tau ))}{(m_{i}+\sqrt{m_{i}^{2}+\check{\vec{\kappa}}_{i}^{2}(\tau )%
})\sqrt{m_{i}^{2}+\check{\vec{\kappa}}_{i}^{2}(\tau )}}-  \nonumber \\
&-&i\sum_{i\neq j}{\frac{{Q_{i}Q_{j}\check{\vec{\kappa}}_{i}(\tau )\cdot
\vec{\xi}_{i}(\tau )}}{{(m_{i}+\sqrt{m_{i}^{2}+\check{\vec{\kappa}}%
_{i}^{2}(\tau )})\sqrt{m_{i}^{2}+\check{\vec{\kappa}}_{i}^{2}(\tau )}}}}{%
\vec{\xi}}_{i}(\tau )\cdot {\frac{{\partial }}{{\partial {\vec{\eta}}_{i}}}}%
\int d^{3}\sigma _{1}\,\vec{c}({\vec{\sigma}}_{1}-{\vec{\eta}}_{i}(\tau
))\cdot \vec{c}({\vec{\sigma}}_{1}-{\vec{\eta}}_{j}(\tau ))+  \nonumber \\
&+&\int d^{3}\sigma {\frac{1}{2}}[\check{\vec{\pi}}_{\perp }^{2}+\check{\vec{%
B}}^{2}](\tau ,\vec{\sigma})\Big)\approx 0.  \label{b17}
\end{eqnarray}

Eq.(\ref{b16}) suggests the following modification of the ten constraints (
\ref{III15}) describing the isolated system on arbitrary hyperplanes (namely
not on the Wigner hyperplanes)

\begin{eqnarray}
{\cal H}_{\mu }(\tau ) &=&\int {d^{3}\sigma {\cal H}_{\mu }(\tau ,\vec{\sigma%
})}={\ p_{s}}_{\mu }-b_{\mu \tau }\Big\{{\frac{1}{2}}\int d^{3}\sigma \Big(%
\vec{\pi}^{2}(\tau ,\vec{\sigma})+\vec{B}^{2}(\tau ,\vec{\sigma})\Big)+
\nonumber \\
&+&\sum_{i=1}^{N}\sqrt{m_{i}^{2}-iQ_{i}\xi _{i}^{r}(\tau )\xi _{i}^{s}(\tau
)F_{rs}(\tau ,{\vec{\eta}}_{i}(\tau ))++(\check{\vec{\kappa}}_{i}(\tau
)-Q_{i}{\check{\vec{A}}}_{\perp }(\tau ,\vec{\eta}_{i}(\tau )))^{2}}+
\nonumber \\
&+&\frac{1}{2}\sum_{i,j=1}^{N}\delta ^{3}(\vec{\eta}_{i}(\tau )-\vec{\eta}%
_{j}(\tau ))\cdot  \nonumber \\
&\cdot &\frac{Q_{i}(\tau )Q_{j}(\tau )\xi _{i}^{\gamma }(\tau )\xi
_{i}^{\delta }(\tau )}{\sqrt{m_{i}^{2}+\vec{\kappa}_{i}^{2}(\tau )}\sqrt{%
m_{j}^{2}+\vec{\kappa}_{j}^{2}(\tau )}}b_{\tau \gamma }\eta _{\delta \beta
}\xi _{j}^{\alpha }(\tau )\xi _{j}^{\beta }(\tau )b_{\tau \alpha }\Big]+
\nonumber \\
&+&i\sum_{i=1}^{N}\frac{Q_{i}(\tau )\xi _{i}^{\alpha }(\tau )\xi _{i}^{\beta
}(\tau )}{\sqrt{m_{i}^{2}+\vec{\kappa}_{i}^{2}(\tau )}}b_{\tau \alpha }b_{%
\breve{s}\beta }(\tau )\pi ^{\breve{s}}(\tau ,\vec{\eta}_{i}(\tau ))+
\nonumber \\
&-&\frac{i}{2}\sum_{i=1}^{N}\frac{Q_{i}(\tau )\xi _{i}^{\alpha }(\tau )\xi
_{i}^{\beta }(\tau )}{\sqrt{m_{i}^{2}+\vec{\kappa}_{i}^{2}(\tau )}}b_{\breve{%
u}\alpha }(\tau )b_{\breve{v}\beta }(\tau )F_{\breve{u}\breve{v}}(\tau ,\vec{%
\eta}_{i}(\tau ))+  \nonumber \\
&+&\sum_{i=1}^{N}\int d^{3}\sigma \delta ^{3}(\vec{\sigma}-{\vec{\eta}}%
_{i}(\tau ))  \nonumber \\
&&{\frac{{\ iQ_{i}(\tau )\kappa _{i\check{r}}(\tau )\xi _{i}^{\alpha }(\tau
)b_{\check{r}\alpha }(\tau )\xi _{i}^{\beta }(\tau )b_{\check{s}\beta }(\tau
)\Big[\pi ^{\check{s}}(\tau ,\vec{\sigma})+{\frac{{\partial ^{\check{s}}}}{{%
\triangle }}}\sum_{k=1}^{N}Q_{k}(\tau )\delta ^{3}(\vec{\sigma}-{\vec{\eta}}%
_{k}(\tau ))\Big]}}{{(m_{i}+\sqrt{m_{i}^{2}+{\vec{\kappa}}_{i}^{2}(\tau )})%
\sqrt{m_{i}^{2}+{\vec{\kappa}}_{i}^{2}(\tau )}}}}-  \nonumber \\
&-&i\sum_{i\not=j}^{1..N}{\frac{{Q_{i}(\tau )Q_{j}(\tau )\kappa _{i\check{r}%
}(\tau )\xi _{i}^{\alpha }(\tau )b_{\check{r}\alpha }(\tau )}}{{(m_{i}+\sqrt{%
m_{i}^{2}+{\vec{\kappa}}_{i}^{2}(\tau )})\sqrt{m_{i}^{2}+{\vec{\kappa}}%
_{i}^{2}(\tau )}}}}  \nonumber \\
&&\xi _{i}^{\beta }(\tau )b_{\check{s}\beta }(\tau )\int d^{3}\sigma \Big[{%
\frac{{\partial }}{{\partial \sigma ^{\check{s}}}}}{\frac{{\vec{\partial}}}{{%
\triangle }}}\delta (\vec{\sigma}-{\vec{\eta}}_{i}(\tau ))\Big]\cdot {\frac{{%
\vec{\partial}}}{{\triangle }}}\delta ^{3}(\vec{\sigma}-{\vec{\eta}}%
_{j}(\tau ))\Big\}+  \nonumber \\
&&{}  \nonumber \\
&+&b_{\breve{r}\mu }(\tau )\Big\{\int {d^{3}\sigma \big[\vec{\pi}\times \vec{%
B}\big]_{\breve{r}}(\tau ,\vec{\sigma})+\sum_{i=1}^{N}\big[\kappa _{i\breve{r%
}}(\tau )-Q_{i}(\tau )A_{\breve{r}}(\tau ,\vec{\eta}_{i}(\tau ))}\big]\Big\}%
\approx 0,  \nonumber \\
&&{}  \nonumber \\
{\cal H}^{\mu \nu }(\tau ) &=&b_{\breve{r}}^{\mu }(\tau )\int {d^{3}\sigma
\sigma ^{\breve{r}}{\cal H}^{\nu }(\tau ,\vec{\sigma})}-b_{\breve{r}}^{\nu
}(\tau )\int {d^{3}\sigma \sigma ^{\breve{r}}{\cal H}^{\mu }(\tau ,\vec{%
\sigma})}=  \nonumber \\
&=&S_{s}^{\mu \nu }(\tau )-\big(b_{\breve{r}}^{\mu }(\tau )b_{\tau }^{\nu
}-b_{\breve{r}}^{\nu }(\tau )b_{\tau }^{\mu }\big)\Big\{{\frac{1}{2}}\int
d^{3}\sigma \sigma ^{\breve{r}}\Big(\vec{\pi}^{2}(\tau ,\vec{\sigma})+\vec{B}%
^{2}(\tau ,\vec{\sigma})\Big)+  \nonumber \\
&+&\frac{1}{2}\sum_{i,j=1}^{N}\delta ^{3}(\vec{\eta}_{i}(\tau )-\vec{\eta}%
_{j}(\tau ))\cdot  \nonumber \\
&\cdot &\frac{\eta _{i}^{\check{r}}Q_{i}(\tau )Q_{j}(\tau )\xi _{i}^{\gamma
}(\tau )\xi _{i}^{\delta }(\tau )}{\sqrt{m_{i}^{2}+\vec{\kappa}_{i}^{2}(\tau
)}\sqrt{m_{j}^{2}+\vec{\kappa}_{j}^{2}(\tau )}}b_{\tau \gamma }\eta _{\delta
\beta }\xi _{j}^{\alpha }(\tau )\xi _{j}^{\beta }(\tau )b_{\tau \alpha }+
\nonumber \\
&+&\sum_{i=1}^{N}\eta _{i}^{\breve{r}}\sqrt{m_{i}^{2}+\big[\vec{\kappa}%
_{i}(\tau )-Q_{i}(\tau )\vec{A}(\tau ,\vec{\eta}_{i}(\tau ))\big]^{2}}+
\nonumber \\
&+&i\sum_{i=1}^{N}\eta _{i}^{\breve{r}}(\tau )\frac{Q_{i}(\tau )\xi
_{i}^{\alpha }(\tau )\xi _{i}^{\beta }(\tau )}{\sqrt{m_{i}^{2}+\vec{\kappa}%
_{i}^{2}(\tau )}}b_{\tau \alpha }b_{\breve{s}\beta }(\tau )\pi ^{\breve{s}%
}(\tau ,\vec{\eta}_{i}(\tau ))+  \nonumber \\
&-&\frac{i}{2}\sum_{i=1}^{N}\eta _{i}^{\breve{r}}(\tau )\frac{Q_{i}(\tau
)\xi _{i}^{\alpha }(\tau )\xi _{i}^{\beta }(\tau )}{\sqrt{m_{i}^{2}+\vec{%
\kappa}_{i}^{2}(\tau )}}b_{\breve{u}\alpha }(\tau )b_{\breve{v}\beta }(\tau
)F_{\breve{u}\breve{v}}(\tau ,\vec{\eta}_{i}(\tau ))+  \nonumber \\
&+&i\sum_{i=1}^{N}\eta _{i}^{\check{r}}(\tau )  \nonumber \\
&&{\frac{{\ iQ_{i}(\tau )\kappa _{i\check{r}}(\tau )\xi _{i}^{\alpha }(\tau
)b_{\check{r}\alpha }(\tau )\xi _{i}^{\beta }(\tau )b_{\check{s}\beta }(\tau
)\Big[\pi ^{\check{s}}(\tau ,\vec{\sigma})+{\frac{{\partial ^{\check{s}}}}{{%
\triangle }}}\sum_{k=1}^{N}Q_{k}(\tau )\delta ^{3}(\vec{\sigma}-{\vec{\eta}}%
_{k}(\tau ))\Big]}}{{(m_{i}+\sqrt{m_{i}^{2}+{\vec{\kappa}}_{i}^{2}})\sqrt{%
m_{i}^{2}+{\vec{\kappa}}_{i}^{2}}}}}-  \nonumber \\
&-&i\sum_{i\not=j}^{1..N}{\frac{{Q_{i}(\tau )Q_{j}(\tau )\kappa _{i\check{r}%
}(\tau )\xi _{i}^{\alpha }(\tau )b_{\check{r}\alpha }(\tau )}}{{(m_{i}+\sqrt{%
m_{i}^{2}+{\vec{\kappa}}_{i}^{2}})\sqrt{m_{i}^{2}+{\vec{\kappa}}_{i}^{2}}}}}
\nonumber \\
&&\xi _{i}^{\beta }(\tau )b_{\check{s}\beta }(\tau )\int d^{3}\sigma \sigma
^{\check{r}}\Big[{\frac{{\partial }}{{\partial \sigma ^{\check{s}}}}}{\frac{{%
\vec{\partial}}}{{\triangle }}}\delta (\vec{\sigma}-{\vec{\eta}}_{i}(\tau ))%
\Big]\cdot {\frac{{\vec{\partial}}}{{\triangle }}}\delta ^{3}(\vec{\sigma}-{%
\vec{\eta}}_{j}(\tau ))\Big\}+  \nonumber \\
&&{}  \nonumber \\
&+&\big(b_{\breve{r}}^{\mu }(\tau )b_{\breve{s}}^{\nu }(\tau )-b_{\breve{r}%
}^{\nu }(\tau )b_{\breve{s}}^{\mu }(\tau )\big)\Big\{\int {\ d^{3}\sigma
\sigma ^{\breve{r}}\big[\vec{\pi}\times \vec{B}\big]_{\breve{s}}(\tau ,\vec{%
\sigma})}+  \nonumber \\
&+&\sum_{i=1}^{N}\eta _{i}^{\breve{r}}(\tau )\big[\kappa _{i\breve{s}}(\tau
)-Q_{i}(\tau )A_{\breve{s}}(\tau ,\vec{\eta}_{i}(\tau ))\big]\Big\}\approx 0.
\label{b18}
\end{eqnarray}

However, there could be extra terms containing $\xi^{\mu}_ib_{\tau \mu}$
[like the spin-spin and spin-electric field terms in Eq.(\ref{III15})],
which vanish on the Wigner hyperplanes. To check whether they are needed,
one should verify that the Poisson brackets of the ten modified constraints
among themselves close on the constraints as it happens with Eq.(\ref{III15}%
). We are not going to do this check here, since we are only interested in
the rest-frame instant form of dynamics

Again modulo terms in $\xi^{\mu}_i z_{\tau \mu}(\tau , \vec \sigma )$ the
modification of the original constraints of Eq.(\ref{III3}) on arbitrary
spacelike hypersurfaces is

\begin{eqnarray}
{\cal H}_\mu (\tau, \vec{\sigma}) &=& \rho_\mu (\tau, \vec{\sigma}) - l_\mu
(\tau, \vec{\sigma}) \Big[ -\frac{1}{2\sqrt{\gamma(\tau, \vec{\sigma})}}
\pi^{\breve{r}}(\tau, \vec{\sigma}) g_{\breve{r} \breve{s}}(\tau, \vec{\sigma%
}) \pi^s (\tau, \vec{\sigma})+  \nonumber \\
&+& \frac{\sqrt{\gamma(\tau, \vec{\sigma})}}{4} \gamma^{\breve{r} \breve{s}%
}(\tau, \vec{\sigma}) \gamma^{\breve{u} \breve{v}}(\tau, \vec{\sigma}) F_{%
\breve{r} \breve{u}}(\tau, \vec{\sigma}) F_{\breve{s} \breve{v}}(\tau, \vec{%
\sigma}) +  \nonumber \\
&+& \sum_{i=1}^N \delta^3 (\vec{\sigma} -\vec{\eta}_i(\tau)) \cdot  \nonumber
\\
&\cdot &\sqrt{m_i^2 - \gamma^{\breve{r} \breve{s}}(\tau, \vec{\sigma}) \big( %
\kappa_{i \breve{r}}(\tau) - Q_i(\tau) A_{\breve{r}}(\tau, \vec{\sigma}) %
\big) \big( \kappa_{i \breve{s}}(\tau) - Q_i(\tau) A_{\breve{s}}(\tau, \vec{%
\sigma}) \big)} +  \nonumber \\
&+& \frac{1}{2\sqrt{\gamma(\tau, \vec{\sigma})}} \sum_{i,j=1}^N \delta^3 (%
\vec{\sigma} -\vec{\eta}_i(\tau)) \delta^3 (\vec{\sigma} -\vec{\eta}%
_j(\tau)) \cdot  \nonumber \\
&\cdot &\frac{ Q_i(\tau) Q_j(\tau) \xi^\gamma_i(\tau)\xi^\delta_i(\tau)
l_\gamma (\tau, \vec{\sigma}) \eta_{\delta \beta}
\xi^\alpha_j(\tau)\xi^\beta_j(\tau) l_\alpha (\tau, \vec{\sigma})} {\sqrt{%
m_i^2 - \gamma^{\breve{r} \breve{s}}(\tau, \vec{\sigma}) \kappa_{i\breve{r}%
}(\tau) \kappa_{i \breve{s}}(\tau)} \sqrt{m_j^2 - \gamma^{\breve{r} \breve{s}%
}(\tau, \vec{\sigma}) \kappa_{j \breve{r}}(\tau) \kappa_{j \breve{s}}(\tau)}}
+  \nonumber \\
&+& \frac{i}{\sqrt{\gamma(\tau, \vec{\sigma})}} \sum_{i=1}^N \delta^3 (\vec{%
\sigma} -\vec{\eta}_i(\tau)) \cdot  \nonumber \\
&\cdot & \frac{ Q_i(\tau) \xi^\alpha_i(\tau)\xi^\beta_i(\tau)} {\sqrt{m_i^2
- \gamma^{\breve{r} \breve{s}}(\tau, \vec{\sigma}) \kappa_{i\breve{r}}(\tau)
\kappa_{i \breve{s}}(\tau)}} l_\alpha (\tau, \vec{\sigma}) z_{\breve{s}
\beta}(\tau, \vec{\sigma}) \pi^{\breve{s}} (\tau, \vec{\sigma}) -  \nonumber
\\
&-& \frac i2 \sum_{i=1}^N \delta^3 (\vec{\sigma} -\vec{\eta}_i(\tau)) \frac{
Q_i(\tau) \xi^\alpha_i(\tau)\xi^\beta_i(\tau)} {\sqrt{m_i^2 - \gamma^{\breve{%
r} \breve{s}}(\tau, \vec{\sigma}) \kappa_{i\breve{r}}(\tau) \kappa_{i \breve{%
s}}(\tau)}}\cdot  \nonumber \\
&\cdot & z_{\breve{u} \alpha}(\tau, \vec{\sigma}) z_{\breve{v} \beta}(\tau,
\vec{\sigma}) \gamma^{\breve{r} \breve{u}}(\tau, \vec{\sigma}) \gamma^{%
\breve{s} \breve{v}}(\tau, \vec{\sigma}) F_{\breve{r} \breve{s}}(\tau, \vec{%
\sigma}) +  \nonumber \\
&+&i \delta^3(\vec \sigma -{\vec \eta}_i(\tau ))  \nonumber \\
&&{\frac{{Q_i(\tau ) \kappa_{i\check r}(\tau ) \xi_i^{\alpha}(\tau )z_{%
\check r\alpha}(\tau ,\vec \sigma ) \xi^{\beta}_i(\tau ) z_{\check s%
\beta}(\tau ,\vec \sigma ) \Big[ \pi^{\check s}(\tau ,\vec \sigma )+{\frac{{%
\partial^{\check s}}}{{\triangle}}} \sum_{k=1}^NQ_k(\tau )\delta^3(\vec %
\sigma -{\vec \eta}_k(\tau ))\Big]}}{{(m_i+\sqrt{m^2_i-\gamma^{\check u%
\check v}(\tau ,\vec \sigma ) \kappa_{i\check u}(\tau )\kappa_{i\check v%
}(\tau )}) \sqrt{m^2_i-\gamma^{\check u\check v}(\tau ,\vec \sigma )
\kappa_{i\check u}(\tau )\kappa_{i\check v}(\tau )} }}} +  \nonumber \\
&+&i\sum_{i\not= j}^{1..N}Q_iQ_j \gamma^{\check u\check v}(\tau ,\vec \sigma
)  \nonumber \\
&&{\frac{{\partial_{\check u}}}{{\triangle}}} \Big[ {\frac{{\kappa_{i\check r%
}(\tau ) \xi^{\alpha}_i(\tau )z_{\check r\alpha}(\tau ,\vec \sigma )
\xi^{\beta}_i(\tau )z_{\check s\beta}(\tau ,\vec \sigma )}}{{(m_i+\sqrt{%
m^2_i-\gamma^{\check u\check v}(\tau ,\vec \sigma ) \kappa_{i\check u}(\tau
)\kappa_{i\check v}(\tau )}) \sqrt{m^2_i-\gamma^{\check u\check v}(\tau ,%
\vec \sigma ) \kappa_{i\check u}(\tau )\kappa_{i\check v}(\tau )} }}}
\nonumber \\
&&{\frac{{\partial}}{{\partial \sigma^{\check s}}}}\delta^3(\vec \sigma -{%
\vec \eta}_i(\tau )) \Big] {\frac{{\partial_{\check v}}}{{\triangle}}}
\delta^3(\vec \sigma -{\vec \eta}_j(\tau )) \Big] -  \nonumber \\
&&{}  \nonumber \\
&-&\gamma^{\breve{r} \breve{s}}(\tau, \vec{\sigma}) z_{\breve{s} \mu}(\tau,
\vec{\sigma}) \Big[F_{\breve{r} \breve{u}}(\tau, \vec{\sigma}) \pi^{\breve{u}%
}(\tau, \vec{\sigma}) +  \nonumber \\
&+&\sum_{i=1}^N \delta^3 (\vec{\sigma} -\vec{\eta}_i(\tau)) \big(\kappa_{%
\breve{r} i} - Q_i(\tau) A_{\breve{r}}(\tau, \vec{\sigma})\big) \Big] %
\approx 0.  \label{b19}
\end{eqnarray}

\vfill\eject

\section{ Computation of the Field Energy and Momentum Integrals of Section
VI.}

Here we carry out the details in the computation of the field energy and
momentum for the case $N=2$. \ The general $N$ results obtained in the text
are an immediate generalization. From Eq.(\ref{VI41}) and Eq.(\ref{VI43}) we
find that

\begin{eqnarray}
{\vec{E}}_{\perp S}^{2}(\tau ,\sigma ) &=&{\frac{Q_{1}Q_{2}}{8\pi ^{2}}}%
\sum_{m=0}^{\infty }\sum_{n=0}^{\infty }{\frac{1}{(2n)!(2m)!}}\,{\bf \vec{V}}%
_{1}\cdot {\bf \vec{V}}_{2}\times  \nonumber \\
&&\Big[(\dot{\vec{\eta}}_{1}\cdot \vec{\partial}_{\sigma })^{2m+1}|\vec{%
\sigma}-\vec{\eta}_{1}|^{2m-1}\Big]\Big[(\dot{\vec{\eta}}_{2}\cdot \vec{%
\partial}_{\sigma })^{2n+1}|\vec{\sigma}-\vec{\eta}_{2}|^{2n-1}\Big]-
\nonumber \\
&&-{\frac{Q_{1}Q_{2}}{8\pi ^{2}}}\sum_{m=0}^{\infty }\sum_{n=0}^{\infty }{\
\frac{1}{(2n+2)!(2m)!}}({\bf \vec{V}}_{1}\cdot \vec{\partial}_{\sigma })(%
{\bf \vec{U}}_{2}\cdot \vec{\partial}_{\sigma })  \nonumber \\
&&\Big[(\dot{\vec{\eta}}_{1}\cdot \vec{\partial}_{\sigma })^{2m+1}|\vec{%
\sigma}-\vec{\eta}_{1}|^{2m-1}\Big]\Big[(\dot{\vec{\eta}}_{2}\cdot \vec{%
\partial}_{\sigma })^{2n+1}|\vec{\sigma}-\vec{\eta}_{2}|^{2n+1}\Big]-
\nonumber \\
&&-{\frac{Q_{1}Q_{2}}{8\pi ^{2}}}\sum_{m=0}^{\infty }\sum_{n=0}^{\infty }{\
\frac{1}{(2m+2)!(2n)!}}({\bf \vec{V}}_{2}\cdot \vec{\partial}_{\sigma })(%
{\bf \vec{U}}_{1}\cdot \vec{\partial}_{\sigma })  \nonumber \\
&&\Big[(\dot{\vec{\eta}}_{2}\cdot \vec{\partial}_{\sigma })^{2n+1})|\vec{%
\sigma}-\vec{\eta}_{2}|^{2n-1}\Big]\Big[(\dot{\vec{\eta}}_{1}\cdot \vec{%
\partial}_{\sigma })^{2m+1})|\vec{\sigma}-\vec{\eta}_{1}|^{2m+1}\Big]+
\nonumber \\
&&+{\frac{Q_{1}Q_{2}}{8\pi ^{2}}}\sum_{m=0}^{\infty }\sum_{n=0}^{\infty }{%
\frac{1}{(2m+2)!(2n+2)!}}({\bf \vec{U}}_{2}\cdot \vec{\partial}_{\sigma })(%
{\bf \vec{U}}_{1}\cdot \vec{\partial}_{\sigma })  \nonumber \\
&&\Big[\vec{\partial}_{\sigma }(\dot{\vec{\eta}}_{2}\cdot \vec{\partial}%
_{\sigma })^{2n+1})|\vec{\sigma}-\vec{\eta}_{2}|^{2n+1}\Big]\cdot \Big[ \vec{%
\partial}_{\sigma }(\dot{\vec{\eta}}_{1}\cdot \vec{\partial}_{\sigma
})^{2m+1})|\vec{\sigma}-\vec{\eta}_{1}|^{2m+1}\Big],  \label{c1}
\end{eqnarray}
\begin{eqnarray}
{\vec{B}}_{S}^{2}(\tau ,\vec{\sigma}) &=&{\frac{Q_{1}Q_{2}}{8\pi ^{2}}}%
\sum_{m=0}^{\infty }\sum_{n=0}^{\infty }{\frac{1}{(2n)!(2m)!}}({\bf \vec{V}}%
_{1}\cdot {\bf \vec{V}}_{2})  \nonumber \\
&&\Big[(\vec{\partial}_{\sigma }(\dot{\vec{\eta}}_{1}\cdot \vec{\partial}%
_{\sigma })^{2m})|\vec{\sigma}-\vec{\eta}_{1}|^{2m-1}\Big]\cdot \Big[ (\vec{%
\partial}_{\sigma }(\dot{\vec{\eta}}_{2}\cdot \vec{\partial}_{\sigma
})^{2n})|\vec{\sigma}-\vec{\eta}_{2}|^{2n-1}\Big]-  \nonumber \\
&&-{\frac{Q_{1}Q_{2}}{8\pi ^{2}}}\sum_{m=0}^{\infty }\sum_{n=0}^{\infty }{%
\frac{1}{(2n)!(2m)!}}({\bf \vec{V}}_{1}\cdot \vec{\partial}_{\sigma })({\bf
\vec{V}}_{2}\cdot \vec{\partial}_{\sigma })  \nonumber \\
&&\Big[(\dot{\vec{\eta}}_{1}\cdot \vec{\partial}_{\sigma })^{2m})|\vec{\sigma%
}-\vec{\eta}_{1}|^{2m-1}\Big]\cdot \Big[ (\dot{\vec{\eta}}_{2}\cdot \vec{%
\partial}_{\sigma })^{2n})|\vec{\sigma}-\vec{\eta}_{2}|^{2n-1}\Big],
\label{c2}
\end{eqnarray}

\begin{eqnarray}
&&(\vec{E}_{\perp S}(\tau ,\vec{\sigma})\times \vec{B}_{S}(\tau ,\vec{\sigma}%
))_{k}=  \nonumber \\
&=&{\frac{Q_{1}Q_{2}}{16\pi ^{2}}}\sum_{m=0}^{\infty }\sum_{n=0}^{\infty }{%
\frac{1}{(2n)!(2m)!}}\,{\bf \vec{V}}_{1}\cdot {\bf \vec{V}}_{2}\times
\nonumber \\
&&\Big[(\dot{\vec{\eta}}_{1}\cdot {\vec{\partial}}_{\sigma })^{2m+1}|\vec{%
\sigma}-\vec{\eta}_{1}|^{2m-1}\Big]\Big[{\vec{\partial}}_{\sigma }(\dot{\vec{%
\eta}}_{2}\cdot {\vec \partial} _{\sigma })^{2n}|\vec{\sigma}-\vec{\eta}%
_{2}|^{2n-1}\Big]-  \nonumber \\
&&-{\frac{Q_{1}Q_{2}}{16\pi ^{2}}}\sum_{m=0}^{\infty }\sum_{n=0}^{\infty }{%
\frac{1}{(2n)!(2m)!}}\,{\bf \vec{V}}_{2} \times  \nonumber \\
&&\Big[({\bf \vec{V}}_{1}\cdot {\vec{\partial}}_{\sigma })(\dot{\vec{\eta}}%
_{2}\cdot {\vec{\partial}}_{\sigma })^{2n}|\vec{\sigma}-\vec{\eta}%
_{2}|^{2n-1}\Big]\Big[(\dot{\vec{\eta}}_{1}\cdot {\vec{\partial}}_{\sigma
})^{2m+1}|\vec{\sigma}-\vec{\eta}_{1}|^{2m-1}\Big]-  \nonumber \\
&&-{\frac{Q_{1}Q_{2}}{16\pi ^{2}}}\sum_{m=0}^{\infty }\sum_{n=0}^{\infty }{%
\frac{1}{(2n)!(2m+2)!}}({\bf \vec{V}}_{2}\cdot \vec{\partial}_{\sigma })(%
{\bf \vec{U}}_{1}\cdot \vec{\partial}_{\sigma })\times  \nonumber \\
&&\Big[{\vec{\partial}}_{\sigma }(\dot{\vec{\eta}}_{2}\cdot {\vec{\partial}}%
_{\sigma })^{2n}|\vec{\sigma}-\vec{\eta}_{2}|^{2n-1}\Big]\Big[(\dot{\vec{\eta%
}}_{1}\cdot {\vec{\partial}}_{\sigma })^{2m+1}|\vec{\sigma}-\vec{\eta}%
_{1}|^{2m+1}\Big]+  \nonumber \\
&&+{\frac{Q_{1}Q_{2}}{16\pi ^{2}}}\sum_{m=0}^{\infty }\sum_{n=0}^{\infty }{%
\frac{1}{(2n)!(2m+2)!}}\,{\bf \vec{V}}_{2}\times  \nonumber \\
&&\Big[{\vec{\partial}}_{\sigma }({\bf \vec{U}}_{1}\cdot \vec{\partial}%
_{\sigma })(\dot{\vec{\eta}}_{1}\cdot {\vec{\partial}}_{\sigma })^{2m+1}|%
\vec{\sigma}-\vec{\eta}_{1}|^{2m+1}\Big]\cdot \Big[ {\vec{\partial}}_{\sigma
}(\dot{\vec{\eta}}_{2}\cdot {\vec{\partial}}_{\sigma })^{2n}|\vec{\sigma}-%
\vec{\eta}_{2}|^{2n-1}\Big]+  \nonumber \\
&&+(1\longleftrightarrow 2).  \label{c3}
\end{eqnarray}

Our aim here is to compute ${{\frac{1}{2}}}\int d^{3}\sigma (\vec{E}_{\perp
S}^{2}+\vec{B}_{S}^{2})(\tau ,\vec{\sigma})$ and $\int d^{3}\sigma (\vec{E}%
_{\perp S}\times \vec{B}_{S})(\tau ,\vec{\sigma})$ . \ Essentially these
integrals were computed in detail in Appendix B of \cite{ap} for scalar
particles which correspond here to the result obtained by replacing ${\bf
\vec{V}}_{i},{\bf \vec{U}}_{i}\rightarrow \dot{\vec{\eta}}_{i}$ . \
Following steps in that appendix similar to ones that lead to Eqs.(6.4) and
(6.5) in \cite{ap} gives the result Eq.(\ref{VI46}), (\ref{VI47})

\vfill\eject

\section{\protect\bigskip Comparison of $U_{1}(\protect\tau )$ with $U(%
\protect\tau )$.}

Let us compare the terms (\ref{VII7}), (\ref{VII8}) and (\ref{VII9}) of the
potential $U_{1}(\tau )$ with the terms of the potential $U(\tau )$ of Eq.(%
\ref{VI46}) and let us try to combine them together so that, after the
addition of $V_{DSO}(\tau )$ of Eq. (\ref{VII5}), we can find the expression
of the Darwin and spin-dependent terms of the potential $V_{DS}(\tau )$ of
Eq.(\ref{VII2}). We rewrite $U(\tau )$ of Eq.(\ref{VI46}) in the following
form

\begin{eqnarray}
&&U(\tau )={\frac{1}{2}}\int d^{3}\sigma ({\vec{E}}_{\perp S}^{2}+{\vec{B}}%
_{S}^{2})(\tau ,\vec{\sigma}) =  \nonumber \\
&&  \nonumber \\
&=&\sum_{i<j}^{1..N}{\frac{Q_{i}Q_{j}}{4\pi }}\sum_{m=0}^{\infty
}\sum_{n=0}^{\infty }\Big[{\bf \vec{V}}_{i}\cdot {\bf \vec{V}}_{j}(\frac{%
\vec{\kappa}_{i}}{\sqrt{m_{i}^{2}+\vec{\kappa}_{i}{}^{2}}}\cdot \vec{\partial%
}_{ij})^{2m+1}(\frac{\vec{\kappa}_{j}}{\sqrt{m_{j}^{2}+\vec{\kappa}_{j}{}^{2}%
}}\cdot \vec{\partial}_{ij})^{2n+1}\frac{\eta _{ij}^{2n+2m+1}}{(2n+2m+2)!} -
\nonumber \\
&-&({\bf \vec{V}}_{i}\cdot \vec{\partial}_{ij})({\bf \vec{U}}_{j}\cdot \vec{%
\partial}_{ij})(\frac{\vec{\kappa}_{i}}{\sqrt{m_{i}^{2}+\vec{\kappa}%
_{i}{}^{2}}}\cdot \vec{\partial}_{ij})^{2m+1}(\frac{\vec{\kappa}_{j}}{\sqrt{%
m_{j}^{2}+\vec{\kappa}_{j}{}^{2}}}\cdot \vec{\partial}_{ij})^{2n+1}\frac{%
\eta _{ij}^{2n+2m+3}}{(2n+2m+4)!}-  \nonumber \\
&&-({\bf \vec{U}}_{i}\cdot \vec{\partial}_{ij})({\bf \vec{V}}_{j}\cdot \vec{%
\partial}_{ij})(\frac{\vec{\kappa}_{i}}{\sqrt{m_{i}^{2}+\vec{\kappa}%
_{i}{}^{2}}}\cdot \vec{\partial}_{ij})^{2m+1}(\frac{\vec{\kappa}_{j}}{\sqrt{%
m_{j}^{2}+\vec{\kappa}_{j}{}^{2}}}\cdot \vec{\partial}_{ij})^{2n+1}\frac{%
\eta _{ij}^{2n+2m+3}}{(2n+2m+4)!}+  \nonumber \\
&&+({\bf \vec{U}}_{i}\cdot \vec{\partial}_{ij})({\bf \vec{U}}_{j}\cdot \vec{%
\partial}_{ij})(\frac{\vec{\kappa}_{i}}{\sqrt{m_{i}^{2}+\vec{\kappa}%
_{i}{}^{2}}}\cdot \vec{\partial}_{ij})^{2m+1}(\frac{\vec{\kappa}_{j}}{\sqrt{%
m_{j}^{2}+\vec{\kappa}_{j}{}^{2}}}\cdot \vec{\partial}_{ij})^{2n+1}\frac{%
\eta _{ij}^{2n+2m+3}}{(2n+2m+4)!}+  \nonumber \\
&+&{\bf \vec{V}}_{i}\cdot {\bf \vec{V}}_{j}(\frac{\vec{\kappa}_{i}}{\sqrt{%
m_{i}^{2}+\vec{\kappa}_{i}{}^{2}}}\cdot \vec{\partial}_{ij})^{2m}(\frac{\vec{%
\kappa}_{j}}{\sqrt{m_{j}^{2}+\vec{\kappa}_{j}{}^{2}}}\cdot \vec{\partial}%
_{ij})^{2n}\frac{\eta _{ij}^{2n+2m-1}}{(2n+2m)!}-  \nonumber \\
&-&({\bf \vec{V}}_{i}\cdot \vec{\partial}_{ij})({\bf \vec{V}}_{j}\cdot \vec{%
\partial}_{ij})(\frac{\vec{\kappa}_{i}}{\sqrt{m_{i}^{2}+\vec{\kappa}%
_{i}{}^{2}}}\cdot \vec{\partial}_{ij})^{2m}(\frac{\vec{\kappa}_{j}}{\sqrt{%
m_{j}^{2}+\vec{\kappa}_{j}{}^{2}}}\cdot \vec{\partial}_{ij})^{2n}\frac{\eta
_{ij}^{2n+2m+1}}{(2n+2m+2)!}\Big],  \label{d1}
\end{eqnarray}

\noindent where we have rewritten the operators ${\vec {{\bf V}}}_i$ of Eq.(%
\ref{V13}) and ${\vec {{\bf U}}}_i$ of Eq.(\ref{VI42})in the form [${\vec {%
{\bf k}}}_i$ and ${\vec {{\bf l}}}_i$ are operators]

\begin{eqnarray}
{\bf \vec{V}}_{i} &=&\frac{\vec{\kappa}_{i}}{\sqrt{m_{i}^{2}+\vec{\kappa}%
_{i}{}^{2}}}-\frac{i\vec{\xi}_{i}\vec{\xi}_{i}\cdot \vec{\partial}_{ij}}{%
\sqrt{m_{i}^{2}+\vec{\kappa}_{i}{}^{2}}}+\frac{i\vec{\kappa}_{i}\cdot \vec{%
\partial}_{ij}\vec{\kappa}_{i}\cdot \vec{\xi}_{i}\vec{\xi}_{i}}{(m_{i}+\sqrt{%
m_{i}^{2}+\vec{\kappa}_{i}{}^{2}})(m_{i}^{2}+\vec{\kappa}_{i}{}^{2})}:=\frac{%
{\vec {{\bf k}}}_{i}}{\sqrt{m_{i}^{2}+\vec{\kappa}_{i}{}^{2}}},  \nonumber \\
{\bf \vec{U}}_{i}&=&\frac{\vec{\kappa}_{i}}{\sqrt{m_{i}^{2}+\vec{\kappa}%
_{i}{}^{2}}}+\frac{i\vec{\kappa}_{i}\cdot \vec{\partial}_{ij}\vec{\kappa}%
_{i}\cdot \vec{\xi}_{i}\vec{\xi}_{i}}{(m_{i}+\sqrt{m_{i}^{2}+\vec{\kappa}%
_{i}{}^{2}})(m_{i}^{2}+\vec{\kappa}_{i}{}^{2})}:=\frac{{\vec {{\bf l}}}_{i}}{%
\sqrt{m_{i}^{2}+\vec{\kappa}_{i}{}^{2}}} .  \label{d2}
\end{eqnarray}

We consider the last two lines in the expression for $U(\tau )$ given in Eq.(%
\ref{VI46})

\begin{eqnarray}
&&\sum_{i<j}^{1..N}{\frac{Q_{i}Q_{j}}{4\pi }}\sum_{m=0}^{\infty
}\sum_{n=0}^{\infty }\Big[\frac{{\bf {\vec{k}}_{i}}}{\sqrt{m_{i}^{2}+\vec{%
\kappa}_{i}{}^{2}}}\cdot \frac{{\bf {\vec{k}}_{j}}}{\sqrt{m_{j}^{2}+\vec{%
\kappa}_{j}{}^{2}}}  \nonumber \\
&& (\frac{\vec{\kappa}_{i}}{\sqrt{m_{i}^{2}+\vec{\kappa}_{i}{}^{2}}}\cdot
\vec{\partial}_{ij})^{2m}(\frac{\vec{\kappa}_{j}}{\sqrt{m_{j}^{2}+\vec{\kappa%
}_{j}{}^{2}}}\cdot \vec{\partial}_{ij})^{2n}\frac{\eta _{ij}^{2n+2m-1}}{%
(2n+2m)!} -  \nonumber \\
&&-(\frac{{\bf {\vec{k}}_{i}}}{\sqrt{m_{i}^{2}+\vec{\kappa}_{i}{}^{2}}}\cdot
\vec{\partial}_{ij})(\frac{{\bf {\vec{k}}_{j}}}{\sqrt{m_{j}^{2}+\vec{\kappa}%
_{j}{}^{2}}}\cdot \vec{\partial}_{ij})  \nonumber \\
&& (\frac{\vec{\kappa}_{i}}{\sqrt{m_{i}^{2}+\vec{\kappa}_{i}{}^{2}}}\cdot
\vec{\partial}_{ij})^{2m}(\frac{\vec{\kappa}_{j}}{\sqrt{m_{j}^{2}+\vec{\kappa%
}_{j}{}^{2}}}\cdot \vec{\partial}_{ij})^{2n}\frac{\eta _{ij}^{2n+2m+1}}{%
(2n+2m+2)!}\Big]=  \nonumber \\
&=&\sum_{i\neq j}^{N}{\frac{Q_{i}Q_{j}}{8\pi }}\Big[\frac{{\bf {\vec{k}}_{i}}%
}{\sqrt{m_{i}^{2}+\vec{\kappa}_{i}{}^{2}}}\cdot \frac{{\bf {\vec{k}}_{j}}}{%
\sqrt{m_{j}^{2}+\vec{\kappa}_{j}{}^{2}}}{\large (}\sum_{m=0}^{\infty }\Big([(%
\frac{\vec{\kappa}_{i}}{\sqrt{m_{i}^{2}+\vec{\kappa}_{i}{}^{2}}}\cdot \vec{%
\partial}_{ij})^{2m}+  \nonumber \\
&&+(\frac{\vec{\kappa}_{j}}{\sqrt{m_{j}^{2}+\vec{\kappa}_{j}{}^{2}}}\cdot
\vec{\partial}_{ij})^{2m}]\frac{\eta _{ij}^{2m-1}}{(2m)!}\Big)-\eta
_{ij}^{-1}{\large )} -  \nonumber \\
&&-(\frac{{\bf {\vec{k}}_{i}}}{\sqrt{m_{i}^{2}+\vec{\kappa}_{i}{}^{2}}}\cdot
\vec{\partial}_{ij})(\frac{{\bf {\vec{k}}_{j}}}{\sqrt{m_{j}^{2}+\vec{\kappa}%
_{j}{}^{2}}}\cdot \vec{\partial}_{ij}){\large (}\sum_{m=0}^{\infty }\Big([(%
\frac{\vec{\kappa}_{i}}{\sqrt{m_{i}^{2}+\vec{\kappa}_{i}{}^{2}}}\cdot \vec{%
\partial}_{ij})^{2m} +  \nonumber \\
&&+(\frac{\vec{\kappa}_{j}}{\sqrt{m_{j}^{2}+\vec{\kappa}_{j}{}^{2}}}\cdot
\vec{\partial}_{ij})^{2m}]\frac{\eta _{ij}^{2m+1}}{(2m+2)!}\Big)-\frac{1}{2}%
\eta _{ij}{\large )} +  \nonumber \\
&&+\sum_{m=0}^{\infty }\sum_{n=0}^{\infty }[\frac{{\bf {\vec{k}}_{i}}}{\sqrt{%
m_{i}^{2}+\vec{\kappa}_{i}{}^{2}}}\cdot \frac{{\bf {\vec{k}}_{j}}}{\sqrt{%
m_{j}^{2}+\vec{\kappa}_{j}{}^{2}}}  \nonumber \\
&&(\frac{\vec{\kappa}_{i}}{\sqrt{m_{i}^{2}+\vec{\kappa}_{i}{}^{2}}}\cdot
\vec{\partial}_{ij})^{2m+2}(\frac{\vec{\kappa}_{j}}{\sqrt{m_{j}^{2}+\vec{%
\kappa}_{j}{}^{2}}}\cdot \vec{\partial}_{ij})^{2n+2}\frac{\eta
_{ij}^{2n+2m+3}}{ (2n+2m+4)!} -  \nonumber \\
&&-(\frac{{\bf {\vec{k}}_{i}}}{\sqrt{m_{i}^{2}+\vec{\kappa}_{i}{}^{2}}}\cdot
\vec{\partial}_{ij})(\frac{{\bf {\vec{k}}_{j}}}{\sqrt{m_{j}^{2}+\vec{\kappa}%
_{j}{}^{2}}}\cdot \vec{\partial}_{ij})  \nonumber \\
&&(\frac{\vec{\kappa}_{i}}{\sqrt{m_{i}^{2}+\vec{\kappa}_{i}{}^{2}}}\cdot
\vec{\partial}_{ij})^{2m+2}(\frac{\vec{\kappa}_{j}}{\sqrt{m_{j}^{2}+\vec{%
\kappa}_{j}{}^{2}}}\cdot \vec{\partial}_{ij})^{2n+2}\frac{\eta
_{ij}^{2n+2m+5}}{ (2n+2m+6)!}]\Big].  \label{d3}
\end{eqnarray}

The first three lines\ on the right hand side contains two separate types of
summations, each of which can be summed to close forms. \ The first type
involves

\begin{equation}
\sum_{m=0}^{\infty }(\frac{\vec{\kappa}_{i}}{\sqrt{m_{i}^{2}+\vec{\kappa}%
_{i}{}^{2}}}\cdot \vec{\partial}_{ij})^{2m}\frac{\eta _{ij}^{2m-1}}{(2m)!}=%
\frac{1}{\eta _{ij}}\frac{\sqrt{m_{i}^{2}+\vec{\kappa}_{i}{}^{2}}}{\sqrt{%
m_{i}^{2}+(\vec{\kappa}_{i}\cdot {\hat{\eta}}_{ij})^{2}}}.  \label{d4}
\end{equation}

In order to perform the second type we use

\begin{eqnarray}
&&\frac{\vec{\partial}_{ij}\vec{\partial}_{ij}}{\sqrt{m_{i}^{2}+\vec{\kappa}%
_{i}{}^{2}}}(\frac{\vec{\kappa}_{i}}{\sqrt{m_{i}^{2}+\vec{\kappa}_{i}{}^{2}}}%
\cdot \vec{\partial}_{ij})^{2m}\frac{\eta _{ij}^{2m+1}}{(2m+2)!}  \nonumber
\\
&=&\frac{(2m+1)!!(2m-1)!!}{\eta _{ij}(2m+2)!(\sqrt{m_{i}^{2}+\vec{\kappa}%
_{i}{}^{2}})^{2m+1}}\Big({\bf I}[\vec{\kappa}_{i}^{2}-(\vec{\kappa}_{i}\cdot
{\hat{\eta}}_{ij})^{2}]^{m} +  \nonumber \\
&&+2m[\vec{\kappa}_{i}\vec{\kappa}_{i}-(\vec{\kappa}_{i}\cdot {\hat{\eta}}%
_{ij})(\vec{\kappa}_{i}{\hat{\eta}}_{ij}+{\hat{\eta}}_{ij}\vec{\kappa}_{i})](%
\vec{\kappa}_{i}^{2}-(\vec{\kappa}_{i}\cdot {\hat{\eta}}_{ij})^{2})^{m-1} -
\nonumber \\
&&-(1-\delta _{m0}){\hat{\eta}}_{ij}{\hat{\eta}}_{ij}(\vec{\kappa}_{i}^{2}-(%
\vec{\kappa}_{i}\cdot {\hat{\eta}}_{ij})^{2})^{m-1}(\vec{\kappa}%
_{i}^{2}-(2m+1)(\vec{\kappa}_{i}\cdot {\hat{\eta}}_{ij})^{2})\Big) .
\label{d5}
\end{eqnarray}

\noindent where ${\bf I}$ is the unit dyad. \ This complex form was obtained
by examples for $m=0,1,2,3,4$. \ As a simple check on this result consider
(for $m\neq 0$) the contraction Eq.( \ref{d5}) with the dyad $\vec{\kappa}%
_{i}\vec{\kappa}_{j}$ . This gives

\begin{eqnarray}
&&(\frac{\vec{\kappa}_{i}}{\sqrt{m_{i}^{2}+\vec{\kappa}_{i}{}^{2}}}\cdot
\vec{\partial}_{ij})^{2m+2}\frac{\eta _{ij}^{2m+1}}{(2m+2)!}=  \nonumber \\
&=&\frac{(2m+1)!!(2m-1)!!}{\eta _{ij}(2m+2)!(\sqrt{m_{i}^{2}+\vec{\kappa}%
_{i}{}^{2}})^{2m+2}}  \nonumber \\
&&\Big(\vec{\kappa}_{i}^{2}(\vec{\kappa}_{i}^{2}-(\vec{\kappa}_{i}\cdot {%
\hat{\eta}}_{ij})^{2})^{m}+2m[(\vec{\kappa}_{i}^{2})^{2}-2\vec{\kappa}%
_{i}^{2}(\vec{\kappa}_{i}\cdot {\hat{\eta}}_{ij})^{2}](\vec{\kappa}_{i}^{2}-(%
\vec{\kappa}_{i}\cdot {\hat{\eta}}_{ij})^{2})^{m-1} -  \nonumber \\
&&-(1-\delta _{m0})(\vec{\kappa}_{i}\cdot {\hat{\eta}}_{ij})^{2}(\vec{\kappa}%
_{i}^{2}-(\vec{\kappa}_{i}\cdot {\hat{\eta}}_{ij})^{2})^{m-1}(\vec{\kappa}%
_{i}^{2}-(2m+1)(\vec{\kappa}_{i}\cdot {\hat{\eta}}_{ij})^{2})\Big) =
\nonumber \\
&=&\frac{(2m+1)!!(2m-1)!!(\vec{\kappa}_{i}^{2}-(\vec{\kappa}_{i}\cdot {\hat{%
\eta}}_{ij})^{2})^{m-1}}{\eta _{ij}(2m+2)!(\sqrt{m_{i}^{2}+\vec{\kappa}%
_{i}{}^{2}})^{2m+2}}  \nonumber \\
&&\times \Big(\vec{\kappa}_{i}^{2}(\vec{\kappa}_{i}^{2}-(\vec{\kappa}%
_{i}\cdot {\hat{\eta}}_{ij})^{2})+2m[(\vec{\kappa}_{i}^{2})^{2}-2\vec{\kappa}%
_{i}^{2}(\vec{\kappa}_{i}\cdot {\hat{\eta}}_{ij})^{2}] -  \nonumber \\
&&-(\vec{\kappa}_{i}\cdot {\hat{\eta}}_{ij})^{2}(\vec{\kappa}_{i}^{2}-(2m+1)(%
\vec{\kappa}_{i}\cdot {\hat{\eta}}_{ij})^{2})\Big) =  \nonumber \\
&=&\frac{[(2m+1)!!]^{2}(\vec{\kappa}_{i}^{2}-(\vec{\kappa}_{i}\cdot {\hat{%
\eta}}_{ij})^{2})^{m+1}}{\eta _{ij}(2m+2)!(\sqrt{m_{i}^{2}+\vec{\kappa}%
_{i}{}^{2}})^{2m+2}},  \label{d6}
\end{eqnarray}

\noindent and it agrees with the equation described above Eq.(\ref{VI32}).
Using Eq.(\ref{d5}), the second sum (we separate out the $m=0$ term and let $%
x=\frac{\vec{\kappa}_{i}^{2}-(\vec{\kappa}_{i}\cdot {\hat{\eta}}_{ij})^{2}}{(%
\sqrt{m_{i}^{2}+\vec{\kappa}_{i}{}^{2}})^{2}}$) has three parts, the first
of which is

\begin{eqnarray}
&&\sum_{m=1}^{\infty }\frac{(2m+1)!!(2m-1)!!}{\eta _{ij}(2m+2)!(\sqrt{%
m_{i}^{2}+\vec{\kappa}_{i}{}^{2}})^{2m+1}}{\bf I}[\vec{\kappa}_{i}^{2}-(\vec{%
\kappa}_{i}\cdot {\hat{\eta}}_{ij})^{2}]^{m} =  \nonumber \\
&=&x\frac{{\bf I}}{\eta _{ij}\sqrt{m_{i}^{2}+\vec{\kappa}_{i}{}^{2}}}%
\sum_{m=0}^{\infty }\frac{(-)^{m}x^{m}}{2m+3}%
{-1/2 \choose m+2}%
=  \nonumber \\
&=&\frac{{\bf I}}{\eta _{ij}\sqrt{m_{i}^{2}+\vec{\kappa}_{i}{}^{2}}}%
x^{1/2}\int_{0}^{x^{1/2}}dz\sum_{m=0}^{\infty }(-)^{m}z^{2m+2}%
{-1/2 \choose m+2}%
=  \nonumber \\
&=&\frac{{\bf I}}{\eta _{ij}\sqrt{m_{i}^{2}+\vec{\kappa}_{i}{}^{2}}}\frac{%
1-x/2-\sqrt{1-x}}{x}  \nonumber \\
&=&\frac{{\bf I}}{\eta _{ij}\sqrt{m_{i}^{2}+\vec{\kappa}_{i}{}^{2}}}\Big[-%
\frac{1}{2}-\frac{\sqrt{m_{i}^{2}+\vec{\kappa}_{i}{}^{2}}\sqrt{m_{i}^{2}+(%
\vec{\kappa}_{i}{}\cdot \hat{\eta}_{ij})^{2}}}{\vec{\kappa}_{i}^{2}-(\vec{%
\kappa}_{i}\cdot {\hat{\eta}}_{ij})^{2}}+\frac{(\sqrt{m_{i}^{2}+\vec{\kappa}%
_{i}{}^{2}})^{2}}{\vec{\kappa}_{i}^{2}-(\vec{\kappa}_{i}\cdot {\hat{\eta}}%
_{ij})^{2}}\Big] =  \nonumber \\
&=&\frac{{\bf I}}{2\eta _{ij}\sqrt{m_{i}^{2}+\vec{\kappa}_{i}{}^{2}}}\frac{%
\sqrt{m_{i}^{2}+\vec{\kappa}_{i}{}^{2}}-\sqrt{m_{i}^{2}+(\vec{\kappa}%
_{i}{}\cdot \hat{\eta}_{ij})^{2}}}{\sqrt{m_{i}^{2}+\vec{\kappa}_{i}{}^{2}}+%
\sqrt{m_{i}^{2}+(\vec{\kappa}_{i}{}\cdot \hat{\eta}_{ij})^{2}}},  \label{d7}
\end{eqnarray}

\noindent while the second part is

\begin{eqnarray}
&&\sum_{m=1}^{\infty }\frac{(2m+1)!!(2m-1)!!}{\eta _{ij}(2m+2)!(\sqrt{%
m_{i}^{2}+\vec{\kappa}_{i}{}^{2}})^{2m+1}}  \nonumber \\
&&2m\Big[\vec{\kappa}_{i}\vec{\kappa}_{i}-(\vec{\kappa}_{i}\cdot {\hat{\eta}}%
_{ij})(\vec{\kappa}_{i}{\hat{\eta}}_{ij}+{\hat{\eta}}_{ij}\vec{\kappa}_{i})%
\Big](\vec{\kappa}_{i}^{2}-(\vec{\kappa}_{i}\cdot {\hat{\eta}}%
_{ij})^{2})^{m-1} =  \nonumber \\
&=&\frac{[\vec{\kappa}_{i}\vec{\kappa}_{i}-(\vec{\kappa}_{i}\cdot {\hat{\eta}%
}_{ij})(\vec{\kappa}_{i}{\hat{\eta}}_{ij}+{\hat{\eta}}_{ij}\vec{\kappa}_{i})]%
}{\eta _{ij}(\vec{\kappa}_{i}^{2}-(\vec{\kappa}_{i}\cdot {\hat{\eta}}%
_{ij})^{2})\sqrt{m_{i}^{2}+\vec{\kappa}_{i}{}^{2}}}\sum_{m=1}^{\infty }\frac{%
(2m+1)!!(2m-1)!!}{(2m+2)!}2mx^{m}.  \label{d8}
\end{eqnarray}

Using

\begin{eqnarray}
&&\sum_{m=1}^{\infty }\frac{(2m+1)!!(2m-1)!!}{(2m+2)!}2mx^{m}=  \nonumber \\
&=&x\sum_{m=0}^{\infty }(-)^{m}x^{m}%
{-1/2 \choose m+2}%
\frac{2m+2}{2m+3} =  \nonumber \\
&=&x\sum_{m=0}^{\infty }(-)^{m}x^{m}%
{-1/2 \choose m+2}%
-x\sum_{m=0}^{\infty }(-)^{m}x^{m}%
{-1/2 \choose m+2}%
\frac{1}{2m+3}=  \nonumber \\
&=&\frac{1}{x}[\frac{1}{\sqrt{1-x}}-1-\frac{x}{2}-1+x/2+\sqrt{1-x}] =
\nonumber \\
&=&\frac{1}{x}[\frac{1}{\sqrt{1-x}}-2+\sqrt{1-x}].  \label{d9}
\end{eqnarray}

Eq.(\ref{d8}) becomes

\begin{eqnarray}
&=&\frac{[\vec{\kappa}_{i}\vec{\kappa}_{i}-(\vec{\kappa}_{i}\cdot {\hat{\eta}%
}_{ij})(\vec{\kappa}_{i}{\hat{\eta}}_{ij}+{\hat{\eta}}_{ij}\vec{\kappa}_{i})]%
\sqrt{m_{i}^{2}+\vec{\kappa}_{i}{}^{2}}}{\eta _{ij}(\vec{\kappa}_{i}^{2}-(%
\vec{\kappa}_{i}\cdot {\hat{\eta}}_{ij})^{2})^{2}}[\frac{\sqrt{m_{i}^{2}+(%
\vec{\kappa}_{i}{}\cdot \hat{\eta}_{ij})^{2}}}{\sqrt{m_{i}^{2}+\vec{\kappa}%
_{i}{}^{2}}}+\frac{\sqrt{m_{i}^{2}+\vec{\kappa}_{i}{}^{2}}}{\sqrt{m_{i}^{2}+(%
\vec{\kappa}_{i}{}\cdot \hat{\eta}_{ij})^{2}}}-2]=  \nonumber \\
&=&\frac{[\vec{\kappa}_{i}\vec{\kappa}_{i}-(\vec{\kappa}_{i}\cdot {\hat{\eta}%
}_{ij})(\vec{\kappa}_{i}{\hat{\eta}}_{ij}+{\hat{\eta}}_{ij}\vec{\kappa}_{i})]%
}{\eta _{ij}\sqrt{m_{i}^{2}+(\vec{\kappa}_{i}{}\cdot \hat{\eta}_{ij})^{2}}(%
\sqrt{m_{i}^{2}+\vec{\kappa}_{i}{}^{2}}+\sqrt{m_{i}^{2}+(\vec{\kappa}%
_{i}{}\cdot \hat{\eta}_{ij})^{2}})^{2}}.  \label{d10}
\end{eqnarray}

Finally our third part of the sum is

\begin{eqnarray}
&&-\frac{{\hat{\eta}}_{ij}{\hat{\eta}}_{ij}}{\eta _{ij}}\sum_{m=1}^{\infty }%
\frac{(2m+1)!!(2m-1)!!}{(2m+2)!\Big(\sqrt{m_{i}^{2}+\vec{\kappa}_{i}{}^{2}}%
)^{2m+1}}(\vec{\kappa}_{i}^{2}-(\vec{\kappa}_{i}\cdot {\hat{\eta}}_{ij})^{2}%
\Big)^{m-1}\Big(\vec{\kappa}_{i}^{2}-(2m+1)(\vec{\kappa}_{i}\cdot {\hat{\eta}%
}_{ij})^{2}\Big) =  \nonumber \\
&=&-\frac{{\hat{\eta}}_{ij}{\hat{\eta}}_{ij}}{\eta _{ij}}\frac{1}{\sqrt{%
m_{i}^{2}+\vec{\kappa}_{i}{}^{2}}(\vec{\kappa}_{i}^{2}-(\vec{\kappa}%
_{i}\cdot {\hat{\eta}}_{ij})^{2})}x  \nonumber \\
&&\Big[ \vec{\kappa}_{i}^{2}\sum_{m=0}^{\infty }\frac{(-)^{m}x^{m}}{2m+3}%
{-1/2 \choose m+2}%
-(\vec{\kappa}_{i}\cdot {\hat{\eta}}_{ij})^{2}\sum_{m=0}^{\infty
}(-)^{m}x^{m}%
{-1/2 \choose m+2}%
\Big].  \label{d11}
\end{eqnarray}

Using the results above we get

\begin{eqnarray}
&&x\Big[\vec{\kappa}_{i}^{2}\sum_{m=0}^{\infty }\frac{(-)^{m}x^{m}}{2m+3}%
{-1/2 \choose m+2}%
-(\vec{\kappa}_{i}\cdot {\hat{\eta}}_{ij})^{2}\sum_{m=0}^{\infty
}(-)^{m}x^{m}%
{-1/2 \choose m+2}%
\Big] =  \nonumber \\
&=&-\frac{(\vec{\kappa}_{i}\cdot {\hat{\eta}}_{ij})^{2}}{x}[\frac{1}{\sqrt{%
1-x}}-1-\frac{x}{2}]+\frac{\vec{\kappa}_{i}^{2}}{x}[1-x/2-\sqrt{1-x}] =
\nonumber \\
&=&\frac{\sqrt{m_{i}^{2}+\vec{\kappa}_{i}{}^{2}}-\sqrt{m_{i}^{2}+(\vec{\kappa%
}_{i}{}\cdot \hat{\eta}_{ij})^{2}}}{\sqrt{m_{i}^{2}+\vec{\kappa}_{i}{}^{2}}+%
\sqrt{m_{i}^{2}+(\vec{\kappa}_{i}{}\cdot \hat{\eta}_{ij})^{2}}}\Big[\frac{%
\vec{\kappa}_{i}^{2}}{2}-\frac{(\vec{\kappa}_{i}\cdot {\hat{\eta}}%
_{ij})^{2}(2\sqrt{m_{i}^{2}+\vec{\kappa}_{i}{}^{2}}+\sqrt{m_{i}^{2}+(\vec{%
\kappa}_{i}{}\cdot \hat{\eta}_{ij})^{2}})}{2\sqrt{m_{i}^{2}+(\vec{\kappa}%
_{i}{}\cdot \hat{\eta}_{ij})^{2}}}\Big],  \label{d12}
\end{eqnarray}

\noindent so that our third part becomes

\begin{eqnarray}
&&-\frac{{\hat{\eta}}_{ij}{\hat{\eta}}_{ij}}{\eta _{ij}}\sum_{m=1}^{\infty }%
\frac{(2m+1)!!(2m-1)!!}{(2m+2)!(\sqrt{m_{i}^{2}+\vec{\kappa}_{i}{}^{2}}%
)^{2m+1}}(\vec{\kappa}_{i}^{2}-(\vec{\kappa}_{i}\cdot {\hat{\eta}}%
_{ij})^{2})^{m-1}(\vec{\kappa}_{i}^{2}-(2m+1)(\vec{\kappa}_{i}\cdot {\hat{%
\eta}}_{ij})^{2})=  \nonumber \\
&=&-\frac{{\hat{\eta}}_{ij}{\hat{\eta}}_{ij}}{\eta _{ij}\sqrt{m_{i}^{2}+\vec{%
\kappa}_{i}{}^{2}}}\frac{1}{{\large (}\sqrt{m_{i}^{2}+\vec{\kappa}_{i}{}^{2}}%
+\sqrt{m_{i}^{2}+(\vec{\kappa}_{i}{}\cdot \hat{\eta}_{ij})^{2}}{\large )}^{2}%
}  \nonumber \\
&&\Big[ \frac{\vec{\kappa}_{i}^{2}}{2}-\frac{(\vec{\kappa}_{i}\cdot {\hat{%
\eta}}_{ij})^{2}(2\sqrt{m_{i}^{2}+\vec{\kappa}_{i}{}^{2}}+\sqrt{m_{i}^{2}+(%
\vec{\kappa}_{i}{}\cdot \hat{\eta}_{ij})^{2}})}{2\sqrt{m_{i}^{2}+(\vec{\kappa%
}_{i}{}\cdot \hat{\eta}_{ij})^{2}}}\Big]-  \nonumber \\
&=&-\frac{{\hat{\eta}}_{ij}{\hat{\eta}}_{ij}}{\eta _{ij}\sqrt{m_{i}^{2}+\vec{%
\kappa}_{i}{}^{2}}}\Big[\frac{\sqrt{m_{i}^{2}+\vec{\kappa}_{i}{}^{2}}-\sqrt{%
m_{i}^{2}+(\vec{\kappa}_{i}{}\cdot \hat{\eta}_{ij})^{2}}}{2{\large (}\sqrt{%
m_{i}^{2}+\vec{\kappa}_{i}{}^{2}}+\sqrt{m_{i}^{2}+(\vec{\kappa}_{i}{}\cdot
\hat{\eta}_{ij})^{2}}{\large )}} -  \nonumber \\
&&-\frac{(\vec{\kappa}_{i}\cdot {\hat{\eta}}_{ij})^{2}\sqrt{m_{i}^{2}+\vec{%
\kappa}_{i}{}^{2}}}{{\large (}\sqrt{m_{i}^{2}+\vec{\kappa}_{i}{}^{2}}+\sqrt{%
m_{i}^{2}+(\vec{\kappa}_{i}{}\cdot \hat{\eta}_{ij})^{2}}{\large )}^{2}\sqrt{%
m_{i}^{2}+(\vec{\kappa}_{i}{}\cdot \hat{\eta}_{ij})^{2}}}\Big].  \label{d13}
\end{eqnarray}

Combining these three parts of \ $U(\tau )$ gives us

\begin{eqnarray}
&&\sum_{i\neq j}^{N}{\frac{Q_{i}Q_{j}}{8\pi }}\Big[\frac{{\bf {\vec{k}}_{i}}%
}{\sqrt{m_{i}^{2}+\vec{\kappa}_{i}{}^{2}}}\cdot \frac{{\bf {\vec{k}}_{j}}}{%
\sqrt{m_{j}^{2}+\vec{\kappa}_{j}{}^{2}}}  \nonumber \\
&&{\large (}\sum_{m=0}^{\infty }\Big([(\frac{\vec{\kappa}_{i}}{\sqrt{%
m_{i}^{2}+\vec{\kappa}_{i}{}^{2}}}\cdot \vec{\partial}_{ij})^{2m}+(\frac{%
\vec{\kappa}_{j}}{\sqrt{m_{j}^{2}+\vec{\kappa}_{j}{}^{2}}}\cdot \vec{\partial%
}_{ij})^{2m}]\frac{\eta _{ij}^{2m-1}}{(2m)!}-\eta _{ij}^{-1}\Big){\large )} -
\nonumber \\
&&-(\frac{{\bf {\vec{k}}_{i}}}{\sqrt{m_{i}^{2}+\vec{\kappa}_{i}{}^{2}}}\cdot
\vec{\partial}_{ij})(\frac{{\bf {\vec{k}}_{j}}}{\sqrt{m_{j}^{2}+\vec{\kappa}%
_{j}{}^{2}}}\cdot \vec{\partial}_{ij}){\large (}\sum_{m=0}^{\infty }\Big([(%
\frac{\vec{\kappa}_{i}}{\sqrt{m_{i}^{2}+\vec{\kappa}_{i}{}^{2}}}\cdot \vec{%
\partial}_{ij})^{2m}+  \nonumber \\
&&+(\frac{\vec{\kappa}_{j}}{\sqrt{m_{j}^{2}+\vec{\kappa}_{j}{}^{2}}}\cdot
\vec{\partial}_{ij})^{2m}]\frac{\eta _{ij}^{2m+1}}{(2m+2)!}-\frac{1}{2}\eta
_{ij}\Big)\Big] =  \nonumber \\
&=&\sum_{i\neq j}^{N}{\frac{Q_{i}Q_{j}}{8\pi }}\Big[\frac{{\bf {\vec{k}}_{i}}%
}{\sqrt{m_{i}^{2}+\vec{\kappa}_{i}{}^{2}}}\cdot \frac{{\bf {\vec{k}}_{j}}}{%
\sqrt{m_{j}^{2}+\vec{\kappa}_{j}{}^{2}}}\frac{1}{\eta _{ij}}\Big(\frac{\sqrt{%
m_{i}^{2}+\vec{\kappa}_{i}{}^{2}}}{\sqrt{m_{i}^{2}+(\vec{\kappa}_{i}\cdot {%
\hat{\eta}}_{ij})^{2}}}+\frac{\sqrt{m_{j}^{2}+\vec{\kappa}_{j}{}^{2}}}{\sqrt{%
m_{j}^{2}+(\vec{\kappa}_{j}\cdot {\hat{\eta}}_{ij})^{2}}}-1\Big)-  \nonumber
\\
&&-\frac{{\bf {\vec{k}}_{i}}}{2\sqrt{m_{i}^{2}+\vec{\kappa}_{i}{}^{2}}}\cdot
\frac{{\bf {\vec{k}}_{j}}}{\sqrt{m_{j}^{2}+\vec{\kappa}_{j}{}^{2}}}\frac{1}{%
\eta _{ij}}+\frac{{\bf {\vec{k}}_{i}}}{2\sqrt{m_{i}^{2}+\vec{\kappa}%
_{i}{}^{2}}}\cdot (\frac{{\bf {\vec{k}}_{j}}}{\sqrt{m_{j}^{2}+\vec{\kappa}%
_{j}{}^{2}}}\,\cdot \frac{{\hat{\eta}}_{ij}}{\eta _{ij}}{\hat{\eta}}_{ij}) -
\nonumber \\
&&-\frac{{\bf {\vec{k}}_{i}\cdot {\vec{k}}_{j}}}{\sqrt{m_{i}^{2}+\vec{\kappa}%
_{i}{}^{2}}\sqrt{m_{j}^{2}+\vec{\kappa}_{j}{}^{2}}}\Big(\frac{1}{2\eta _{ij}}%
\frac{\sqrt{m_{i}^{2}+\vec{\kappa}_{i}{}^{2}}-\sqrt{m_{i}^{2}+(\vec{\kappa}%
_{i}{}\cdot \hat{\eta}_{ij})^{2}}}{\sqrt{m_{i}^{2}+\vec{\kappa}_{i}{}^{2}}+%
\sqrt{m_{i}^{2}+(\vec{\kappa}_{i}{}\cdot \hat{\eta}_{ij})^{2}}}+  \nonumber
\\
&&+\frac{1}{2\eta _{ij}(\sqrt{m_{j}^{2}+\vec{\kappa}_{j}{}^{2}})^{2}}\frac{%
\sqrt{m_{j}^{2}+\vec{\kappa}_{j}{}^{2}}-\sqrt{m_{j}^{2}+(\vec{\kappa}%
_{j}{}\cdot \hat{\eta}_{ij})^{2}}}{\sqrt{m_{j}^{2}+\vec{\kappa}_{j}{}^{2}}+
\sqrt{m_{j}^{2}+(\vec{\kappa}_{j}{}\cdot \hat{\eta}_{ij})^{2}}}\Big)-
\nonumber \\
&&-\frac{{\bf {\vec{k}}_{i}\cdot \Big({\vec{k}}_{j}\cdot }}{\sqrt{m_{j}^{2}+%
\vec{\kappa}_{j}{}^{2}}}\frac{[\vec{\kappa}_{i}\vec{\kappa}_{i}-(\vec{\kappa}%
_{i}\cdot {\hat{\eta}}_{ij})(\vec{\kappa}_{i}{\hat{\eta}}_{ij}+{\hat{\eta}}%
_{ij}\vec{\kappa}_{i})]}{\eta _{ij}\sqrt{m_{i}^{2}+(\vec{\kappa}_{i}{}\cdot
\hat{\eta}_{ij})^{2}}}\frac{1}{(\sqrt{m_{i}^{2}+\vec{\kappa}_{i}{}^{2}}+%
\sqrt{m_{i}^{2}+(\vec{\kappa}_{i}{}\cdot \hat{\eta}_{ij})^{2}})^{2}}\Big) -
\nonumber \\
&&-\frac{{\bf {\vec{k}}_{i}\cdot \Big({\vec{k}}_{j}\cdot }}{\sqrt{m_{i}^{2}+%
\vec{\kappa}_{i}{}^{2}}}\frac{[\vec{\kappa}_{j}\vec{\kappa}_{j}-(\vec{\kappa}%
_{j}\cdot {\hat{\eta}}_{ij})(\vec{\kappa}_{j}{\hat{\eta}}_{ij}+{\hat{\eta}}%
_{ij}\vec{\kappa}_{j})]}{\eta _{ij}\sqrt{m_{j}^{2}+(\vec{\kappa}_{j}{}\cdot
\hat{\eta}_{ij})^{2}}}\frac{1}{(\sqrt{m_{j}^{2}+\vec{\kappa}_{j}{}^{2}}+%
\sqrt{m_{j}^{2}+(\vec{\kappa}_{j}{}\cdot \hat{\eta}_{ij})^{2})^{2}}}\Big) +
\nonumber \\
&&+\frac{{\bf {\vec{k}}_{i}\cdot \Big({\vec{k}}_{j}\cdot }}{\sqrt{m_{i}^{2}+%
\vec{\kappa}_{i}{}^{2}}\sqrt{m_{j}^{2}+\vec{\kappa}_{j}{}^{2}}}\frac{{\hat{%
\eta}}_{ij}{\hat{\eta}}_{ij}}{\eta _{ij}}  \nonumber \\
&&\Big( \frac{\sqrt{m_{i}^{2}+\vec{\kappa}_{i}{}^{2}}-\sqrt{m_{i}^{2}+(\vec{%
\kappa}_{i}{}\cdot \hat{\eta}_{ij})^{2}}}{2{\large (}\sqrt{m_{i}^{2}+\vec{%
\kappa}_{i}{}^{2}}+\sqrt{m_{i}^{2}+(\vec{\kappa}_{i}{}\cdot \hat{\eta}%
_{ij})^{2}}{\large )}}-  \nonumber \\
&&-\frac{(\vec{\kappa}_{i}\cdot {\hat{\eta}}_{ij})^{2}\sqrt{m_{i}^{2}+\vec{%
\kappa}_{i}{}^{2}}}{{\large (}\sqrt{m_{i}^{2}+\vec{\kappa}_{i}{}^{2}}+\sqrt{%
m_{i}^{2}+(\vec{\kappa}_{i}{}\cdot \hat{\eta}_{ij})^{2}}{\large )}^{2}\sqrt{
m_{i}^{2}+(\vec{\kappa}_{i}{}\cdot \hat{\eta}_{ij})^{2}}}\Big)\Big) +
\nonumber \\
&&+\frac{{\bf {\vec{k}}_{i}\cdot \Big({\vec{k}}_{j}\cdot }}{\sqrt{m_{i}^{2}+%
\vec{\kappa}_{i}{}^{2}}\sqrt{m_{j}^{2}+\vec{\kappa}_{j}{}^{2}}}\frac{{\hat{%
\eta}}_{ij}{\hat{\eta}}_{ij}}{\eta _{ij}}  \nonumber \\
&&\Big( \frac{\sqrt{m_{j}^{2}+\vec{\kappa}_{j}{}^{2}}-\sqrt{m_{j}^{2}+(\vec{%
\kappa}_{j}{}\cdot \hat{\eta}_{ij})^{2}}}{2{\large (}\sqrt{m_{j}^{2}+\vec{%
\kappa}_{j}{}^{2}}+\sqrt{m_{j}^{2}+(\vec{\kappa}_{j}{}\cdot \hat{\eta}%
_{ij})^{2}}{\large )}}-  \nonumber \\
&&-\frac{(\vec{\kappa}_{j}\cdot {\hat{\eta}}_{ij})^{2}\sqrt{m_{j}^{2}+\vec{%
\kappa}_{j}{}^{2}}}{{\large (}\sqrt{m_{j}^{2}+\vec{\kappa}_{j}{}^{2}}+\sqrt{%
m_{j}^{2}+(\vec{\kappa}_{j}{}\cdot \hat{\eta}_{ij})^{2}}{\large )}^{2}\sqrt{%
m_{j}^{2}+(\vec{\kappa}_{j}{}\cdot \hat{\eta}_{ij})^{2}}}\Big)\Big)\Big].
\label{d14}
\end{eqnarray}

Rearrangement leads to

\begin{eqnarray}
&&\sum_{i\neq j}^{N}{\frac{Q_{i}Q_{j}}{8\pi }}\Big[\frac{{\bf {\vec{k}}%
_{i}\cdot {\vec{k}}_{j}}}{\sqrt{m_{i}^{2}+\vec{\kappa}_{i}{}^{2}}\sqrt{%
m_{j}^{2}+\vec{\kappa}_{j}{}^{2}}}\frac{1}{\eta _{ij}}\Big(\frac{\sqrt{%
m_{i}^{2}+\vec{\kappa}_{i}{}^{2}}}{\sqrt{m_{i}^{2}+(\vec{\kappa}_{i}\cdot {%
\hat{\eta}}_{ij})^{2}}}+\frac{\sqrt{m_{j}^{2}+\vec{\kappa}_{j}{}^{2}}}{\sqrt{%
m_{j}^{2}+(\vec{\kappa}_{j}\cdot {\hat{\eta}}_{ij})^{2}}}-1\Big)+  \nonumber
\\
&&+\frac{{\bf {\vec{k}}_{i}\cdot {\vec{k}}_{j}-{\vec{k}}_{i}\cdot ({\vec{k}}%
_{j}\cdot ({\hat{\eta}}_{ij}{\hat{\eta}}_{ij}))}}{2\sqrt{m_{i}^{2}+\vec{%
\kappa}_{i}{}^{2}}\sqrt{m_{j}^{2}+\vec{\kappa}_{j}{}^{2}}}\frac{1}{\eta _{ij}%
}-  \nonumber \\
&&-\frac{({\bf {\vec{k}}}_{i}\cdot {\bf {\vec{k}}}_{j}-{\bf {\vec{k}}}%
_{i}\cdot ({\bf {\vec{k}}}_{j}\cdot ({\hat{\eta}}_{ij}{\hat{\eta}}_{ij})))}{%
\sqrt{m_{j}^{2}+\vec{\kappa}_{j}{}^{2}}}\frac{1}{\eta _{ij}{\large (}\sqrt{%
m_{i}^{2}+\vec{\kappa}_{i}{}^{2}}+\sqrt{m_{i}^{2}+(\vec{\kappa}_{i}{}\cdot
\hat{\eta}_{ij})^{2}}{\large )}}-  \nonumber \\
&&-\frac{({\bf {\vec{k}}}_{i}\cdot {\bf {\vec{k}}}_{j}-{\bf {\vec{k}}}%
_{i}\cdot ({\bf {\vec{k}}}_{j}\cdot ({\hat{\eta}}_{ij}{\hat{\eta}}_{ij})))}{%
\sqrt{m_{i}^{2}+\vec{\kappa}_{i}{}^{2}}}\frac{1}{\eta _{ij}{\large (}\sqrt{%
m_{j}^{2}+\vec{\kappa}_{j}{}^{2}}+\sqrt{m_{j}^{2}+(\vec{\kappa}_{j}{}\cdot
\hat{\eta}_{ij})^{2}}{\large )}}-  \nonumber \\
&&-\Big(({\bf {\vec{k}}}_{i}\cdot \vec{\kappa}_{i})({\bf {\vec{k}}}_{j}\cdot
\vec{\kappa}_{i}-({\bf {\vec{k}}}_{j}\cdot {\hat{\eta}}_{ij})(\vec{\kappa}%
_{i}\cdot {\hat{\eta}}_{ij}))-{\bf {\vec{k}}}_{j}\cdot \lbrack (\hat{\vec{k}}%
_{i}\cdot {\hat{\eta}}_{ij})(\vec{\kappa}_{i}-{\hat{\eta}}_{ij}(\vec{\kappa}%
_{i}\cdot {\hat{\eta}}_{ij}))](\vec{\kappa}_{i}\cdot {\hat{\eta}}_{ij})\Big)
\nonumber \\
&& \frac{1}{\eta _{ij}\sqrt{m_{j}^{2}+\vec{\kappa}_{j}{}^{2}}\sqrt{%
m_{i}^{2}+(\vec{\kappa}_{i}{}\cdot \hat{\eta}_{ij})^{2}}}\frac{1}{(\sqrt{%
m_{i}^{2}+\vec{\kappa}_{i}{}^{2}}+\sqrt{m_{i}^{2}+(\vec{\kappa}_{i}{}\cdot
\hat{\eta}_{ij})^{2}})^{2}}-  \nonumber \\
&&-\Big(({\bf {\vec{k}}}_{j}\cdot \vec{\kappa}_{j})({\bf {\vec{k}}}_{i}\cdot
\vec{\kappa}_{j}-({\bf {\vec{k}}}_{i}\cdot {\hat{\eta}}_{ij})(\vec{\kappa}%
_{j}\cdot {\hat{\eta}}_{ij}))-{\bf {\vec{k}}}_{i}\cdot \lbrack (\hat{\vec{k}}%
_{j}\cdot {\hat{\eta}}_{ij})(\vec{\kappa}_{j}-{\hat{\eta}}_{ij}(\vec{\kappa}%
_{j}\cdot {\hat{\eta}}_{ij}))](\vec{\kappa}_{j}\cdot {\hat{\eta}}_{ij})\Big)
\nonumber \\
&& \frac{1}{\eta _{ij}\sqrt{m_{i}^{2}+\vec{\kappa}_{i}{}^{2}}\sqrt{%
m_{j}^{2}+(\vec{\kappa}_{j}{}\cdot \hat{\eta}_{ij})^{2}}}\frac{1}{(\sqrt{%
m_{j}^{2}+\vec{\kappa}_{j}{}^{2}}+\sqrt{m_{j}^{2}+(\vec{\kappa}_{j}{}\cdot
\hat{\eta}_{ij})^{2})^{2}}}\Big].  \label{d15}
\end{eqnarray}
Note that the double dyadic dot products are defined such that the right
most $k^{\prime }s$ are contracted with the right most $\hat{\eta}^{\prime
}s $.

Now we combine this with the expression (\ref{VII7}) for the corresponding
portion of $U_{1}$ (involving ${\vec{A}}_{\perp }$). First note that

\begin{eqnarray}
&&-\sum_{i=1}^{N}\frac{Q_{i}\vec{\kappa}_{i}(\tau )\cdot {\vec{A}}_{\perp
}(\tau ,\vec{\eta}_{i}(\tau ))}{\sqrt{m_{i}^{2}+\check{\vec{\kappa}}%
_{i}(\tau )^{2}}} =  \nonumber \\
&=&-\sum_{i\neq j}^{N}{\frac{Q_{j}Q_{i}}{4\pi \sqrt{m_{i}^{2}+\vec{\kappa}%
_{i}{}^{2}}}\Big[\Big(}\vec{\kappa}_{i}{+\frac{i\vec{\kappa}_{j}\cdot \vec{%
\xi}_{j}\vec{\kappa}_{i}\cdot \vec{\xi}_{j}\vec{\partial}_{ij}}{\sqrt{%
m_{j}^{2}+\vec{\kappa}_{j}^{2}}(m_j+\sqrt{m_{j}^{2}+\vec{\kappa}_{j}^{2}})}%
\Big)\cdot \lbrack }\vec{\kappa}_{j}+\hat{\eta}_{ij}\vec{\kappa}_{j}\cdot
\hat{\eta}_{ij}]  \nonumber \\
&& {\frac{1}{|\vec{\eta}_{i}-\vec{\eta}_{j}|(\sqrt{m_{j}^{2}+{\vec{\kappa}%
_{j}}^{2}}+\sqrt{m_{j}^{2}+(\vec{\kappa}_{j}\cdot {\hat{\eta}}_{ij})^{2}})}-%
\vec{\kappa}_{i}\cdot \vec{\xi}_{j}\vec{\xi}_{j}\cdot \vec{\partial}_{ij}}%
\frac{1}{|\vec{\eta}_{i}-\vec{\eta}_{j}|\sqrt{m_{j}^{2}+(\vec{\kappa}%
_{j}\cdot {\hat{\eta}}_{ij})^{2}}}\Big]=  \nonumber \\
&&{}  \nonumber \\
&=&-\sum_{i\neq j}^{N}{\frac{Q_{j}Q_{i}}{4\pi }\Big[ }\vec{\kappa}_{i}\cdot
\vec{ k}_{j}\frac{1}{\eta _{ij}\sqrt{m_{i}^{2}+\vec{\kappa}_{i}{}^{2}}\sqrt{
m_{j}^{2}+(\vec{\kappa}_{j}\cdot {\hat{\eta}}_{ij})^{2}}}-  \nonumber \\
&&-\vec{\kappa}_{i}\cdot \Big( {\bf {\vec{l}}}_{j}\cdot ({\bf I}-{\hat{\eta }%
}_{ij}{\hat{\eta}}_{ij})\frac{1}{\eta _{ij}\sqrt{m_{i}^{2}+\vec{\kappa}%
_{i}{}^{2}}\sqrt{m_{j}^{2}+(\vec{\kappa}_{j}\cdot {\hat{\eta}}_{ij})^{2}}}({%
\frac{\sqrt{m_{j}^{2}+\vec{\kappa}_{j}^{2}}}{\sqrt{m_{j}^{2}+{\vec{\kappa}%
_{j}}^{2}}+\sqrt{m_{j}^{2}+(\vec{\kappa}_{j}\cdot {\hat{\eta}}_{ij})^{2}}})%
\Big)\Big]},  \nonumber \\
&&{}  \label{d16}
\end{eqnarray}

\noindent or

\begin{eqnarray}
&=&-\sum_{i\neq j}^{N}{\frac{Q_{j}Q_{i}}{8\pi }\Big[}\frac{\vec{\kappa}%
_{i}\cdot {\bf {\vec{k}}}_{j}}{\sqrt{m_{i}^{2}+\vec{\kappa}_{i}{}^{2}}}\frac{%
1}{\eta _{ij}\sqrt{m_{j}^{2}+(\vec{\kappa}_{j}\cdot {\hat{\eta}}_{ij})^{2}}}
+  \nonumber \\
&&+\frac{\vec{\kappa}_{j}\cdot {\bf {\vec{k}}}_{i}}{\sqrt{m_{j}^{2}+{\vec{%
\kappa}_{j}}^{2}}}\frac{1}{\eta _{ij}\sqrt{m_{i}^{2}+(\vec{\kappa}_{i}\cdot {%
\hat{\eta}}_{ij})^{2}}} -  \nonumber \\
&&-[\vec{\kappa}_{i}\cdot {\bf {\vec{l}}}_{j}-\vec{\kappa}_{i}\cdot (\hat{%
\vec{l}}_{j}\cdot ({\hat{\eta}}_{ij}{\hat{\eta}}_{ij}))]  \nonumber \\
&& \frac{1}{\eta _{ij}\sqrt{m_{i}^{2}+\vec{\kappa}_{i}{}^{2}}\sqrt{%
m_{j}^{2}+(\vec{\kappa}_{j}\cdot {\hat{\eta}}_{ij})^{2}}}({\frac{\sqrt{%
m_{j}^{2}+\vec{\kappa}_{j}^{2}}}{\sqrt{m_{j}^{2}+{\vec{\kappa}_{j}}^{2}}+
\sqrt{m_{j}^{2}+(\vec{\kappa}_{j}\cdot {\hat{\eta}}_{ij})^{2}}})}-  \nonumber
\\
&&-[\vec{\kappa}_{j}\cdot {\bf {\vec{l}}}_{i}-\vec{\kappa}_{j}\cdot (\hat{%
\vec{l}}_{i}\cdot ({\hat{\eta}}_{ij}{\hat{\eta}}_{ij}))]  \nonumber \\
&& \frac{1}{\eta _{ij}\sqrt{m_{j}^{2}+{\vec{\kappa}_{j}}^{2}}\sqrt{%
m_{i}^{2}+(\vec{\kappa}_{i}\cdot {\hat{\eta}}_{ij})^{2}}}({\frac{\sqrt{%
m_{i}^{2}+\vec{\kappa}_{i}^{2}}}{\sqrt{m_{i}^{2}+{\vec{\kappa}_{i}}^{2}}+%
\sqrt{m_{i}^{2}+(\vec{\kappa}_{i}\cdot {\hat{\eta}}_{ij})^{2}}})\Big]}.
\label{d17}
\end{eqnarray}

Now combine Eq.(\ref{d17}) with the above term Eq.(\ref{d15}) from $U(\tau )$%
. \ In order to clarify the cancellations that will take place consider the
case without spin for which the spin operator containing terms ${\bf {\vec{k}%
}}_{i}$ and ${\bf {\vec{l}}}_{i}$ reduce to the simple momentum factor $\vec{%
\kappa}_{i}$. In that case Eq.(\ref{d15}) becomes

\begin{eqnarray}
&=&\sum_{i\neq j}^{N}{\frac{Q_{i}Q_{j}}{8\pi }}\Big[\frac{\vec{\kappa}%
_{i}\cdot \vec{\kappa}_{j}}{\sqrt{m_{i}^{2}+\vec{\kappa}_{i}{}^{2}}\sqrt{%
m_{j}^{2}+\vec{\kappa}_{j}{}^{2}}}\frac{1}{\eta _{ij}}\Big(\frac{\sqrt{%
m_{i}^{2}+\vec{\kappa}_{i}{}^{2}}}{\sqrt{m_{i}^{2}+(\vec{\kappa}_{i}\cdot {%
\hat{\eta}}_{ij})^{2}}}+\frac{\sqrt{m_{j}^{2}+\vec{\kappa}_{j}{}^{2}}}{\sqrt{%
m_{j}^{2}+(\vec{\kappa}_{j}\cdot {\hat{\eta}}_{ij})^{2}}}-1\Big) +  \nonumber
\\
&&+\frac{\vec{\kappa}_{i}\cdot \vec{\kappa}_{j}-\vec{\kappa}_{i}\cdot {\hat{%
\eta}}_{ij}\vec{\kappa}_{j}\cdot {\hat{\eta}}_{ij}}{\eta _{ij}}\Big(\frac{1}{%
2\sqrt{m_{i}^{2}+\vec{\kappa}_{i}{}^{2}}\sqrt{m_{j}^{2}+\vec{\kappa}%
_{j}{}^{2}}} -  \nonumber \\
&&-\frac{\sqrt{m_{i}^{2}+\vec{\kappa}_{i}{}^{2}}}{\sqrt{m_{j}^{2}+\vec{\kappa%
}_{j}{}^{2}}\sqrt{m_{i}^{2}+(\vec{\kappa}_{i}{}\cdot \hat{\eta}_{ij})^{2}}(%
\sqrt{m_{i}^{2}+\vec{\kappa}_{i}{}^{2}}+\sqrt{m_{i}^{2}+(\vec{\kappa}%
_{i}{}\cdot \hat{\eta}_{ij})^{2}})}-  \nonumber \\
&&-\frac{\sqrt{m_{j}^{2}+\vec{\kappa}_{j}{}^{2}}}{\sqrt{m_{i}^{2}+\vec{\kappa%
}_{i}{}^{2}}\sqrt{m_{j}^{2}+(\vec{\kappa}_{j}{}\cdot \hat{\eta}_{ij})^{2}}(%
\sqrt{m_{j}^{2}+\vec{\kappa}_{j}{}^{2}}+\sqrt{m_{j}^{2}+(\vec{\kappa}%
_{j}{}\cdot \hat{\eta}_{ij})^{2}})}\Big) \Big],  \label{d18}
\end{eqnarray}

\noindent while\ the corresponding portion of $U_{1}(\tau )$ becomes\

\begin{eqnarray}
&=&-\sum_{i\neq j}^{N}{\frac{Q_{j}Q_{i}}{8\pi }\Big[}\frac{\vec{\kappa}%
_{i}\cdot \vec{\kappa}_{j}}{\sqrt{m_{j}^{2}+{\vec{\kappa}_{j}}^{2}}}\frac{1}{%
\eta _{ij}\sqrt{m_{j}^{2}+(\vec{\kappa}_{j}\cdot {\hat{\eta}}_{ij})^{2}}} +
\nonumber \\
&&+\frac{\vec{\kappa}_{j}\cdot \vec{\kappa}_{j}}{\sqrt{m_{i}^{2}+\vec{\kappa}%
_{i}{}^{2}}}\frac{1}{\eta _{ij}\sqrt{m_{i}^{2}+(\vec{\kappa}_{i}\cdot {\hat{%
\eta}}_{ij})^{2}}}-  \nonumber \\
&&-[\vec{\kappa}_{i}\cdot \vec{\kappa}_{j}-\vec{\kappa}_{i}\cdot {\hat{\eta}}%
_{ij}\vec{\kappa}_{j}\cdot {\hat{\eta}}_{ij}]\frac{1}{\eta _{ij}\sqrt{%
m_{i}^{2}+\vec{\kappa}_{i}^{2}}\sqrt{m_{j}^{2}+(\vec{\kappa}_{j}\cdot {\hat{%
\eta}}_{ij})^{2}}}  \nonumber \\
&&({\frac{\sqrt{m_{j}^{2}+\vec{\kappa}_{j}^{2}}}{\sqrt{m_{j}^{2}+{\vec{\kappa%
}_{j}}^{2}}+\sqrt{m_{j}^{2}+(\vec{\kappa}_{j}\cdot {\hat{\eta}}_{ij})^{2}}})
} -  \nonumber \\
&&-[\vec{\kappa}_{i}\cdot \vec{\kappa}_{j}-\vec{\kappa}_{i}\cdot {\hat{\eta}}%
_{ij}\vec{\kappa}_{j}\cdot {\hat{\eta}}_{ij}]\frac{1}{\eta _{ij}\sqrt{%
m_{j}^{2}+{\vec{\kappa}_{j}}^{2}}\sqrt{m_{i}^{2}+(\vec{\kappa}_{i}\cdot {%
\hat{\eta}}_{ij})^{2}}}  \nonumber \\
&&({\frac{\sqrt{m_{i}^{2}+\vec{\kappa}_{i}^{2}}}{\sqrt{m_{i}^{2}+{\vec{\kappa%
}_{i}}^{2}}+\sqrt{m_{i}^{2}+(\vec{\kappa}_{i}\cdot {\hat{\eta}}_{ij})^{2}}})%
\Big]} .  \nonumber \\
&&{}  \label{d19}
\end{eqnarray}

Adding the two gives

\begin{equation}
V_{LDO}^{(spinless)}(\tau ) =-\sum_{i\neq j}^{N}{\frac{Q_{i}Q_{j}}{16\pi }}%
\frac{\vec{\kappa}_{i}\cdot \vec{\kappa}_{j}+\vec{\kappa}_{i}\cdot {\hat{\eta%
}}_{ij}\vec{\kappa}_{j}\cdot {\hat{\eta}}_{ij}}{\sqrt{m_{i}^{2}+\vec{\kappa}%
_{i}{}^{2}}\sqrt{m_{j}^{2}+\vec{\kappa}_{j}{}^{2}}}\,\,\, \frac{1}{\eta _{ij}%
},  \label{d20}
\end{equation}

\noindent which is our relativistic extension of the standard Darwin
interaction for spinless particles \cite{ap}

\bigskip Now that we have checked our results without spin, we combine the
spin-dependent terms of Eqs.(\ref{d15}), (\ref{d17}) to get (${\bf {\vec{k}}}%
:=\vec{\kappa}+{\bf {\vec{f}}};{\bf {\vec{l}}}:=\vec{\kappa}+{\bf {\vec{g}}}$%
)

\begin{eqnarray}
&&\sum_{i\neq j}^{N}{\frac{Q_{i}Q_{j}}{8\pi }}\Big[-\frac{\vec{\kappa}%
_{i}\cdot \vec{\kappa}_{j}+\vec{\kappa}_{i}\cdot {\hat{\eta}}_{ij}\vec{\kappa%
}_{j}\cdot {\hat{\eta}}_{ij}}{2\sqrt{m_{i}^{2}+\vec{\kappa}_{i}{}^{2}}\sqrt{%
m_{j}^{2}+\vec{\kappa}_{j}{}^{2}}}\frac{1}{\eta _{ij}} +  \nonumber \\
&&+\Big(\frac{{\bf {\vec{f}}_{i}\cdot (\vec{\kappa}_{j}+{\vec{f}}_{j})}}{%
\sqrt{m_{j}^{2}+\vec{\kappa}_{j}{}^{2}}\eta _{ij}\sqrt{m_{i}^{2}+(\vec{\kappa%
}_{i}\cdot {\hat{\eta}}_{ij})^{2}}}+\frac{{\bf {\vec{f}}_{j}\cdot (\vec{%
\kappa}_{i}+{\vec{f}}_{i})}}{\sqrt{m_{i}^{2}+\vec{\kappa}_{i}{}^{2}}\eta
_{ij}\sqrt{m_{j}^{2}+(\vec{\kappa}_{j}\cdot {\hat{\eta}}_{ij})^{2}}} -
\nonumber \\
&&-\frac{\vec{\kappa}_{i}\cdot {\bf {\vec{f}}}_{j}+\vec{\kappa}_{j}\cdot
\hat{\vec{f}}_{i}+{\bf {\vec{f}}}_{i}\cdot {\bf {\vec{f}}}_{j}}{\sqrt{%
m_{i}^{2}+\vec{\kappa}_{i}{}^{2}}\sqrt{m_{j}^{2}+\vec{\kappa}_{j}{}^{2}}}%
\frac{1}{\eta _{ij}}\Big) +  \nonumber \\
&&+\Big(\vec{\kappa}_{i}\cdot {\bf {\vec{f}}}_{j}+\vec{\kappa}_{j}\cdot \hat{%
\vec{f}}_{i}+{\bf {\vec{f}}}_{i}\cdot {\bf {\vec{f}}}_{j}-({\bf {\vec{f}}}%
_{j}\vec{\kappa}_{i}\cdot {\hat{\eta}}_{ij}+{\bf {\vec{f}}}_{i}\vec{\kappa}%
_{j}\cdot {\hat{\eta}}_{ij}+{\bf {\vec{f}}}_{j}{\bf {\vec{f}}}_{i}\cdot {%
\hat{\eta}}_{ij})\cdot {\hat{\eta}}_{ij}\Big)  \nonumber \\
&&\Big(\frac{1}{2\sqrt{m_{i}^{2}+\vec{\kappa}_{i}{}^{2}}\sqrt{m_{j}^{2}+\vec{%
\kappa}_{j}{}^{2}}\eta _{ij}}-  \nonumber \\
&&-\frac{1}{\eta _{ij}\sqrt{m_{j}^{2}+\vec{\kappa}_{j}{}^{2}}{\large (}\sqrt{%
m_{i}^{2}+\vec{\kappa}_{i}{}^{2}}+\sqrt{m_{i}^{2}+(\vec{\kappa}_{i}{}\cdot
\hat{\eta}_{ij})^{2}}{\large )}}-  \nonumber \\
&&-\frac{1}{\sqrt{m_{i}^{2}+\vec{\kappa}_{i}{}^{2}}\eta _{ij}{\large (}\sqrt{%
m_{j}^{2}+\vec{\kappa}_{j}{}^{2}}+\sqrt{m_{j}^{2}+(\vec{\kappa}_{j}{}\cdot
\hat{\eta}_{ij})^{2}}{\large )}}\Big)-  \nonumber \\
&&-\Big((\vec{\kappa}_{i}\cdot \vec{\kappa}_{i})({\bf {\vec{f}}}_{j}\cdot
\vec{\kappa}_{i}-({\bf {\vec{f}}}_{j}\cdot {\hat{\eta}}_{ij})(\vec{\kappa}%
_{i}\cdot {\hat{\eta}}_{ij}))-{\bf {\vec{f}}}_{j}\cdot [ (\vec{\kappa}%
_{i}\cdot {\hat{\eta}}_{ij})(\vec{\kappa}_{i}-{\hat{\eta}}_{ij}(\vec{\kappa}%
_{i}\cdot {\hat{\eta}}_{ij}))](\vec{\kappa}_{i}\cdot {\hat{\eta}}_{ij}) +
\nonumber \\
&&+({\bf {\vec{f}}}_{i}\cdot \vec{\kappa}_{i})(\vec{\kappa}_{j}\cdot \vec{%
\kappa}_{i}-(\vec{\kappa}_{j}\cdot {\hat{\eta}}_{ij})(\vec{\kappa}_{i}\cdot {%
\hat{\eta}}_{ij}))-\vec{\kappa}_{j}\cdot [ ({\bf {\vec{f}}}_{i}\cdot {\hat{%
\eta}}_{ij})(\vec{\kappa}_{i}-{\hat{\eta}}_{ij}(\vec{\kappa}_{i}\cdot {\hat{%
\eta}}_{ij}))](\vec{\kappa}_{i}\cdot {\hat{\eta}}_{ij})+  \nonumber \\
&&+({\bf {\vec{f}}}_{i}\cdot \vec{\kappa}_{i})({\bf {\vec{f}}}_{j}\cdot \vec{%
\kappa}_{i}-({\bf {\vec{f}}}_{j}\cdot {\hat{\eta}}_{ij})(\vec{\kappa}%
_{i}\cdot {\hat{\eta}}_{ij}))-{\bf {\vec{f}}}_{j}\cdot [ ({\bf {\vec{f}}}%
_{i}\cdot {\hat{\eta}}_{ij})(\vec{\kappa}_{i}-{\hat{\eta}}_{ij}(\vec{\kappa}%
_{i}\cdot {\hat{\eta}}_{ij}))](\vec{\kappa}_{i}\cdot {\hat{\eta}}_{ij})\Big)
\nonumber \\
&& \frac{1}{\eta _{ij}\sqrt{m_{j}^{2}+\vec{\kappa}_{j}{}^{2}}\sqrt{%
m_{i}^{2}+(\vec{\kappa}_{i}{}\cdot \hat{\eta}_{ij})^{2}}}\frac{1}{(\sqrt{%
m_{i}^{2}+\vec{\kappa}_{i}{}^{2}}+\sqrt{m_{i}^{2}+(\vec{\kappa}_{i}{}\cdot
\hat{\eta}_{ij})^{2}})^{2}} -  \nonumber \\
&&-\Big((\vec{\kappa}_{j}\cdot \vec{\kappa}_{j})({\bf {\vec{f}}}_{i}\cdot
\vec{\kappa}_{j}-({\bf {\vec{f}}}_{i}\cdot {\hat{\eta}}_{ij})(\vec{\kappa}%
_{j}\cdot {\hat{\eta}}_{ij}))-{\bf {\vec{f}}}_{i}\cdot [ (\vec{\kappa}%
_{j}\cdot {\hat{\eta}}_{ij})(\vec{\kappa}_{j}-{\hat{\eta}}_{ij}(\vec{\kappa}%
_{j}\cdot {\hat{\eta}}_{ij}))](\vec{\kappa}_{j}\cdot {\hat{\eta}}_{ij}) +
\nonumber \\
&&+{\large (}{\bf {\vec{f}}}_{j}\cdot \vec{\kappa}_{j})(\vec{\kappa}%
_{i}\cdot \vec{\kappa}_{j}-(\vec{\kappa}_{i}\cdot {\hat{\eta}}_{ij})(\vec{%
\kappa}_{j}\cdot {\hat{\eta}}_{ij}))-\vec{\kappa}_{i}\cdot [ ({\bf {\vec{f}}}%
_{j}\cdot {\hat{\eta}}_{ij})(\vec{\kappa}_{j}-{\hat{\eta}}_{ij}(\vec{\kappa}%
_{j}\cdot {\hat{\eta}}_{ij}))](\vec{\kappa}_{j}\cdot {\hat{\eta}}_{ij}) +
\nonumber \\
&&+{\large (}{\bf {\vec{f}}}_{j}\cdot \vec{\kappa}_{j})({\bf {\vec{f}}}%
_{i}\cdot \vec{\kappa}_{j}-({\bf {\vec{f}}}_{i}\cdot {\hat{\eta}}_{ij})(\vec{%
\kappa}_{j}\cdot {\hat{\eta}}_{ij}))-{\bf {\vec{f}}}_{i}\cdot [ ({\bf {\vec{f%
}}}_{j}\cdot {\hat{\eta}}_{ij})(\vec{\kappa}_{j}-{\hat{\eta}}_{ij}(\vec{%
\kappa}_{j}\cdot {\hat{\eta}}_{ij}))](\vec{\kappa}_{j}\cdot {\hat{\eta}}%
_{ij})\Big)  \nonumber \\
&& \frac{1}{\eta _{ij}\sqrt{m_{i}^{2}+\vec{\kappa}_{i}{}^{2}}\sqrt{%
m_{j}^{2}+(\vec{\kappa}_{j}{}\cdot \hat{\eta}_{ij})^{2}}}\frac{1}{(\sqrt{%
m_{j}^{2}+\vec{\kappa}_{j}{}^{2}}+\sqrt{m_{j}^{2}+(\vec{\kappa}_{j}{}\cdot
\hat{\eta}_{ij})^{2})^{2}}}+  \nonumber \\
&&+\Big(\vec{\kappa}_{i}\cdot {\bf {\vec{g}}}_{j}-\vec{\kappa}_{i}\cdot (%
\hat{\vec{g}}_{j}\cdot ({\hat{\eta}}_{ij}{\hat{\eta}}_{ij}))\Big)  \nonumber
\\
&& \frac{1}{\eta _{ij}\sqrt{m_{i}^{2}+{\vec{\kappa}_{i}}^{2}}\sqrt{%
m_{j}^{2}+(\vec{\kappa}_{j}\cdot {\hat{\eta}}_{ij})^{2}}}({\frac{\sqrt{%
m_{j}^{2}+\vec{\kappa}_{j}^{2}}}{\sqrt{m_{j}^{2}+{\vec{\kappa}_{j}}^{2}}+
\sqrt{m_{j}^{2}+(\vec{\kappa}_{j}\cdot {\hat{\eta}}_{ij})^{2}}})} +
\nonumber \\
&&+\Big(\vec{\kappa}_{j}\cdot {\bf {\vec{g}}}_{i}-\vec{\kappa}_{j}\cdot (%
\hat{\vec{g}}_{i}\cdot ({\hat{\eta}}_{ij}{\hat{\eta}}_{ij}))\Big)  \nonumber
\\
&& \frac{1}{\eta _{ij}\sqrt{m_{j}^{2}+{\vec{\kappa}_{j}}^{2}}\sqrt{%
m_{i}^{2}+(\vec{\kappa}_{i}\cdot {\hat{\eta}}_{ij})^{2}}}({\frac{\sqrt{%
m_{i}^{2}+\vec{\kappa}_{i}^{2}}}{\sqrt{m_{i}^{2}+{\vec{\kappa}_{i}}^{2}}+%
\sqrt{m_{i}^{2}+(\vec{\kappa}_{i}\cdot {\hat{\eta}}_{ij})^{2}}})}\Big],
\label{d21}
\end{eqnarray}

\noindent in which

\begin{eqnarray}
{\bf {\vec{f}}}_{i} &=&-i\vec{\xi}_{i}\vec{\xi}_{i}\cdot \vec{\partial}_{ij}+%
\frac{i\vec{\kappa}_{i}\cdot \vec{\partial}_{ij}\vec{\kappa}_{i}\cdot \vec{%
\xi}_{i}\vec{\xi}_{i}}{(m_{i}+\sqrt{m_{i}^{2}+\vec{\kappa}_{i}{}^{2}})\sqrt{%
m_{i}^{2}+\vec{\kappa}_{i}{}^{2}}},  \nonumber \\
{\bf {\vec{g}}}_{i} &=&\frac{i\vec{\kappa}_{i}\cdot \vec{\partial}_{ij}\vec{%
\kappa}_{i}\cdot \vec{\xi}_{i}\vec{\xi}_{i}}{(m_{i}+\sqrt{m_{i}^{2}+\vec{%
\kappa}_{i}{}^{2}})\sqrt{m_{i}^{2}+\vec{\kappa}_{i}{}^{2}}}.  \label{d22}
\end{eqnarray}

To this we must add the remaining parts of $U_{1}(\tau )$ and $U(\tau )$. To
complete $U_{1}(\tau )$ we need the expression for ${\cal R}_{i}$ in Eq.(\ref
{VII4}). The magnetic portion (\ref{VII8}) is

\begin{equation}
\sum_{i=1}^{N}\frac{iQ_{i}\vec{\xi}_{i}\times \vec{\xi}_{i}\cdot \check{\vec{%
B}}(\tau ,\vec{\eta}_{i})}{2\sqrt{m_{i}^{2}+\check{\vec{\kappa}}_{i}^{2}}}%
=-\sum_{i\neq j}^{N}{\frac{iQ_{i}Q_{j}\vec{\xi}_{i}\times \vec{\xi}_{i}\cdot
({\bf {\vec{k}}}_{j}\times \vec{\partial}_{ij})}{8\pi \sqrt{m_{i}^{2}+\vec{%
\kappa}_{i}{}^{2}}}}\Big({\frac{1}{|\vec{\eta}_{i}-\vec{\eta}_{j}|}}\frac{1}{%
\sqrt{m_{j}^{2}+(\vec{\kappa}_{j}\cdot {\hat{\eta}}_{ij})^{2}}}\Big),
\label{d23}
\end{equation}

\noindent while the transverse electric part (\ref{VII9}) is

\begin{eqnarray}
&&-i\sum_{i=1}^{N}\frac{Q_{i}\check{\vec{\kappa}}_{i}\cdot \vec{\xi}_{i}\vec{%
\xi}_{i}\cdot \check{\vec{\pi}}_{\perp }(\tau ,\vec{\eta}_{i})}{\sqrt{%
m_{i}^{2}+\check{\vec{\kappa}}_{i}{}^{2}}(\sqrt{m_{i}^{2}+\check{\vec{\kappa}%
}_{i}^{2}}+m_{i})}=  \nonumber \\
&=&i\sum_{i\neq j}^{N}{\frac{Q_{i}Q_{j}\vec{\kappa}_{i}\cdot \vec{\xi}_{i}}{%
4\pi \sqrt{m_{i}^{2}+\vec{\kappa}_{i}{}^{2}}(\sqrt{m_{i}^{2}+\vec{\kappa}%
_{i}^{2}}+m_{i})}}\Big[\Big( \vec{\xi}_{i}{+\frac{i\vec{\kappa}_{j}\cdot
\vec{\xi}_{j}\vec{\xi}_{i}\cdot \vec{\xi}_{j}\vec{\partial}_{ij}}{\sqrt{%
m_{j}^{2}+\vec{\kappa}_{j}^{2}}(\sqrt{m_{j}^{2}+\vec{\kappa}_{j}^{2}}+m_{j})}%
}\Big)\cdot \frac{\hat{\eta}_{ij}}{\eta _{ij}^{2}}  \nonumber \\
&&\times {(}\frac{m_{j}^{2}\sqrt{m_{j}^{2}+\vec{\kappa}_{j}^{2}}}{%
[m_{j}^{2}+(\vec{\kappa}_{j}\cdot \hat{\eta}_{ij})^{2}]^{3/2}}-1)-{i\vec{\xi}
}_{i}\cdot {\vec{\xi}_{j}\vec{\xi}_{j}\cdot \vec{\partial}_{ij}\vec{\kappa}%
_{j}\cdot \vec{\partial}}_{ij}\frac{1}{\eta _{ij}}\frac{1}{\sqrt{m_{j}^{2}+(%
\vec{\kappa}_{j}\cdot \hat{\eta}_{ij})^{2}}}\Big].  \label{d24}
\end{eqnarray}

Combining Eq.(\ref{d23}) and Eq.(\ref{d24}) with Eq.(\ref{d21}), with Eq.(%
\ref{VII5}) for $V_{DSO}$ and with the first four lines of Eq.(\ref{d1})\
produces Eq.(\ref{VII11}).

\vfill\eject

\bigskip

\section{Summation of Two-Body Rest Energy to Closed Form}

\bigskip

\ In the special case of the two body system we \ can obtain a closed form
if we use the rest frame condition $\widetilde{\vec{\kappa}}_{1}+\widetilde{%
\vec{\kappa}}_{2}=0$ . \ The expression we get in this way may be used with
the Dirac brackets associated with $\widetilde{\vec{\kappa}}_{1}+\widetilde{%
\vec{\kappa}}_{2}\approx 0$, \ \ \ so that the final reduced phase contains
only $\tilde{\eta}=|\widetilde{\vec{\eta}_{1}}-\widetilde{\vec{\eta}_{2}}|$
and $\widetilde{\vec{\kappa}}:=\widetilde{\vec{\kappa}}_{1}=-\widetilde{\vec{%
\kappa}}_{2}.$ Using the identity below Eq.(\ref{VI32})

\begin{equation}
(\widetilde{\vec{\kappa}}_{1}\cdot \vec{\partial}_{12})^{2m+1}(\widetilde{%
\vec{\kappa}}_{2}\cdot \vec{\partial}_{12})^{2n+1}\tilde{\eta}^{2(m+n)+1}=-%
\frac{[(2m+2n+1)!!]^{2}}{\tilde{\eta}}(\widetilde{\vec{\kappa}}^{2}-(%
\widetilde{\vec{\kappa}}\cdot \hat{\eta})^{2})^{m+n+1},  \label{e1}
\end{equation}

\noindent and the identity Eq.(\ref{d5}) in the form

\begin{eqnarray}
&&\frac{\vec{\partial}_{12}\vec{\partial}_{12}}{\sqrt{m_{1}^{2}+{\widetilde{%
\vec{\kappa}}}{}^{2}}\sqrt{m_{2}^{2}+{\widetilde{\vec{\kappa}}}{}^{2}}}({%
\frac{\widetilde{\vec{\kappa}}}{\sqrt{m_{1}^{2}+{\widetilde{\vec{\kappa}}}%
^{2}}}}\cdot \vec{\partial}_{12})^{2m+2}({\frac{\widetilde{\vec{\kappa}}}{%
\sqrt{m_{2}^{2}+{\widetilde{\vec{\kappa}}}^{2}}}}\cdot \vec{\partial}%
_{12})^{2n}\frac{\tilde{\eta}^{2n+2m+3}}{(2m+2n+4)!}=  \nonumber \\
&=&\frac{(2m+2n+3)!!(2m+2n+1)!!}{\tilde{\eta}(2m+2n+4)!(\sqrt{m_{1}^{2}+{%
\widetilde{\vec{\kappa}}}{}^{2}})^{2m+3}(\sqrt{m_{2}^{2}+{\widetilde{\vec{%
\kappa}}}{}^{2}})^{2n+1}}\Big[{\bf I}[\widetilde{\vec{\kappa}}^{2}-(%
\widetilde{\vec{\kappa}}\cdot \hat{\eta})^{2}]^{m+n+1} +  \nonumber \\
&&+(2m+2n+1)[\widetilde{\vec{\kappa}}\widetilde{\vec{\kappa}}-(\widetilde{%
\vec{\kappa}}\cdot {\hat{\eta}})(\widetilde{\vec{\kappa}}{\hat{\eta}}+{\hat{%
\eta}}\widetilde{\vec{\kappa}})](\widetilde{\vec{\kappa}}^{2}-(\widetilde{%
\vec{\kappa}}\cdot \hat{\eta})^{2})^{m+n} -  \nonumber \\
&&-{\hat{\eta}\hat{\eta}}(\widetilde{\vec{\kappa}}^{2}-(\widetilde{\vec{%
\kappa}}\cdot \hat{\eta})^{2})^{m+n}(\widetilde{\vec{\kappa}}^{2}-(2m+2n+3)(%
\widetilde{\vec{\kappa}}\cdot {\hat{\eta}})^{2})\Big].  \label{e2}
\end{eqnarray}

\noindent the higher order Darwin plus spin part $V_{HDS}$ $=U_{HDS}+U_{HDS%
\text{ }}^{\prime }$ becomes

\begin{eqnarray}
&&V_{HDS}=-\sum_{i<j}{\frac{Q_{1}Q_{2}}{8\pi }}\sum_{m=0}^{\infty
}\sum_{n=0}^{\infty }\Big[\widetilde{{\bf \vec{k}}}_{1}\cdot \widetilde{{\bf
\vec{k}}}_{2}\frac{[(2m+2n+1)!!]^{2}}{\tilde{\eta}(2n+2m+2)!}[\widetilde{%
\vec{\kappa}}^{2}-(\widetilde{\vec{\kappa}}\cdot \hat{\eta})^{2}]^{n+m+1}
\nonumber \\
&&({\frac{1}{\sqrt{m_{1}^{2}+{\widetilde{\vec{\kappa}}}^{2}}}})^{2m+1}({%
\frac{1}{\sqrt{m_{2}^{2}+{\widetilde{\vec{\kappa}}}^{2}}}})^{2n+1}({\frac{1}{%
m_{1}^{2}+{\widetilde{\vec{\kappa}}}^{2}}}+{\frac{1}{m_{2}^{2}+{\widetilde{%
\vec{\kappa}}}^{2}}})-  \nonumber \\
&-&\frac{(2m+2n+3)!!(2m+2n+1)!!}{(2m+2n+4)!(\sqrt{m_{1}^{2}+{\widetilde{\vec{%
\kappa}}}{}^{2}})^{2m+1}(\sqrt{m_{2}^{2}+{\widetilde{\vec{\kappa}}}{}^{2}}%
)^{2n+1}}\Big[\Big(\widetilde{{\bf \vec{k}}}_{1}\cdot \widetilde{{\bf \vec{l}%
}}_{2}[\widetilde{\vec{\kappa}}^{2}-(\widetilde{\vec{\kappa}}\cdot \hat{\eta}%
)^{2}]^{m+n+1}+  \nonumber \\
&+&(2m+2n+1)\Big(\widetilde{{\bf \vec{k}}}_{1}\cdot {\widetilde{\vec{\kappa}}%
}\widetilde{{\bf \vec{l}}}_{2}\cdot \widetilde{\vec{\kappa}}-\widetilde{{\bf
\vec{k}}}_{1}\cdot (\widetilde{{\bf \vec{l}}}_{2}\cdot (\widetilde{\vec{%
\kappa}}{\hat{\eta}}+{\hat{\eta}}\widetilde{\vec{\kappa}}))(\widetilde{\vec{%
\kappa}}\cdot {\hat{\eta}})\Big)(\widetilde{\vec{\kappa}}^{2}-(\widetilde{%
\vec{\kappa}}\cdot \hat{\eta})^{2})^{m+n}-  \nonumber \\
&-&\widetilde{{\bf \vec{k}}}_{1}\cdot (\widetilde{{\bf \vec{l}}}_{2}\cdot {%
\hat{\eta}\hat{\eta})}(\widetilde{\vec{\kappa}}^{2}-(\widetilde{\vec{\kappa}}%
\cdot \hat{\eta})^{2})^{m+n}(\widetilde{\vec{\kappa}}^{2}-(2m+2n+3)(%
\widetilde{\vec{\kappa}}\cdot {\hat{\eta}})^{2})\Big)  \nonumber \\
&&({\frac{1}{m_{1}^{2}+{\widetilde{\vec{\kappa}}}^{2}}}-{\frac{1}{\sqrt{%
m_{1}^{2}+{\widetilde{\vec{\kappa}}}{}^{2}}\sqrt{m_{2}^{2}+{\widetilde{\vec{%
\kappa}}}{}^{2}}}})+  \nonumber \\
&+&\Big(\widetilde{{\bf \vec{k}}}_{2}\cdot \widetilde{{\bf \vec{l}}}_{1}[%
\widetilde{\vec{\kappa}}^{2}-(\widetilde{\vec{\kappa}}\cdot \hat{\eta}%
)^{2}]^{m+n+1}+  \nonumber \\
&+&(2m+2n+1)\Big(\widetilde{{\bf \vec{k}}}_{2}\cdot {\widetilde{\vec{\kappa}}%
}\widetilde{{\bf \vec{l}}}_{1}\cdot \widetilde{\vec{\kappa}}-\widetilde{{\bf
\vec{k}}}_{2}\cdot (\widetilde{{\bf \vec{l}}}_{1}\cdot (\widetilde{\vec{%
\kappa}}{\hat{\eta}}+{\hat{\eta}}\widetilde{\vec{\kappa}}))(\widetilde{\vec{%
\kappa}}\cdot {\hat{\eta}})\Big)(\widetilde{\vec{\kappa}}^{2}-(\widetilde{%
\vec{\kappa}}\cdot \hat{\eta})^{2})^{m+n}-  \nonumber \\
&-&\widetilde{{\bf \vec{k}}}_{2}(\widetilde{{\bf \vec{l}}}_{1}\cdot {\hat{%
\eta}\hat{\eta})}(\widetilde{\vec{\kappa}}^{2}-(\widetilde{\vec{\kappa}}%
\cdot \hat{\eta})^{2})^{m+n}(\widetilde{\vec{\kappa}}^{2}-(2m+2n+3)(%
\widetilde{\vec{\kappa}}\cdot {\hat{\eta}})^{2})\Big)  \nonumber \\
&&({\frac{1}{m_{2}^{2}+{\widetilde{\vec{\kappa}}}^{2}}}-{\frac{1}{\sqrt{%
m_{1}^{2}+{\widetilde{\vec{\kappa}}}{}^{2}}\sqrt{m_{2}^{2}+{\widetilde{\vec{%
\kappa}}}{}^{2}}}})+  \nonumber \\
&+&\Big(\widetilde{{\bf \vec{l}}}_{2}\cdot \widetilde{{\bf \vec{l}}}_{1}[%
\widetilde{\vec{\kappa}}^{2}-(\widetilde{\vec{\kappa}}\cdot \hat{\eta}%
)^{2}]^{m+n+1}+  \nonumber \\
&+&(2m+2n+1)\Big(\widetilde{{\bf \vec{l}}}_{2}\cdot {\widetilde{\vec{\kappa}}%
}\widetilde{{\bf \vec{l}}}_{1}\cdot \widetilde{\vec{\kappa}}-\widetilde{{\bf
\vec{l}}}_{2}\cdot (\widetilde{{\bf \vec{l}}}_{1}\cdot (\widetilde{\vec{%
\kappa}}{\hat{\eta}}+{\hat{\eta}}\widetilde{\vec{\kappa}}))(\widetilde{\vec{%
\kappa}}\cdot {\hat{\eta}})\Big)(\widetilde{\vec{\kappa}}^{2}-(\widetilde{%
\vec{\kappa}}\cdot \hat{\eta})^{2})^{m+n}-  \nonumber \\
&-&\widetilde{{\bf \vec{l}}}_{2}\cdot (\widetilde{{\bf \vec{l}}}_{1}\cdot {%
\hat{\eta}\hat{\eta})}(\widetilde{\vec{\kappa}}^{2}-(\widetilde{\vec{\kappa}}%
\cdot \hat{\eta})^{2})^{m+n}(\widetilde{\vec{\kappa}}^{2}-(2m+2n+3)(%
\widetilde{\vec{\kappa}}\cdot {\hat{\eta}})^{2})\Big)({\frac{1}{\sqrt{%
m_{1}^{2}+{\widetilde{\vec{\kappa}}}{}^{2}}\sqrt{m_{2}^{2}+{\widetilde{\vec{%
\kappa}}}{}^{2}}}})\Big]\frac{1}{\tilde{\eta}}+  \nonumber \\
&+&\frac{[(2m+2n+1)!!]^{2}}{(2n+2m+2)!(\sqrt{m_{1}^{2}+{\widetilde{\vec{%
\kappa}}}{}^{2}})^{2m+1}(\sqrt{m_{2}^{2}+{\widetilde{\vec{\kappa}}}{}^{2}}%
)^{2n+1}}  \nonumber \\
&&\Big(\widetilde{\vec{\kappa}}\cdot \widetilde{{\bf \vec{k}}}_{2}\frac{(%
\widetilde{{\bf \vec{k}}}_{1}-\widetilde{{\bf \vec{l}}}_{1})\cdot \widetilde{%
\vec{\kappa}}-(\widetilde{{\bf \vec{k}}}_{1}-\widetilde{{\bf \vec{l}}}%
_{1})\cdot {\hat{\eta}}\widetilde{\vec{\kappa}}\cdot {\hat{\eta}}}{m_{1}^{2}+%
{\widetilde{\vec{\kappa}}}{}^{2}}-  \nonumber \\
&-&\widetilde{\vec{\kappa}}\cdot \widetilde{{\bf \vec{k}}}_{1}\frac{(%
\widetilde{{\bf \vec{k}}}_{2}-\widetilde{{\bf \vec{l}}_{2}})\cdot \widetilde{%
\vec{\kappa}}-(\widetilde{{\bf \vec{k}}}_{2}-\widetilde{{\bf \vec{l}}}%
_{2})\cdot {\hat{\eta}}\widetilde{\vec{\kappa}}\cdot {\hat{\eta}}}{m_{2}^{2}+%
{\widetilde{\vec{\kappa}}}{}^{2}}\Big)\frac{(\widetilde{\vec{\kappa}}^{2}-(%
\widetilde{\vec{\kappa}}\cdot \hat{\eta})^{2})^{m+n}}{\tilde{\eta}}-
\nonumber \\
&-&\widetilde{{\bf \vec{k}}}_{1}\cdot \widetilde{{\bf \vec{k}}}_{2}\frac{%
[(2m+2n+3)!!]^{2}}{\tilde{\eta}(2n+2m+4)!}[\widetilde{\vec{\kappa}}^{2}-(%
\widetilde{\vec{\kappa}}\cdot \hat{\eta})^{2}]^{n+m+2}({\frac{1}{\sqrt{%
m_{1}^{2}+{\widetilde{\vec{\kappa}}}^{2}}}})^{2m+3}({\frac{1}{\sqrt{%
m_{2}^{2}+{\widetilde{\vec{\kappa}}}^{2}}}})^{2n+3}+  \nonumber \\
&+&\frac{(2m+2n+5)!!(2m+2n+3)!!}{(2m+2n+6)!(\sqrt{m_{1}^{2}+{\widetilde{\vec{%
\kappa}}}{}^{2}})^{2m+3}(\sqrt{m_{2}^{2}+{\widetilde{\vec{\kappa}}}{}^{2}}%
)^{2n+3}}\Big(\widetilde{{\bf \vec{k}}}_{1}\cdot \widetilde{{\bf \vec{k}}}%
_{2}[\widetilde{\vec{\kappa}}^{2}-(\widetilde{\vec{\kappa}}\cdot \hat{\eta}%
)^{2}]^{m+n+2}+  \nonumber \\
&+&(2m+2n+3)\Big(\widetilde{{\bf \vec{k}}}_{1}\cdot {\widetilde{\vec{\kappa}}%
}\widetilde{{\bf \vec{k}}}_{2}\cdot \widetilde{\vec{\kappa}}-\widetilde{{\bf
\vec{k}}}_{1}\widetilde{{\bf \vec{k}}}_{2}\cdot \cdot (\widetilde{\vec{\kappa%
}}\cdot {\hat{\eta}})(\widetilde{\vec{\kappa}}{\hat{\eta}}+{\hat{\eta}}%
\widetilde{\vec{\kappa}})\Big)(\widetilde{\vec{\kappa}}^{2}-(\widetilde{\vec{%
\kappa}}\cdot \hat{\eta})^{2})^{m+n+1}-  \nonumber \\
&-&\widetilde{{\bf \vec{k}}}_{1}\cdot (\widetilde{{\bf \vec{k}}}_{2}\cdot {%
\hat{\eta}\hat{\eta})}(\widetilde{\vec{\kappa}}^{2}-(\widetilde{\vec{\kappa}}%
\cdot \hat{\eta})^{2})^{m+n+1}(\widetilde{\vec{\kappa}}^{2}-(2m+2n+5)(%
\widetilde{\vec{\kappa}}\cdot {\hat{\eta}})^{2})\Big)\frac{1}{\tilde{\eta}}%
\Big].  \label{e3}
\end{eqnarray}

We use

\begin{equation}
\frac{\lbrack (2m+2n+1)!!]^{2}}{(2n+2m+2)!}=\frac{(-)^{n+m}}{2(n+m+1)}%
{-3/2 \choose n+m}%
,  \label{e4}
\end{equation}

\noindent and we put $m+n=l$, so that $0\leq m\leq l$ and $0\leq l<\infty $.
Then we perform the $m$ sum using

\begin{equation}
\sum_{m=0}^{l}\left( \frac{x}{y}\right) ^{m}=\frac{y^{l+1}-x^{l+1}}{%
y^{l}(y-x)},  \label{e5}
\end{equation}

\noindent and we obtain

\begin{eqnarray}
&&V_{HDS}=-\sum_{i<j}{\frac{Q_{1}Q_{2}}{8\pi }}\sum_{l=0}^{\infty }\Big[[%
\widetilde{\vec{\kappa}}^{2}-(\widetilde{\vec{\kappa}}\cdot \hat{\eta}%
)^{2}]^{l}\Big(\frac{({\frac{1}{\sqrt{m_{2}^{2}+{\widetilde{\vec{\kappa}}}%
^{2}}}})^{2l+2}-({\frac{1}{\sqrt{m_{1}^{2}+{\widetilde{\vec{\kappa}}}^{2}}}}%
)^{2l+2}}{m_{1}^{2}-m_{2}^{2}}\Big)\Big]  \nonumber \\
&&\Big(\widetilde{{\bf \vec{k}}}_{1}\cdot \widetilde{{\bf \vec{k}}}_{2}\frac{%
(-)^{l}}{2(l+1)}%
{-3/2 \choose l}%
\frac{[\widetilde{\vec{\kappa}}^{2}-(\widetilde{\vec{\kappa}}\cdot \hat{\eta}%
)^{2}]}{\tilde{\eta}}(\sqrt{{\frac{m_{2}^{2}+{\widetilde{\vec{\kappa}}}^{2}}{%
m_{1}^{2}+{\widetilde{\vec{\kappa}}}^{2}}}}+\sqrt{{\frac{m_{1}^{2}+{%
\widetilde{\vec{\kappa}}}^{2}}{m_{2}^{2}+{\widetilde{\vec{\kappa}}}^{2}}}})+
\nonumber \\
&&+\frac{(-)^{l}}{2(l+2)(2l+3)}%
{-3/2 \choose l+1}%
[\Big(\widetilde{{\bf \vec{k}}}_{1}\cdot \widetilde{{\bf \vec{l}}}_{2}[%
\widetilde{\vec{\kappa}}^{2}-(\widetilde{\vec{\kappa}}\cdot \hat{\eta})^{2}]+
\nonumber \\
&&+(2l+1)\Big(\widetilde{{\bf \vec{k}}}_{1}\cdot {\widetilde{\vec{\kappa}}}%
\widetilde{{\bf \vec{l}}}_{2}\cdot \widetilde{\vec{\kappa}}-\widetilde{{\bf
\vec{k}}}_{1}(\widetilde{{\bf \vec{l}}}_{2}\cdot (\widetilde{\vec{\kappa}}{%
\hat{\eta}}+{\hat{\eta}}\widetilde{\vec{\kappa}}))(\widetilde{\vec{\kappa}}%
\cdot {\hat{\eta}})\Big)-  \nonumber \\
&&-\widetilde{{\bf \vec{k}}}_{1}\cdot (\widetilde{{\bf \vec{l}}}_{2}\cdot {%
\hat{\eta}\hat{\eta})}(\widetilde{\vec{\kappa}}^{2}-(2l+3)(\widetilde{\vec{%
\kappa}}\cdot {\hat{\eta}})^{2})\Big)(\sqrt{{\frac{m_{2}^{2}+{\widetilde{%
\vec{\kappa}}}^{2}}{m_{1}^{2}+{\widetilde{\vec{\kappa}}}^{2}}}}-{1})+
\nonumber \\
&&+\Big(\widetilde{{\bf \vec{k}}}_{2}\cdot \widetilde{{\bf \vec{l}}}_{1}[%
\widetilde{\vec{\kappa}}^{2}-(\widetilde{\vec{\kappa}}\cdot \hat{\eta}%
)^{2}]+(2l+1)\Big(\widetilde{{\bf \vec{k}}}_{2}\cdot {\widetilde{\vec{\kappa}%
}}\widetilde{{\bf \vec{l}}}_{1}\cdot \widetilde{\vec{\kappa}}-\widetilde{%
{\bf \vec{k}}}_{2}\widetilde{{\bf \vec{l}}}_{1}\cdot \cdot (\widetilde{\vec{%
\kappa}}\cdot {\hat{\eta}})(\widetilde{\vec{\kappa}}{\hat{\eta}}+{\hat{\eta}}%
\widetilde{\vec{\kappa}})\Big)-  \nonumber \\
&&-\widetilde{{\bf \vec{k}}}_{2}\cdot (\widetilde{{\bf \vec{l}}}_{1}\cdot {%
\hat{\eta}\hat{\eta})}(\widetilde{\vec{\kappa}}^{2}-(\widetilde{\vec{\kappa}}%
\cdot \hat{\eta})^{2})^{l}(\widetilde{\vec{\kappa}}^{2}-(2l+3)(\widetilde{%
\vec{\kappa}}\cdot {\hat{\eta}})^{2})\Big)(\sqrt{{\frac{m_{1}^{2}+{%
\widetilde{\vec{\kappa}}}^{2}}{m_{2}^{2}+{\widetilde{\vec{\kappa}}}^{2}}}}-{1%
})+  \nonumber \\
&&+\Big(\widetilde{{\bf \vec{l}}}_{2}\cdot \widetilde{{\bf \vec{l}}}_{1}[%
\widetilde{\vec{\kappa}}^{2}-(\widetilde{\vec{\kappa}}\cdot \hat{\eta}%
)^{2}]+(2l+1)\Big(\widetilde{{\bf \vec{l}}}_{2}\cdot {\widetilde{\vec{\kappa}%
}}\widetilde{{\bf \vec{l}}}_{1}\cdot \widetilde{\vec{\kappa}}-\widetilde{%
{\bf \vec{l}}}_{2}\cdot (\widetilde{{\bf \vec{l}}}_{1}\cdot (\widetilde{\vec{%
\kappa}}{\hat{\eta}}+{\hat{\eta}}\widetilde{\vec{\kappa}}))(\widetilde{\vec{%
\kappa}}\cdot {\hat{\eta}})\Big)-  \nonumber \\
&&-\widetilde{{\bf \vec{l}}}_{2}\cdot (\widetilde{{\bf \vec{l}}}_{1}\cdot {%
\hat{\eta}\hat{\eta})}(\widetilde{\vec{\kappa}}^{2}-(2l+3)(\widetilde{\vec{%
\kappa}}\cdot {\hat{\eta}})^{2}\Big))\Big]\Big)\frac{1}{\tilde{\eta}}+
\nonumber \\
&&+\frac{(-)^{l}}{2(l+1)}%
{-3/2 \choose l}%
\Big(\widetilde{\vec{\kappa}}\cdot \widetilde{{\bf \vec{k}}}_{2}\frac{(%
\widetilde{{\bf \vec{k}}}_{1}-\widetilde{{\bf \vec{l}}}_{1})\cdot \widetilde{%
\vec{\kappa}}-(\widetilde{{\bf \vec{k}}}_{1}-\widetilde{{\bf \vec{l}}}%
_{1})\cdot {\hat{\eta}}\widetilde{\vec{\kappa}}\cdot {\hat{\eta}}}{m_{1}^{2}+%
{\widetilde{\vec{\kappa}}}{}^{2}}-  \nonumber \\
&&-\widetilde{\vec{\kappa}}\cdot \widetilde{{\bf \vec{k}}}_{1}\frac{(%
\widetilde{{\bf \vec{k}}}_{2}-\widetilde{{\bf \vec{l}}_{2}})\cdot \widetilde{%
\vec{\kappa}}-(\widetilde{{\bf \vec{k}}}_{2}-\widetilde{{\bf \vec{l}}}%
_{2})\cdot {\hat{\eta}}\widetilde{\vec{\kappa}}\cdot {\hat{\eta}}}{m_{2}^{2}+%
{\widetilde{\vec{\kappa}}}{}^{2}}\Big)\frac{\sqrt{m_{1}^{2}+{\widetilde{\vec{%
\kappa}}}{}^{2}}\sqrt{m_{2}^{2}+{\widetilde{\vec{\kappa}}}{}^{2}}}{\tilde{%
\eta}}+  \nonumber \\
&&+\widetilde{{\bf \vec{k}}}_{1}\cdot \widetilde{{\bf \vec{k}}}_{2}\frac{%
(-)^{l}}{2(l+2)}%
{-3/2 \choose l+1}%
\frac{[\widetilde{\vec{\kappa}}^{2}-(\widetilde{\vec{\kappa}}\cdot \hat{\eta}%
)^{2}]}{\tilde{\eta}\sqrt{m_{1}^{2}+{\widetilde{\vec{\kappa}}}{}^{2}}\sqrt{%
m_{2}^{2}+{\widetilde{\vec{\kappa}}}{}^{2}}}-  \nonumber \\
&&-\frac{(-)^{l}}{2(l+3)(2l+5)}%
{-3/2 \choose l+2}%
\Big(\widetilde{{\bf \vec{k}}}_{1}\cdot \widetilde{{\bf \vec{k}}}_{2}[%
\widetilde{\vec{\kappa}}^{2}-(\widetilde{\vec{\kappa}}\cdot \hat{\eta}%
)^{2}]^{2}+  \nonumber \\
&&+(2l+3)\Big(\widetilde{{\bf \vec{k}}}_{1}\cdot {\widetilde{\vec{\kappa}}}%
\widetilde{{\bf \vec{k}}}_{2}\cdot \widetilde{\vec{\kappa}}-\widetilde{{\bf
\vec{k}}}_{1}\cdot (\widetilde{{\bf \vec{k}}}_{2}\cdot (\widetilde{\vec{%
\kappa}}{\hat{\eta}}+{\hat{\eta}}\widetilde{\vec{\kappa}}))(\widetilde{\vec{%
\kappa}}\cdot {\hat{\eta}})\Big)(\widetilde{\vec{\kappa}}^{2}-(\widetilde{%
\vec{\kappa}}\cdot \hat{\eta})^{2})-  \nonumber \\
&&-\widetilde{{\bf \vec{k}}}_{1}\cdot (\widetilde{{\bf \vec{k}}}_{2}\cdot {%
\hat{\eta}\hat{\eta})}(\widetilde{\vec{\kappa}}^{2}-(\widetilde{\vec{\kappa}}%
\cdot \hat{\eta})^{2})(\widetilde{\vec{\kappa}}^{2}-(2l+5)(\widetilde{\vec{%
\kappa}}\cdot {\hat{\eta}})^{2})\Big)\frac{1}{\sqrt{m_{1}^{2}+{\widetilde{%
\vec{\kappa}}}{}^{2}}\sqrt{m_{2}^{2}+{\widetilde{\vec{\kappa}}}{}^{2}}}\frac{%
1}{\tilde{\eta}}\Big].  \label{e6}
\end{eqnarray}

We use
\[
\sum_{l=0}^{\infty }\frac{(-)^{l}}{2(l+1)}%
{-3/2 \choose l}%
x^{l}=\frac{1}{x}((1-x)^{-1/2}-1)
\]
and similar identities applied in Appendix D and we obtain Eq.(\ref{VIII3}).

\vfill\eject

\end{document}